\documentclass[12pt,a4paper,francais]{article}
\usepackage[francais]{babel}
\usepackage[utf8]{inputenc}
\usepackage{amsmath}
\usepackage{amsfonts}
\usepackage{amssymb}
\usepackage{graphicx}
\usepackage{slashbox}
\usepackage{wrapfig}

\newcommand{\be}{\begin{equation}}
\newcommand{\ee}{\end{equation}}

\newcommand{\aaa}{\alpha}
\newcommand{\insc}{N_{insc}}
\newcommand{\expr}{N_{expr}}
\newcommand{\vot}{N_{vot}}
\newcommand{\gagn}{N_{gagn}}
\newcommand{\inscaa}{N_{insc}^\alpha}
\newcommand{\inscour}{N_{insc,\:cour}}
\newcommand{\inscoa}{N_{insc,\:cour}^\alpha}
\newcommand{\pio}{\pi_{insc,\:0}}
\newcommand{\piinsc}{\pi_{insc,\:0}^\alpha}
\newcommand{\res}{\rho}
\newcommand{\resa}{\rho^\alpha}
\newcommand{\taa}{\tau^\alpha}
\newcommand{\tbb}{\tau^\beta}
\newcommand{\toa}{\tau^\alpha_0}
\newcommand{\saa}{\sigma^\alpha_0}
\newcommand{\vaa}{\mathcal{V}^\alpha_{16}}

\newcommand{\saam}{\sigma'^{\,\alpha}_0}
\newcommand{\cva}{c_v^\alpha}

\newcommand{\poa}{p_0^\alpha}
\newcommand{\pcour}{p_{cour}^\alpha}

\newcommand{\nra}{n_r^{\alpha}}
\newcommand{\tra}{\tau_r^\alpha}

\newcommand{\naa}{N^\alpha}
\newcommand{\pab}{p_{\beta\rightarrow\alpha}}
\newcommand{\mva}{m^{\mathcal{V}_\alpha}}
\newcommand{\pcum}{\mathcal{P}_{>}}
\newcommand{\fva}{F^{\mathcal{V}_\alpha}}
\newcommand{\iva}{I^{\mathcal{V}_\alpha}}
\newcommand{\dd}{\mathrm{d}}

\newcommand{\meq}{m^*}

\newcommand{\rr}{\mathbf{r}}
\newcommand{\kk}{\mathbf{k}}
\newcommand{\qq}{\mathbf{q}}
\newcommand{\x}{\,.\,}

\newcommand{\ffa}{\tilde{f}^\alpha}

\newcommand{\ti}{-}

\begin{document}


\renewcommand{\thepage}{\arabic{page}}
\setcounter{page}{1}
\vspace*{5cm}
\hfill
\begin{minipage}[r]{0.5\linewidth}
A mon ami, mon frère, Pierre Villard\\
$\ti$et j'ai eu la chance d'être son ami$\ti$,\\
grand voyageur,\\
esprit libre,\\
intelligent, trop intelligent~?\\
Mais mort trop tôt.
\end{minipage}%

\newpage

\vspace*{5cm}
\hfill
\begin{minipage}[r]{0.8\linewidth}
\textit{Voyant aussi qu'au milieu des troubles qui divisaient la ville, nombre de citoyens, par indifférence, s'en remettaient au hasard, [Solon] porta contre eux cette loi singulière~:} Quiconque, en temps de trouble, ne prendra pas les armes pour l'un des deux partis, sera frappé d'atimie et exclu de la cité.\\
Aristote, \textit{Constitution d'Athènes}~\cite{aristote}
\end{minipage}
\vspace*{3cm}

\hfill
\begin{minipage}[r]{0.8\linewidth}
\textit{Douter de tout ou tout croire, ce sont deux solutions également commodes, qui l'une et l'autre nous dispensent de réfléchir.}\\
Henri Poincaré, \textit{La science et l'hypothèse}~\cite{poincare}
\end{minipage}

\clearpage
\begin{Large}
\textbf{Avertissement}\\
\end{Large}
\vspace*{0.5cm}\\

Le texte que l'on trouve ci-dessous correspond, à de mineures modifications près, au corps principal de la thèse de doctorat~\footnote{Thèse de l'Université Paris Diderot (Paris 7), soutenue le 29/09/2009 et de jury constitué par Jean-Philippe Bouchaud (CFM et École polytechnique, Paris), Jean Chiche (CEVIPOF, Sciences Po, Paris), Santo Fortunato (Institute for Scientific Interchange, Turin), Silvio Franz (LPTMS, Université Paris-Sud), Serge Galam (CRÉA, École Polytechnique, Paris), Matteo Marsili (The Abdus Salam International Centre for Theoretical Physics, Trieste) et Jean-Pierre Nadal (LPS, ENS et CAMS, EHESS, Paris).} menée au Service de Physique de l'État Condensé (au CEA Saclay) et de directeur de thèse Jean-Philippe Bouchaud.\\

Contrairement à la version remise à l'Université pour la soutenance de thèse, et principalement pour des raisons d'unité, d'autres travaux ne figurent pas dans cette version~:
\begin{enumerate}
\item Un article $\ti$réalisé avec Jean-Philippe Bouchaud$\ti$ modélisait une expérience~\cite{salganik_exp_final} qui mesurait comment la connaissance des choix des autres agents pouvait modifier le choix d'un agent~\cite{of_songs_and_men}.
\item Un développement supplémentaire accompagnait l'article précédent (mais qui se trouve sur la version 1 dans arXiv). 
\item Un papier $\ti$écrit avec Serge Galam$\ti$ poursuivait le travail du mémoire de Master intitulé \textit{Une distraction mathématique sur l'évolution d'opinion}, et se préoccupait des conséquences théoriques des contrariants lors d'une campagne électorale~\cite{chaotic}.
\item Une publication $\ti$développée principalement par Matteo Marsili et Salvatore Miccichè$\ti$ abordait l'éconophysique et le temps minimal pour l'émergence d'une structure dans les systèmes financiers~\cite{emergence}.\vspace{0.5cm}
\end{enumerate}

En revanche on trouvera une postface, écrite en octobre 2010, complétant quelque peu le travail principal réalisé en thèse.

\clearpage
\begin{Large}
\textbf{Remerciements}\\
\end{Large}
\vspace*{0.5cm}

Si la science est une belle école de pensée, la pratiquer l'est encore davantage. Et c'est un plaisir, une chance, un luxe, que de pouvoir se sentir bénéficiaire d'une longue tradition de réflexion.

Je remercie donc toutes les personnes qui, de près ou de loin, directement ou indirectement, m'ont permis de m'engager dans ce voyage intellectuel. Je pense à toutes les personnes très intelligentes que cette thèse m'a permis de côtoyer, et notamment Jean-Philippe Bouchaud. Je pense de plus aux discussions critiques et aux encouragements $\ti$qui m'ont bien aidé$\ti$ venant de Jean-Pierre Nadal et de Bertrand Roehner. Je pense aussi aux quelques personnes $\ti$et elles sont rares$\ti$ qui ont abordé leurs responsabilités administratives avec hauteur, comme Mme Turpin au lycée de Châtenay-Malabry, et Éric Vincent au SPEC du CEA de Saclay. Je n'oublie pas les membres du labo pour leur aide et pour les bons moments passés à boire un café ou à manger ensemble $\ti$avec en exergue, la fameuse et plus que lapidaire \og réduction d'Ivan \fg. Un grand merci à Laurent Barry, Frédéric Nowacki, et Philippe-Alexandre Pouille, notamment pour leur vision pertinente en rapport à ce travail de recherche. Enfin, cette thèse doit beaucoup, par leurs stimulantes réflexions, à Arnaud Laroche et à Lionel Tabourier, qui de plus ont eu la lourde tâche de supporter, et mes discussions alambiquées, et leurs conseils très souvent délaissés.

Cette thèse doit aussi beaucoup à Brigitte Hazart qui a rassemblé, collecté, comblé et envoyé $\ti$avec toute la sympathie qui la définit$\ti$ les données électorales, tout en y joignant à l'occasion des informations connexes.

Un très grand merci à Catherine Quilliet, maîtresse incontestée de la juste virgule, pour sa relecture du manuscrit, ainsi qu'à Josiane Tabourier pour sa longue relecture attentive et minutieuse, expurgeant les trop de fautes résiduelles de français.\\

Et comment ne pas remercier celui qui m'a initié à \og penser en physique \fg, et qui fut $\ti$au risque d'en devenir prétentieux$\ti$ mon maître à penser~: Georges Oberlechner.

Enfin, tant qu'à être dingue allons-y. Il y a un plaisir certain de se sentir héritier d'une longue tradition intellectuelle $\ti$à laquelle je me sens, en partie, redevable. Mais ce travail n'aurait pu se réaliser sans ce nouvel et formidable outil~: l'informatique. Et encore moins, sans la possibilité de disposer des données et de pouvoir se confronter à elles. Et là, je serai tenté de les remercier, car pour qui se permet de changer de point de vue, de théorie, voire d'idéologie, les données sont heureusement têtues.

\clearpage

\tableofcontents

\clearpage
\section{Introduction}
\hfill
\begin{minipage}[r]{0.8\linewidth}
\textit{Ils font partie de la race des petits qui rendent toute chose petite.}\\
D'après \textit{Ainsi parlait Zarathoustra} de Friedrich Nietzsche~\cite{nietzsche}
\end{minipage}
\vspace{0.75cm}

Que peut dire la physique sur les élections~?

Voilà la question centrale qui animera toute cette thèse. Une question tellement vaste, qu'elle nécessite quelques précisions.

Les élections se réduisent, ici, aux seules données électorales~: résultats électoraux et taux de participation, par commune, voire par bureau de vote. Les données électorales sont considérées comme des mesures physiques, irréversibles et nettes, du choix de chaque électeur qui s'exprime lors d'un vote, ou de chaque électeur inscrit sur les listes électorales qui participe ou s'abstient à l'élection. Bref, les données électorales seront manipulées comme toute autre grandeur ou mesure physique, sans leur assigner une quelconque particularité $\ti$i.e. sans se préoccuper de savoir pour qui ou pour quoi les électeurs sont appelés aux urnes, et en considérant donc toutes les élections de la même façon, sans faire aucune distinction.

Les réponses que nous apporterons, l'angle d'attaque que nous suivrons, la grille d'investigation que nous adopterons, ne prétendent nullement à l'exhaustivité. (Comment pourrait-il en être autrement en sciences~?)

Enfin, les sciences sociales ne contribuent que très $\ti$trop~?$\ti$ peu à ce travail, ce qui constitue peut-être l'une des plus grandes lacunes de cette étude. Pour se disculper quelque peu, je dirai que cette thèse... est une thèse de physique. Je pourrai ensuite évoquer le temps limité d'une thèse (deux années et demie pour cette partie consacrée aux élections), ma trop faible connaissance des sciences sociales au commencement de la thèse, le fait de n'avoir pas trouvé (mais peut-être ai-je mal cherché~?) de travaux dans ce domaine qui conviennent à l'objectif de cette thèse. (Mais, plutôt qu'une simple pétition de principe nous affinerons cette critique, une fois notre travail développé, à la section~\ref{pt-conclusion-critique} de la Conclusion.)\\

Précisons donc l'objectif de cette thèse.

Nous ne prétendons nullement comprendre, et encore moins expliquer, le comportement de l'électeur en rapport à ses caractéristiques historiques ou socio-culturelles, ou bien en rapport aux divers évènements économiques, sociaux ou politiques survenus lors de, ou avant, la campagne électorale. Nous cherchons à comprendre les données électorales, comme le ferait un physicien face à une somme de mesures fiables, d'un champ encore en friche.

Cette compréhension se décompose en deux étapes. Tout d'abord, une recherche empirique des régularités présentes dans les données électorales. Ensuite, une investigation d'ordre théorique~: retrouver les principaux phénomènes empiriques à partir d'un modèle reposant à la fois sur un faible nombre d'ingrédients, et sur une description microscopique plausible.

Et tout ceci, en se tenant uniquement, et volontairement, aux données électorales par commune (ou par bureau de vote), et à leur localisation spatiale (la position géographique de la mairie de chaque commune)\footnote{L'idée d'étudier les résultats électoraux par commune et par bureau de vote revient entièrement à Jean-Philippe Bouchaud, ainsi que l'idée forte de cette étude~: la prise en compte de leur localisation spatiale.}. C'est l'un des partis pris de cette étude, et constitue en quelque sorte un défi à relever.\\

Les deux points ci-dessus, l'aspect empirique et l'aspect théorique de cette thèse, nécessitent encore des éclaircissements.

Nous nous efforcerons de trouver des régularités dans les données électorales, et des régularités seulement. Nous bannirons, notamment, le contexte dans lequel se situe l'élection, les particularités et caractéristiques des candidats, des partis politiques, des choix proposés aux votes des électeurs, le comportement d'un même électeur face à diverses élections, etc. Bref, afin de parvenir aux régularités, nous nous intéresserons aux questions dites simples, au détriment des questions dites intéressantes $\ti$i.e. les questions qui attirent généralement l'attention du citoyen.

Comme de coutume en physique, en éliminant les  \og détails \fg, nous espérons atteindre des traits saillants des phénomènes~; traits sur lesquels pourront s'appuyer ensuite d'autres investigations, tant théoriques qu'empiriques. Nous n'affirmons pas que ces \og détails \fg{} sont inintéressants, ou ne méritent aucune attention scientifique, nous disons seulement que nous ne voulons pas en tenir compte ici. Nous prenons délibérément le pari que les seules données électorales, conjuguées à leur localisation géographique, permettront d'obtenir des régularités, voire des traits pertinents, du phénomène électoral.

Prenons un exemple. Avant d'étudier convenablement la chute des corps, et en particulier la chute des feuilles $\ti$aussi belles et poétiques soient-elles$\ti$, il semble préférable de commencer l'étude par la chute des graves dans le vide. Non pas que les \og détails \fg{} (ici la forme et la rigidité de la feuille, la présence de l'air, etc.) n'aient aucune importance, mais, sans que l'on puisse s'en abstraire, le début d'une construction scientifique de la chute des corps paraîtrait bien hasardeuse et bien difficile.

Faisons une digression. Nous constatons \textit{a posteriori} que l'une des plus grandes erreurs des premiers penseurs grecs présocratiques, notamment les milésiens Thalès, Anaximandre et Anaximène~\cite{grecs} $\ti$appelés par la suite, à tort au sens étymologique, philosophes$\ti$ fut de s'intéresser à des questions dites intéressantes (comme le principe originel, \textit{archê}, ou l'éclair, le tonnerre, les ouragans, la grêle, l'arc-en-ciel, les tremblements de terre, etc.), plutôt qu'à des questions dites $\ti$\textit{a posteriori} encore$\ti$ simples. (Ce qui n'enlève rien à l'immense mérite qui leur est dû, ainsi qu'à celui de leur civilisation. La critique \textit{a posteriori}, une fois \og hissé sur les épaules des géants \fg{}, étant toujours facile.) Encore faudrait-il, au temps présent, savoir quels problèmes simples se poser, et ce sans faire intervenir de particularités historiques, spatiales, téléologiques, d'essence, de nature, etc. Sous cette optique partielle, le génie de Galilée~\cite{galilee_1, galilee_2} fut d'arriver à trouver ces \og bons \fg{} problèmes, outrepassant par là même les détails inhérents à une vision commune, et d'y avoir répondu. A l'écheveau par la suite de pouvoir être dénoué. Ou déjoué.

Signalons dans cette jeune branche $\ti$à l'avenir, espérons-le, bien prometteur$\ti$ de la physique, nommée physique sociale~\cite{quetelet} ou sociophysique, l'existence de très belles études, étayées par une multiplicité de sources différentes, sur la recherche des régularités empiriques dans les phénomènes sociaux (notamment, \cite{roehner_speculation, roehner_driving}).\\

Nous nous efforcerons ensuite de trouver des modèles qui, bien évidemment, s'accordent le mieux possible aux données, ou plus précisément aux principaux traits tirés de l'analyse empirique précédente. De plus, les modèles utilisés doivent traduire une dynamique plausible, à l'échelle microscopique (à l'échelle des agents, ici), et au regard de la grandeur mesurée (les choix des agents d'une commune). Précisons que nous cherchons à élaborer des modèles, et non une théorie. Nous en sommes encore à un stade, semble-t-il, bien prématuré pour pouvoir l'envisager. En revanche chaque modèle développe, sous une forme qui lui est propre, une même thématique commune, une même question centrale posée aux données.

Sous l'angle épistémologique, nous comprenons ici les modèles comme des questions posées aux données. Les modèles que nous utiliserons reposent en effet sur peu d'ingrédients, et leurs traits saillants s'identifient aisément. La signification probable, tirée d'un accord ou non du modèle aux données, s'en trouve dès lors facilitée.

Enfin, signalons que nous adopterons sciemment l'hypothèse zéro des phénomènes sociaux traités par la physique, et notamment par la physique statistique~: l'individu comme un spin qui opine, un atome social.\\

Avant de commencer l'étude à proprement parler, faisons deux remarques préalables.

Nous emploierons sans scrupule les mots d'influence sociale, d'imitation, d'interaction entre agents, et ce sans tenir compte d'une riche littérature (voir par exemple~\cite{moscovici, psy_soc_exp}) qui les analyse, les distingue et les teste expérimentalement.

La possibilité de manipuler et de se confronter aux données électorales (une mesure précise et irréversible du choix des électeurs) possède un avantage supplémentaire~: esquiver au maximum toute référence à cet \og objet \fg{} encore trop flou, ou trop inconsistant à nos yeux, nommé opinion\footnote{Je remercie Stéphane Laurens et Birgitta Orfali pour avoir appuyé ce point lors d'une discussion orale.}~\cite{laurens_opinion, bourdieu}.

\clearpage
\section{Enquête : des généralités aux prédictions}
\label{section-enquete}
\hfill
\begin{minipage}[r]{0.8\linewidth}
\textit{Le risque le plus certain est de ne pas prendre de risque.}\\
D'après Arnaud Laroche~\cite{arnaud}
\end{minipage}
\vspace{0.75cm}

La plupart des études de physique consacrées aux résultats électoraux concernent les élections à choix multiple, où l'on cherche essentiellement à établir, puis à justifier, les distributions statistiques en loi de puissance ou en log-normale de la répartition des votes sur les nombreux candidats~; comme au Brésil~\cite{costa_filho_scaling_vot, costa_filho_bresil_el2, lyra_bresil_el, bernardes_bresil_el}, au Brésil et en Inde~\cite{gonzalez_bresil_inde_el}, en Inde et au Canada~\cite{hit_is_born}, au Mexique~\cite{baez_mexiq_el, morales_mexiq_el}, en Indonésie~\cite{situngkir_indonesie_el}, et enfin avec une recherche de loi universelle~\cite{fortunato_universality}. Araripe et al.~\cite{araripe_plurality} étudient les statistiques des élections des maires, en se focalisant sur les cas à petit nombre de candidats. Sadovsky et Gliskov~\cite{sadovsky_russie_el} se préoccupent d'une typologie des élections russes. Schneider et al.~\cite{schneider_impact} mettent en évidence une corrélation entre les résultats électoraux et le nombre d'adhérents des partis politiques. Hern\'andez-Salda\~na~\cite{hernandez_bvot_mexique} étudie la distribution des votes de trois élections mexicaines par cabine de vote et l'espacement de leur résultat. Berger et al.~\cite{berger_contextual} montrent comment la présence d'écoles, d'églises, etc., influence le vote, notamment aux élections générales d'Arizona (USA). 

Pour une approche théorique de l'évolution d'opinion, on pourra consulter~\cite{fortunato_stat_phys, stauffer_sociophys, bettencourt_epidemiological, schweitzer_brownian} avec leurs nombreuses références.

\subsection{Les résultats des élections nationales binaires}
\label{section-res}

\subsubsection{Vue d'ensemble}
\label{pt-vue-res}
Pour des raisons de simplicité nous étudions les élections portant, d'une part, sur un choix binaire, et d'autre part, sur le même choix à l'échelle nationale, française en l'occurrence. Nous restreignons de plus cette étude à la France métropolitaine. Un choix binaire semble en effet plus facile à analyser $\ti$pour une première étude$\ti$ qu'un choix multiple. Nous justifierons par la suite pourquoi étudier des élections nationales qui portent sur le même choix, et pourquoi se limiter aux données de la France métropolitaine. Nous ne faisons plus ensuite de discrimination, i.e. nous analyserons sans aucune distinction toute élection répondant aux critères ci-dessus, dès lors que nous possédons leurs données sur l'ensemble des communes, voire sur l'ensemble des bureaux de vote. La table~\ref{tres} indique les élections traitées~: les seconds tours d'une présidentielle et les référendums $\ti$dans lesquels, l'électeur qui s'exprime, dépose dans l'urne un bulletin de vote au nom de l'une des deux alternatives proposées sur l'ensemble du pays.

\begin{table}[h]
\begin{tabular}{|c|c|c|c|c|c|}
\hline
Année & type d'élection & choix binaire & exprim & moy\\
\hline
1992 & référendum, traité de Maastricht & \textbf{oui}-non & 25,2 $10^6$ & 0,508\\ 
1995 & présidentielle, second tour & \textbf{Chirac}-Jospin & 29,0 $10^6$ & 0,525\\ 
2000 & référendum, quinquennat présidentiel& \textbf{oui}-non & 9,84 $10^6$ & 0,729\\ 
2002 & présidentielle, second tour & \textbf{Chirac}-Le Pen & 30,1 $10^6$ & 0,820\\ 
2005 & référendum, traité constitutionnel européen& oui-\textbf{non} & 27,5 $10^6$ & 0,550\\ 
2007 & présidentielle, second tour & Royal-\textbf{Sarkozy} & 34,2 $10^6$ & 0,533\\
\hline
\end{tabular}
\caption{\small Élections nationales portant sur un même choix binaire à l'échelle nationale, et analysées pour leurs résultats. Le choix en gras désigne le gagnant du scrutin avec sa moyenne (moy) sur le nombre de votes exprimés (exprim) pour la France métropolitaine.}
\label{tres}
\end{table}

Les données électorales proviennent\footnote{Mme Brigitte Hazart, du \textit{bureau des élections et des études politiques} m'a aimablement fourni les données, suite parfois, à un véritable travail de sa part. Cette thèse lui doit beaucoup, et je la remercie grandement. Un remerciement à Frédéric Debever de la préfecture de l'Aube qui m'a envoyé par courrier les résultats du second tour des présidentielles de $1995$, des communes de son département.} du \textit{bureau des élections et des études politiques} du \textit{ministère de l'intérieur}. Pour chacune des élections, figurent au minimum, le nom de la commune (et éventuellement le numéro du bureau de vote), les nombres de personnes inscrites sur la liste électorale, de votants, de bulletins blancs ou nuls, de suffrages exprimés, et enfin, le nombre de voix pour chacun des choix proposés aux électeurs.

Considérons une élection donnée. $\expr$ désigne le nombre de bulletins exprimés, et $\gagn$, le nombre de voix du résultat gagnant, tous deux à l'échelle nationale. Le choix gagnant au niveau national sert ensuite de référence aux résultats de toutes les communes (ou de tous les bureaux de vote) de la France. Ensuite, $\expr^\aaa$ et $\gagn^\aaa$, désignent respectivement pour la commune (ou le bureau de vote) $\aaa$, le nombre de bulletins exprimés et le nombre de votes en faveur du choix nationalement gagnant (même si $\frac{\gagn^\aaa}{\expr^\aaa}<50\%$).

Afin de pouvoir comparer plus aisément les résultats électoraux des différentes élections, notamment en diminuant l'influence de leur résultat moyen, il convient de changer de variable, i.e. de passer du résultat en pourcentage usuellement donné
\be \label{eres} \res = \frac{\gagn}{\expr}~,\ee
borné entre $0$ et $1$, à une variable non bornée et symétrique, comme par exemple 
\be\label{etau} \tau = \ln\big(\frac{\gagn}{\expr-\gagn}\big) = \ln\big(\frac{\res}{1-\res}\big)~.\ee
De manière générale, $\tau=\ln\big(\frac{N_+}{N_-}\big)$, où $N_+$ et $N_-$ dénotent les nombres de personnes qui ont réalisé chacun des deux choix binaires \og + \fg{} et \og - \fg, sur un total de $N=N_-+N_+$ personnes. Avec cette notation générale, $\rho$ s'écrirait comme $\rho=\frac{N_+}{N}$. Noter que $\ln\big(\frac{N_-}{N_+}\big)=-\tau$, ce qui procure l'aspect symétrique de $\tau$. Bref, un résultat $\res$ de $50\%$ équivaut à $\tau = 0$, ceux supérieurs à $50\%$ donnent $\tau > 0$ et inversement. Afin d'éviter un $\tau$ infini nous transformons dans les deux cas extrêmes $\gagn$ d'une demi-voix, soit plus précisément $\gagn=\expr\rightarrow\gagn=\expr-0,5$ et $\gagn=0 \rightarrow\gagn=0,5$.

\begin{figure}
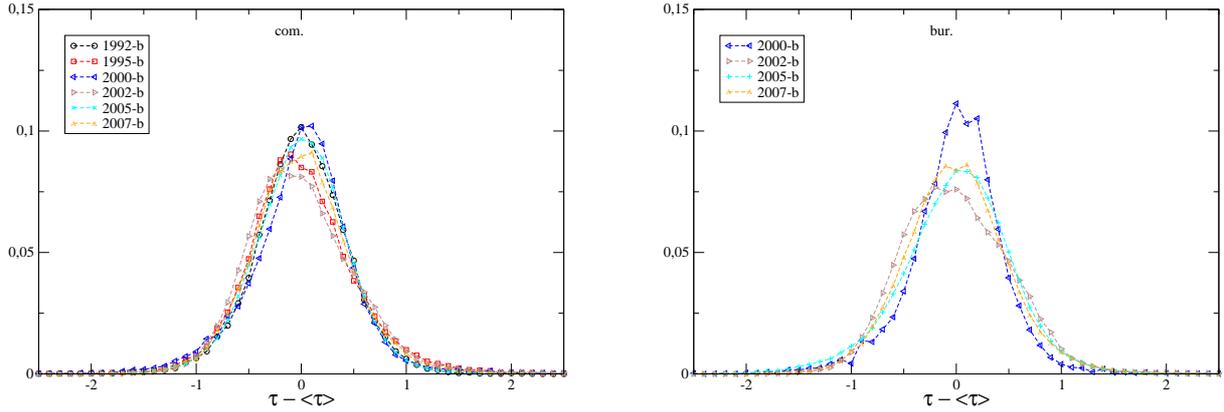

\includegraphics[scale = 0.32]{histo-log-res.eps}\hfill
\includegraphics[scale = 0.32]{histo-log-bvot-res.eps}
\caption{\small Histogrammes des résultats $\tau$ des communes à gauche et des bureaux de vote à droite, ramenés à une moyenne nulle. La lettre $b$ ajoutée à l'année précise une l'élection de type binaire. Ces légendes s'appliqueront aux élections équivalentes des figures qui suivront.}
\label{fhisto-log-res}
\end{figure}

Les figures~\ref{fhisto-log-res} représentent les histogrammes, centrés sur leur moyenne, de l'ensemble des résultats $\taa$ par commune (ou par bureau de vote). La table~\ref{tstat-log-res} fournit des statistiques élémentaires de la distribution des résultats par commune $\taa$ et, si possible, par bureau de vote. Les histogrammes et statistiques en grandeur $\res$ se trouvent en appendice (cf. Fig.~\ref{fhisto-dens-res} et Tab.~\ref{tstat-dens-res}). Nous travaillerons par la suite préférentiellement avec la grandeur $\tau$, puisqu'elle laisse apparaître bien plus de régularités entre différentes élections des histogrammes et des statistiques, qu'avec la variable usuelle $\res$. Remarquons néanmoins que l'on peut choisir d'autres grandeurs symétriques et non bornées obtenues à partir de $\res$, comme par exemple $-erfc^{(-1)}(2\res)$, ou toute autre fonction cumulative inverse d'une distribution de probabilité symétrique et définie sur $\mathbb{R}$.

\begin{table}[h]
\begin{tabular}{|c||c|c|c|c|c||c|c|c|c|c|}
\hline
Année & \textbf{com.} & moy & éc-typ & skew & kurt & \textbf{bur.} & moy & éc-typ & skew & kurt\\
\hline
1992 & 36186 & -0,164 & 0,447 & -0,159 & 2,48 & & & & &\\ 
1995 & 36197 & 0,187 & 0,524 & 0,357 & 2,71 & & & & &\\
2000 & 36202 & 0,874 & 0,498 & -0,116 & 2,77 & 62253 & 0,948 & 0,454 & -0,289 & 3,33\\
2002 & 36215 & 1,48 & 0,521 & 0,776 & 2,26 & 62352 & 1,54 & 0,521 & 0,476 & 1,17\\
2005 & 36220 & 0,377 & 0,443 & -0,021 & 1,37 & 62775 & 0,295 & 0,530 & -0,224 & 0,989\\
2007 & 36219 & 0,257 & 0,487 & 0,174 & 2,31 & 63516 & 0,185 & 0,499 & 0,185 & 1,72\\
\hline
\end{tabular}
\caption{\small Moyennes (moy), écarts-types (éc-typ), skewness (skew) et kurtosis (kurt) des résultats $\tau$ sur l'ensemble des communes (com.), et si possible sur l'ensemble des bureaux de vote (bur.).}
\label{tstat-log-res}
\end{table}

\subsubsection{La distribution binomiale d'un vote aléatoire, indépendant et identiquement distribué}
\label{pt-binomial-res}
Pour commencer, vérifions l'hypothèse la plus naïve possible, autrement dit vérifions si le vote représente un processus aléatoire indépendant et identiquement distribué sur l'ensemble du territoire.

\begin{figure}
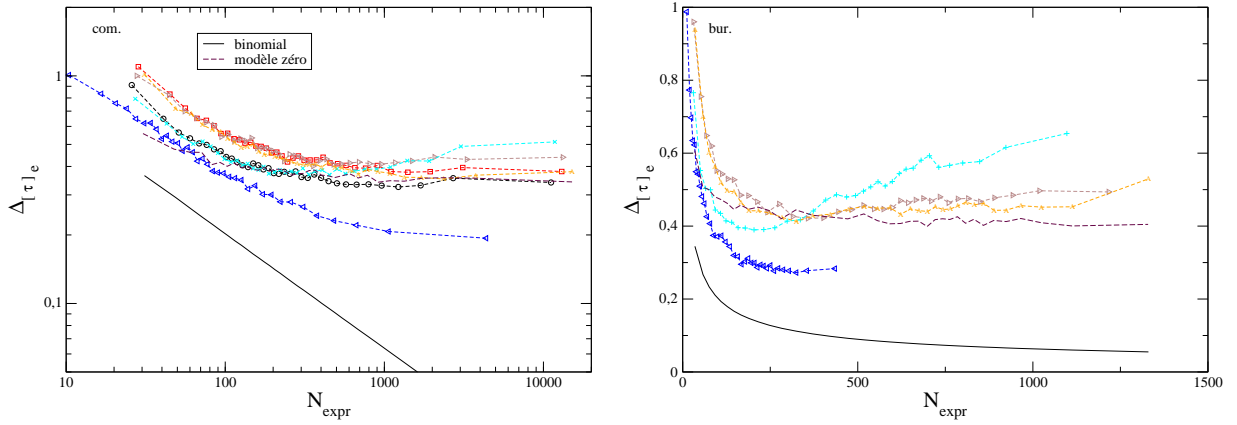

\includegraphics[scale=0.32]{sigma-taille-log-res.eps}\hfill
\includegraphics[scale=0.32]{sigma-taille-log-bvot-res.eps}
\caption{\small Écart-type des résultats $\tau$ en fonction du nombre de suffrages exprimés $\expr$, calculé avec $36$ intervalles ; pour les communes à gauche et pour les bureaux de votes à droite. Le \og modèle zéro \fg{} (voir Eq.~(\ref{emod0})) et les tirages binomiaux indépendants (voir Eq.~(\ref{ebinomial})) s'effectuent à partir des valeurs réelles de l'élection 2007-b. La valeur moyenne $\expr$ de l'échantillon des communes de plus grande taille doit se considérer avec précaution puisqu'interviennent des valeurs extrêmes. La notation $\Delta_{[...]_{\,e}}$ signifie l'écart-type d'une variable au sein d'un échantillon $e$ (répertorié sur l'axe des abscisses), et s'appliquera aux figures similaires.}
\label{fsigma-taille-res}
\end{figure}

Pour cela, les figures~\ref{fsigma-taille-res} tracent pour chaque élection les écarts-types des résultats des communes (ou des bureaux de vote) en fonction du nombre de suffrages exprimés $\expr$ par commune (ou par bureau de vote), et les comparent au cas d'une distribution binomiale des votes. Les $36$ intervalles, à l'intérieur desquels s'évaluent les écarts-types, contiennent sensiblement le même nombre de communes ($\simeq 1000$), ou de bureaux de vote ($\simeq 1700$). Autrement dit, les $36$ boîtes qui regroupent les communes (ou les bureaux de vote) n'ont pas la même largeur, mais contiennent toutes le même nombre environ de communes (ou de bureaux de vote).

L'écart-type d'une variable indépendante et identiquement distribuée, réalisée dans des ensembles de taille $N$, décroît $\ti$avec des conditions peu restrictives$\ti$ en $\frac{1}{\sqrt{N}}$ d'après la loi normale. Et dans le cas qui nous concerne, le processus binomial décrit plus spécifiquement un vote aléatoire et indépendant. En effet, un processus de vote aléatoire et indépendant à l'intérieur d'une commune (ou d'un bureau de vote), de nombre de bulletins exprimés $\expr$, correspond à $\expr$ tirages aléatoires indépendants, de probabilité de réussite $\overline{\res}$ en faveur du choix gagnant à l'échelle nationale, et ($1-\overline{\res}$) en faveur de l'autre choix. Il faut ensuite imposer la même valeur $\overline{\res}$ à l'ensemble des communes (ou des bureaux de vote) pour que ce processus devienne identiquement distribué. Dans ce cas, $\overline{\res}$ provient du résultat moyen de l'élection à l'échelle nationale (cf. Tabs.~\ref{tres} ou \ref{tstat-dens-res}). En résumé, selon l'hypothèse d'un vote aléatoire, indépendant et identiquement distribué sur l'ensemble du territoire national, le résultat $\resa$ d'une commune (ou d'un bureau de vote) de bulletins exprimés $\expr^\aaa$, dérive d'une loi binomiale de probabilité $\overline{\res}$ sur $\expr^\aaa$ essais, soit~:
\be \label{ebinomial} \resa = Binomial(\overline{\res}\; ;\expr^\aaa)~.\ee

Notons $\Delta_{[...]}\big|_{_{\expr}}$, l'écart-type d'une grandeur pour des communes (ou des bureaux de vote) de même $\expr$. Il vient alors pour des tirages binomiaux provenant de l'équation~(\ref{ebinomial})~: $\Delta_{[\res,\,binomial]}\big|_{_{\expr}} = \sqrt{\frac{\overline{\res}(1-\overline{\res})}{\expr}}$. Ce qui s'écrit en variable $\tau$~: $\Delta_{[\tau,\,binomial]}\big|_{_{\expr}} \simeq \frac{1}{\sqrt{\overline{\res}(1-\overline{\res})\expr}}$, avec de nouveau la décroissance en $1/\sqrt{\expr}$ de la loi normale.

Puisque le résultat moyen d'une élection varie en fonction de la taille des communes ou des bureaux de vote (voir les figures~\ref{fmoy-taille-res} en annexe), la probabilité $\overline{\res}$ de la formule précédente devient $\overline{\res}_N$, la moyenne des $\res$ de l'échantillon où se situe $\expr$. La Fig.~\ref{fsigma-taille-res} reporte cette expression pour l'élection 2007-b où $\expr$ correspond à la moyenne des suffrages exprimés de l'échantillon. (Les figures~\ref{fhisto-insc-expr} en annexe représentent les histogrammes du nombre de suffrages exprimés et du nombre d'électeurs inscrits par commune et par bureau de vote.)

Assez pinaillé avec ces variations de moyenne. Les figures~\ref{fsigma-taille-res} réfutent sans équivoque l'hypothèse naïve testée. Une élection diffère donc d'un processus aléatoire où chaque électeur dépose indépendamment un bulletin de vote, avec la même probabilité à l'échelle nationale.

\subsubsection{\og Modèle zéro \fg}
\label{pt-mod0-res}
A ce stade, les figures~\ref{fsigma-taille-res} peuvent suggérer un rapprochement avec un type de modèle simpliste, noté ici \og modèle zéro\fg. Ces figures laissent en effet apparaître une même tendance générale pour les $6$ élections analysées : un écart-type $\Delta_{[\tau]}\big|_{_{\expr}}$ décroissant en $1/\sqrt{\expr}$ pour les faibles $\expr$, et stable pour les grands $\expr$. Le \og modèle zéro\fg{} complexifie un peu plus l'hypothèse naïve d'un tirage binomial uniforme $\ti$précédemment testée et invalidée$\ti$ en ajoutant à la probabilité $\overline{\res}$, sur laquelle opère le tirage binomial, un bruit supplémentaire, $\eta$, par commune.

Plus précisément, selon le \og modèle zéro \fg, la probabilité sur laquelle repose les $\expr^\aaa$ tirages binomiaux $\ti$aléatoires et indépendants$\ti$ fluctue d'une commune à l'autre, et vaut $(\overline{\rho}_N + \eta^\aaa)$ pour une commune $\aaa$ ayant $\expr^\aaa$ bulletins exprimés. $\eta^\aaa$ dénote la réalisation sur la commune $\aaa$ d'une variable aléatoire, $\eta$, indépendante, et donc non corrélée entre différentes communes. $\overline{\res}_N$ désigne, comme précédemment, la moyenne des $\res$ de l'échantillon en $\expr$ auquel appartient $\expr^\aaa$. Le résultat $\resa$ d'une commune $\aaa$ s'écrit alors comme 
\be \label{emod0} \resa = Binomial(\overline{\res}_{N} + \eta^\aaa\; ; \expr^\aaa) ~,\ee
où $\eta^\aaa$, indépendant d'une commune à l'autre, provient d'une distribution de moyenne nulle, d'écart-type $\sigma_\eta$, et de nature gaussienne pour des raisons de généralité et de simplicité. L'amplitude des fluctuations, $\sigma_\eta$, provient alors de l'échantillon $e$ en $\expr$ qui minimise l'écart-type en grandeur $\res$ des résultats, i.e. $\sigma_\eta = (\Delta_{[\res]_e})_{_{min}}$. Cet échantillon se trouve parmi les échantillons de grands $\expr$, i.e. avec un bruit binomial faible.

Attardons-nous sur l'amplitude ou écart-type $\sigma_\eta$ des fluctuations $\eta$. Les figures~\ref{fsigma-taille-res} montrent pour des échantillons $e$ de grands $\expr$, un écart-type des $\taa$ qui tend vers une constante valant approximativement $\Delta_{[\tau]_e} \simeq 0,4$  pour les $6$ élections traitées, exceptée celle de $2000$. Or, si $\tau = \overline{\tau} + \xi$, avec $\overline{\tau}=\ln(\frac{\overline{\rho}}{1-\overline{\rho}})$, il vient au premier ordre en $\xi$ de l'équation~(\ref{etau})~: $\rho \simeq \overline{\rho} + \overline{\rho}(1-\overline{\rho})\,\xi$. Il en découle que $\Delta_{[\res]_{\,e}} \simeq \overline{\res}(1-\overline{\res})\Delta_{[\tau]_{\,e}}$, et toujours pour des échantillons, $e$, de grands $\expr$. Or, dans les échantillons à forts $\expr$, les fluctuations de $\eta$ prédominent sur celles occasionnées par un bruit binomial, i.e. $\Delta_{[\res]_{\,e}}\simeq\sigma_\eta$. En d'autres termes, le bruit binomial proportionnel à $1\big/\sqrt{\expr}$ devient, dans les échantillons à grands $\expr$, négligeable devant les fluctuations de $\eta$. Fait notable~? donc, l'intensité des fluctuations de $\eta$ s'écrit comme, $\sigma_\eta \simeq \overline{\res}\x(1-\overline{\res})\x A$, avec $A = (\Delta_{[\tau]_{\,e}})_{_{min}} \simeq 0,4$ pour $5$ des $6$ élections traitées.\\

\begin{wrapfigure}[7]{r}{0.33 \textwidth}
  \centering
  \includegraphics[scale=0.2]{mod0.eps}
\end{wrapfigure}

Voyons maintenant par des arguments heuristiques, pourquoi l'écart-type des $\tau$ dans des échantillons à forte population ($\Delta_{[\tau]_e}$) ne dépend pas du résultat moyen ($\overline{\tau}$ ou $\overline{\rho}$, cf. Tabs.~\ref{tstat-log-res} et \ref{tstat-dens-res}) de l'élection considérée. (On écarte dans ce paragraphe le cas de l'élection de $2000$ qui semble représenter une exception.)

Plaçons-nous pour cela dans dans les communes à grand $\expr$, i.e. dans des communes que le bruit binomial affecte peu, et que nous négligerons ici. Prenons une image assez classique en physique qui consiste à interpréter l'existence d'un équilibre, ici la moyenne $\rho$ de la commune, par l'action mécanique des ressorts. Considérons alors un équilibre, $\rho$, résultant de la tension entre deux ressorts coaxiaux différents~: le premier, de constante de raideur $K_+$ tend à déplacer $\rho$ vers $1$ ; le second, de constante de raideur $K_-$, tend au contraire à attirer $\rho$ vers $0$. A l'équilibre $K_+(1-\rho)=K_-\rho$, d'où $\tau=\ln(\frac{\rho}{1-\rho})=\ln(\frac{K_+}{K_-})$.

Considérons maintenant un ensemble de couples de ressorts ($K_+,~K_-$) caractérisés par une moyenne respective $\overline{K_+}$ et $\overline{K_-}$ avec, chose importante, des fluctuations relatives d'intensité indépendante des valeurs moyennes. Une commune donnée aura alors $K_+ = \overline{K_+} + \delta~K_+$ telle que $\frac{\delta~K_+}{\overline{K_+}}=\varepsilon_+$, et de façon similaire, $K_-=\overline{K_-}+\delta~K_-$ avec $\frac{\delta~K_-}{\overline{K_-}}=\varepsilon_-$. Ici, $\langle\varepsilon_+\rangle=\langle\varepsilon_-\rangle=0$ et $\langle(\varepsilon_+)^2\rangle=\langle(\varepsilon_-)^2\rangle=(\sigma_\varepsilon)^2$. Avec un développement limité au premier ordre en $\varepsilon$, $\tau\simeq \ln(\frac{\overline{K_+}}{\overline{K_-}})+(\varepsilon_+-\varepsilon_-)$. En conclusion, $\tau$ fluctue, comme nous le cherchions, autour de sa moyenne, $\overline{\tau}\simeq\ln(\frac{\overline{K_+}}{\overline{K_-}})$ avec une variance ($2(\sigma_\varepsilon)^2$) indépendante de la moyenne $\overline{\tau}$.

Nous avons développé ces petits calculs pour un résultat somme toute assez classique $\ti$dès lors que l'intensité relative des fluctuations ne dépend pas de la valeur moyenne et qu'interviennent des grandeurs logarithmiques$\ti$ afin de mieux comprendre les fluctuations de $\tau$ dans les zones à forte population, approximativement identiques pour $5$ des $6$ élections analysées ici (voir Fig.~\ref{fsigma-taille-res}).

Notons pour terminer que cette vision simpliste, correspondante au dénommé \og modèle zéro \fg, a aussi le mérite de retrouver des statistiques assez comparables à celles des résultats réels. Mais nous y reviendrons ultérieurement.

\subsubsection{Prise en compte des coordonnées spatiales des communes}
\label{pt-coord-spatiales-res}
Bref, devrions-nous en rester là et nous contenter d'une variante du \og modèle zéro \fg{} ? Non, bien évidemment~! Mais quel questionnement pourrait enrichir cette étude~? Sur quelle méthode d'investigation se fonder~?

Parmi l'ensemble des grilles d'analyses qu'arbore la physique, nous adopterons l'une des plus classiques en physique statistique : l'influence due au voisinage.

Nous disposons via le Répertoire Géographique des Communes~\cite{ign} de l'IGN (l'Institut Géographique National) des coordonnées spatiales XY (longitude et latitude selon l'ellipsoïde de Lambert 2 étendu\footnote{Un remerciement à Alain Borg de l'IGN pour m'avoir fourni les informations nécessaires à l'exploitation du fichier téléchargeable~\cite{ign}.}) de la mairie de chaque commune à une précision de $100~m$ près. Cette base de données, établie pour l'année $1999$, servira pour toutes les élections traitées. Autrement dit, dans le cas des fusions ou des scissions des communes par rapport à cette date $\ti$ce qui ne dépasse pas la vingtaine de cas sur plus de $36200$ communes métropolitaines$\ti$ nous nous ramènerons à leur recensement de $1999$. (Les Figures~\ref{frepartition-taille} indiquent les coordonnées XY des communes ou des bureaux de vote, avec de plus, un classement selon leur population.)

Le Répertoire Géographique des Communes~\cite{ign} fournit aussi pour chaque commune, l'altitude, la superficie et une estimation de la population.

Justifions les restrictions que nous avions volontairement imposées, en début de ce chapitre (cf. section~\ref{pt-vue-res}), à cette étude portant sur les données électorales. Une prise en compte aisée des effets dus à la localisation spatiale des résultats électoraux nécessite de considérer des élections portant sur un même choix à l'échelle nationale. Pour s'en convaincre, il suffit d'envisager la difficulté d'appréhender les effets spatiaux des résultats électoraux sur des élections communales $\ti$qui portent sur un ensemble de choix, propre à chaque commune. D'autre part, l'analyse des données électorales cantonnées à la France métropolitaine permet d'amoindrir les effets de bord d'une étude à caractère spatial. Les effets de bords deviennent en effet d'autant plus prépondérants que les surfaces considérées deviennent petites.

\subsubsection{Corrélations spatiales}
\label{pt-cor-spatial-res}
Calculons les corrélations spatiales des résultats, $\tau$, d'une même élection en fonction de la distance séparant les communes.

Mais auparavant introduisons un détail permettant d'amoindrir l'effet de l'hétérogénéité de la répartition spatiale des communes. Notons $r_{\aaa\beta}$ la distance séparant les communes $\alpha$ et $\beta$, puis $D = \sqrt{\frac{S}{2\,n_{com}}}$ la distance caractéristique du département (où se situe la commune $\aaa$) de superficie $S$ et de nombre de communes $n_{com}$. ($D \simeq 2,7~km$ en moyenne sur l'ensemble des départements.) La grandeur sans dimension, $R_{\aaa\beta} = \frac{r_{\aaa\beta}}{D}$, évalue alors la distance séparant la commune $\beta$ de $\aaa$, en tenant compte de la densité des communes du département de $\aaa$.

La corrélation spatiale $C(R)$ peut s'écrire comme
\be \label{ecorrel-spatiale} C(R) = \frac{\frac{1}{N_{\aaa\beta}}\sum_{\aaa}\sum_{\beta}\taa\tbb - \langle\taa\rangle\:\langle\tbb\rangle}{\sqrt{\langle(\taa)^2\rangle - \langle\taa\rangle^2}\  \sqrt{\langle(\tbb)^2\rangle - \langle\tbb\rangle^2 }},\ee
où le calcul s'effectue sur les $N_{\aaa\beta}$ paires de communes appartenant à des couronnes $R_{\aaa\beta} \in ]2n; 2(n+1)]$. Les moyennes comme $\langle\taa\rangle = \frac{1}{N_{\aaa\beta}} \sum_{\aaa,\beta} \taa$ et $\langle\tbb\rangle = \frac{1}{N_{\aaa\beta}} \sum_{\aaa,\beta} \tbb$ s'obtiennent à partir des communes $\aaa$ et $\beta$ ainsi considérées. (Ces deux moyennes se distinguent l'une de l'autre à cause de la dissymétrie engendrée par les différentes distances caractéristiques départementales.) Enfin signalons que la corrélation spatiale peut se calculer différemment comme par exemple dans~\cite{solomon_micro_poland}.\\

\begin{figure}
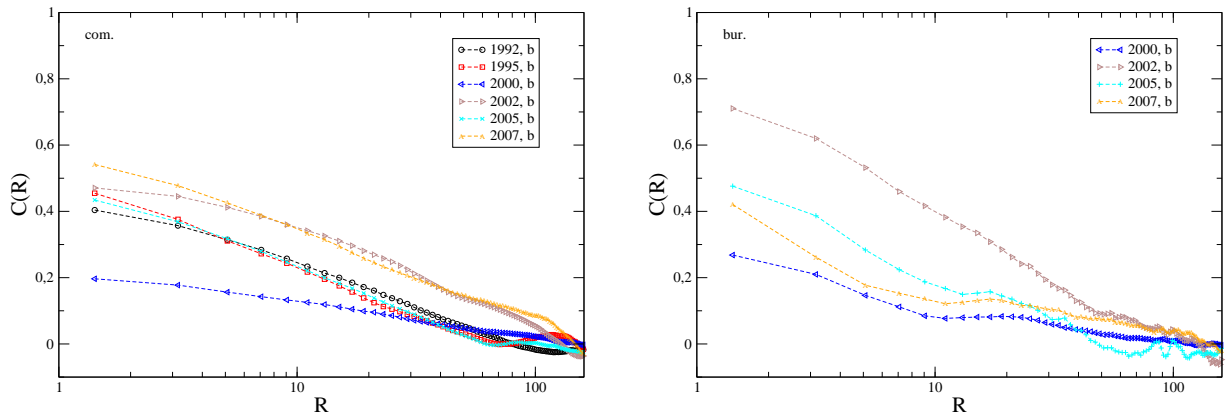

\includegraphics[scale = 0.32]{correl-log-res.eps}\hfill
\includegraphics[scale = 0.32]{correl-log-bvot-res.eps}
\caption{\small Corrélation spatiale des résultats $\tau$, des communes à gauche et des bureaux de vote à droite, où $R_{\aaa\beta}=\frac{r_{\aaa\beta}}{D} \in ]2n; 2(n+1)]$. La distance départementale caractéristique $D$ (environ $2,7~km$) pour les communes s'applique également à celle des bureaux de vote.}
\label{fcorrel-res}
\end{figure}

Les figures~\ref{fcorrel-res} montrent pour les corrélations $C(R)$ entre communes~:
\begin{itemize}
\item une corrélation spatiale très étendue à l'échelle de la France (supérieure à environ $80\,R$, i.e. à des distances supérieures à $200~km$ environ)~;
\item une décroissance quasi-logarithmique en distance~;
\item une corrélation relativement importante (environ $0,5$) entre proches communes, pour toutes les élections hormis celle de $2000$.
\end{itemize}
Ces courbes permettront par la suite de retenir, voire de calibrer, différents modèles.\\

Mais dans l'immédiat, l'étude des corrélations spatiales des résultats infirme le \og modèle zéro \fg{} pour lequel $C(R) < 0,06$ ici~; bien loin donc des valeurs obtenues pour les résultats réels. (Peut-être aurions-nous pu écrire un bel article avec de belles figures à l'appui et force commentaires sociologiques si nous n'avions pu adjoindre aux données électorales leur information spatiale~?)

\subsubsection{Effets de l'environnement}
\label{pt-environnement-res}
Puisque l'étude des corrélations spatiales $C(R)$ des résultats par commune a montré une forte dépendance du résultat d'une commune par rapport à ses communes voisines (voir Fig.~\ref{fcorrel-res}), scrutons plus finement cette influence locale. Voyons donc les relations existantes entre les résultats d'une commune et les résultats des communes de son environnement.

Définissons l'environnement $\vaa$ d'une commune centrale $\aaa$, par ses $n_p=16$ plus proches communes. Bien qu'arbitraire, le nombre $n_p=16$ concilie à la fois, la proximité des communes faisant partie de l'environnement de la commune centrale, à des fluctuations $\ti$dues aux tailles des échantillons$\ti$ pas trop importantes. Tous les autres choix $n_p = 4, 8, 32, 64$ et $128$ que nous avons testés, vérifient les conclusions tirées de $n_p=16$. Autrement dit, la convention arbitraire, $n_p=16$, du nombre des communes les plus voisines d'une commune centrale $\aaa$, et considérées comme l'environnement de $\aaa$, a un caractère robuste. Sauf mention contraire, toutes les tables et figures qui suivront faisant intervenir l'environnement autour d'une commune centrale, auront celui-ci défini par ses $n_p = 16$ plus proches communes. Remarque~: à l'échelle des bureaux de vote, l'environnement $\vaa$ d'une commune centrale $\aaa$ comprend l'ensemble des bureaux de vote que contiennent les $n_p=16$ plus proches communes de la commune centrale $\aaa$.

L'environnement $\vaa$ d'une commune centrale $\aaa$ peut se caractériser relativement correctement à l'aide des deux premiers moments de sa distribution des résultats $\tau$. 
\begin{itemize}
\item la moyenne de ses résultats électoraux~:
\be \label{etoa} \toa = \frac{1}{n_p} \sum_{\beta \in \vaa} \tbb ~;\ee
\item la déviation standard de ses résultats électoraux~:
\be \label{esaa} \saa = \sqrt{\frac{1}{n_p-1} \sum_{\beta \in \vaa} (\tbb - \toa)^2} ~,\ee
\end{itemize}
où $\beta$ désigne l'une des $n_p$ communes les plus proches de la commune centrale $\aaa$. Quand $\toa$ et $\saa$ s'évalueront non pas à partir des données par commune, mais à partir des données par bureau de vote, il suffira de remplacer $n_p$ dans les deux équations ci-dessus, par le nombre de bureaux de vote à l'intérieur de $\vaa$.

Noter que nous optons pour la moyenne définie ci-dessus plutôt que par $\ln(\frac{\sum_{\beta \in \vaa} \gagn^\beta}{\sum_{\beta \in \vaa} \expr^\beta-\gagn^\beta})$ afin d'éluder la variation des résultats des communes en fonction de leur taille (voir Fig.~\ref{fmoy-taille-res}). De même, nous préférons définir l'environnement d'une commune centrale par un nombre, $n_p$, fixe de ses plus proches communes, plutôt que par un nombre variable de communes dont la proximité à la commune centrale n'excède pas une distance limite fixe, et ce afin d'écrire proprement la dispersion standard $\saa$. Ainsi, $\saa$ ne fluctue pas plus ou moins fortement selon le nombre plus ou moins grand de communes voisines prises en compte. Enfin, tout au long de cette étude nous amalgamons la valeur de l'écart-type, à son estimation statistique~\cite{proba-stat}.

Les figures~\ref{fcorrel-toa-saa-res} en annexe tracent les corrélations spatiales des $\toa$ et des $\saa$, calculés à partir des résultats électoraux par commune pour chaque élection. Ces courbes peuvent se comparer aux corrélations spatiales des $\toa$ et $\saa$ issus d'un mélange aléatoire des résultats des communes, ainsi que ceux calculés à partir du \og modèle zéro \fg. De plus, les figures~\ref{frepartition-saa-res} reportent, en fonction de leur classement, les positions géographiques des $\saa$ de l'élection 2007-b, issus des résultats par commune et par bureau de vote.\\

L'effet du voisinage sur le résultat de la commune centrale peut s'analyser en traçant la moyenne et l'écart-type des $\taa$ en fonction de $\toa$ et de $\saa$. Nous calculons donc la moyenne et l'écart-type des résultats $\taa$ de toutes les communes centrales ayant les mêmes valeurs de $\toa$ (ou de $\saa$), ou plus précisément à l'intérieur des échantillons définis en $\toa$ (ou en $\saa$). Les valeurs $\toa$ et $\saa$ caractérisant l'environnement se répartissent dans cette étude en $36$ intervalles, contenant chacun environ $1000$ communes centrales.

\begin{figure}[t!]
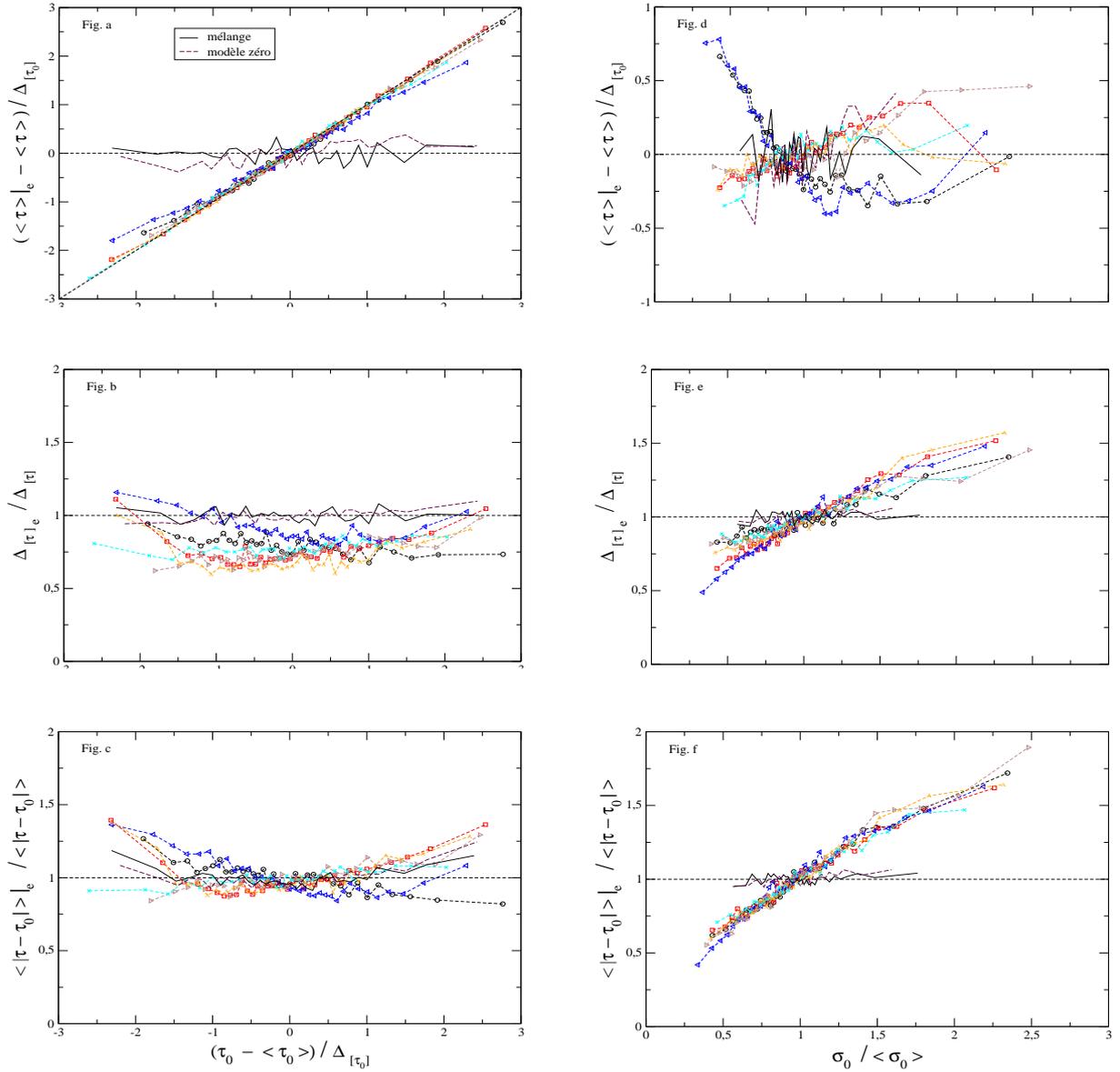

\includegraphics[width=7.5cm, height=5cm]{1-res.eps}\hfill
\includegraphics[width=7.5cm, height=5cm]{3-res.eps}\vspace{0.25cm}
\includegraphics[width=7.5cm, height=5cm]{2-res.eps}\hfill
\includegraphics[width=7.5cm, height=5cm]{4-res.eps}\vspace{0.25cm}
\includegraphics[width=7.5cm, height=5cm]{5-res.eps}\hfill
\includegraphics[width=7.5cm, height=5cm]{6-res.eps}
\caption{\small Effet de l'environnement sur le résultat de la commune centrale. En trait continu, les résultats de l'élection 2007-b mélangés aléatoirement~; et en tirets, le \og modèle zéro \fg{} appliqué au données de l'élection 2007-b. Cette légende vaudra pour les figures \ref{fhisto-environnement-res} et \ref{fs0t0-res} en annexe.}
\label{fenvironnement-res}
\end{figure}

Afin d'étudier de concert toutes ces élections nous normalisons les valeurs caractérisant l'environnement $\vaa$, soit $\saa\rightarrow\frac{\saa}{\langle \sigma_0 \rangle}$ et $\toa\rightarrow\frac{\toa - \langle \tau_0 \rangle}{\Delta_{[\tau_0]}}$ (où $\langle ... \rangle$ et $\Delta_{[...]}$ désignent respectivement la moyenne et la déviation standard de la grandeur considérée sur toutes les communes de la France métropolitaine). L'écart-type des résultats $\tau$ par échantillon $e$, $\Delta_{[\tau]}\big|_{_e}$ (noté $\Delta_{[\tau]_{\,e}}$ aux figures~\ref{fenvironnement-res}), se normalise en le divisant par l'écart-type à l'échelle nationale (i.e. $\Delta_{[\tau]}|_{_e}\rightarrow\frac{\Delta_{[\tau]}|_{_e}}{\Delta_{[\tau]}}$), et quant à la moyenne des résultats $\tau$ dans un échantillon $e$~: $\langle\tau\rangle\big|_{_e}\rightarrow\frac{\langle\tau\rangle|_{_e} - \langle \tau \rangle}{\Delta_{[\tau_0]}}$. Nous utilisons volontairement le même facteur correctif $\frac{1}{\Delta_{[\tau_0]}}$ pour l'abscisse et l'ordonnée afin de rendre explicite la droite $y = x$.

Les courbes ainsi obtenues peuvent se voir aux figures~\ref{fenvironnement-res}, ainsi que la moyenne de l'écart absolu $|\taa - \toa|$ en fonction de $\toa$ et de $\saa$. Les figures en annexe, Fig.~\ref{fhisto-environnement-res} et Fig.~\ref{fs0t0-res} représentent respectivement les histogrammes de $\toa$, de $\saa$ et de $(\taa-\toa)$ ainsi qu'une éventuelle connexion entre les deux grandeurs caractérisant l'environnement, soit $\saa = f(\toa)$ et $\toa = f(\saa)$. Afin de mieux distinguer les spécificités des résultats réels, et plus particulièrement des spécificités liées à leur localisation spatiale, nous adjoignons deux autres courbes à ces courbes réelles. La première provient d'un mélange aléatoire des résultats de l'élection de $2007$, où chaque commune se voit attribuer le résultat d'une autre commune choisie aléatoirement. La seconde dérive du \og modèle zéro \fg, appliqué encore aux données de l'élection $2007$.

Remarquons la connexion entre corrélation d'une part, et, d'autre part, le type de courbes tracées ici. Considérons pour cela le cas simpliste de deux variables aléatoires $X$ et $Y$ telles que $Y_i = a X_i + b + \eta_i$, où $\eta$ désigne un bruit blanc gaussien de moyenne nulle et indépendant de $X$. Le coefficient $a$ peut se voir directement sur la courbe donnant la moyenne de $Y$ en fonction de $X$, i.e. de la moyenne des $Y$ en fonction de l'intervalle des $X$. Pour simplifier, supposons que l'écart-type, $\Delta_{[Y]}\big|_{_e}$, des $Y$ calculé à l'intérieur d'un échantillon $e$ en $X$, prenne la même valeur pour tous les échantillons, égale à une constante $c$ multipliée par l'écart-type des $Y$ sur l'ensemble des réalisations (i.e. $\Delta_{[Y]}\big|_{_e}=c.\Delta_{[Y]}$). La corrélation $C_{XY} = \frac{\langle XY \rangle - \langle X \rangle.\langle Y \rangle}{\Delta_{[X]}.\Delta_{[Y]}}$ entre $X$ et $Y$ sur l'ensemble des réalisations peut alors s'écrire comme $C_{XY} \simeq \pm \sqrt{1-c^2}$. Cette équation convient à un nombre d'intervalles suffisamment grand (pour que $X\simeq cste$ à l'intérieur d'un intervalle défini en $X$) mais pas trop grand (pour que l'écart-type des $Y$ calculé dans un échantillon ait un sens). Le coefficient $c$, directement visible sur la courbe reportant l'écart-type de $Y$ en fonction de $X$, indique donc un degré de corrélation entre $X$ et $Y$. Plus $c$ devient petit devant $1$, plus le taux de corrélation entre $X$ et $Y$ augmente.\\

Interprétons les courbes de la figure~\ref{fenvironnement-res}. Les courbes de Fig.~\ref{fenvironnement-res}-a permettent de savoir qu'en moyenne la commune centrale vote comme la moyenne des communes de son voisinage. Cette régularité s'applique à chaque élection, et ne dépend donc, ni du résultat global de l'élection à l'échelle nationale, ni de la moyenne des résultats locaux, $\toa$. La figure~\ref{fenvironnement-res}-b indique l'intensité de cette polarisation qui, en outre, ne dépend quasiment pas de la valeur de $\toa$, sauf à ses valeurs extrêmes. La figure~\ref{fenvironnement-res}-c montre que l'écart absolu $|\taa - \toa|$ en fonction de $\toa$ des résultats réels s'apparente à celui issu d'un mélange aléatoire des résultats. Nous reviendrons ultérieurement sur la différence de comportement des courbes réelles de la figure~\ref{fenvironnement-res}-d. Par contre, les courbes des figures \ref{fenvironnement-res}-e et \ref{fenvironnement-res}-f dévoilent une nouvelle forme de régularité qui fait intervenir $\saa$. En effet l'écart en valeur absolue, $|\taa - \toa|$, entre le résultat $\taa$ de la commune centrale $\aaa$ et la moyenne $\toa$ des résultats des communes du voisinage de $\aaa$, augmente avec $\saa$ de façon similaire pour chacune des $6$ élections (voir Fig.~\ref{fenvironnement-res}-f). L'accroissement de $|\taa - \toa|$ avec la dispersion $\saa$ implique ensuite, comme la figure~\ref{fenvironnement-res}-e le visualise, une augmentation de l'écart-type des $\taa$ avec $\saa$. Bref, l'écart des résultats entre celui de la commune centrale et ceux de son voisinage augmente avec la dispersion locale des résultats, et de la même manière pour toutes les élections.

Ces figures viennent, notamment, de mettre en évidence un phénomène quelque peu nouveau à nos yeux. L'influence de l'environnement sur une commune centrale ne se réduit pas à sa moyenne. L'hétérogénéité des résultats des communes comprises dans l'environnement importe aussi. La dispersion, des résultats $\tau$ de l'environnement, se trouve de plus connectée au résultat de la commune centrale de la même manière pour chacune des élections analysées.\\

Efforçons nous maintenant de comprendre l'augmentation de $|\taa - \toa|$ en fonction de $\saa$, qui se reproduit quasiment à l'identique pour les résultats des $6$ élections analysées (voir Fig.~\ref{fenvironnement-res}-f). Cherchons l'origine de cette permanence de la façon la plus simple possible $\ti$i.e. nécessitant le moins d'interprétation ou d'information possible$\ti$ avec des arguments d'ordre statistique. L'hypothèse la plus naïve attribuerait à l'écart entre $\taa$ et $\toa$, la valeur de la dispersion $\saa$ des résultats du voisinage autour de la commune $\aaa$, en moyenne évidemment. Dit autrement, la commune centrale $\aaa$, plongée dans son environnement, n'introduit ni plus ni moins de dispersion des résultats que celle propre à son environnement, $\saa$. Dit encore autrement, la même \og agitation \fg{} des résultats (comme celle dérivant en physique d'un bain thermique local) prévaut aussi bien pour les communes de l'environnement que pour la commune centrale.

Pour vérifier cette hypothèse, nous traçons directement et sans normalisation l'écart-type de ($\taa - \toa$) en fonction de la dispersion $\saa$ des $6$ élections, et les comparons à deux courbes issues des simulations reposant sur l'hypothèse testée. La première simulation procure de façon synchrone à chaque commune $\aaa$ une valeur $\taa$ tirée aléatoirement suivant une distribution gaussienne issue des paramètres réels : la moyenne $\toa$ et l'écart-type $\saa$ mesurés pour l'élection de $2007$. Quant à la seconde simulation, commençant par des résultats $\taa$ distribués aléatoirement, chaque itération attribue à une commune $\aaa$ tirée au hasard, un résultat $\taa$ provenant d'une variable aléatoire gaussienne centrée sur $\toa$ et d'écart-type $\saa$ (avec, $\toa$ calculé d'après les résultats simulés et $\saa$ issu des résultats réels de l'élection de $2007$). Dans les deux cas, la distribution de $(\taa-\toa)$ correspond, conformément à l'hypothèse testée, à une gaussienne de moyenne nulle et d'écart-type égal à $\saa$. La figure~\ref{fdelta-saa-res} en annexe, atteste assez correctement la validité de cette hypothèse, même si en y regardant de près dans les zones à fortes dispersions $\saa$, les résultats réels échappent légèrement à cette règle.\\

En résumé, les figures~\ref{fenvironnement-res} mettent en lumière deux régularités différentes qui engagent le résultat d'une commune centrale $\aaa$  aux résultats des communes de son environnement~:
\begin{itemize}
\item en moyenne, une commune centrale $\aaa$ vote comme les communes environnantes, i.e. comme $\toa$~;
\item l'écart entre $\taa$ et $\toa$ augmente régulièrement avec l'hétérogénéité locale, $\saa$, des résultats des communes environnantes.
\end{itemize}

\subsubsection{Corrélations temporelles}
\label{pt-tempo-res}
Après avoir constaté des régularités à caractère spatial de $\saa$ $\ti$que nous ignorions jusqu'alors$\ti$, recherchons l'existence d'une éventuelle régularité à caractère temporel de $\saa$. Plus précisément, voyons comment varie $\saa$ d'une élection à l'autre pour l'ensemble des communes $\aaa$.

Afin de répondre quantitativement à cette question, il convient de calculer la corrélation de $\saa$ d'une élection à l'autre sur l'ensemble des communes. La corrélation temporelle d'une grandeur $X$, prenant pour chaque commune $\aaa$ les valeurs $X^\aaa(t_i)$ et $X^\aaa(t_j)$ aux deux élections aux temps $t_i$ et $t_j$, s'exprime comme
\be \label{ecorrel-temporelle} C_{t_i,t_j}(X) = \frac{\langle X^\aaa(t_i)X^\aaa(t_j) \rangle - \langle X^\aaa(t_i)\rangle \: \langle X^\aaa(t_j)\rangle}{\sqrt{\langle X^\aaa(t_i)^2\rangle - \langle X^\aaa(t_i)\rangle^2}\ \sqrt{\langle X^\aaa(t_j)^2\rangle - \langle X^\aaa(t_j)\rangle^2}}~, \ee
où les moyennes $\langle ...\rangle$ s'effectuent sur l'ensemble des communes. Ensuite, la moyenne des corrélations temporelles et leur déviation standard sur l'ensemble des $n_{ij}$ couples d'élections s'évaluent respectivement comme $\overline{C}_t(X)=\frac{1}{n_{ij}}\sum_{i,j\neq i} C_{t_i,t_j}(X)$ et $\Delta_{[C_t(X)]} = \sqrt{\frac{1}{n_{ij}-1}\sum_{i,j\neq i} (C_{t_i,t_j}(X)- \overline{C}_t(X))^2}$.

La table~\ref{ttempo-saa-res} fournit les corrélations de $\saa$ pour chaque couple d'élections, puis leur moyenne et leur écart-type sur l'ensemble des couples d'élections, et enfin le rapport de ces deux dernières grandeurs (une sorte de rapport \textit{bruit/signal}). Elle indique des $C_{t_i,t_j}(\sigma_0)$, premièrement, de valeurs relativement élevées (environ $0,6$), et deuxièmement, assez régulières sur tous les couples d'élections différentes. $\saa$ montre par là un caractère relativement permanent, ou régulier, au cours des différentes élections.

\begin{table}[h]
\begin{minipage}[c]{0.56\linewidth}
\begin{tabular}{|c|c|c|c|c|c|}
\hline
\backslashbox{$t_i$}{$t_j$} & 1995 & 2000 & 2002 & 2005 & 2007\\
\hline
1992 & 0,577 & 0,608 & 0,606 & 0,527 & 0,576\\
1995 & & 0,548 & 0,541 & 0,606 & 0,748\\
2000 & & & 0,616 & 0,487 & 0,557\\
2002 & & & & 0,523 & 0,558\\
2005 & & & & & 0,614\\
\hline
\end{tabular}
\end{minipage}\hfill
\begin{minipage}[c]{0.4\linewidth}
\caption{\small Corrélation temporelle $C_{t_i,t_j}(\sigma_0)$ des $\saa$ sur chaque couple d'élections $(t_i,t_j)$. Moyenne $\overline{C}_t(\sigma_0) = 0,579$ ; écart-type $\Delta_{[C_t(\sigma_0)]} = 0,060$ ; $\frac{\Delta_{[C_t(\sigma_0)]}}{\overline{C}_t(\sigma_0)} = 0,104$.}
\label{ttempo-saa-res}
\end{minipage}
\end{table}

\begin{figure}
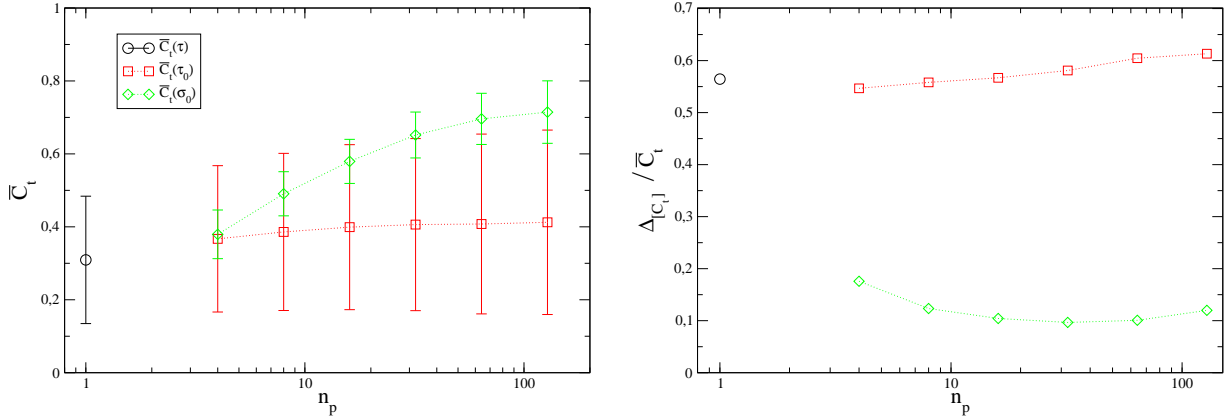

\includegraphics[scale = 0.32]{bilan-cor-tempo-res-multippv.eps}\hfill
\includegraphics[scale = 0.32]{ratio-multippv-res.eps}
\caption{\small Moyenne des corrélations temporelles $\overline{C}_t$ sur l'ensemble des couples d'élections, avec leur déviation standard $\Delta_{[C_t]}$ en barre d'erreur, évalués pour $\taa$, $\toa$ et $\saa$ des résultats électoraux. La figure de droite visualise leur rapport, $\frac{\Delta_{[C_t]}}{\overline{C}_t}$, une sorte de rapport \textit{bruit/signal}. Le nombre de communes, $n_p$, intervenant dans les grandeurs $\toa$ et $\saa$ (voir Eqs.~(\ref{etoa},~\ref{esaa})), égal à 16 jusqu'alors varie ici de 4 à 128. Les valeurs correspondantes à $\tau$ se situent, par extension, à $n_p=1$.}
\label{fbilan-tempo-res}
\end{figure}

Une fois de plus, la régularité des valeurs mesurées à la table~\ref{ttempo-saa-res} ne prend son sens que placée dans son contexte. Cette table doit se comparer aux corrélations temporelles des résultats par commune $\taa$, et des moyennes $\toa$. La table~\ref{ttempo-res} en annexe indique pour les résultats $\taa$, d'une part de fortes variations selon les couples d'élections et d'autre part une plus faible corrélation en moyenne. (Plus précisément, $\overline{C}_t(\tau) = 0,309$, $\Delta_{[C_t(\tau)]} = 0,175$ et $\frac{\Delta_{[C_t(\tau)]}}{\overline{C}_t(\tau)} = 0,564$.) Les corrélations temporelles des $\toa$ examinées à la table~\ref{ttempo-toa-res} (pour lesquelles $\overline{C}_t(\tau_0) = 0,399$, $\Delta_{[C_t(\tau_0)]} = 0,226$ et $\frac{\Delta_{[C_t(\tau_0)]}}{\overline{C}_t(\tau_0)} = 0,567$) infirment l'hypothèse d'une grande stabilité et d'une forte corrélation temporelle des $\saa$ qui dérivent d'un simple effet de lissage (par le fait de prendre en compte la valeur associée, non pas à une seule commune, mais à $n_p=16$ communes). Afin de s'en convaincre pleinement, la figure~\ref{fbilan-tempo-res} rapporte pour $\toa$ et $\saa$, définies avec un nombre variable $n_p$ de communes voisines allant de $4$ à $128$, la moyenne $\overline{C}_t$ et la déviation standard $\Delta_{[C_t]}$ de leurs corrélations temporelles sur l'ensemble des couples d'élections, ainsi que leur rapport. (La moyenne et la dispersion standard des corrélations des résultats $\taa$ ou des $\toa$ s'estiment à partir de leur valeur absolue $|C_{t_i,t_j}|$ afin de ne pas se fourvoyer dans la convention adoptée ici, qui revient à considérer le résultat $\tau$ ou $\res$ d'une élection en fonction du choix gagnant à l'échelle nationale, et ce indépendamment d'une des polarisations possibles comme les distinctions droite-gauche, pro ou anti-européaniste, etc., diverses et sujettes, semble-t-il, à bien des controverses.)

Enfin, nous constatons aussi de plus grandes et régulières valeurs des corrélations temporelles des $\saa$ comparées à celles des $(\taa - \toa)$ ou des $|\taa - \toa|$, (avec $\overline{C}_t(\taa - \toa)\simeq 0,24$, $\Delta_{[C_t(\taa -\toa)]}\simeq 0,15$, et  $\overline{C}_t(|\taa - \toa|)\simeq 0,27$, $\Delta_{[C_t(|\taa - \toa|)]}\simeq 0,08$). 

\subsubsection{Une étude singulière}
\label{pt-sarko}
Contrairement à la règle que nous avions volontairement adoptée au début de cette étude $\ti$celle de traiter toutes les élections de façon identique sans faire intervenir les spécificités inhérentes à chacune d'entre elles$\ti$, nous allons nous arrêter momentanément au second tour de la présidentielle de $2007$. Cette partie peut se voir comme une digression divertissante qui renforce l'importance, si le besoin se fait encore sentir, de $\saa$. Le lecteur peut sauter sans encombre cette section, et se référer à l'annexe~\ref{annexe-croiss-pop} pour une analyse plus approfondie du lien entre croissance de population et $\saa$.\\

J'avais l'impression lorsque je collectais \og à la main \fg{} les résultats des 20 arrondissements de Paris de cette élection $\ti$pour compléter les données informatisées par commune$\ti$, que le vote pour Sarkozy ou Royal de l'arrondissement dépendait de son taux d'augmentation de population. Plus la population $\ti$ramenée au nombre d'électeurs inscrits sur les listes électorales$\ti$ par arrondissement s'accroissait, plus l'arrondissement votait en moyenne pour Ségolène Royal.

Éprouvons donc cette assertion pour l'ensemble des communes de France métropolitaine, en utilisant une fois de plus la corrélation $C_{XY}$ entre deux grandeurs : $X^\aaa=\taa$, le résultat de la commune, et $Y^\aaa=p^\aaa=\ln\left(\frac{\insc^\aaa(2007)}{\insc^\aaa(1992)}\right)$, le taux d'augmentation du nombre d'inscrits de la commune entre $1992$ et $2007$. (Rappelons l'expression $C_{XY} = \frac{\langle XY \rangle - \langle X \rangle.\langle Y \rangle}{\sqrt{\langle X^2 \rangle - \langle X \rangle ^2}.\sqrt{\langle Y^2 \rangle - \langle Y \rangle ^2}}$, où la moyenne $\langle ...\rangle$ s'effectue sur l'ensemble des communes $\aaa$.)

La corrélation sur toutes les communes du résultat $\tau$ et du taux d'augmentation de population $p$, aboutit à une corrélation $C_{\tau p} = 0,034$. Une corrélation non seulement faible, voire négligeable, mais opposée à celle que j'avais entrevue. Qu'en conclure~?\begin{itemize}
\item Qu'il ne faut jamais se fier à ses premières impressions, y compris dans la chose politique~?
\item Qu'en politique comme ailleurs, les explications ne sauraient se satisfaire d'un vague discours fondé sur quelques données glanées au passage~?
\item Que la France métropolitaine ne se résume pas à Paris~?
\item Qu'il ne faut jamais prendre au sérieux les élucubrations émanant des sciences dites dures qui empiètent sur les sciences dites sociales~? puisque, bien évidemment ces deux sciences étudient deux \og objets \fg{} hermétiques et complètement distincts à l'instar des deux mondes sublunaire et supralunaire d'Aristote.
\end{itemize}
\vspace{0.25cm}

Bref, plus sérieusement poursuivons notre questionnement. Situons de nouveau une grandeur dans son contexte, autrement dit calculons différentes corrélations faisant intervenir non seulement la valeur $\taa$ ou $p^\aaa$ de la commune centrale, mais aussi celles de son environnement. Notons $p_0^\aaa$ et $\sigma_{p,0}^\aaa$, la moyenne et la déviation standard de $p$ dans le voisinage $\vaa$ d'une commune centrale $\aaa$. Elles s'obtiennent de la même façon que $\toa$ et $\saa$ par rapport à $\tau$, i.e. en transposant dans les équations.~(\ref{etoa},~\ref{esaa}), $\tau$ par $p$. La table~\ref{tsarko} fournit quelques-unes des corrélations sur l'ensemble des communes~; soit entre le résultat d'une commune centrale $\taa$ et une grandeur liée à l'accroissement de population dans son voisinage, soit entre le taux d'augmentation d'une commune centrale $p^\aaa$ et une grandeur associée aux résultats $\tau$ dans son voisinage.

\begin{table}[h!]
\begin{tabular}{|c||c|c|c|c|}
\hline
$X^\aaa=\taa$ et $Y^\aaa=$ & $p_0^\aaa$ : 0,072 & $\sigma_{p,0}^\aaa$ : -0,026 & $(p^\alpha - p_0^\aaa)$ : -0,019 & $|p^\alpha - p_0^\aaa|$ : 0,014\\
\hline
$Y^\aaa=p^\alpha$ et $X^\aaa=$ & $\toa$ : 0,069 & $\saa$ : \textbf{-0,290} & $(\taa - \toa)$ : -0,023 & $|\taa - \toa|$ : -0,162\\
\hline
\end{tabular}
\caption{\small Corrélations $C_{XY}$ entre les grandeurs, $X$ liée au résultat du second tour de la présidentielle de $2007$, et $Y$ liée au taux d'accroissement du nombre d'inscrits entre $1992$ et $2007$. Pour mémoire~: avec $X^\aaa=\taa$ et $Y^\aaa=p^\alpha = \ln\left(\frac{\insc^\aaa(2007)}{\insc^\aaa(1992)}\right)$, $C_{XY} = 0,034$.}
\label{tsarko}
\end{table}

Avec $X^\aaa=\saa$ et $Y^\aaa=p^\alpha$, $C_{XY} = -0,290$ dépasse nettement, en valeur absolue, toutes les autres manières de calculer les autres corrélations de la table~\ref{tsarko}. Cette corrélation signifie qu'en moyenne, plus l'hétérogénéité des résultats des communes voisines à une commune centrale $\aaa$ diminue, plus le taux d'augmentation du nombre d'inscrits de la commune $\aaa$ entre $1992$ et $2007$ augmente. En d'autres termes, l'homogénéité des résultats augmente, en moyenne, avec l'augmentation du taux de croissance de la population. Autrement dit, les zones à faible taux de croissance de population ont, en moyenne, une forte dispersion des résultats à l'élection 2007-b. Notons aussi que la deuxième valeur de corrélation, en valeur absolue, de la table~\ref{tsarko} concerne $X^\aaa=|\taa - \toa|$ et $Y^\aaa=p^\alpha$ (avec $C_{XY} = -0,162$). Ce dernier point confirme la connexion entre $\saa$ et $|\taa - \toa|$ entrevue aux figures~\ref{fenvironnement-res}, et \ref{fdelta-saa-res}, tout en accordant de nouveau (cf. les corrélations temporelles de $\saa$ comparées à celles des $|\taa - \toa|$, ou des $(\taa - \toa)$ vues à la section~\ref{pt-tempo-res}) la primauté de $\saa$ sur $|\taa - \toa|$ (ou sur $(\taa-\toa)$) pour caractériser ou étudier des résultats électoraux.

En résumé, cette petite étude croisant les résultats du second tour de la présidentielle de 2007 aux taux d'augmentation de la population, renforce, à ce stade, l'importance $\ti$ici quelque peu inattendue, ou non intuitive$\ti$ de $\saa$. Importance, précisons-le, ici relative au taux d'augmentation de population.

Notons que l'importance de $\saa$ par rapport au taux d'accroissement de la population, peut provenir d'une relation entre ces deux grandeurs, soit directe, soit indirecte $\ti$car médiée par une autre grandeur intermédiaire. L'annexe~\ref{annexe-croiss-pop} étudie plus rigoureusement, certes avec les données disponibles, la relation entre $\saa$ de l'élection 2007 et le taux d'augmentation de la population de 1992 à 2007.\\

Cette digression renforce, y compris indirectement, l'importance de $\saa$, et s'ajoute aux régularités précédemment observées associées à $\saa$. Régularités qui concernent aussi bien un rapport d'une commune centrale à son environnement (voir Fig.~\ref{fenvironnement-res}) qu'un aspect temporel (voir Tab.~\ref{ttempo-saa-res}). Ce qui amène naturellement à se demander pourquoi.

\subsubsection{Comprendre $\mathbf{\saa}$}
\label{pt-comprendre-res}
Il y aurait deux approches possibles qui permettraient de mieux comprendre les régularités associées à $\saa$.

La première se base sur une permanence du comportement qui régit le vote de l'électeur à chaque élection. Ces permanences de comportement pourraient peut-être s'apparenter à une généralisation, ou à une extrapolation, des observations faites en psychologie sociale $\ti$réalisées notamment par Sherif, Asch, Milgram et Moscovici$\ti$ concernant le conformisme d'un individu face à un groupe plus ou moins homogène~\cite{moscovici, psy_soc_exp}. Dans le même ordre d'idée, M\"{u}ller et al.~\cite{stauffer_inhomogeneous-T} attribuent au spin central, une règle qui dépend d'une sorte de dispersion des spins incluant son voisinage, et ce pour modéliser théoriquement la ségrégation des habitations selon un genre de modèle de Schelling.

La seconde approche possible se fonde quant à elle sur une connexion de $\saa$ à une \og texture \fg, à un \og substrat \fg{} géographique, à des grandeurs lentement variables aux échelles de temps rencontrées. Les régularités spatiales et temporelles associées à $\saa$ pourraient ensuite s'expliquer de deux manières différentes : soit $\saa$ contiendrait des informations à caractère politique, historique, sociologique, etc., relatives à sa zone géographique d'évaluation, soit au contraire, $\saa$ brillerait par son manque total de signification politique. Autrement dit, $\saa$ pourrait alors se rattacher à des données non électorales de deux façons possibles : soit $\saa$ révélerait des propriétés positives (d'ordre politique et sociologique~\cite{baudoin-socio-po, chiche-party}) du milieu géographique, soit à l'inverse, $\saa$ mesurerait un bruit (sans aucune signification politique), dû par exemple aux tailles finies des communes, de la façon de collecter les votes via une répartition spatiale en communes, etc. 

Inutile de dire que cette étude n'aspire pas à trancher entre ces deux approches possibles $\ti$qui ne s'excluent d'ailleurs pas obligatoirement$\ti$, mais cherche à mieux appréhender $\saa$ par les questions qu'elle soulève. Quelle démarche adopter alors~?\\

Pour commencer, nous privilégions la seconde approche, puis chercherons à la tester avec les données disponibles. Si les données dont nous disposons ne permettent pas de valider la seconde approche, alors nous nous tournerons vers la première. Autrement dit, une fois soulevée la question de l'origine plausible de $\saa$ $\ti$qui dépasse le cadre de cette étude$\ti$, nous cherchons, comme de coutume en science, une relation forte entre $\saa$ et une autre grandeur mesurable. Et nous nous en contenterons si nous parvenons à établir cette relation.

Il reste donc à savoir quelles données non électorales, variant peu à l'échelle de temps considéré (entre 1992 et 2007), utiliser. Actuellement nous ne possédons que la population, la surface, l'altitude et la position géographique, comme seules données non électorales par commune. La table~\ref{tsarko}, la figure~\ref{fmoy-taille-res}, et bien évidemment la figure~\ref{fsigma-taille-res}, attestent d'un lien entre la population et les résultats électoraux, ce qui nous incite à poursuivre dans cette voie. En outre, rechercher un éventuel lien entre $\saa$ et la population permet de se confronter à l'une des sources potentielles du bruit : celle que $\saa$ contiendrait sous forme de bruit statistique dû aux effets de tailles finies des communes.\\

Remarquons au préalable que la population se ramène tout le long de cette étude au nombre d'inscrits sur les listes électorales, et ce sans se préoccuper, dans l'immédiat, de questions connexes. Puisque $\saa$ fait intervenir $n_p$ communes de l'environnement $\vaa$ d'une commune centrale $\aaa$ (voir Eq.~(\ref{esaa})), la population de $\vaa$ se définit directement comme la population sur la couronne des $n_p$ communes~:
\be \label{einscoa} \inscoa = \sum_{\beta\in\vaa} \insc^\beta ~.\ee

\begin{figure}
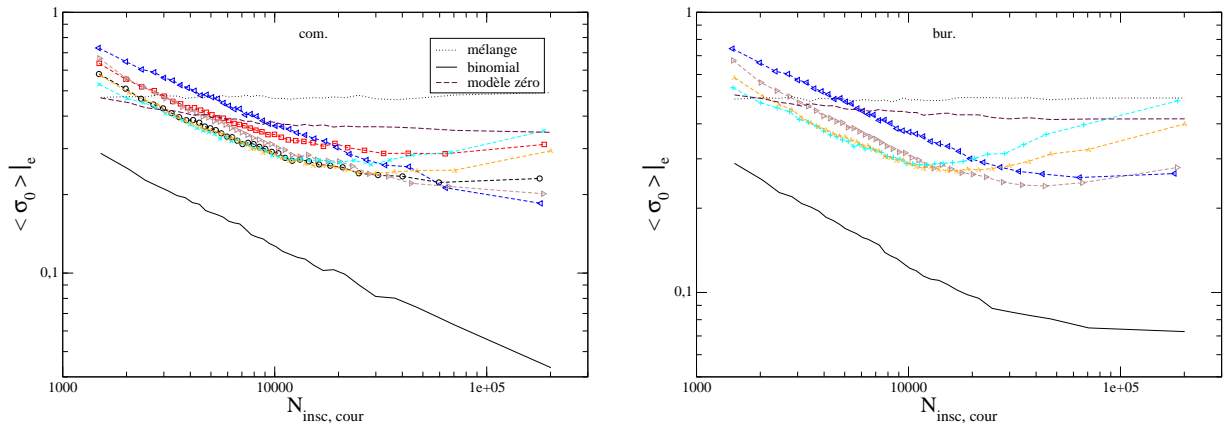

\includegraphics[scale = 0.32]{s0-com-taille-inscrit-res.eps}\hfill
\includegraphics[scale = 0.32]{s0-bvot-taille-inscrit-res.eps}
\caption{\small Moyenne des $\saa$ en fonction de $\inscoa$, calculé avec $36$ intervalles. $\saa$ s'évalue à partir des résultats par commune, à gauche, et des résultats par bureau de vote, à droite. Les tirages aléatoires binomiaux, le \og modèle zéro \fg{} et le mélange aléatoire des résultats s'effectuent à partir des valeurs réelles de l'élection 2007-b.}
\label{fsaa-taille-res}
\end{figure}

Les figures~\ref{fsaa-taille-res} reportent $\saa$ en fonction de la taille de la population, et permettent donc de voir facilement la relation entre ces deux grandeurs. Afin de jauger la présence d'un bruit dû aux effets de tailles finies des communes (ou des bureaux de vote), nous comparons les courbes issues des résultats réels à celle dérivant d'une simulation binomiale (cf. Eq.~(\ref{ebinomial}) et section~\ref{pt-binomial-res}), et accessoirement, à celle provenant du \og modèle zéro \fg{} (cf. Eq.~(\ref{emod0}) et section~\ref{pt-mod0-res}). Voyons maintenant les réponses apportées par les figures~\ref{fsaa-taille-res} aux questions posées plus haut.

Ces figures fournissent deux éléments de réponse puisqu'elles montrent clairement~:
\begin{itemize}
\item une forte correspondance entre la taille de la population et $\saa$~; 
\item la présence de bruit dû aux tailles finies dans $\saa$, excepté dans les zones à forte population.
\end{itemize}
Nous nous satisfaisons de ces réponses et ne poursuivrons donc pas nos investigations à la recherche de règles comportementales des électeurs, associées à la dispersion des résultats dans le voisinage d'une commune centrale $\ti$sans pour autant se positionner sur l'existence ou non de ces règles.\\

Mais revenons aux figures~\ref{fsaa-taille-res}. La présence du bruit dû aux tailles finies, hormis les zones des plus fortes populations, se décèle par l'allure similaire des courbes réelles et celles issues d'un tirage binomial de même probabilité $\overline{\res}$. Les courbes réelles et celle provenant du tirage binomial ont en effet sensiblement les mêmes exposants des lois de puissance de $\saa$ en fonction de $\inscoa$. Quelques remarques néanmoins. La loi de puissance, $\saa \propto \frac{1}{(\inscoa)^{0,42}}$, obtenue pour la distribution binomiale, diffère de celle attendue, $\saa \propto \frac{1}{\sqrt{\inscoa}}$, à cause de l'inhomogénéité en taille de la répartition spatiale des communes, et à plus forte raison des bureaux de vote. Notons aussi que la courbe obtenue avec le \og modèle zéro \fg{} se distingue nettement des courbes réelles, à cause, probablement, de l'état de corrélation des votes à l'intérieur des communes d'un même voisinage. (Le \og modèle zéro \fg{} part en effet de l'hypothèse des votes indépendants à l'intérieur d'une commune, et aussi, entre différentes communes.) La courbe relative à l'élection de $2000$ se situe au-dessus des autres, à cause de son faible nombre de suffrages exprimés ($\expr \simeq 0,25\:\insc$) à l'aune des autres élections. Mais cet argument ne tient pas pour comparer les autres élections entre elles. Bien d'autres informations peuvent s'extraire des figures~\ref{fsaa-taille-res} $\ti$notamment l'irrégularité des comportements dans les zones à forte population$\ti$, mais comme nous le disions en préambule, la présente investigation concerne les régularités, et les régularités seulement. Que cela nous suffise pour l'instant.

\subsubsection{Des régularités expliquées}
\label{pt-explique-res}
Voyons pourquoi les précédentes régularités associées à $\saa$ s'expliquent par la forte connexion entre la taille de la population et $\saa$. 

Puisque la population des communes varie faiblement d'une élection à l'autre, $\saa$ garde d'une élection à l'autre, approximativement la même valeur $\ti$à un décalage global près$\ti$ sur sa zone d'évaluation (voir Fig.~\ref{fsaa-taille-res}). Ainsi, se comprennent la corrélation temporelle assez forte des $\saa$, ainsi que leur faible variation d'une élection à l'autre (voir Tab.~\ref{ttempo-saa-res}).

Puisque autour d'une commune $\aaa$ de taille donnée se trouvent en moyenne d'autres communes de même taille~\cite{repartition_pop} (voir aussi Figs.~\ref{frepartition-taille} et \ref{fenvironnement-inscrit}), le bruit statistique dû aux tailles finies possède la même intensité moyenne, aussi bien pour la commune centrale $\aaa$ que pour celles de son environnement. Ainsi, se justifie plus aisément la variation régulière de $|\taa - \toa|$ en fonction de $\saa$ (voir Fig.~\ref{fenvironnement-res}-f).

Enfin, puisque le taux de croissance de la population augmente avec la taille de la population~\cite{repartition_pop} (voir aussi l'annexe~\ref{annexe-croiss-pop}), $\saa$ se trouve donc indirectement $\ti$et donc dans une moindre mesure$\ti$ relié au taux de croissance de la population. L'anti-corrélation entre $\saa$ et le taux d'augmentation de population (voir Tab.~\ref{tsarko}) s'éclaire de la même façon, à savoir~: plus la taille de la population augmente (et donc plus $\saa$ diminue, en moyenne), plus le taux de croissance de population augmente. (L'annexe~\ref{annexe-croiss-pop} étudie plus scrupuleusement la relation entre $\saa$ et le taux d'augmentation de population, purgée de leur connexion indirecte via la taille de la population.)

En outre et à titre anecdotique, se clarifie l'apparente irrégularité des courbes de la figure~\ref{fenvironnement-res}-d, où les élections des années 1992 et 2000 se différencient nettement des autres. La figure~\ref{fmoy-taille-res} indique un comportement similaire pour les élections de $1992$ et de $2000$~: à la différence des autres élections, le résultat $\taa$ d'une commune croît en moyenne quand le nombre de suffrages exprimés, et donc sa taille de population, augmente. ($\inscoa$ et le nombre d'inscrits $\insc^\aaa$ par commune centrale $\aaa$ varient de concert, à cause de la répartition spatiale de la population~\cite{repartition_pop} (voir aussi Figs.~\ref{frepartition-taille} et \ref{fenvironnement-inscrit})). Ainsi, $\taa$ et $\saa$ peuvent de nouveau se retrouver indirectement connectés, et ce via la taille de la population. Et pour les élections de $2000$ et de $2005$, quand la population augmente, $\taa$ augmente et $\saa$ diminue, ce qui implique la décroissance de $\taa$ pour la croissance de $\saa$ comme le montre la figure~\ref{fenvironnement-res}-d.

\subsubsection{Du bruit dans $\mathbf{\saa}$}
\label{pt-bruit-res}
Nous voulons ici quantifier l'accointance de $\saa$ avec la taille de la population (voir Fig.~\ref{fsaa-taille-res}) et, plus précisément encore, avec le bruit statistique dû aux tailles finies des communes (ou des bureaux de vote).

Nous commençons pour cela, par rechercher l'expression mathématique du bruit statistique occasionné par les tailles finies de population. N'ayant pas de modèle de vote à l'échelle de la commune, ou entre plusieurs communes avoisinantes, nous nous référons au bruit statistique binomial dû aux tailles finies. Le rapprochement d'allure entre les courbes réelles et celles issues du tirage binomial dans les figures~\ref{fsaa-taille-res} conforte notre position, à savoir~: se référer au bruit statistique binomial pour décrire le bruit statistique dû aux tailles finies des résultats réels contenu dans $\saa$.\\

Selon un tirage binomial (cf. Eq.~(\ref{ebinomial}) et section~\ref{pt-binomial-res}), l'écart-type des réalisations possibles pour une commune (ou pour un bureau de vote) avec $\expr^\aaa$ suffrages exprimés, s'écrit comme $\frac{\sqrt{\overline{\res}(1-\overline{\res})}}{\sqrt{\expr^\aaa}}$. En considérant des petites fluctuations, et en négligeant l'asymétrie (ou skewness) de la distribution binomiale, le résultat $\resa$ d'une commune (ou d'un bureau de vote) $\aaa$ avec $\expr^\aaa$ suffrages exprimés devient selon une simulation binomiale~: $\resa \simeq \overline{\res} + \frac{\sqrt{\overline{\res}(1-\overline{\res})}}{\sqrt{\expr^\aaa}}\:\xi^\aaa$, où $\xi^\aaa$ dénote une réalisation d'une distribution $\xi$, de moyenne nulle et de variance unité, gaussienne par simplicité. Le changement de variable $\res$ à $\tau$, donné par l'équation~(\ref{etau}), implique $\taa \simeq \overline{\tau} + \frac{1}{\sqrt{\overline{\res}(1-\overline{\res})}\ \sqrt{\expr^\aaa}}\:\xi^\aaa$, où $\overline{\tau}=\ln(\frac{\overline{\res}}{1-\overline{\res}})$. La variance engendrée par les communes (ou bureaux de vote) du voisinage $\vaa$ d'une commune centrale devient selon l'hypothèse d'un tirage binomial, $\langle \frac{(\xi^\beta)^2}{\overline{\res}(1-\overline{\res})}\frac{1}{\expr^\beta}\rangle_{_{\beta\in\vaa}}\simeq\frac{1}{\overline{\res}(1-\overline{\res})}\langle\frac{1}{\expr^\beta}\rangle_{_{\beta\in\vaa}}$.

Il convient alors d'introduire la grandeur $\piinsc$ définie par~:
\be \label{epiinsc} \piinsc = \sqrt{\langle\frac{1}{\insc^\beta}\rangle_{_{\beta\in\vaa}}} = \sqrt{\frac{1}{n_p} \sum_{\beta\in\vaa} \frac{1}{\insc^\beta}}~,\ee
qui relie la façon dont la population se répartit sur l'environnement $\vaa$ autour d'une commune centrale $\aaa$, au bruit statistique binomial dû aux tailles finies des populations. (Quand $\piinsc$ se calculera non pas à partir des populations des communes, mais à partir des populations des bureaux de vote, il suffira de remplacer dans le membre de droite, $n_p$ par le nombre de bureaux de vote contenus dans la zone $\vaa$.) Bien que nous aurions dû par souci de rigueur définir une grandeur relative aux $\expr^\beta$ (et non pas aux $\insc^\beta$), nous avons néanmoins opté pour $\piinsc$ associée au nombre d'inscrits, puisque, d'une part, cette grandeur fait directement intervenir la population des communes qui varie peu d'une élection à l'autre, et d'autre part, elle engendre une différence négligeable sur les résultats des corrélations qui suivront.

A titre indicatif, un \og modèle zéro \fg{} $\ti$bâti sur l'hypothèse d'un vote, ni corrélé à l'intérieur d'une commune, ni entre communes avoisinantes$\ti$ appliqué aux inscrits et non pas aux nombres de suffrages exprimés, donnerait~:
\be \label{esaa-mod0} (\saa)^2 \simeq \sigma_\eta^2 + \frac{1}{\overline{\rho}(1-\overline{\rho})}\x(\piinsc)^2 - \sigma_\eta^2\x(\piinsc)^2,\ee

La figure~\ref{fcorrel-piinsc} en annexe montre les corrélations spatiales de $\pio$, et de son équivalent pour une seule commune, à savoir $\frac{1}{\sqrt{\insc}}$, toutes deux pour les données de l'élection 2007-b. Il ressort une étonnante similarité des corrélations spatiales des résultats réels (cf. Fig.~\ref{fcorrel-res}) avec celle de $\frac{1}{\sqrt{\insc}}$, corroborée par la similarité des corrélations spatiales des $\toa$ des résultats électoraux (cf. Fig.~\ref{fcorrel-toa-saa-res}) avec celles des $\piinsc$.\\

\begin{figure}
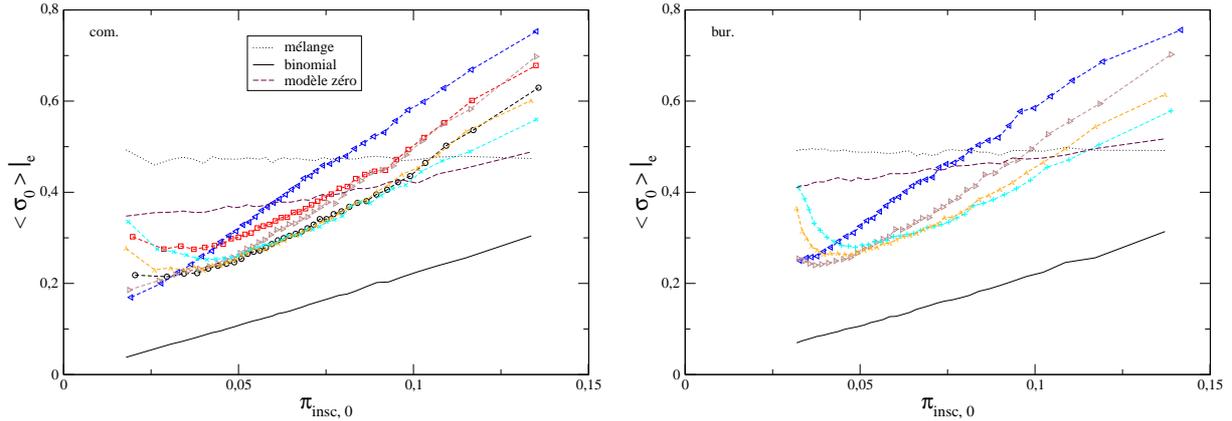

\includegraphics[scale = 0.32]{s0-com-piinsc-res.eps}\hfill
\includegraphics[scale = 0.32]{s0-bvot-piinsc-res.eps}
\caption{\small Moyenne des $\saa$ des résultats électoraux en fonction de $\piinsc$ calculé avec $36$ intervalles, évalués conjointement par commune à gauche, et par bureau de vote à droite. Les tirages aléatoires binomiaux, le \og modèle zéro \fg{} et le mélange aléatoire des résultats s'effectuent à partir des valeurs réelles de l'élection 2007-b.}
\label{fsaa-piinsc-res}
\end{figure}

Les figures~\ref{fsaa-piinsc-res} représentent, pour chaque élection, $\saa$ en fonction de $\piinsc$, calculés tous deux à partir des données par commune, et si possible, par bureau de vote. Comme de coutume, s'y ajoutent les courbes issues des tirages binomiaux, du \og modèle zéro \fg{} et enfin du mélange aléatoire des résultats, dérivés tous trois des données réelles des résultats de l'élection 2007-b. Nous y voyons une nette confirmation de la relation entre $\saa$ et le bruit statistique binomial dû aux tailles finies. L'écart entre les courbes réelles et le \og modèle zéro \fg{} devient patent (comme aux figures~\ref{fsaa-taille-res}), ce qui accrédite de nouveau l'existence des corrélations intercommunales des communes avoisinantes. Notons enfin que la pente des $\saa$ pour l'année $2007$ prend une valeur supérieure à celles provenant du tirage binomial correspondant. Ce dernier constat provient, probablement, de la présence des corrélations intercommunales, mais nous ne savons pas l'expliquer davantage.

Afin de quantifier l'accord entre $\saa$ et le bruit statistique binomial dû aux tailles finies des communes (ou des bureaux de vote), il suffit, d'après ce qui précède, de corréler $\saa$ à $\piinsc$. Les valeurs correspondantes (obtenues à la table~\ref{tsaa-piinsc-res}) peuvent se confronter au cas d'un bruit binomial pur affectant à chaque commune $\aaa$ un résultat $\resa = Binomial(\overline{\res} ; \expr^\aaa)$ (où $\overline{\res}$ provient de la moyenne des résultats réels $\resa$ de l'élection considérée, et calculés à la table~\ref{tstat-dens-res}, comme le montrait la section~\ref{pt-binomial-res}). Cette table indique les valeurs moyennes et les dispersions standards évaluées sur l'ensemble des élections traitées. (Même s'il semble ridicule de s'échiner à calculer des moyennes et des écarts-types sur un ensemble de $6$ valeurs à peine, ces calculs prendront plus de signification ultérieurement.) 

La table~\ref{tsaa-piinsc-res} relative aux données par commune, corrobore les précédentes observations tirées des figures~\ref{fsaa-taille-res} et \ref{fsaa-piinsc-res}, à savoir une connexion forte entre $\saa$ et la taille de la population, et plus particulièrement des corrélations entre $\saa$ et $\piinsc$~:
\begin{itemize}
\item d'assez fortes valeurs (aux alentours de $0,7$)~;
\item relativement permanentes d'une élection à l'autre.
\end{itemize}
Les données associées aux bureaux de vote témoignent, quant à elles, d'une plus grande irrégularité, probablement fort intéressante et significative, mais volontairement délaissée dans cette section.
 
\begin{table}
\begin{tabular}{|c|c|c|c|c|c|c|}
\cline{2-7}
\multicolumn{1}{c|}{}
 & 1992 & 1995 & 2000 & 2002 & 2005 & 2007\\
\hline
com. & 0,722 & 0,632 & 0,783 & 0,719 & 0,591 & 0,664\\
bur. & & & 0,775 & 0,719 & 0,472 & 0,602\\
\hline
\end{tabular}
\caption{\small Corrélations $C_{XY}$ pour chaque élection sur l'ensemble des communes, entre la dispersion des résultats $X^\aaa=\saa$ et $Y^\aaa=\piinsc$. $\saa$ et $\piinsc$ se calculent soit à partir des valeurs par commune (com.), soit à partir des valeurs par bureaux de vote (bur.). Pour les communes et sur l'ensemble des élections : $C_{XY} = 0,685 \pm 0,069$. (Par comparaison, avec un tirage binomial (voir texte) sur les données par commune, $C_{XY} = 0,827 \pm 0,013$ ; et sur les données par bureau de vote, $\overline{C}_{XY} = 0,829 \pm 0,025$.)}
\label{tsaa-piinsc-res}
\end{table}

\subsubsection{\og Séparer le bon grain de l'ivraie \fg}
\label{pt-separer-res}
Faut-il conclure de la présence d'un bruit dans $\saa$ (comme l'attestent les figures~\ref{fsaa-taille-res}, \ref{fsaa-piinsc-res} et Tab.~\ref{tsaa-piinsc-res}) que $\saa$ mesure uniquement un bruit dû aux tailles finies des communes (ou des bureaux de vote), et ne reflète par là même aucune information locale à caractère positif~? Nous appelons information locale à caractère positif, une propriété inhérente à la zone géographique où se calcule $\saa$, qui ressort des considérations d'ordre politique, sociologique, historique, comportementale, etc. Bref, qui ressort de toute spécificité locale différente de celle que lui confère la répartition de la population dans cette zone au travers d'un bruit dû aux tailles finies.

Nous partons du présupposé qu'une information spécifiquement locale, à caractère dit positif, se retrouve identiquement exprimée dans $\saa$ à chaque élection. Mais qu'importe le présupposé, nous cherchons à proprement parler, non pas la présence d'une information locale à caractère positif telle qu'elle puisse se manifester dans $\saa$ à une élection donnée, mais plutôt une information locale $\ti$dite, faute de mieux, à caractère positif$\ti$ que manifeste la régularité temporelle de $\saa$ à diverses élections. En d'autres termes, nous cherchons à déceler la présence d'une propriété locale autour d'une commune centrale $\aaa$, qui se répercute par les régularités de $\saa$ à différentes élections.

Bien évidemment, nous ne nous interrogeons nullement sur la nature de cette information locale, mais nous nous contenterons uniquement d'essayer de la mesurer, y compris \textit{a minima}. Nous tenterons du mieux possible d'évaluer dans quelle proportion $\ti$minimale$\ti$ cette information locale contribue à la permanence temporelle des $\saa$ (voir Tab.~\ref{ttempo-saa-res}). Nous essaierons ensuite de comprendre pourquoi, de par son expression mathématique, $\saa$ contient une information locale. Enfin, à titre indicatif, nous illustrerons par une carte les zones à plus ou moins forte régularité temporelle des $\saa$. 

Ces développements se trouvent à l'annexe~\ref{annexe-extraire}. Mais auparavant, expliquons mieux le problème et notre façon d'opérer.\\

De par sa définition même (cf. Eq.~(\ref{esaa})), $\saa$ mesure directement la dispersion des résultats électoraux à l'intérieur de sa zone géographique $\vaa$ $\ti$les $n_p=16$ plus proches communes de la commune centrale $\aaa$. Les faibles valeurs de $\saa$ indiquent des zones à forte homogénéité des résultats électoraux, tandis que les fortes valeurs de $\saa$ pointent les zones de forte fracture, ou d'hétérogénéité, des résultats électoraux. Mais nous savons également que $\saa$ dépend fortement de la taille des communes. L'existence d'un bruit statistique dû aux tailles finies des communes implique à lui seul, des fortes valeurs de $\saa$ dans les zones à faible population, et inversement dans les zones fortement peuplées. Cela se traduit par une permanence temporelle des $\saa$, mesurée par la corrélation des $\saa$ à différentes élections (cf. Eq.~(\ref{ecorrel-temporelle}) et section~\ref{pt-tempo-res}), telle que $\overline{C}_t(\sigma_0)\simeq 0,69$ avec un tirage binomial des résultats, et $\overline{C}_t(\sigma_0)\simeq 0,14$ avec des résultats issus du \og modèle zéro \fg. (Pour mémoire, $\overline{C}_t(\sigma_0)\simeq 0,58$ (voir Tab.~\ref{ttempo-saa-res}) pour les résultats réels des six élections étudiées.) Bref, tout ceci appelle à se poser la question suivante~: comment parvenir à identifier d'une valeur de $\saa$ ou, mieux, d'une série de valeurs de $\saa$, une propriété à caractère positif et non entachée de bruit statistique dû aux tailles finies des communes (ou des bureaux de vote)~?

Prenons un exemple pour mieux comprendre. Supposons que les données électorales s'accordent avec le \og modèle zéro \fg $\ti$traité à la section~\ref{pt-mod0-res} et donné par l'équation~(\ref{emod0}). Connaissant $\piinsc$ et en utilisant l'équation~(\ref{esaa-mod0}), nous pourrions en déduire pour chaque élection l'amplitude $\sigma_\eta^\aaa$ du bruit à l'intérieur de la zone $\vaa$. $\saa$ contient du bruit dû aux tailles finies des communes, à la différence de $\sigma_\eta^\aaa$ qui n'en contient pas. $\sigma_\eta^\aaa$ caractérise donc une information ou grandeur locale lors de l'élection traitée. Une information locale à une élection donnée que nous pourrions ensuite exploiter. Il y a maintenant un pas à franchir entre, d'une part, une information locale, à une élection donnée, non entachée de bruit statistique dû aux tailles finies des communes, et d'autre part, une information locale pérenne. Dans le cas du \og modèle zéro \fg, $\sigma_\eta^\aaa$ provient des fluctuations d'un bruit, et ne possède donc pas le caractère d'information locale pérenne que nous recherchons. Autrement dit, $\sigma_\eta^\aaa$ varie aléatoirement d'une élection à une autre, et ne reflète aucune particularité locale à caractère dit positif. Concrètement, nous vérifierions le caractère non permanent de $\sigma_\eta^\aaa$ au travers de ses corrélations temporelles, $\overline{C}_t(\sigma_\eta)$, sur tous les couples d'élections.

L'exemple du \og modèle zéro\fg{} ayant permis d'éclaircir nos propos, passons aux résultats électoraux réels qui, comme nous le savons, ne s'accordent pas à ce modèle. Les résultats électoraux à l'intérieur de la zone géographique $\vaa$ des $n_p$ communes ne constituent pas en effet des variables indépendantes les unes des autres (cf. notamment la moyenne des $\taa$ en fonction $\toa$ de Fig~\ref{fenvironnement-res}-a et les corrélations entre proches communes $C(R\simeq 1)$ de Fig.~\ref{fcorrel-res}), ni à l'intérieur des villes pour leurs bureaux de vote (cf. l'annexe~\ref{annexe-communautes}). L'état de la corrélation entre les communes à l'intérieur de la zone pourrait fournir une $\ti$parmi tant d'autres$\ti$ information locale associée à cette zone géographique. Mais il subsiste un problème majeur~: nous ne savons pas remonter de $\saa$ à ces corrélations en l'absence de modèle de vote. En d'autres termes, en l'absence de modèle de vote à l'échelle des communes, voire infra-communale, nous ne savons pas extirper de $\saa$, le bruit dû aux tailles finies. (A l'annexe~\ref{annexe-extraire}, nous tenterons néanmoins d'estimer indirectement les corrélations intercommunales, par comparaison des $\saa$ réels avec des $\saa$ virtuellement produits lors d'une absence de corrélation intercommunale.)\\

Nous allons tenter de contourner le problème et l'attaquer par un autre biais, en se focalisant sur son aspect temporel. Nous partons pour cela du présupposé suivant~: une fois détachée du bruit résiduel, l'information locale se manifeste par une permanence temporelle. Nous voulons donc nous intéresser à la partie pérenne que témoigne à toutes les élections $\saa$ ôté du bruit statistique des tailles finies, et qui exprime ce que nous appelons, peut-être à tort, information locale à caractère positif. Par exemple, si $\saa$ d'une commune $\aaa$ donnée présente pour chaque élection une valeur bien supérieure (ou inférieure) comparée aux autres valeurs des zones à population similaire, alors, nous en déduisons que la zone géographique sur laquelle s'évalue $\saa$ recèle une information locale et pérenne. Cet exemple nous montre comment une information partielle, et pérenne, peut se déceler au travers des $\saa$~: 
\begin{itemize}
\item en se référant à d'autres zones comparables $\ti$en terme de bruit dû aux tailles finies$\ti$~;
\item en tenant compte d'une série d'élections.
\end{itemize}

Une façon de filtrer le bruit statistique dû aux tailles finies consiste alors à comparer $\ti$ou plus précisément à faire des corrélations$\ti$, à différentes élections, des $\saa$ ayant entre eux la même intensité de ce bruit statistique. Ce bruit statistique, d'une égale valeur $\ti$aux fluctuations près$\ti$ pour le groupe de communes centrales considérées, n'aura alors plus d'effet sur les corrélations temporelles de leur $\saa$ à différentes élections. Nous pouvons alors accéder par ce biais à une information locale, à caractères positif et pérenne.\\

L'annexe~\ref{annexe-extraire} essaie de détecter, puis de mesurer, la présence de l'information locale à caractère positif que contient $\saa$, et ce de différentes façons. Elle développe pour cela, l'argumentation qui précède. Il ressort que la partie pérenne de $\saa$ filtré du bruit statistique des tailles finies, contribue pour un minimum de $15\%$ à la permanence temporelle globale des $\saa$ (mesurée par $\overline{C}_t(\sigma_0)$). Ensuite, elle tente de mesurer l'état des corrélations des résultats $\tau$ des communes (ou des bureaux de vote) à une hypothétique grandeur locale, puisque, selon la discussion précédente, l'information locale positive peut aussi se comprendre par l'état des corrélations intercommunales. L'annexe indique aussi que, globalement, ladite information locale positive se manifeste davantage dans les zones à forte population. Enfin, et à titre anecdotique, l'annexe représente la position des communes centrales $\aaa$ ayant une plus ou moins grande variation relative de leurs $\saa$ à différentes élections.

\subsubsection{Critique inspirée de \og la possibilité de la mesure crée le phénomène \fg}
\label{pt-critique-res}
Avant de conclure ce chapitre, nous nous servirons de cette belle assertion de Lionel Tabourier~\cite{lionel} $\ti$sans nous questionner sur ce qui, à son tour, initie la possibilité de la mesure$\ti$ pour critiquer les régularités observées liées à $\saa$, dont l'origine se situe essentiellement dans leur imbrication avec des valeurs de population, $\piinsc$ en l'occurrence.

Premièrement, une lapalissade où il n'y a pas de quoi véritablement s'exciter. Nous avons pu croiser les valeurs $\saa$ à $\piinsc$ parce que nous disposions des données permettant de le faire. Nous n'affirmons pas en revanche que $\piinsc$ contient la plus grande affinité avec $\saa$. Dit autrement, nous n'affirmons pas, par exemple, qu'il ne puisse exister de meilleure corrélation entre $\saa$ et une autre grandeur, que celle entre $\saa$ et $\piinsc$. Nous n'affirmons pas non plus que la corrélation entre $\saa$ et $\piinsc$ fournisse le plus d'information à caractère politique, sociologique, etc. $\ti$dit en un mot, positif$\ti$, sur la zone géographique où s'évalue $\saa$. Néanmoins, les tables~\ref{tsarko} et \ref{tsaa-piinsc-res} montrent une meilleure connivence de $\saa$ avec $\piinsc$, qu'avec le taux d'augmentation de population. Nous aurions pu aussi croiser $\saa$ à la densité de population, mais cette dernière grandeur capture moins clairement le bruit statistique dû aux tailles finies, que celle provenant directement des populations. De futures études cherchant à savoir comment $\saa$ se corrèle à des grandeurs directement issues des champs économiques, sociaux, etc., enrichiraient très certainement l'analyse des résultats électoraux $\ti$surtout dans les zones à forte population, dans lesquelles les propriétés dites positives se manifestent davantage dans $\saa$ (voir Figs.~\ref{ftempo-piinsc-res} et \ref{fraa-res}). L'annexe~\ref{annexe-croiss-pop} montre une façon de déceler une information dite positive qui, de plus, se lie en partie au bruit statistique des tailles finies. (Elle cherche à déceler une relation directe entre le taux de croissance de la population et $\saa$, et ce sans la connexion indirecte avec le bruit dû aux tailles finies exprimé par $\piinsc$.)

Mais revenons à la possibilité de mesurer $\saa$ et $\piinsc$ et plus particulièrement au rôle des appareils de mesure, ou mieux, des échelles de mesure. Les figures~\ref{fsaa-taille-res} et \ref{fsaa-piinsc-res} ainsi que la table~\ref{tsaa-piinsc-res} permettent de confronter l'effet occasionné par deux différents \og appareils de mesure \fg{} des choix des électeurs et de la taille des populations~: les communes et les bureaux de vote. (La différence entre ces deux appareils de mesure n'intervient bien évidemment que pour les plus grandes communes, i.e. celles qui possèdent plus d'un bureau de vote.) L'effet majeur introduit par ces deux échelles de mesure différentes, consiste à amplifier l'écart par rapport au comportement bruité \og normal \fg. Dit autrement, l'échelle du bureau de vote, par rapport à celle de la commune, permet de voir plus finement si les $\saa$ d'une élection se démarquent du bruit statistique dû aux tailles finies $\ti$et qui témoignent de la sorte d'une plus grande richesse, en terme de signification politique, sociologique, etc.

Enfin, plus intéressant encore, la possibilité de comparer entre elles des élections de natures différentes et portant sur des questions différentes (voir Tab.~\ref{tres}) nous incite à réfléchir un peu plus sur la signification à accorder aux grandeurs électorales. Les résultats $\taa$ ou $\toa$ auraient pu de façon bien surprenante exhiber de fortes et régulières valeurs de corrélations temporelles entre de si disparates élections (cf. Tabs.~\ref{ttempo-res} et \ref{ttempo-toa-res}). Il eût fallu pour cela, éventuellement, ne considérer que des élections portant sur des choix similaires, ou ne traiter que des résultats d'un même parti politique aux différentes élections. (Notons néanmoins que la table~\ref{ttempo-toa-res}, relative aux corrélations temporelles des $\toa$, contient des informations à caractère politique. Elle fournit en effet pour chaque couple d'élections, outre le signe d'une corrélation ou d'anticorrélation (avec la convention d'attribuer au résultat électoral, la valeur du choix vainqueur à l'échelle nationale), mais aussi des valeurs absolues généralement, soit grandes (supérieures à environ $1/2$), soit faibles (inférieures à environ $1/10$, voire proches de zéro).) Comme nous pouvions nous y attendre, la régularité temporelle des $\saa$ (cf. Tab.~\ref{ttempo-saa-res}) à des élections différentes en nature et en choix provient alors non pas de leurs récurrentes propriétés politiques, mais, au contraire, de leur absence de signification politique, bref de leur bruit statistique attaché aux tailles finies de population. Il faut cependant nuancer cette affirmation, puisque les régularités temporelles de $\saa$ procèdent aussi de la spécificité à caractère positif (i.e. ne traduisant pas le bruit statistique dû aux tailles finies) inhérente à leur zone géographique (cf. annexe~\ref{annexe-extraire}).

\subsubsection{Résumé}
\label{pt-resume-res}
En résumé, cette partie a mis en évidence les principaux points et régularités suivants.
\begin{itemize}
\item Il convient d'utiliser une variable non bornée pour mesurer les résultats électoraux binaires, comme par exemple $\tau$ (voir Eq.~(\ref{etau})), afin d'étudier leurs régularités (voir Tab.~\ref{tstat-log-res} et Fig.~\ref{fhisto-log-res} à comparer avec Tab.~\ref{tstat-dens-res} et Fig.~\ref{fhisto-dens-res}).
\item Les résultats électoraux par commune présentent une corrélation spatiale, d'une part relativement forte à courte distance, et d'autre part à longue portée avec une décroissance quasi-logarithmique (voir Fig.~\ref{fcorrel-res}).
\item Une analyse plus fine de la connexion entre le résultat d'une commune centrale $\aaa$ et les communes de son voisinage révèle deux régularités : la commune centrale vote en moyenne comme son environnement ($\taa \simeq \toa$), et l'écart entre ces deux valeurs, $|\taa - \toa|$, augmente régulièrement avec la dispersion $\saa$ des résultats des communes de son environnement (voir Fig.~\ref{fenvironnement-res} et Eqs.~(\ref{etoa},~\ref{esaa})).
\item $\saa$ possède une forte régularité temporelle, à la différence d'autres grandeurs liées aux résultats électoraux comme $\taa$ et $\toa$ (voir Tabs.~\ref{ttempo-saa-res}, \ref{ttempo-res} et \ref{ttempo-toa-res}).
\item $\saa$ atteste une forte connexion avec la taille de la population (voir Fig.\ref{fsaa-taille-res}), ce qui explique, en partie, les diverses régularités associées à $\saa$.
\item $\saa$ traduit non seulement la présence d'un bruit dû aux tailles finies des communes ou des bureaux de vote (voir Fig.~\ref{fsaa-piinsc-res} et Tab.~\ref{tsaa-piinsc-res}), mais aussi une information positive inhérente à sa zone géographique, vraisemblablement à caractère politique ou sociologique, et dite positive (voir annexe~\ref{annexe-extraire}).
\end{itemize}
\vspace{0.5cm}

Signalons aussi que l'annexe~\ref{annexe-communautes} étudie l'indépendance des résultats électoraux des bureaux de vote au sein d'une même ville, ou dans les zones urbaines, et montre $\ti$en accord avec ce qui précède$\ti$ leur forte corrélation.\\

Enrichissons cette étude en se confrontant à une autre forme de choix binaire du domaine électoral, et virtuellement plus simple que le précédent~: s'abstenir ou non à une élection. Nous chercherons notamment à vérifier si les points ci-dessus restent encore valables pour les taux de participation aux élections.

\clearpage
\subsection{Le taux de participation des élections nationales}
\label{section-abst}
\hfill
\begin{minipage}[r]{0.8\linewidth}
\textit{Où l'on voit que, pressé par le temps, le lecteur avisé pouvait survoler cette partie.}
\end{minipage}

\subsubsection{Vue d'ensemble}
\label{pt-vue-abst}
Poursuivons ce travail par l'étude d'une autre attitude binaire associée aux élections : ou bien l'électeur inscrit sur les listes électorales se déplace à son bureau de vote et opine en faveur de l'un des choix proposés $\ti$voire appose un bulletin blanc ou nul$\ti$, ou bien il ne participe pas à l'élection.

Les raisons de simplicité entrevues précédemment dictent les mêmes restrictions~: premièrement, ne considérer que les élections portant sur le même choix à l'échelle nationale, et deuxièmement, se limiter aux données de la France métropolitaine. Ainsi, les élections analysées pour leurs taux de participation (voir Tab.~\ref{tabst}) concernent aussi bien celles à choix multiple\footnote{Depuis 2004, les élections européennes présentent sept listes différentes pour la France métropolitaine. Les différences entre ces sept listes régionales portent uniquement sur la présence des plus petits partis et le nom des tête de listes des autres partis. Sans trop commettre d'erreur, nous assimilons donc leur taux de participation à ceux issus d'élections portant sur le même choix à l'échelle nationale.} que celles à choix binaire $\ti$précédemment analysées quant à leurs résultats.

Outre, d'un côté, l'engagement physique de l'individu mesuré par le taux de participation, et de l'autre côté, la mesure d'une opinion potentiellement labile~\cite{laurens_opinion}, ces deux mesures possèdent une différence essentielle liée à un point déjà soulevé à la section~\ref{pt-critique-res}. Les taux de participation se réfèrent à la même question (participer ou non à l'élection donnée) pour les $12$ élections traitées, tandis que les résultats des $6$ élections binaires analysées précédemment se calculent à partir de questions différentes (cf. les choix proposés aux électeurs de la table.~\ref{tres}).

\begin{table}[hd]
\begin{tabular}{|c|c|c|c|c|c}
\hline
Année & choix & type d'élection & inscrits & moyenne\\
\hline
1992 & binaire & référendum, traité de Maastricht & 36,6 $10^6$ & 0,713\\
1994 & multiple & européennes & 37,6 $10^6$ & 0,539\\
1995 & multiple & présidentielle, premier tour & 38,4 $10^6$ & 0,795\\
1995 & binaire & présidentielle, second tour & 38,4 $10^6$ & 0,805\\
1999 & multiple & européennes & 38,4 $10^6$ & 0,478\\
2000 & binaire & référendum, quinquennat présidentiel & 38,2 $10^6$ & 0,308\\
2002 & multiple & présidentielle, premier tour & 39,2 $10^6$ & 0,729\\
2002 & binaire & présidentielle, second tour & 39,2 $10^6$ & 0,810\\
2004 & multiple & européennes & 39,9 $10^6$ & 0,434\\
2005 & binaire & référendum, traité constitutionnel européen& 39,7 $10^6$ & 0,711\\
2007 & multiple & présidentielle, premier tour & 41,9 $10^6$ & 0,854\\
2007 & binaire & présidentielle, second tour & 41,9 $10^6$ & 0,853\\
\hline
\end{tabular}
\caption{\small Élections nationales analysées pour leurs taux de participation. Le nombre d'inscrits et la moyenne du taux de participation se réfèrent à la France métropolitaine.}
\label{tabst}
\end{table}

En résumé, l'étude des taux de participation semble plus facile que celle des résultats électoraux pour les raisons suivantes~:
\begin{itemize}
\item un plus grand nombre de données ($12$ élections connues pour leurs taux de participation, comparé à $6$, pour leurs résultats électoraux)~;
\item les choix, de s'abstenir ou non lors des différentes élections, paraissent plus homogènes entre eux que les différents choix associés aux expressions des votes~;
\item un choix à prendre par l'électeur, semble-t-il, plus simple $\ti$car un choix qui engage moins l'électeur, ou qui suscite moins de clivage, de positionnement, d'attente, etc., idéologique$\ti$ pour la participation à une élection, comparé à celui de l'expression du vote.
\end{itemize}
Bien que l'étude des régularités paraisse plus facile pour les taux de participation par rapport aux résultats électoraux, la précédente étude a eu néanmoins le grand mérite de détacher assez nettement certaines spécificités électorales $\ti$comme l'importance de l'hétérogénéité des résultat $\saa$ sur une zone géographique.

Bref, il importe maintenant de se pencher sur cette autre source de données électorales $\ti$les taux de participation aux élections$\ti$ et de savoir ce qu'elle apporte de nouveau. Et notamment, infirmera-t-elle, ou confirmera-t-elle les précédentes régularités dégagées de l'étude des résultats électoraux~?\\

Le taux de participation de la commune (ou du bureau de vote) $\aaa$ s'écrit comme $\resa = \frac{\vot^\aaa}{\inscaa}$, et en grandeur $\tau$ (voir Eq.~(\ref{etau})) comme $\taa = \ln(\frac{\vot^\aaa}{\inscaa - \vot^\aaa})$, où $\inscaa$ et $\vot^\aaa$ désignent respectivement le nombre d'électeurs inscrits sur la liste électorale de la commune (ou du bureau de vote) $\aaa$, et le nombre d'électeurs de $\aaa$ qui ont voté à l'élection considérée. Comme précédemment, nous convenons de modifier $\vot^\aaa$ d'une demi-voix dans les deux cas extrêmes ($\vot^\aaa = \inscaa \rightarrow \vot^\aaa=\inscaa-0,5$ et $\vot^\aaa = 0 \rightarrow \vot^\aaa = 0,5$) pour éviter que $\taa$ devienne infini.

\begin{figure}
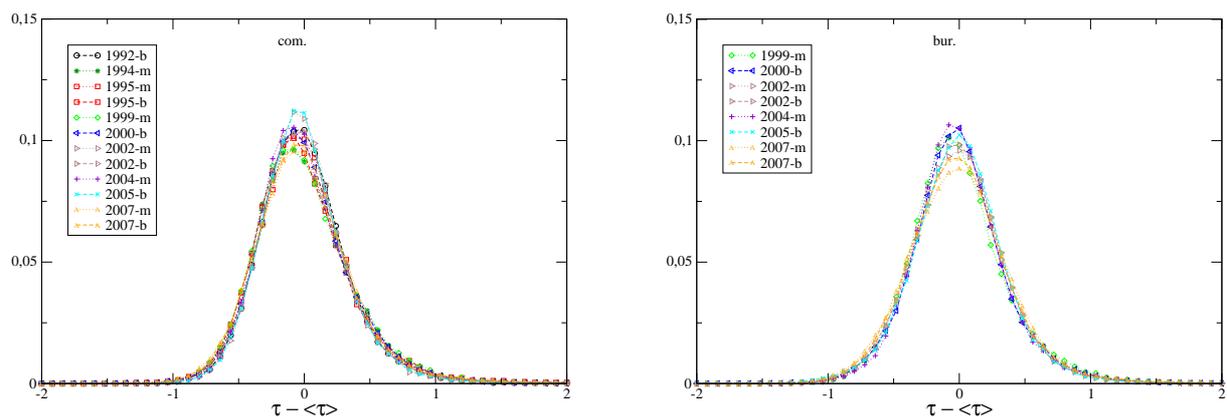

\includegraphics[scale = 0.32]{histo-log-abst.eps}\hfill
\includegraphics[scale = 0.32]{histo-log-bvot-abst.eps}
\caption{\small Histogrammes des taux de participation $\tau$ ramenés à une moyenne nulle, des communes à gauche et des bureaux de vote à droite. La lettre $b$ (ou $m$) ajoutée à l'année précise la nature binaire (ou multiple) de l'élection. Ces légendes s'appliqueront aux élections équivalentes des figures qui suivront.}
\label{fhisto-log-abst}
\end{figure}

Les figures~\ref{fhisto-log-abst} représentent les histogrammes, centrés sur leur moyenne, de l'ensemble des taux de participation $\taa$ par commune (ou par bureau de vote). La table~\ref{tstat-log-abst} fournit des statistiques élémentaires de la distribution des taux de participation par commune $\taa$, et si possible, par bureau de vote. Les histogrammes et statistiques en grandeur $\res$ se trouvent en appendice (cf. Fig.~\ref{fhisto-dens-abst} et Tab.~\ref{tstat-dens-abst}). De nouveau, les histogrammes et statistiques associés aux taux de participation confirment l'avantage d'utiliser une grandeur non bornée, comme $\tau$, si l'on se préoccupe de régularités. 

Mais plus important encore, la distribution des taux de participation $\taa$ par commune (ou par bureau de vote) semble particulièrement permanente, similaire, d'une élection à une autre. Ceci se traduit quantitativement par des écarts-types de $\tau$ par commune (ou bien par bureau de vote), sensiblement égaux pour chacune des $12$ élections, et ce indépendamment des taux de participation moyen $\ti$qui s'étend de $31\%$ à $86\%$ (voir Tabs.~\ref{tabst}, \ref{tstat-log-abst}). Noter que les écarts-types en grandeur $\tau$ des résultats ($\simeq 0,49$) dépassent ceux des taux de participation ($\simeq 0,38$), ce qui tend à accréditer le type de classement retenu ici des choix binaires ~: entre d'un côté, celui d'un vote entre deux choix, et, d'un autre côté, participer ou non à l'élection $\ti$et dans les deux cas, sans tenir compte ni de l'enjeu ni de la nature des élections.

Notons enfin une dissymétrie (skewness) en grandeur $\tau$ toujours positive, sur laquelle nous reviendrons ultérieurement.\\

\begin{table}[h]
\begin{tabular}{|c||c|c|c|c|c||c|c|c|c|c|}
\hline
Élection & \textbf{com.} & moy & éc-typ & skew & kurt & \textbf{bur.} & moy & éc-typ & skew & kurt\\
\hline
1992-b & 36186 & 1,13 & 0,355 & 1,05 & 5,48 & & & & &\\
1994-m & 36196 & 0,358 & 0,398 & 0,837 & 9,23 & & & & &\\
1995-m & 36195 & 1,60 & 0,375 & 0,928 & 5,37 & & & & &\\
1995-b & 36197 & 1,72 & 0,398 & 1,35 & 5,54 & & & & &\\
1999-m & 36209 & 0,146 & 0,392 & 1,15 & 7,50 & 62005 & 0,031 & 0,386 & 1,01 & 6,10\\
2000-b & 36212 & -0,626 & 0,377 & 0,858 & 8,27 & 62267 & -0,729 & 0,374 & 0,681 & 6,63\\
2002-m & 36217 & 1,24 & 0,347 & 1,25 & 9,46 & 62356 & 1,11 & 0,369 & 0,856 & 6,73\\
2002-b & 36217 & 1,67 & 0,367 & 1,26 & 6,40 & 62358 & 1,56 & 0,379 & 0,861 & 5,25\\
2004-m & 36221 & -0,095 & 0,366 & 1,45 & 9,82 & 62775 & -0,182 & 0,363 & 1,03 & 8,37\\
2005-b & 36223 & 1,13 & 0,351 & 1,58 & 12,0 & 62778 & 1,01 & 0,373 & 0,805 & 7,16\\
2007-m & 36219 & 1,98 & 0,396 & 1,06 & 8,02 & 63514 & 1,88 & 0,405 & 0,718 & 5,37\\
2007-b & 36219 & 1,99 & 0,394 & 1,22 & 5,28 & 63516 & 1,88 & 0,397 & 0,883 & 4,19\\
\hline
\end{tabular}
\caption{\small Moyennes (moy), écarts-types (éc-typ), skewness (skew) et kurtosis (kurt) des taux de participation $\tau$ sur l'ensemble des communes (com.) et si possible sur l'ensemble des bureaux de vote (bur.).}
\label{tstat-log-abst}
\end{table}

\begin{figure}
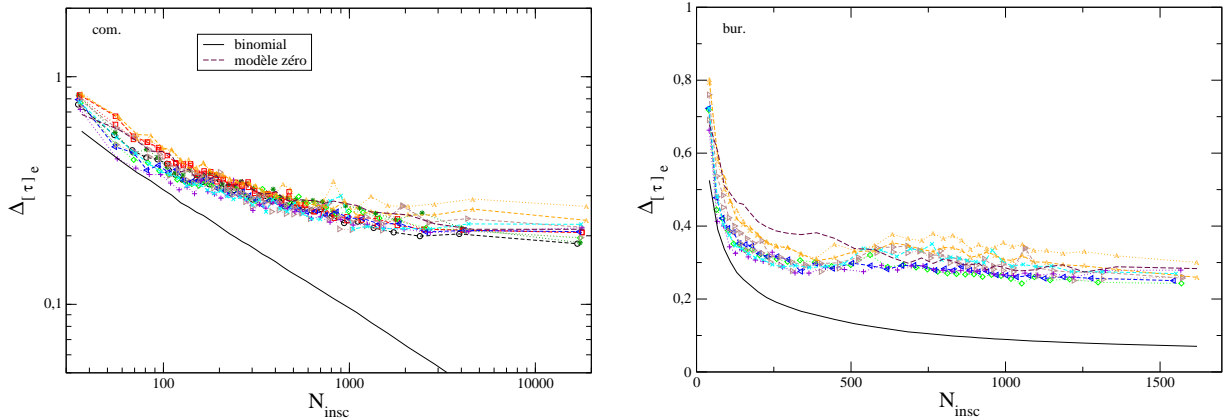

\includegraphics[scale=0.32]{sigma-taille-log-abst.eps}\hfill
\includegraphics[scale=0.32]{sigma-taille-log-bvot-abst.eps}
\caption{\small Écart-type des taux de participation $\tau$ en fonction du nombre d'inscrits $\insc$, calculé avec $36$ intervalles ; pour les communes à gauche et pour les bureaux de votes à droite. Le \og modèle zéro \fg{} (voir section~\ref{pt-mod0-res}) et les tirages binomiaux uniformes (voir section~\ref{pt-binomial-res}) s'effectuent à partir des valeurs réelles de l'élection 2007-b. La valeur moyenne $\insc$ de l'échantillon des communes de plus grande taille doit se considérer avec précaution puisqu'interviennent des valeurs extrêmes.}
\label{fsigma-taille-abst}
\end{figure}

Les figures~\ref{fsigma-taille-abst} et \ref{fmoy-taille-abst} tracent respectivement l'écart-type et la moyenne de $\taa$ à l'intérieur de $36$ intervalles en $\inscaa$ et peuvent se comparer à leurs homologues (Figs.~\ref{fsigma-taille-res} et \ref{fmoy-taille-res}) associés aux résultats des élections. Les conclusions de la section~\ref{pt-binomial-res} s'appliquent également aux taux de participation, soit plus précisément~: d'une part les taux de participations diffèrent d'un processus aléatoire, indépendant et identique à l'échelle nationale (i.e. un tirage binomial à probabilité $\overline{\res}$ uniforme), et d'autre part un assez bon accord, à ce stade, du \og modèle zéro \fg{} aux données réelles. Noter au passage le meilleur accord du \og modèle zéro \fg{} avec les taux de participation réels, comparés aux résultats électoraux réels (cf. Figs.~\ref{fsigma-taille-res} et \ref{fsigma-taille-abst}).

Par contre, la nouveauté $\ti$en terme de régularité$\ti$ apportée par l'étude des taux de participation concerne leur décroissance en fonction de la taille de la population. Les inscrits participent plus, en moyenne, et de façon récurrente, aux élections dans les communes à faible population, que dans celles à forte population (cf. Fig~\ref{fmoy-taille-abst}).\\

Comment comprendre ce nouveau point~? La première façon consiste à invoquer des raisons d'ordre politique, sociologique, ou comportementale, pour expliquer pourquoi, les électeurs participent en moyenne plus aux élections dans les plus petites communes que dans les plus grandes villes $\ti$pour un aperçu d'une investigation de nature sociologico-politique du taux d'abstention en France, voir~\cite{muxel_mobilisation_electorale}. (Nous pouvons évoquer dans les communes faiblement peuplées, une plus grande capacité de la mairie à mobiliser l'électorat, une plus grande proportion de personnes âgées $\ti$qui votent plus en moyenne$\ti$, une élection vue comme une incitation à se rencontrer, etc.~\footnote{Je remercie une nouvelle fois Brigitte Hazart, pour m'avoir indiqué toutes ces raisons, ainsi que la référence~\cite{muxel_mobilisation_electorale}.}) La deuxième façon nous pousse à suspecter les données traitées. Le nombre d'électeurs inscrits sur les listes électorales, ne capture pas parfaitement le nombre d'électeurs qui vivent dans la commune. Pourquoi ne pas émettre l'hypothèse que ce biais dépend de la taille de la commune~? Usant de l'\textit{épochê} chère aux Sceptiques grecs, nous accepterons telle quelle la décroissance du taux de participation avec l'augmentation de la taille de la population des communes. De plus, nous acceptons tout au long de ce travail que le nombre d'inscrits sur les listes électorales reflète fidèlement la population de la commune (ou du bureau de vote) $\ti$comme nous l'avions fait au chapitre précédent, sans l'avoir précisé.

\subsubsection{Heurs et malheurs du \og modèle zéro \fg}
\label{pt-mod0-abst}
\begin{figure}
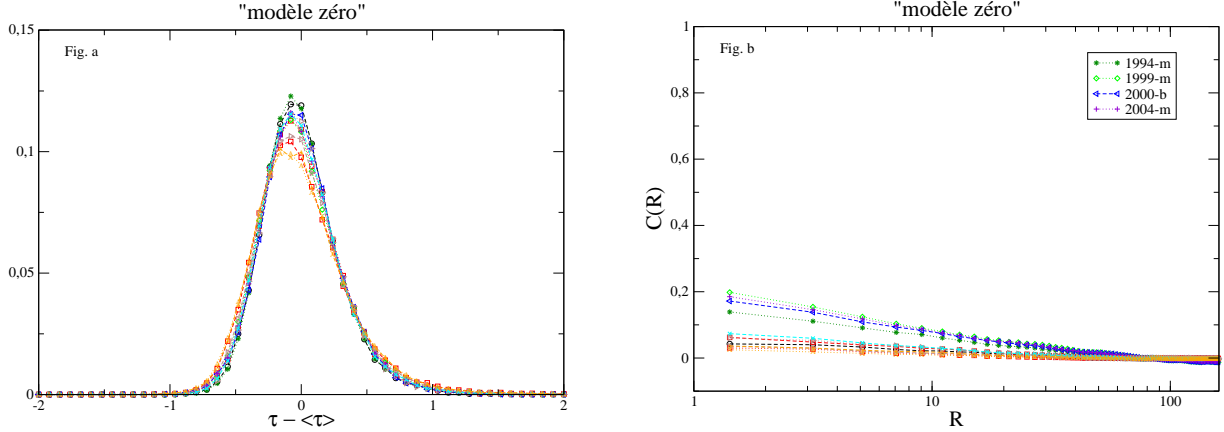

\includegraphics[scale = 0.32]{histo-log-mod0.eps}\hfill
\includegraphics[scale = 0.32]{correl-log-mod0.eps}
\caption{\small \og Modèle zéro \fg{} (cf. Eq.~(\ref{emod0})) des taux de participation par commune. A gauche, histogrammes des $\tau$, et à droite, corrélation spatiale des $\tau$.}
\label{fmod0}
\end{figure}

Revenons au \og modèle zéro \fg{} (cf. Eq.(~\ref{emod0}) et section~\ref{pt-mod0-res}) qui, de par sa simplicité, pourra mettre en lumière certaines caractéristiques des données réelles.

Fait encore notable~? l'amplitude du bruit $\sigma_\eta=\sqrt{< \eta^2 >}$ (qui provient de l'échantillon en $\insc$ minimisant l'écart-type en grandeur $\res$ des taux de participation) vaut $\sigma_\eta \simeq \overline{\res}\x(1-\overline{\res})\x A$, avec $A = (\Delta_{[\tau]_{\,e}})_{_{min}} \simeq 0,2$ pour les $12$ élections analysées. (Cette valeur de $A$, qui minimise l'écart-type des $\tau$ à l'intérieur d'un échantillon $e$ en $\insc$, se retrouve directement à la figure~\ref{fsigma-taille-abst}, et dans les échantillons à forte population.)

Les histogrammes de la figure~\ref{fmod0}-a représentent la distribution sur l'ensemble des communes des $\taa$ obtenus à partir de ce modèle simpliste $\ti$dont les paramètres s'ajustent aux taux de participation réels de chacune des $12$ élections analysées. Nous y distinguons une apparente similarité de ces $12$ distributions entre elles, ainsi qu'une permanente dissymétrie (skewness). Vu le caractère rudimentaire du modèle, la dissymétrie des distributions des $\tau$ obtenus par ce modèle ne peut résulter que de la décroissance moyenne du taux de participation en fonction de la population de la commune, et ce pour chacune des élections (voir Fig.~\ref{fmoy-taille-abst}). (De façon légèrement plus quantitative, les écarts-types et les skewness dérivés du \og modèle zéro \fg{} se trouvent respectivement environ $(13 \pm 6)\%$ et $(23 \pm 15)\%$ inférieurs à ceux provenant des valeurs réelles.) Ainsi, la diminution permanente du taux de participation en fonction de la population de la commune implique, en partie du moins, la skewness permanente constatée sur les valeurs réelles (voir Fig.~\ref{fhisto-log-abst} et Tab.~\ref{tstat-log-abst}). 

Notons enfin des valeurs non négligeables de la corrélation spatiale $C(R)$ des $\taa$ obtenus à partir du \og modèle zéro \fg{} (voir Fig.~\ref{fmod0}-b) pour les taux de participation des $4$ élections, 1994-m, 1999-m, 2000-b et 2004-m. Noter que ces quatre élections ont la particularité d'avoir la plus forte variation des $\tau$ en fonction de la taille de la population des communes (voir Fig.~\ref{fmoy-taille-abst}). Obtenir de la sorte des corrélations spatiales convenables au regard des corrélations réelles $\ti$sensiblement de même portée, de forme quasi-logarithmique et de valeurs acceptables entre proches voisins$\ti$ nous semble particulièrement remarquable, et appelle à commentaires.

Les corrélations spatiales, qui témoignent usuellement de l'existence d'interactions locales à l'échelle microscopique (des agents, ici), peuvent aussi se concevoir différemment. Les corrélations spatiales peuvent aussi résulter d'un même comportement moyen des agents, en fonction des caractéristiques mesurables des communes dans lesquelles ils résident. Le \og modèle zéro \fg{} montre en effet par son aspect rudimentaire, qu'une simple dépendance du taux de participation en fonction de la population de la commune peut entraîner des corrélations spatiales assez acceptables.

En résumé, le caractère simpliste du \og modèle zéro \fg{} permet de mettre en évidence les deux points suivants~:
\begin{itemize}
\item la dissymétrie permanente de la distribution des taux de participation $\tau$ sur l'ensemble des communes semble dériver de la variation du taux de participation en fonction de la population de la commune~;
\item la corrélation spatiale, qui traduit généralement la présence d'interactions locales à l'échelle microscopique (ici, à l'échelle des agents), peut aussi se comprendre comme un comportement commun, en moyenne, des agents qui résident dans des communes de mêmes caractéristiques (en l'occurrence ici, la taille en population de la commune).
\end{itemize}

\subsubsection{Corrélations spatiales}
\label{pt-cor-spatial-abst}
\begin{figure}
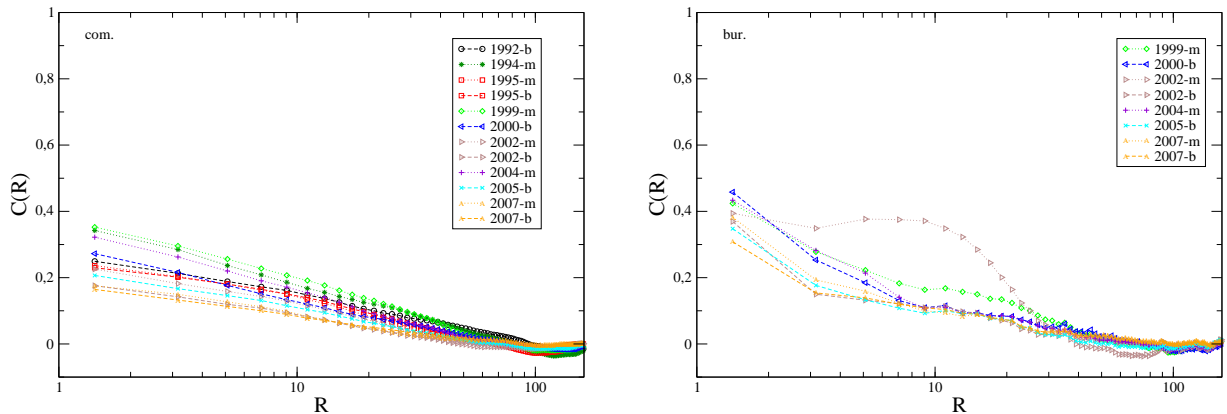

\includegraphics[scale = 0.32]{correl-log-abst.eps}\hfill
\includegraphics[scale = 0.32]{correl-log-bvot-abst.eps}
\caption{\small Corrélation spatiale des taux de participation $\tau$ des communes à gauche et des bureaux de vote à droite, où $R_{\aaa\beta}=\frac{r_{\aaa\beta}}{D} \in ]2n; 2(n+1)]$. La distance départementale caractéristique $D$ (environ $2,7~km$) pour les communes s'applique également à celle des bureaux de vote.}
\label{fcorrel-abst}
\end{figure}

Les figures~\ref{fcorrel-abst} tracent les corrélations spatiales, $C(R)$, des taux de participation par commune et par bureau de vote. Les corrélations spatiales entre communes possèdent une décroissance à longue portée (environ égale à $80\,R$, soit environ $200~km$) et une forme quasi-logarithmique, comme celles calculées à partir des résultats des élections (voir Fig.~\ref{fcorrel-res}).

Par contre, les corrélations entre proches communes des résultats électoraux dépassent (sauf pour l'élection de $2000$) celles des taux de participation. Noter aussi que les trois plus fortes valeurs des corrélations entre proches communes $\ti$qui se détachent relativement par rapport aux autres élections$\ti$ concernent trois des quatre élections qui procurent au \og modèle zéro \fg{} les plus fortes corrélations, à savoir les trois élections européennes.

\subsubsection{Effets de l'environnement}
\label{pt-environnement-abst}
\begin{figure}[t!]
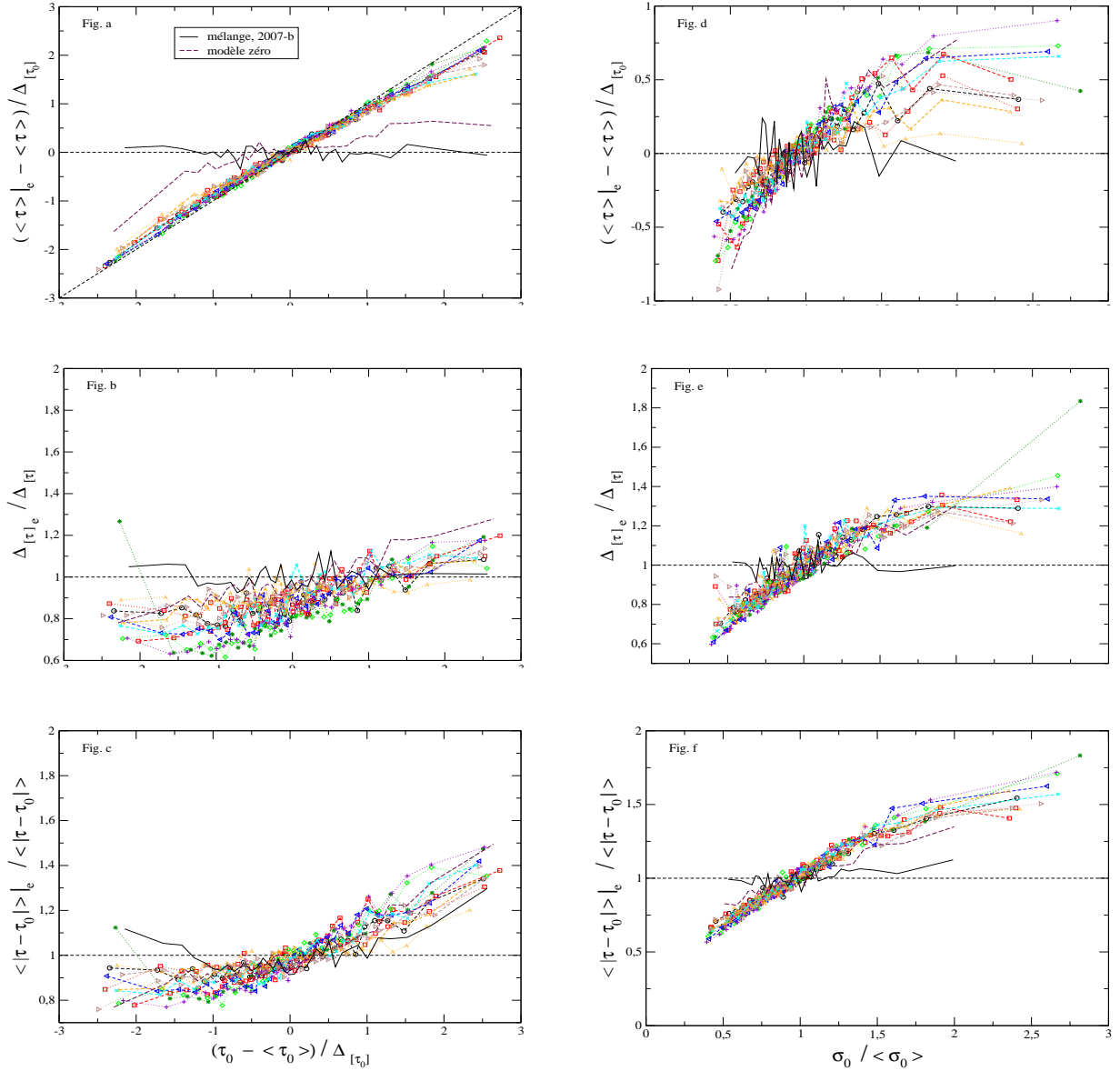

\includegraphics[width=7.5cm, height=5cm]{1-abst.eps}\hfill
\includegraphics[width=7.5cm, height=5cm]{3-abst.eps}\vspace{0.25cm}
\includegraphics[width=7.5cm, height=5cm]{2-abst.eps}\hfill
\includegraphics[width=7.5cm, height=5cm]{4-abst.eps}\vspace{0.25cm}
\includegraphics[width=7.5cm, height=5cm]{5-abst.eps}\hfill
\includegraphics[width=7.5cm, height=5cm]{6-abst.eps}
\caption{\small Effet de l'environnement sur le taux de participation de la commune centrale.  En trait continu, les taux de participation de l'élection 2007-b mélangés aléatoirement~; et en tirets, le \og modèle zéro \fg{} appliqué au données de l'élection 2007-b. Cette légende vaudra pour les figures~\ref{fhisto-environnement-abst} et \ref{fs0t0-abst} en annexe.}
\label{fenvironnement-abst}
\end{figure}

L'effet des taux d'abstention de l'environnement $\vaa$ d'une commune centrale $\aaa$ $\ti$défini de nouveau par les $n_p=16$ plus proches communes de $\aaa$\--- sur le taux d'abstention de la commune centrale se voit aux figures~\ref{fenvironnement-abst}. Ces figures cherchent, comme à la section~\ref{pt-environnement-res}, à déceler la connexion entre le taux de participation $\taa$ d'une commune centrale $\aaa$ et les taux de participation de son environnement $\ti$mesurés par $\toa$ et $\saa$ (cf. Eqs.~(\ref{etoa},~\ref{esaa})). Ces figures ressortent les deux mêmes régularités que celles relatives aux résultats (comparer aux figures~\ref{fenvironnement-res})~: la commune centrale se comporte en moyenne comme son environnement ($\taa \simeq \toa$), et l'écart entre ces deux valeurs, $|\taa - \toa|$, se relie fortement et en moyenne à la dispersion $\saa$ des $\tau$ des communes dans l'environnement de la commune centrale. (Les normalisations effectuées dans ces deux figures (cf. section~\ref{pt-environnement-res}) accentuent leur ressemblance.)

En revanche, toutes les courbes de Fig.~\ref{fenvironnement-abst}-d montrent une croissance de $\taa$ en fonction de $\saa$~; ce qui s'accorde avec la décroissance de $\taa$ en fonction de la taille de la population, jointe à diminution de $\saa$ en fonction de la taille de la population (voir discussion similaire pour les résultats à la section~\ref{pt-explique-res}).

Les figures~\ref{fcorrel-toa-saa-abst} en annexe tracent les corrélations spatiales des $\toa$ et des $\saa$, calculés à partir des taux de participation par commune pour chaque élection. En outre, les figures~\ref{frepartition-saa-abst} reportent, en fonction de leur classement, les positions géographiques des $\saa$ de l'élection 2007-b des taux de participation par commune et par bureau de vote. Enfin, Fig.~\ref{fhisto-environnement-res} et Fig.~\ref{fs0t0-res} représentent respectivement les histogrammes de $\toa$, de $\saa$ et de $(\taa-\toa)$, ainsi qu'une connexion entre les deux grandeurs caractérisant l'environnement, $\saa$ et $\toa$.

\subsubsection{Corrélations temporelles}
\label{pt-tempo-abst}
La table~\ref{ttempo-saa-abst} relate les corrélations temporelles, $C_{t_i,t_j}(\sigma_0)$, des $\saa$ des taux de participation entre des élections différentes. Les valeurs obtenues, et plus précisément leur moyenne $\overline{C}_t(\sigma_0)$ et leur dispersion standard $\Delta_{[C_t(\sigma_0)]}$ sur l'ensemble des couples d'élections, se rapprochent énormément de celles obtenues à partir des résultats (cf. Tab.~\ref{ttempo-saa-res}).

En revanche, les corrélations temporelles des taux de participation $\taa$, ou des $\toa$, exhibent de bien plus grandes et régulières valeurs au regard de leurs homologues provenant des résultats (voir les tables~\ref{ttempo-abst} et \ref{ttempo-toa-abst} comparées aux tables \ref{ttempo-res} et \ref{ttempo-toa-res}). Les valeurs de $C_{t_i,t_j}(\tau_0)$ pour $n_p = 16$ communes dépassent même, en moyenne, celles de $C_{t_i,t_j}(\sigma_0)$. Ceci peut s'expliquer par le fait que les taux de participation des différentes élections, à l'inverse des résultats des élections, répondent à la même question (s'abstenir ou non) $\ti$même si les enjeux diffèrent d'une élection à l'autre. La plus grande homogénéité du choix mesuré (participer ou non à l'élection) permet en effet de donner plus de sens à la comparaison possible des $\taa$, d'une même commune $\aaa$ à différentes élections (voir à ce sujet la discussion à la section~\ref{pt-critique-res}).

\begin{figure}[t]
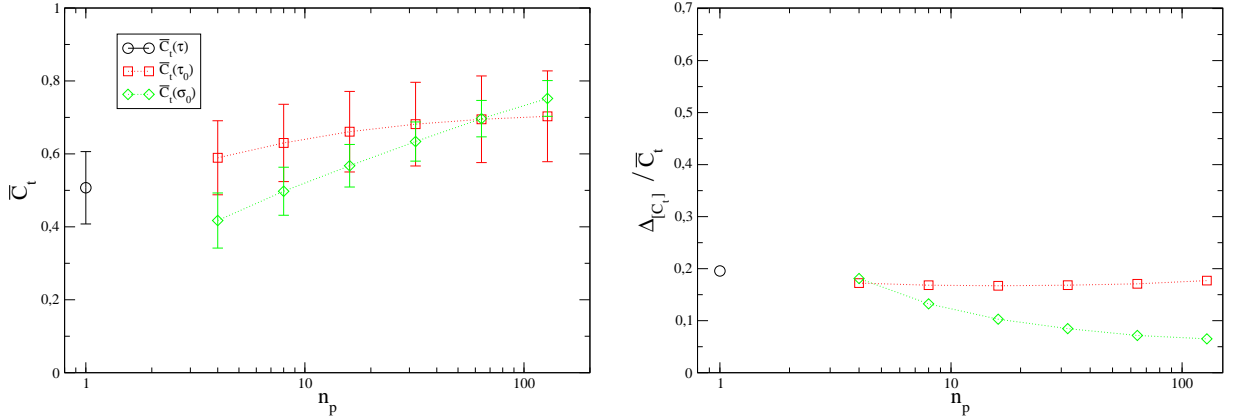

\includegraphics[scale = 0.32]{bilan-cor-tempo-abst-multippv.eps}\hfill
\includegraphics[scale = 0.32]{ratio-multippv-abst.eps}
\caption{\small Moyenne des corrélations temporelles $\overline{C}_t$ sur l'ensemble des couples d'élections, avec leur déviation standard $\Delta_{[C_t]}$ en barre d'erreur, évalués pour $\taa$, $\toa$ et $\saa$ des taux de participation. La figure de droite visualise leur rapport, $\frac{\Delta_{[C_t]}}{\overline{C}_t}$, une sorte de rapport \textit{bruit/signal}. Le nombre de communes, $n_p$, intervenant dans les grandeurs $\toa$ et $\saa$ (voir Eqs.~(\ref{etoa},~\ref{esaa})), égal à 16 jusqu'alors varie ici de 4 à 128. Les valeurs correspondantes à $\tau$ se situent, par extension, à $n_p=1$.}
\label{fbilan-tempo-abst}
\end{figure}

Enfin, la figure~\ref{fbilan-tempo-abst} (comme sa cons\oe{}ur, Fig.~\ref{fbilan-tempo-res}, relative aux résultats électoraux) reporte les moyennes $\overline{C}_t$ et les écarts-types $\Delta_{[C_t]}$ pour les $\saa$, $\toa$ et $\taa$, ainsi que l'influence du nombre $n_p$ de communes prises en compte.

\begin{table}[h]
\small
\begin{tabular}{|c|c|c|c|c|c|c|c|c|c|c|c|}
\hline
\backslashbox{$t_i$}{$t_j$} & 94-m & 95-m & 95-b & 99-m & 00-b & 02-m & 02-b & 04-m & 05-b & 07-m & 07-b\\
\hline
1992-b & 0,636 & 0,602 & 0,600 & 0,597 & 0,559 & 0,501 & 0,521 & 0,543 & 0,515 & 0,484 & 0,520\\ 
1994-m & & 0,609 & 0,595 & 0,686 & 0,607 & 0,524 & 0,509 & 0,603 & 0,535 & 0,463 & 0,505\\
1995-m & & & 0,732 & 0,607 & 0,555 & 0,551 & 0,549 & 0,554 & 0,531 & 0,507 & 0,535\\
1995-b & & & & 0,605 & 0,547 & 0,537 & 0,554 & 0,556 & 0,524 & 0,500 & 0,536\\
1999-m & & & & & 0,697 & 0,589 & 0,570 & 0,700 & 0,592 & 0,510 & 0,551\\ 
2000-b & & & & & & 0,552 & 0,546 & 0,662 & 0,542 & 0,485 & 0,532\\
2002-m & & & & & & & 0,694 & 0,581 & 0,557 & 0,499 & 0,535\\
2002-b & & & & & & & & 0,566 & 0,572 & 0,530 & 0,574\\
2004-m & & & & & & & & & 0,642 & 0,535 & 0,575\\
2005-b & & & & & & & & & & 0,567 & 0,600\\
2007-m & & & & & & & & & & & 0,704\\
\hline
\end{tabular}
\normalsize
\caption{\small Taux de participation : corrélation temporelle $C_{t_i,t_j}(\sigma_0)$ des $\saa$ sur chaque couple d'élections $(t_i,t_j)$. Moyenne $\overline{C}_t(\sigma_0) = 0,567$ ; écart-type $\Delta_{[C_t(\sigma_0)]} = 0,058$ ; $\frac{\Delta_{[C_t(\sigma_0)]}}{\overline{C}_t(\sigma_0)} = 0,103$.}
\label{ttempo-saa-abst}
\end{table}

\subsubsection{$\mathbf{\saa}$ à la croisée des grandeurs}
\label{pt-saa-abst}
\begin{figure}[t]
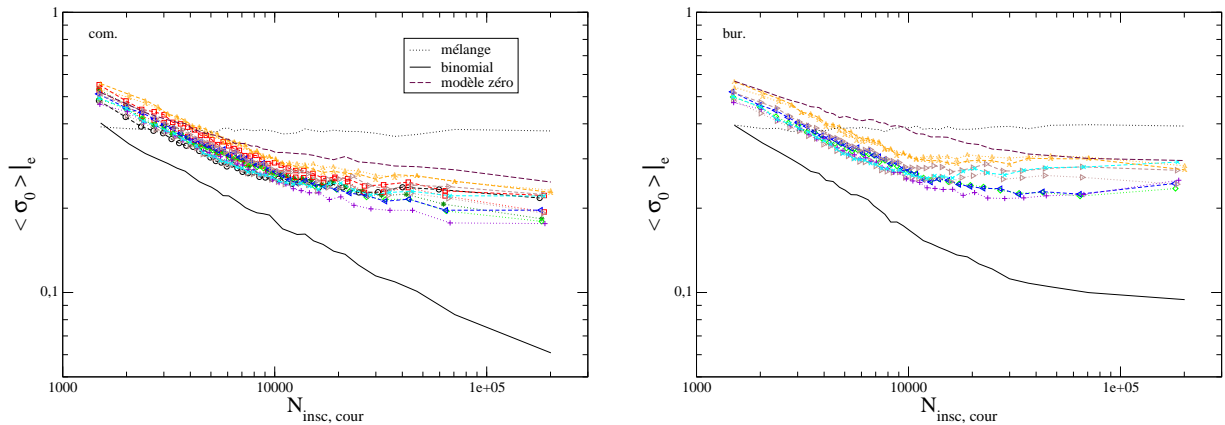

\includegraphics[scale = 0.32]{s0-com-taille-inscrit-abst.eps}\hfill
\includegraphics[scale = 0.32]{s0-bvot-taille-inscrit-abst.eps}
\caption{\small Moyenne des $\saa$ en fonction de $\inscoa$, calculé avec $36$ intervalles. $\saa$ s'évalue à partir des taux de participation par commune, à gauche, et par bureau de vote, à droite. Les tirages aléatoires binomiaux, le \og modèle zéro \fg et le mélange aléatoire des taux de participation s'effectuent à partir des valeurs réelles de l'élection 2007-b.}
\vspace*{0.5cm}
\label{fsaa-taille-abst}
\end{figure}

En se référant à la démarche suivie au chapitre précédent (cf section~\ref{pt-comprendre-res}), les figures~\ref{fsaa-taille-abst} tracent $\saa$ des taux de participation en fonction du nombre d'inscrits $\inscoa = \sum_{\beta\in\vaa} \insc^\beta$. Il s'en déduit les mêmes conclusions que pour les résultats électoraux (voir Fig.\ref{fsaa-taille-res} et la discussion afférente), à savoir une forte connexion de $\saa$ avec la taille de la population. Cette forte connexion explique, comme nous l'avions déjà évoqué à la section~\ref{pt-explique-res}, les diverses régularités associées à $\saa$.

\begin{figure}
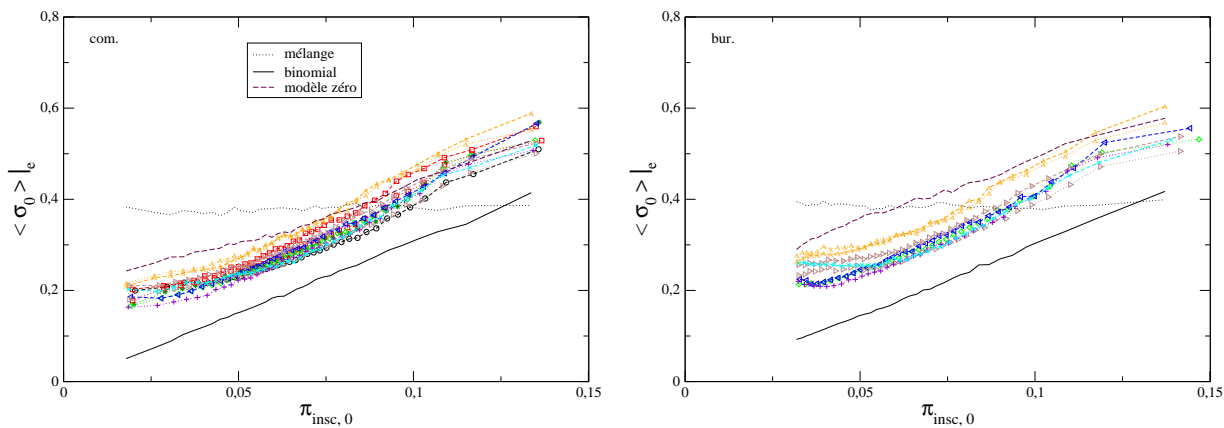

\includegraphics[scale = 0.32]{s0-com-piinsc-abst.eps}\hfill
\includegraphics[scale = 0.32]{s0-bvot-piinsc-abst.eps}
\caption{\small Moyenne des $\saa$ des taux de participation en fonction de $\piinsc$ calculé avec $36$ intervalles, évalués conjointement par commune à gauche, et par bureau de vote à droite. Les tirages aléatoires binomiaux, le \og modèle zéro \fg{} et le mélange aléatoire des taux de participation s'effectuent à partir des valeurs réelles de l'élection 2007-b.}
\label{fsaa-piinsc-abst}
\end{figure}
Comme au chapitre précédent (cf. section~\ref{pt-bruit-res}), la connexion de $\saa$ avec la taille de la population s'établit plus clairement avec la grandeur $\piinsc$ (donnée à l'équation~(\ref{epiinsc})), et procure de plus, une vision plus intime du rôle tenu par le bruit statistique dû aux tailles finies des communes (ou des bureaux de vote). Les figures~\ref{fsaa-piinsc-abst} tracent directement $\saa$ en fonction de $\piinsc$, tandis que la table~\ref{tsaa-piinsc-abst} fournit les corrélations entre $\saa$ et $\piinsc$. Nous pouvons les comparer à leurs homologues relatives aux résultats électoraux (cf. Fig.~\ref{fsaa-piinsc-res}, et Tab.~\ref{tsaa-piinsc-res}) et constater, notamment, les mêmes régularités précédemment discutées.

\begin{table}[h]
\footnotesize
\begin{tabular}{|c|c|c|c|c|c|c|c|c|c|c|c|c|}
\cline{2-13}
\multicolumn{1}{c|}{}
 & 92-b & 94-m & 95-m & 95-b & 99-m & 00-b & 02-m & 02-b & 04-m & 05-b & 07-m & 07-b\\
\hline
com. & 0,636 & 0,633 & 0,665 & 0,662 & 0,661 & 0,677 & 0,615 & 0,635 & 0,679 & 0,609 & 0,610 & 0,663\\
bur. & & & & & 0,640 & 0,664 & 0,563 & 0,586 & 0,656 & 0,555 & 0,558 & 0,628\\
\hline
\end{tabular}
\normalsize
\caption{\small Corrélations $C_{XY}$ pour chaque élection sur l'ensemble des communes entre la dispersion des taux de participation $X^\aaa=\saa$ et $Y^\aaa=\piinsc$. $\saa$ et $\piinsc$ se calculent, soit à partir des valeurs par commune (com.), soit à partir des valeurs par bureaux de vote (bur.). Sur l'ensemble des élections, pour les données par commune, $C_{XY} = 0,645 \pm 0,026$ (et, $C_{XY} = 0,823 \pm 0,013$ avec un tirage binomial)~; et pour les données par bureau de vote, $C_{XY} = 0,606 \pm 0,046$ (et, $C_{XY} = 0,832 \pm 0,008$ avec un tirage binomial).}
\label{tsaa-piinsc-abst}
\end{table}

\subsubsection{L'ivresse de la séparation}
\label{pt-separer-abst}
L'annexe~\ref{annexe-extraire} essaie, par plusieurs méthodes, d'extirper de $\saa$ une part d'information locale dite positive. Par comparaison avec l'information locale positive contenue dans les $\saa$ issus des résultats électoraux, il ressort les deux points suivants~:
\begin{itemize}
\item les $\saa$ des taux de participation, dénués du bruit statistique dû aux tailles finies, ont un aspect permanent plus marqué que ceux des résultats électoraux (parce que, probablement, les taux de participation à différentes élections demandent des choix à faire aux électeurs, plus homogènes que ceux relatifs à l'expression d'un vote)~;
\item l'état des corrélations inter-communales, qui permet de comprendre d'une certaine façon l'information locale dite positive, montre un plus fort degré de corrélation pour les résultats électoraux, comparé aux taux d'abstention (en accord avec les corrélations des $\tau$ entre communes voisines, plus fortes avec les résultats électoraux qu'avec les taux de participation).
\end{itemize}

\subsubsection{Récapitulatif}
\label{pt-recapitulatif-abst}
L'étude portant sur les taux de participation de douze élections nationales confirme donc l'ensemble des conclusions tirées de l'analyse des résultats électoraux (voir section~\ref{pt-resume-res}), à savoir~:
\begin{itemize}
\item La grandeur $\tau$ convient à une étude portant sur la recherche des régularités.
\item Les corrélations des $\taa$ par commune montrent une décroissance à longue portée, de forme quasi-logarithmique (avec une valeur à courte distance relativement importante pour les résultats électoraux, et plus faible pour les taux de participation)~;
\item Une commune centrale $\aaa$ se comporte en moyenne comme son entourage ($\taa \simeq \toa$), et l'écart à son entourage, $|\taa - \toa|$, augmente régulièrement avec la dispersion, $\saa$, de son voisinage.
\item $\saa$ présente une forte permanence temporelle~;
\item $\saa$ se relie fortement à la taille de la population, expliquant par là même les diverses régularités qu'il engage~;
\item $\saa$ traduit une part de bruit dû aux tailles finies des communes ou des bureaux de vote, mais aussi une information à caractère positif sur la zone où il s'évalue.
\end{itemize}
\vspace{0.5cm}

L'étude des taux de participation a, en outre, mis en exergue l'importance de pouvoir comparer des choix similaires (ici, participer ou s'abstenir à une élection) sur différentes élections. Les $\taa$ ou $\toa$ des taux de participation montrent alors une forte stabilité temporelle, et bien plus grande que celles correspondantes aux résultats électoraux.

En revanche, les taux de participation montrent une moins forte corrélation entre communes voisines, que celle des résultats électoraux (élection de 2000, mise à part).

Signalons enfin que l'annexe~\ref{annexe-communautes} étudie l'indépendance des taux de participation des bureaux de vote au sein d'une même ville, ou dans les zones urbaines, et montre $\ti$en accord avec ce qui précède$\ti$ leur moins forte corrélation par rapport aux résultats électoraux.

Enfin, terminons cette partie en rappelant la similarité sur toutes les élections des distributions de probabilité $\mathcal(\taa-\langle\tau\rangle)$ (cf. les histogrammes de Fig.~\ref{fhisto-log-abst} des $\taa$ de toutes les communes ou de tous les bureaux vote, centrés sur leur moyenne)~; avec une asymétrie en partie expliquée par la diminution du taux de participation avec l'augmentation du nombre d'habitants des communes.\\

Il s'impose à présent d'élargir cette étude au risque de la prédiction, ainsi qu'au jeu de la signification.

\clearpage
\subsection{Prédictions et Discussion}
\label{section-predictions}

\subsubsection{Prédictions}
\label{pt-predictions}
Certaines régularités observées sur les résultats et sur les taux de participation des élections étudiées ne peuvent que nous inciter à nous risquer au jeu de la prédiction. Mais auparavant, précisons que les prédictions qui suivront ne se fondent que sur des régularités de certains phénomènes, et non pas sur leur connaissance intime, et encore moins sur une théorie qui les sous-tend. Après avoir observé comment une vingtaine de pierres tombent, un esprit naïf ne se laisserait-il pas porter à prévoir la manière dont la prochaine tombera à son tour~?

Bien entendu, ces prédictions s'appliquent sur des élections comparables\footnote{Un merci à Laurent Barry, auteur du remarquable essai, \textit{La parenté}, qui m'avait conseillé de ne pas me cantonner à la prochaine élection, les européennes de 2009 au moment de la discussion, et ce pour mieux tester la pertinence de la démarche suivie.}, i.e. sur des élections portant sur un même choix à l'échelle de la France, et binaire en ce qui concerne leurs résultats électoraux. Les prédictions portent sur des grandeurs quantitatives et régulières. Et nous outrepassons l'aspect critiquable et ridicule de tenir compte d'échantillons aussi restreints en taille (i.e. des ensembles de valeurs construits sur $6$ résultats électoraux et $12$ taux de participation). Noter au passage que les grandeurs sur lesquelles portent les prédictions ne semblent pas montrer de variation, ou de dérive, globale au cours du temps étudié (de 1992 à 2007). Enfin, les prédictions qui suivront se feront sur une barre d'erreur arbitraire de \textit{2 sigma}, où \textit{sigma} dénote la déviation standard mesurée sur les données correspondantes.\\

\begin{enumerate}
\item \textbf{Écart-type en grandeur $\mathbf{\tau}$} (voir Eq.~(\ref{etau})) sur l'ensemble des communes.\\
D'après les tables~\ref{tstat-log-res} et \ref{tstat-log-abst}~:
\begin{table}[h!]
\begin{center}
\begin{tabular}{|c|c|c|}
\hline
$\Delta_{[\tau]}$ & mesures & attendu\\
\hline
résultats & $0,487 \pm 0,035$ & $[0,417 ; 0,557]$\\
taux de participation & $0,376 \pm 0,019$ & $[0,338 ; 0,414]$\\
\hline
\end{tabular}
\vspace*{0.5cm}\\
\end{center}
\end{table}

\item \textbf{Corrélations $\mathbf{C_{t_i,t_j}(\sigma_0)}$} sur l'ensemble des communes des $\saa$, à différentes élections.\\
$\saa$ se calcule à partir des valeurs par commune (voir Eqs.~(\ref{esaa})). Bien entendu, la valeur attendue porte sur la moyenne des $C_{t_i,t_j}(\sigma_0)$ (cf. Eq.~\ref{ecorrel-temporelle}) entre les élections passées, $t_i$, et l'élection à venir, $t_j$. D'après les tables~\ref{ttempo-saa-res} et \ref{ttempo-saa-abst} :
\begin{table}[h!]
\begin{center}
\begin{tabular}{|c|c|c|}
\hline
$C_{t_i,t_j}(\sigma_0)$ & mesures & attendu\\
\hline
résultats & $0,579 \pm 0,060$ & $[0,459 ; 0,699]$\\
taux de participation & $0,567 \pm 0,058$ & $[0,451 ; 0,683]$\\
\hline
\end{tabular}
\vspace*{0.5cm}\\
\end{center}
\end{table}
\item \textbf{Corrélation entre $\mathbf{\saa}$ et $\mathbf{\piinsc}$} sur l'ensemble des communes.\\
$\saa$ et $\piinsc$ se calculent encore ici à partir des valeurs par commune (voir Eqs.~(\ref{esaa},~\ref{epiinsc})). D'après les tables~\ref{tsaa-piinsc-res} et \ref{tsaa-piinsc-abst}~:
\begin{table}[h!]
\begin{center}
\begin{tabular}{|c|c|c|}
\hline
$C_{\sigma_0\,\pi_{insc,\,0}}$ & mesures & attendu\\
\hline
résultats & $0,685 \pm 0,069$ & $[0,547 ; 0,823]$\\
taux de participation & $0,645 \pm 0,026$ & $[0,593 ; 0,697]$\\
\hline
\end{tabular}
\end{center}
\end{table}

\end{enumerate}

Un désaccord entre les valeurs à venir et celles prédites ci-dessus s'interpréterait, soit comme une mauvaise investigation de notre part, soit comme le signe d'une plus grande richesse, i.e. d'une manifestation d'un phénomène nouveau ou plus intense $\ti$à l'aune des élections analysées dans ce travail.

\subsubsection{\textit{Addenda~: résultats des prédictions des européennes de 2009}}
\label{pt-addenda}
\textit{Bien avant le 7 juin 2009, date de l'élection des européennes en France, j'avais orienté ce travail d'étude des données électorales en vue de lui faire fournir des prédictions. A vrai dire, je pensais que ce travail de thèse aurait abouti plus tôt (j'avais déjà terminé en février 2009, aux annexes près, la rédaction de} Enquête : des généralités aux prédictions). \textit{Le jour des élections approchant, et n'ayant toujours pas mené à terme la partie théorique, j'ai voulu officialiser~\cite{3_predictions} en mai les prédictions précédemment établies.}

\textit{Tout ça pour dire que les prédictions faites sur les taux de participation des élections européennes de 2009 n'ont pas changé ma façon de poser le problème, et que je n'ai pas bâti cette rédaction sur des prédictions vérifiées} a posteriori.\\

\textit{Nous donnons maintenant les résultats des prédictions, tels qu'ils apparaissent dans~\cite{3_results}.}

\renewcommand{\thetable}{\arabic{table}}
\setcounter{table}{0}
\small

\textbf{1. Standard deviation of $\taa$}
\begin{table}[h!]
\small
\begin{center}
\begin{tabular}{|c|c|c|c|}
\cline{2-4}
\multicolumn{1}{c|}{}
 & Previous measures & Expected measure & Real value\\
\hline
Standard deviation of $\taa$ & $0.376 \pm 0.019$ & $[0.338 ; 0.414]$ & 0.360\\
\hline
\end{tabular}\\
\end{center}
\end{table}

\textbf{2. Correlation of $\mathbf{\saa}$ at different elections}
\begin{table}[h!]
\small
\begin{center}
\begin{tabular}{|c|c|c|c|}
\cline{2-4}
\multicolumn{1}{c|}{}
& Previous measures & Expected measure & Real measure\\
\hline
$C_{t_i,t_j}(\sigma_0)$ & $0.567 \pm 0.058$ & $[0.451 ; 0.683]$ & $0.577$\\
\hline
\end{tabular}
\begin{flushleft}(Tab.~\ref{ttempo-saa-abst-09} gives the twelve $C_{t_i,t_j}(\sigma_0)$, where $t_i$ is one of the past twelve elections and, $t_j$, the 2009 European Parliament election.)\end{flushleft}
\end{center}
\end{table}

\textbf{3. Correlation between $\mathbf{\saa}$ and $\mathbf{\piinsc}$}
\begin{table}[h!]
\small
\begin{center}
\begin{tabular}{|c|c|c|c|}
\cline{2-4}
\multicolumn{1}{c|}{}
 & Previous measures & Expected measure & Real measure\\
\hline
Correlation between $\saa$ and $\piinsc$ & $0.645 \pm 0.026$ & $[0.593 ; 0.697]$ & 0.657\\
\hline
\end{tabular}
\end{center}
\end{table}

\begin{table}[h!]
\small
\begin{tabular}{|c|c|c|c|c|c|c|c|c|c|c|c|}
\hline
92-b & 94-m & 95-m & 95-b & 99-m & 00-b & 02-m & 02-b & 04-m & 05-b & 07-m & 07-b\\
\hline
0.534 &  0.585 & 0.533 &  0.529 & 0.648 &  0.622 &  0.539 & 0.537 & 0.679 &  0.596 &  0.532 &  0.589\\
\hline
\end{tabular}
\caption{\small $C_{t_i,t_j}(\sigma_0)$, where $t_i$ is one of the previous twelve elections and $t_j$, the 2009 European Parliament election. The mean value is $0.577$.}
\label{ttempo-saa-abst-09}
\end{table}

\renewcommand{\thetable}{\arabic{table}}
\setcounter{table}{9}

\normalsize
\subsubsection{Discussion}
\label{pt-discussions}
Tout ça pour ça ? Tout ce travail d'enquête pour en arriver à déceler de bien piètres régularités. Qui, de plus pour celles qui engagent $\saa$, doivent leur régularité à la présence d'un bruit statistique dû aux tailles finies.

Devrions-nous nous consoler en arguant de la possibilité d'émettre les trois maigres prédictions\footnote{\textit{Prédire n'est pas expliquer}~: un titre explicite d'un livre de René Thom, laissé en suspens ici.} ci-dessus~? Devrions-nous en arriver à implorer la mansuétude des bons connaisseurs de la chose publique, labile par essence~? Devrions-nous nous justifier et quémander des circonstances atténuantes, pour avoir vu l'importance de $\saa$ et y mesurer une information locale positive, auprès des amateurs et interprétateurs de la glose publique, inépuisable à leurs sens~? Devrions-nous pour autant contenter les grands penseurs opinant que, de toute façon, la chose publique échappe par essence à la science~?

Non~!

Non, et de façon peu surprenante pour plusieurs raisons. Pour plusieurs raisons qui relèvent de certains aspects de la démarche scientifique.

Première raison : pousser au bout autant qu'il se peut une investigation, et ce sans se poser la question dans un premier temps ni de son utilité, ni de sa bonne moralité. Ce travail a en effet mis l'accent sur des questions dites simples, plutôt que sur celles qualifiées d'intéressantes, i.e. qui généralement stimulent de prime abord la curiosité de l'honnête homme. (Remarque~: nous ne parlons ici que d'accent, d'inclination, de notre démarche, puisque de manière générale, l'aspect dit simple d'une question n'exclut pas forcément son aspect dit intéressant, et inversement.)

Deuxième raison : parvenir à étudier un phénomène en éliminant autant qu'il se peut les \og détails \fg{} à l'échelle du phénomène mesuré. Ceci s'insère dans la dynamique à réduire l'information que pratique la science~\cite{solomonoff}. Probablement, avons-nous pu réussir notre challenge $\ti$observer des régularités$\ti$, justement parce que nous avons délaissé au préalable pléthore de particularités, de données sociologiques, politiques, historiques, et n'avons guère usé d'interprétation. Par contre, nous ne prétendons pas qu'il n'existe pas d'autres grilles d'investigation~\cite{siegfried}. De plus, la prise en compte d'autres données pourrait rendre caduque certaines parties de ce travail $\ti$comme la prise en compte des coordonnées spatiales des communes, permettant le calcul des corrélations spatiales, avait entraîné le rejet du \og modèle zéro \fg.

La troisième raison fait intervenir le jeu entre régularité et hors régularité, qui anime l'élaboration scientifique des connaissances~\cite{kuhn, poincare}. Ici, ce jeu donne de la signification aux choses, après les avoir dépouillées de leur apparente irrégularité. Dit autrement, une fois cerné le champ des phénomènes réguliers, ceux qui s'en échappent portent du sens, de l'information. Dit encore autrement~: par-delà les régularités, pointe le sens.\\

Je reprends l'exemple de l'esprit naïf qui regarde tomber des pierres. (Je vous le concède, en terme poétique on peut mieux faire.) Si cet esprit naïf se dote d'une démarche scientifique, à moins que la démarche scientifique ne l'ait déjà pourvu d'un esprit naïf, non seulement il croira que les prochaines pierres à leur tour tomberont, mais entreprendra des mesures. Mieux, il cherchera à voir dans ces phénomènes le plus de régularités possible. Je ne parle pas ici de la possibilité de formuler une théorie sous-jacente, potentiellement bâtie sur les régularités observées. Fort donc des régularités observées, il saura si un objet a un comportement \og normal \fg{} ou non. Il saura alors qu'un comportement qui échappe à la règle possède potentiellement une plus grande richesse que les autres. Mais plus important encore, il s'attendra à ce que cet échappement à la règle lui permette d'accéder à l'existence d'autres propriétés. Par exemple, une feuille qui tombe l'informera de l'existence de l'air, de la présence du vent, de l'importance de la la forme de la feuille, de sa masse volumique, de sa rigidité, etc. Mais cette richesse, ce dévoilement\footnote{\og Nature aime se cacher.\fg{} Héraclite.}~? s'opère plus facilement une fois constaté, par contraste, l'apparente régularité des pierres qui tombent.\\

Que dire alors sur les élections telles que nous les avons étudiées~?

Quel enseignement tirer, notamment, d'une observation minutieuse des figures~\ref{fsigma-taille-res},~\ref{fsigma-taille-abst}, et plus particulièrement encore, des figures~\ref{fsaa-taille-res},~\ref{fsaa-taille-abst},~\ref{fsaa-piinsc-res} et \ref{fsaa-piinsc-abst}~? Une question de vocabulaire au préalable~: les irrégularités se définissent en terme de différence par rapport aux autres élections de même type, puis, en l'absence de modèle de vote, par rapport aux tirages binomiaux. En outre, comme nous l'avions signalé plus haut, l'écart au bruit binomial se distingue plus aisément par l'analyse des données issues des bureaux de vote, plutôt que par celles collectées à l'échelle communale. Enfin, il va sans dire que les irrégularités se réfèrent, implicitement et nécessairement, par rapport à la grille d'analyse retenue pour étudier les données (ici, principalement, l'hétérogénéité locale $\saa$ en fonction de la répartition de la population). Bref, que dire donc des irrégularités observées sur ces courbes~?

Premièrement, les irrégularités ne se manifestent pas pour les taux de participation, mais uniquement pour les résultats électoraux de certaines élections. Deuxièmement, si les régularités apparaissent pour certaines élections, elles ne se manifestent que dans les zones à forte population, i.e. à faible $\piinsc$.

Encore reste-t-il à trouver un moyen autre que visuel pour déceler la présence des irrégularités, ou dit plus simplement, pour les quantifier. Il convient pour cela de se référer aux tables~\ref{tsaa-piinsc-res} et \ref{tsaa-piinsc-abst}, par bureau de vote si possible, et de comparer les valeurs obtenues, soit entre différentes élections, soit avec celles issues des tirages binomiaux. Il ressort que les résultats électoraux du référendum de $2005$ se détachent, et de loin, des autres élections\footnote{Merci à Philippe-Alexandre Pouille pour avoir attiré mon attention sur l'importance des irrégularités et sur la particularité du référendum de 2005~; et aussi pour ses critiques pertinentes sur les questions idéologiques soulevées par cette étude, ainsi que sur le concept d'atome social.}.

Que conclure donc~? Au préalable, j'attribue aux irrégularités précédemment rencontrées la manifestation dans le champ électoral de ce que j'appelle, faute de mieux, le \og fait politique \fg. Que dire donc de l'ainsi dénommé \og fait politique \fg, tel qu'il s'applique dans les élections étudiées~?

La réponse tirée de cette étude, tient en deux points. Ledit \og fait politique \fg{} ne se produit, potentiellement, que~: \begin{itemize}
\item dans les zones à forte population~;
\item pour les résultats électoraux de certaines élections.
\end{itemize}
En outre, avec les élections traitées ici, et selon les critères retenus pour décrypter les données, ce dénommé \og fait politique \fg{} se déploie le plus ostensiblement dans l'expression des votes du référendum de $2005$.

Pourquoi~? Avec son corollaire : comment étudier le \og fait politique \fg~? Comment pouvoir l'aborder après ce travail consacré aux régularités~? Que cache-t-il~? La physique statistique procure-t-elle une bonne grille d'investigation~? Que de questions qui méritent de s'y appesantir... mais pas ici.

\clearpage
\section{D'un modèle l'autre}
\label{section-rfim}
\hfill
\begin{minipage}[r]{0.8\linewidth}
\textit{Méfiez-vous des gens qui regardent le monde par un trou.}\\
D'après le \textit{Pantagruel} de François Rabelais~\cite{rabelais}
\end{minipage}
\vspace{0.75cm}

Jusqu'ici\footnote{Toute la partie qui suit doit beaucoup à Jean-Philippe Bouchaud.} nous nous sommes principalement préoccupés de la recherche des régularités. Nous nous sommes appesantis sur les régularités liées à $\saa$, en essayant de comprendre leur origine ainsi que leur signification. En revanche, nous n'avons pas poussé plus loin l'analyse des régularités associées aux corrélations spatiales. Plus important encore, nous n'avons pas cherché à établir un modèle simpliste à l'échelle microscopique, i.e. à l'échelle des agents, dans lequel s'insèrent les régularités observées. La partie qui suit va tenter d'y remédier.

\subsection{Préambule}
\label{section-modeles-preambule}
La tâche qui dès lors nous incombe consiste, principalement, à vouloir \og sauver les apparences \fg\footnote{Avec un clin d'\oe{}il au livre richement documenté de Pierre Duhem~\cite{duhem}, et à ses réflexions iconoclastes.} à l'aide des modèles très simples, puis, subsidiairement, à vouloir se prêter au jeu de l'interprétation $\ti$i.e. à vouloir tirer quelque enseignement de la concordance ou non des modèles avec les phénomènes.

Précisons la démarche que nous suivrons pour obtenir un modèle dit satisfaisant. Premièrement, nous nous efforcerons à la simplicité. (Un modèle, à nos yeux, doit faire ressortir les traits saillants, sous-jacents aux phénomènes étudiés, i.e. il doit permettre de retrouver les caractéristiques générales des données à partir d'un minimum d'ingrédients, qui deviennent par là même significatifs.) La modélisation se basera sur des prémisses plausibles et facilement formalisables. (Précisons que nous nous placerons au sein du même canevas conceptuel tout au long de cette section.) Ensuite, l'impérative remise en cause, l'irremplaçable confrontation aux données, permettra de retenir ou non les modèles obtenus. (La simplicité du modèle a le mérite de laisser transparaître son parti pris, les traits saillants qu'il engage, et qui auront les données réelles pour réponse.) Et nous nous arrêterons là~: nous n'irons pas à décréter que les choses, les individus, se comportent réellement tels que le modèle l'édicte. Ce qui ne nous empêchera pas pour autant d'analyser le modèle de façon critique, puis de l'interpréter. 

Insistons de nouveau sur le fait que nous nous bornons dans cette thèse à vouloir sauver les phénomènes, et ce au travers d'une question simple que nous soumettons aux données. Nous ne prétendons pas qu'il existe un et un seul modèle qui satisfasse à la confrontation des données, et encore moins, une seule question à poser aux données. (Ce qui nous permettra de garder la même façon de formaliser les phénomènes sociaux tout au long de ce travail, sans pour autant nier l'existence d'autres approches conceptuelles.) Sur un plan personnel, je ne crois pas que la physique doive se borner à sauver les phénomènes, ou faire répondre les données à une question ponctuelle. La physique gagne, et de beaucoup, à poser des théories, i.e. des cadres conceptuels larges et cohérents qui décrivent $\ti$en l'état des connaissances$\ti$ très acceptablement les phénomènes. Mais cela ne peut se produire qu'après un long travail expérimental$\ti$au sens large$\ti$ et théorique. Et je ne crois pas que la physique sociale~\cite{quetelet} ait déjà atteint ces degrés de maturité.

Bref, commençons par passer en revue les diverses observations à retrouver, puis explicitons les bases sur lesquelles reposent les modèles que nous utiliserons.

\subsubsection{Les phénomènes à sauver}
\label{pt-sozein-ta-phainomena}
Il est important de remarquer que les régularités des corrélations spatiales concernent les données électorales (résultats et taux de participation) à l'échelle de la commune, et non pas à l'échelle des bureaux de vote (cf. Figs.~\ref{fcorrel-res} et \ref{fcorrel-abst}). Nous nous concentrerons donc sur les résultats électoraux et sur les taux de participation par commune.\\

Récapitulons les diverses mesures effectuées précédemment $\ti$concernant la distribution des $\taa$ ainsi que la corrélation spatiale des résultats et des taux de participation des communes$\ti$ qui doivent être retrouvées par les modèles.
\begin{itemize}
\item La moyenne de $\taa$ en fonction de $\toa$ (voir Figs.~\ref{fenvironnement-res}-a et \ref{fenvironnement-abst}-a) se représente convenablement bien par une droite de pente sensiblement égale à un.
\item Les corrélations spatiales sont à longue portée. La portée caractéristique vaut approximativement $80\:R\simeq200~km$ pour toutes les élections $\ti$excepté principalement les résultats de 2002-b et de 2007-b.
\item Les corrélations spatiales ont une décroissance en distance quasi-logarithmique (cf. Figs.~\ref{fcorrel-res} et \ref{fcorrel-abst}).
\item La distribution des résultats ou des taux de participation est toujours unimodale (voir Figs.~\ref{fhisto-log-res} et \ref{fhisto-log-abst}).
\end{itemize}
Insistons sur le fait que les régularités ci-dessus ne dépendent~:
\begin{itemize}
\item ni de la moyenne globale du résultat ou du taux de participation (la moyenne sur l'ensemble des communes s'étend en effet de $<\resa>\simeq50\%$ à environ $80\%$ pour les résultats électoraux, et d'environ $35\%$ à $87\%$ pour les taux de participation, comme le montrent les tables~\ref{tstat-dens-res} et \ref{tstat-dens-abst});
\item ni de la valeur de la corrélation entre proches communes ($C(R\simeq 1)$ s'étend ici d'environ $0,16$ à $0,54$, résultats électoraux et taux de participation confondus).
\end{itemize}

Notons au passage que la forme des corrélations spatiales recèle plus d'information, i.e. est plus contraignante, que la moyenne des $\taa$ en fonction des $\toa$.\\

A cela se rajoutent d'autres phénomènes plus fins et d'autres régularités qui font intervenir des écarts-types des $\taa$ à l'échelle nationale, ou locale $\ti$i.e. sur l'environnement $\vaa$ des $n_p=16$ plus proches communes d'une commune centrale. Ces régularités ont déjà été recensées à la section~\ref{pt-predictions} consacrée aux prédictions. Rappelons-les néanmoins succinctement~:
\begin{itemize}
\item l'écart-type des $\taa$ sur l'ensemble des communes vaut dans les alentours de $0,49$ pour les résultats, et $0,37$ pour les taux de participation;
\item la corrélations entre $\saa$ et $\piinsc$ vaut dans les environs de $0,66$ pour les résultats et taux de participation confondus;
\item la corrélation temporelle des $\saa$ entre différentes élections vaut approximativement $0,57$, à la fois pour les résultats et les taux de participation.
\end{itemize}
\vspace{0.25cm}

Présentons maintenant notre façon de poser le problème $\ti$et le parti pris qu'il engage$\ti$ pour retrouver les phénomènes.

\subsubsection{Bagatelles pour un modèle}
\label{pt-bagatelles}
C'est une chance de disposer de tant de données, de mesures nettes du choix effectué en un jour par une quarantaine de millions d'individus. Nous profiterons de cette opportunité pour nous poser une question centrale, simple, et tenterons d'exiger des données leur verdict. 

Répétons de nouveau que nous ne pensons pas qu'il n'y ait qu'une seule grille d'investigation possible, ni que la réponse des données puisse être exhaustive. Une question avalisée par les données laisserait la place à d'autres approches possibles, surtout au stade actuel de développement de la physique sociale. Une réponse négative, quant à elle, serait riche d'enseignement puisqu'elle permettrait de rejeter clairement la question posée. 

La question à poser dépend intrinsèquement des données. Or, les données électorales collectent en un jour les choix (du vote ou de la participation à l'élection) des électeurs d'une commune, voire d'un bureau de vote. D'autre part, nous n'utiliserons que les données électorales conjuguées à la localisation spatiale des communes, à l'exclusion de toutes autres données de nature économique, sociologique, etc. 

D'où le truisme suivant~: la question posée ne peut avoir de réponse qu'à l'échelle collective, et ne concerne que les choix, ou les intentions de choix, des agents au jour de l'élection. Sont exclues de la sorte des questions d'ordre psychologique, sociologique, etc., i.e. des questions liées aux linéaments idéologiques de l'agent, ou aux facteurs socio-économiques, etc. qui conditionnent statistiquement le choix des agents. Sont aussi exclues des questions concernant le comportement d'un même électeur à différentes élections. Sont enfin exclues les questions liées au déroulement de la campagne électorale, et notamment la rétroaction probable des partis politiques sur la dénommée opinion publique.

Parmi la multitude des questions possibles, la question que nous posons est la suivante~: une interaction entre agents, basique et limitée au voisinage géographique, peut-elle expliquer les phénomènes~? Noter que la simplicité de l'interaction va de pair avec une interaction locale, i.e. cantonnée aux endroits où l'agent se situe. L'accent est donc porté ici non pas sur la forme potentiellement complexe de l'interaction, mais sur la possibilité d'une interaction simple $\ti$et donc locale$\ti$ à rendre compte des phénomènes.

L'interaction considérée ici $\ti$au regard du choix à prendre par les agents$\ti$ est sans aucune particularité, et donc la plus simple possible. Autrement dit, elle est par hypothèse symétrique, sans hiérarchie, sans leader, etc., et ne dépend ni de l'état des agents, ni de leur passé, etc. Elle ne dépend que des agents qui peuvent se rencontrer physiquement. En l'absence de données relatives aux divers déplacements des agents, nous restreignons la probabilité de rencontre des agents à leur lieu de résidence $\ti$assimilés aux communes où ils sont inscrits sur les listes électorales. (Bien que l'interaction entre agents soit d'emblée restreinte au voisinage, nous tenterons néanmoins par la suite d'éprouver cet \textit{a priori}.)

Nous essaierons donc de savoir si un magma d'atomes sociaux, interagissant entre eux de façon simple, permet de retrouver les phénomènes précédemment mesurés~? Ou, pour employer un mot à la mode, les phénomènes électoraux émergent-ils d'une interaction basique entre agents voisins~?

Remarquons au passage une autre approche possible entr'aperçue avec le \og modèle zéro \fg{} (cf. Fig.~\ref{fmod0}-b). Une corrélation spatiale des résultats électoraux (ou des taux de participation) peut résulter non pas d'une interaction entre agents, mais d'un comportement statistiquement similaire des agents qui habitent dans des communes ayant le même état, les mêmes conditions. (A la section~\ref{pt-mod0-abst}, l'état d'une commune ne dépendait que de sa population.)

Bref, ce qu'il importe maintenant, n'est pas tant de savoir si la question posée est crédible ou non, si elle correspond ou non à la bienséance idéologique ou intellectuelle du moment, mais plutôt de savoir ce qu'en disent les données.\\

Pour répondre à cette question, nous ferons usage de modèles simplistes. Simplistes afin que les traits saillants du modèle ne soient pas noyés dans un maelström de considérations parasites. Parasites au sens de la question posée.

Bien entendu, un modèle simpliste à faible nombre de paramètres ne peut contenir toute la richesse, toutes les particularités historiques, etc., et à toutes les échelles d'observation du système qu'il prétend décrire. Il doit par contre retrouver certaines grandeurs stylisées du système. En l'occurrence, il doit pouvoir retrouver les régularités précédemment mesurées, et à la bonne échelle d'observation, i.e. à l'échelle de la commune, ici.

De plus, les modèles que nous utiliserons se doivent d'être plausibles. Autrement dit, ils doivent se fonder sur un mécanisme microscopique plausible et cohérent, au détriment d'arguments purement heuristiques déconnectés les uns des autres.

Bref, comme de coutume en science, nous essaierons de retrouver les phénomènes majeurs (les phénomènes à sauver de la section~\ref{pt-sozein-ta-phainomena}) en faisant fi des \og détails \fg{} et en se basant sur une description microscopique qui se veut plausible et pertinente.\\

Tout au long de cette étude, nous allons nous placer à l'intérieur d'un seul cadre conceptuel ouvrant droit à une modélisation possible des phénomènes sociaux. Voyons donc cette grille de travail permettant la formalisation des phénomènes sociaux.

\subsection{RFIM en physique sociale}
\label{section-rfim-phys-soc}

\subsubsection{Précédentes utilisations}
\label{pt-rfim-precedentes}
Parmi la riche boîte à outils~\cite{fortunato_stat_phys, stauffer_sociophys, bettencourt_epidemiological, schweitzer_brownian} dont dispose la physique statistique pour conceptualiser les phénomènes sociaux $\ti$à laquelle il manque encore à mon goût de sérieuses confrontations avec les données, aussi imparfaites ou parcellaires soient-elles, mais des confrontations néanmoins nécessaires pour lui faire acquérir un caractère cumulatif$\ti$, nous utiliserons le modèle d'Ising en champ aléatoire à température nulle, plus connu sous son acronyme anglais RFIM (Random Field Ising Model) à $0~T$. (Tout au long de ce travail, le RFIM sera toujours considéré à température nulle, et nous ne le mentionnerons plus par la suite.)

Initialement conçu pour résoudre les problèmes de type bruit de Barkhausen et autres phénomènes d'hystérésis et d'avalanches, le RFIM~\cite{sethna_1, sethna_2, sethna_3}, dont les lois d'échelle restent valables assez loin du point critique, a été appliqué pour modéliser de façon théorique des phénomènes sociaux~\cite{galam_1, galam_2, galam_3, hit_is_born} et d'interaction entre acheteurs et vendeurs~\cite{nadal_dilemma, nadal_multiple}. L'article~\cite{nadal_generic_properties} passe en revue les différentes propriétés engendrées par ce modèle dans le cadre des phénomènes socio-économiques. Caccioli et al.~\cite{marsili_rfim_memory} enrichissent un modèle d'Ising par la prise en compte d'un effet de mémoire. Q. Michard et J.-Ph. Bouchaud~\cite{collective_shift} analysent trois types de données distinctes les unes des autres $\ti$l'augmentation de l'achat des téléphones portables en Europe, la diminution du taux de natalité en Europe et la fin des applaudissements en salles de spectacle$\ti$ sous l'angle du RFIM, modèle qui inclut selon ses paramètres des discontinuités, l'arrêt brutal des applaudissements en l'occurrence. L'article~\cite{of_songs_and_men} s'appuie sur une belle étude expérimentale~\cite{salganik_exp_final} du téléchargement des chansons en diverses situations, pour étendre le RFIM aux choix non binaires et pour calibrer ensuite de façon semi-quantitative les paramètres du modèle.

Voyons sous quel angle, selon quels concepts, les modèles de type RFIM appréhendent les phénomènes sociaux. En d'autres termes, voyons succinctement les idées sous-jacentes aux phénomènes sociaux sur lesquelles reposent le RFIM.

\subsubsection{Description des phénomènes sociaux par le RFIM}
\label{pt-description-rfim} 
Explicitons brièvement comment le RFIM aborde le choix binaire, $S_i \pm 1$, accompli par un agent $i$. Dans le cadre de cette étude, $+1$ représente, soit le vote en faveur du choix gagnant à l'échelle nationale (cf. chapitre~\ref{section-res}), soit la participation de l'agent à l'élection comme au chapitre~\ref{section-abst}.

Un choix résulte, selon ce modèle, de la superposition de trois composantes identifiables~: l'agent isolé, l'influence \og extérieure \fg{} et enfin, l'interaction entre agents. Détaillons-les succinctement.
\begin{itemize}
\item L'idiosyncrasie, une propension individuelle de l'agent $i$ par rapport au choix proposé. Elle se traduit en physique statistique non seulement par un unique scalaire $h_i$, mais surtout par une variable aléatoire $\ti$à cause de la quantité \textit{a priori} très grande des \og choses \fg{} dont elle dépend. Cette variable aléatoire décrivant la tendance individuelle de l'agent $i$, isolé de l'environnement et des autres agents au moment du choix à prendre, peut être alors considérée comme indépendante et identiquement distribuée (i.i.d.) sur l'ensemble des agents, et fixe au cours du temps analysé.
\item Un champ global, affectant d'une égale façon l'ensemble des agents, représente l'influence extérieure. Il est reporté par une grandeur réelle $F$.
\item L'influence des autres\footnote{\og Je est un autre \fg{} de Rimbaud, certes, mais avec une toute autre conception.}, écrite comme une somme d'influence de chaque agent $j$ sur l'agent $i$, soit $\frac{\sum_{j\in \mathcal{V}_i} J_{ij}S_j}{\sum_{j\in \mathcal{V}_i}1}$ où $\mathcal{V}_i$ représente le voisinage de $i$, $J_{ij}$ l'intensité de l'influence de $j$ sur $i$ et enfin $S_j$ le choix de $j$. (Le dénominateur comptabilise le nombre d'agents dans le voisinage $\mathcal{V}_i$ avec lesquels interagit $i$.) $J_{ij} > 0$ pousse $i$ à se conformer aux choix de $j$, et inversement, $J_{ij} < 0$ incite $i$ à s'opposer aux choix de $j$. L'effet de la plus simple imitation, selon laquelle chaque agent tend à imiter tous les agents de son voisinage d'une égale façon, peut donc se formuler comme $J_{ij} = J > 0$ pour tout $i$ et $j$ du même voisinage. En notant $m_{\mathcal{V}_i} = \frac{\sum_{j\in \mathcal{V}_i} S_j}{\sum_{j\in \mathcal{V}_i}1}$, le choix moyen (la polarisation des spins en physique) dans le voisinage $\mathcal{V}_i$, le terme d'imitation devient $J.m_{\mathcal{V}_i}$.
\end{itemize}
\vspace{0.5cm}
Ainsi, selon le modèle à seuil qu'est le RFIM standard~:
\be \label{erfim} S_i = \operatorname{sign}\big[h_i + F + J \x m_{\mathcal{V}_i} \big]~.\ee
Dans le cas d'une imitation entre agents, l'intensité d'interaction, $J$, est positive.\\

Il est à noter que l'unique hétérogénéité prise en compte par le modèle, réside dans le terme idiosyncratique, et encore, celle-ci dérive d'une variable aléatoire. Bien évidemment, cette description, ou cette formalisation, ne prend son sens qu'à l'échelle des phénomènes collectifs. Autrement dit, le RFIM tente de décrire les phénomènes collectifs, et non pas les linéaments d'ordre psychologique conduisant au choix effectué par l'agent le jour J. En conclusion, le RFIM aborde donc les phénomènes collectifs en mettant l'accent sur l'imitation simple des agents $\ti$au sein d'un environnement décrit par un champ global $F$.

Cette modélisation, de type \og agent-based model \fg, illustre de façon générale l'approche de la physique statistique des phénomènes sociaux. Premièrement, la simplicité prédomine, et les \og détails \fg{} microscopiques à l'échelle de la mesure sont expurgés $\ti$comme souvent en science. Deuxièmement, cette modélisation repose sur une décomposition, une séparation, une non intrication, à la fois des différentes tendances qui interviennent dans le choix de l'agent, et aussi du groupe de personnes pris comme un ensemble d'agents individuels~: un ensemble d'atomes sociaux. 

Terminons enfin cette partie critique en mentionnant l'impossibilité $\ti$connue récemment\footnote{Je remercie Frédéric Nowacki de me l'avoir signalé, expliqué, réexpliqué, et de m'avoir donné ces références.}$\ti$ d'expliquer le noyau en terme d'interaction à deux corps (nucléons), ce qui conduit à introduire phénoménologiquement, actuellement, une force à trois corps~\cite{3nucleons-abinitio, 3nucleons-spectroscopie, 3nucleons-decomposition}.

Revenons maintenant au problème des résultats ou des taux de participations mesurés par commune, et voyons si un modèle issu du RFIM peut s'accorder avec les observations.

\subsection{RFIM avec imitation du choix des agents}
\label{section-rfim-choix}
Commençons par passer en revue chaque terme de l'équation~(\ref{erfim}) $\ti$relative au RFIM standard$\ti$ au regard du problème concret qui se pose ici~: la détermination du résultat ou du taux de participation par commune à une élection donnée. Ceci nous permettra, moyennant quelques hypothèses, d'écrire un modèle basé sur le RFIM standard. Nous confronterons ensuite le modèle obtenu aux données réelles provenant des résultats et des taux de participation.

\subsubsection{Du RFIM standard aux élections dans une commune~: précision sur les termes employés}
\label{pt-rfim-termes}
La commune $\aaa$ contient $\naa$ agents qui effectuent $\naa$ choix binaires $S_i^\aaa$. Le choix binaire se réfère au vote en faveur de l'un des deux candidats en lice du second tour d'une élection présidentielle, ou à choisir entre un oui ou un non lors d'un référendum. Ce type de choix lié au résultat électoral de la commune $\aaa$, implique $\naa = \expr^\aaa$ agents, i.e. un nombre $\expr^\aaa$ de bulletins exprimés. $S_i^\aaa=+1$ si l'agent $i$ de la commune $\aaa$ vote en faveur du choix qui se trouve gagnant à l'échelle nationale, ou $-1$ dans le cas contraire. L'autre type de choix envisagé ici, la participation à une élection, fait intervenir $\naa = \inscaa$ agents inscrits sur les listes électorales de la commune $\aaa$. $S_i^\aaa$ ne dépend plus de la convention précédente d'assigner la valeur $+1$ au choix gagnant à l'échelle nationale de l'élection considérée. Plus simplement, $S_i^\aaa = +1$ si l'agent $i$ de la commune $\aaa$ participe à une élection, et $-1$ s'il ne participe pas, quelle que soit la moyenne nationale du taux de participation de l'élection considérée.

La polarisation de la commune $\aaa$, en terme de résultat (ou de taux de participation) à une élection donnée, s'écrit alors comme 
\be \label{emaa} m^\aaa = 2\resa - 1 = \frac{\sum_{i = 1}^{\naa} S_i^\aaa}{\naa} ~,\ee
où $S_i^\aaa$ représente le choix afférent de l'agent numéro $i$ de la commune $\aaa$.\\

En se basant sur l'Eq.~(\ref{erfim}), il semble naturel d'appliquer à chaque commune $\aaa$ considérée non pas le même champ global $F$ à l'ensemble des communes, mais un champ qui lui est propre, $F^\aaa$. Ensuite, $F^\aaa$ peut se noter comme la somme d'un terme, $f^\aaa$, spécifique à la commune $\aaa$, et d'un champ global $F$, appliqué à toutes les communes, soit $F^\aaa = F + f^\aaa$.

Le terme spécifique $f^\aaa$ peut se comprendre de deux façons différentes. La première façon, en accord avec la manière dont il a été introduit ci-dessus, lui donne la signification d'un champ appliqué qui se surajoute au champ global $F$. La seconde façon lui attribue la signification d'une moyenne~: la moyenne des idiosyncrasies de la commune $\aaa$. En effet, ni $h_i$ seule, ni $F$ seul, ne sont relevants dans Eq.~(\ref{erfim}), mais leur somme $(h_i + F)$ l'est. Prenons le cas d'un champ uniforme $F$ appliqué à chaque commune. Pour un agent $i$ d'une commune $\aaa$, l'expression précédente, relevante, devient $(h_i^\aaa + F)$, où $h_i^\aaa$ dénote l'idiosyncrasie de cet agent. Selon cette écriture, la spécificité de la commune $\aaa$ est transcrite par les particularités de $h_i^\aaa$. En notant maintenant $f^\aaa$ comme la moyenne des idiosyncrasies $h_i^\aaa$ des agents de la commune $\aaa$, i.e. $h_i^\aaa\rightarrow h_i + f^\aaa$, l'expression relevante devient $(h_i+f^\aaa+F)$. Dans cette dernière équation, il importe de voir que $h_i$ représente une idiosyncrasie d'un agent $i$ indépendante de la commune où il réside. La spécificité de la commune $\aaa$ est transcrite de nouveau par le terme $f^\aaa$, et l'agent $i$ perd toute spécificité liée à la commune où il réside.

Bref, nous avons opté, par pure convention, en faveur d'une idiosyncrasie des agents uniforme à l'échelle nationale, i.e. indépendante des lieux de résidence. Elle se note pour un agent $i$ d'une commune $\aaa$ comme $h_i$. La spécificité de la commune $\aaa$ est transcrite par $f^\aaa$ que nous appelons, faute de mieux, tendance spécifique de la commune $\aaa$. Cette tendance propre de la commune $\aaa$ peut ensuite être interprétée, soit comme un champ appliqué qui se surajoute au champ appliqué et global, $F$, soit comme la moyenne des idiosyncrasies des agents qui habitent la commune $\aaa$. 

Voyons maintenant comment écrire le terme d'interaction entre agents.\\

Il semble convenable que la probabilité de rencontre d'un agent avec un autre agent, augmente d'autant plus que la distance qui sépare leur lieu de résidence diminue. Dit autrement, la probabilité que deux agents entrent en relation diminue avec l'éloignement géographique des agents. (N'ayant aucune donnée sur les lieux de travail, de sortie, et autres déplacements des agents, la probabilité de rencontre des agents ne dépend ici que des lieux où ils résident $\ti$assimilés aux communes où ils sont inscrits sur les listes électorales.) En outre, nous pouvons faire l'hypothèse la plus simple possible liée à l'influence potentielle entre deux agents~: l'influence entre deux agents ne dépend que de leur rencontre, ou interaction. Nous écartons par là même des types de modèles dans lesquels la probabilité de s'accorder entre agents dépend de leur opinion préalable, et augmente d'autant plus que les opinions des deux agents sont proches, comme par exemple dans les modèles à opinion continue et confiance limitée (bounded confidence models) à la Deffuant-Weisbuch ou Hegselmann-Krause (voir~\cite{fortunato_stat_phys, lorenz-continuous-opinion} pour une vue d'ensemble). Nous ne prétendons pas par là nier leur validité potentielle, mais nous adoptons l'hypothèse la plus simple possible quant à la manière dont deux agents s'influencent. (Il faudrait entreprendre un tout autre travail pour appliquer aux données électorales les idées qui président aux modèles de type d'opinion continue à confiance limitée. Mais rappelons que le but de cette thèse n'est pas d'affirmer que les  \og choses se passent comme ça \fg, et donc, \og pas autrement \fg, mais de trouver un modèle plausible, et pas nécessairement le meilleur $\ti$terme flou$\ti$, qui puisse rendre compte des phénomènes.) Bref, nous convenons, d'une part qu'un agent puisse être influencé par un autre agent, du moment qu'ils se rencontrent, et d'autre part que la probabilité de rencontre décroît avec la distance géographique qui les sépare. Enfin, selon le RFIM standard, seul le choix (fait ou à faire) d'un agent peut influer sur un autre agent.

Il paraît vraisemblable que la probabilité de rencontre entre un agent $i$, et un autre agent $j$ décroisse exponentiellement en fonction de la distance qui les sépare, et avec une longueur caractéristique $\ell_c$. Comme précédemment, le lieu de résidence d'un agent dans une commune est ramené à la position (les coordonnées XY) de sa mairie. La probabilité de rencontre $\ti$et donc ici, d'interaction$\ti$ d'un agent $i$ dans une commune $\aaa$ avec un agent $j$ dans une commune $\beta$ distantes de $r_{\aaa \beta}$ est alors proportionnelle à 
$\exp(-\frac{r_{\aaa \beta}}{\ell_c})$. 

Ainsi, la somme des influences qu'un agent $i$ d'une commune $\aaa$ reçoit des autres agents $j$ s'écrit comme
\be \label{emva-brut} \mva = \frac{\sum_\beta \sum_{j=1}^{N^\beta} \exp(-\frac{r_{\aaa \beta}}{\ell_c}) \x S_j^\beta}{\sum_\beta \sum_{j=1}^{N^\beta} \exp(-\frac{r_{\aaa \beta}}{\ell_c})}.\ee
(Nous négligerons les termes en $1/\naa$, i.e. la valeur du choix de l'agent $i$ sur l'ensemble des $\naa$ choix de la commune.)

L'équation ci-dessus prend une forme plus simple en notant $\pab$, le poids, ou la contribution, de l'influence globale de la commune $\beta$ sur un agent $i$ d'une commune $\aaa$. Dit autrement, un agent $i$ d'une commune $\aaa$ reçoit l'influence globale des agents d'une commune $\beta$ avec un poids, ou une contribution, $\pab$. Ce poids, ou cette contribution, $\pab$ ne dépend ici que de la population de la commune $\beta$ et de la distance relative $r_{\aaa \beta}$ entre les deux communes, et s'écrit plus explicitement comme
\be \label{epab-pop} \pab = N^\beta \x \exp(-\frac{r_{\aaa \beta}}{\ell_c})~,\ee
ce qui conduit à l'expression générale~:
\be \label{emva} \mva = \frac{\sum_\beta \pab \x m^\beta}{\sum_\beta \pab}~.\ee

Nous aurions pu écrire directement $\mva$ sans faire intervenir $\pab$, mais comme nous le verrons ultérieurement, la forme de $\pab$ recèle une signification ouvrant droit aux interprétations.

Dans cet ordre d'idée, il paraît assez acceptable qu'un agent $i$ interagisse avec un nombre constant d'agents par commune, et donc indépendamment de la taille des communes $\ti$au coefficient d'atténuation près dû à l'éloignement, $\exp(-\frac{r_{\aaa\beta}}{\ell_c})$. Dans ce cas, $\pab$ devient
\be \label{epab} \pab = \exp(-\frac{r_{\aaa \beta}}{\ell_c})~,\ee
à inclure ensuite dans l'équation~(\ref{emva}) conduisant à $\mva$.

Chacun des termes qui intervient à l'intérieur d'une description basée sur le RFIM ayant été abordé, critiquons maintenant notre démarche.\\

Il est important de comprendre notre parti pris et ce que nous voulons faire ressortir. Le modèle utilisé comporte comme paramètres clés, le coefficient d'imitation, $J$, et une longueur caractéristique de prise en compte du voisinage, $\ell_c$. Les tendances spécifiques des communes $f^\aaa$ sont non corrélées, ou mieux, i.i.d. pour l'ensemble des communes $\aaa$, et les idiosyncrasies $h_i$ sont uniformément réparties à l'intérieur du territoire. Remarquons au passage le faible nombre d'ingrédients du modèle, ce qui permet de connaître facilement ses traits saillants. Nous voyons par là que ce modèle se propose de capturer des caractéristiques récurrentes des élections (voir section.~\ref{pt-sozein-ta-phainomena}) à partir d'une interaction microscopique locale~: une simple imitation du choix des agents du voisinage. Autrement dit, ce modèle a pour objectif de faire ressortir certains phénomènes électoraux (les phénomènes à sauver) à partir d'une imitation basique du choix des agents situés dans leur voisinage. C'est peut-être une mauvaise façon d'aborder le problème des résultats électoraux, mais au moins, cet angle d'attaque a le mérite d'être clairement posé, et mieux encore, il permet aux données d'y répondre $\ti$espérons-le$\ti$ clairement. 

La question qui se pose à nous, n'est donc pas de savoir si ce modèle correspond à une bonne façon d'aborder le problème ou non, s'il est plus ou moins plausible, etc., mais plutôt, s'il convient, s'il répond aux données, autrement dit s'il réussit sa confrontation aux données.

Après ce court $\ti$mais nécessaire$\ti$ laïus, écrivons explicitement le modèle du RFIM appliqué aux élections par commune.

\subsubsection{Écriture du modèle}
\label{pt-rfim-ecriture}
Le choix $S_i^\aaa$ d'un agent $i$ d'une commune $\aaa$, de tendance propre $f^\aaa$, de voter en faveur de l'un des deux choix proposés (ou bien de participer ou non à l'élection) s'écrit alors comme
\be \label{esia-choix} S_i^\aaa = \operatorname{sign} \big[h_i + f^\aaa + F + J \x \mva \big]~.\ee

Soient respectivement $p$ et $\pcum$, la densité de probabilité et la probabilité cumulative des idiosyncrasies individuelles $h_i$ $\ti$i.i.d. à l'échelle nationale selon la convention précédemment adoptée. ($\pcum[X] = \mathcal{P}[h_i > X] = \int_X^\infty p(h).\dd h$). Et en ne se préoccupant pas des tailles finies $\ti$question qui comme nous l'avons déjà évoqué affecte $\saa$ principalement$\ti$, le résultat d'une commune $\aaa$, ou son taux de participation à une élection, devient selon le RFIM standard~: 
\be \label{erfim-choix} m^\aaa = 2\:\pcum[-F - f^\aaa -J\x\mva] - 1~.\ee

Voyons maintenant quelles distributions statistiques utiliser pour décrire les idiosyncrasies $h_i$ et les tendances spécifiques communales $f^\aaa$.

Les raisons de généralité et de simplicité vont guider une nouvelle fois notre choix. Sachant que la distribution gaussienne~\cite{phys_stat} est la distribution statistique la plus générale possible (obtenue avec le théorème central limite, sous des conditions peu restrictives) ou la plus simple possible (qui maximise son entropie ou information de Shannon à variance finie), nous l'adopterons pour décrire les distributions $h_i$ et $F^\aaa$. Dans ce cadre, $F^\aaa=F+f^\aaa$ peut être vue comme une distribution gaussienne i.i.d. sur l'ensemble des communes $\aaa$, de moyenne $F$ et d'écart-type $\sigma_f$. Dit autrement, la distribution des $f^\aaa$ qui relate les fluctuations des $F^\aaa$, est une gaussienne de moyenne nulle et d'écart-type $\sigma_f$. Quant à la distribution $h_i$, i.i.d. sur l'ensemble des agents à l'échelle nationale, elle est de moyenne nulle comme nous l'avions précédemment discuté. L'écart-type $\sigma_h$ des $h_i$ est pris comme étant égale à un. Dans l'équation~(\ref{erfim-choix}), $J$ et $F^\aaa$ interviennent de façon relevante dans leur rapport respectif à $\sigma_h$ (i.e. $J/\sigma_h$ et $F^\aaa/\sigma_h$), ce qui incite à opter pour $\sigma_h = 1$ et à ne plus l'écrire par la suite. ($\frac{J}{\sigma_h}\rightarrow J$, $\frac{F}{\sigma_h}\rightarrow F$ et $\frac{\sigma_f}{\sigma_h}\rightarrow \sigma_f$.) L'équation~(\ref{erfim-choix}) s'écrit alors en utilisant la fonction d'erreur complémentaire, $erfc$, comme :
\be \label{erfim-choix-gauss} m^\aaa = \operatorname{erfc}\big[-\frac{F +f^\aaa +J.\mva}{\sqrt{2}}\;\big] - 1~.\ee

Ce modèle comporte donc quatre paramètres. Le premier, $J$, relate l'intensité de l'imitation des choix entre agents. Le deuxième, $F$, relève du champ national appliqué à l'ensemble des agents de toutes les communes. Le troisième, $\sigma_f$, dénote l'intensité des fluctuations, des particularités, des communes $\aaa$. Le quatrième, $\ell_c$, traduit la longueur caractéristique d'interaction entre agents des communes voisines. Notons que les conditions initiales peuvent jouer un rôle important lorsqu'il existe plusieurs équilibres possibles. Et bien entendu, l'analyse du modèle correspond aux solutions stationnaires qu'il procure.

\subsubsection{Réfutation du modèle et question ouverte}
\label{pt-refutation-rfim}
Le modèle RFIM écrit aux équations~(\ref{erfim-choix}) ou (\ref{erfim-choix-gauss}) possède une grave lacune : il ne s'accorde pas avec les données. En d'autres termes, les observations invalident ce modèle qui repose sur l'imitation du choix des agents.

L'argumentation s'appuie sur les deux points suivants : les corrélations des $\taa$ des résultats (ou des taux d'abstention) sont à longue portée, et la distribution des $\taa$ sur l'ensemble des communes est toujours unimodale et non fortement étalée. Le premier point suggère que le système se trouve proche d'un point critique $\ti$ce qui n'est pas difficile à assurer en jouant sur la valeur de $J$. (Le point critique obtenu avec $F^\aaa=F$ uniforme dans l'équation~(\ref{erfim-choix-gauss}), se caractérise par $J=\sqrt{\pi/2}\simeq 1,25$, $F=0$, et une moyenne $<m^\aaa>=0$ sur l'ensemble des communes, soit $<\resa> = 0,5$ et $<\taa>=0$.) A noter que la pente approximativement égale à un de la moyenne des $\taa$ en fonction de $\toa$ conforte ce premier point. Le second point impose une distribution toujours unimodale, y compris pour une moyenne $<\resa>\simeq0,87$ loin du point critique : c'est là que le bât blesse. L'annexe~\ref{annexe-refutation-rfim} établit plus scrupuleusement la réfutation de ce modèle en se servant du cas particulier $<\resa>\simeq 0,8$ rencontré dans les résultats de 2002-b et des taux de participation de plusieurs élections (voir Tabs.~\ref{tstat-dens-res} et \ref{tstat-dens-abst}). Cette annexe montre l'incapacité du modèle à produire simultanément : des corrélations à longue portée des $\resa$, une distribution des $\resa$ non bimodale (ni très étendue), et ce avec une moyenne d'ensemble proche $\res=0,8$.\\

Le principe de réfutabilité pouvant être considéré comme critère de scientificité par certains auteurs\footnote{Avec Karl Popper comme chef de file de ce courant de pensée.}, essayons d'aller plus loin dans notre démarche. Précisons que de notre côté, nous n'avons hélas pu réfuter une théorie\footnote{L'auteur de cette thèse plaide son incapacité à pouvoir l'entreprendre, fortement aidé par sa méconnaissance de théorie quantifiable sur le sujet, s'il y en a.}, mais simplement un modèle. Un modèle fondé sur l'imitation du choix des agents. Ce qui pousse naturellement à se poser la question : quid de l'imitation du choix des agents dans n'importe quel modèle~?

Mais auparavant donnons du corps à la réfutation du modèle écrit à l'équation~(\ref{erfim-choix-gauss}). Rappelons que l'argument principal se base sur un comportement similaire des données (entre tous les résultats électoraux, ou bien entre tous les taux de participation) quelle que soit la moyenne $<\resa>$ sur l'ensemble des communes. Ce qui rentre en contradiction avec la particularité, la singularité, imposée par le point critique du modèle. Avec $<\resa>$ loin du point critique, le modèle ne peut se comporter comme s'il était proche du point critique en $<\resa>=0,5$. (Un point critique néanmoins nécessaire pour faire apparaître des corrélations à longue portée avec ce modèle.)

Cette réfutation devient dès lors robuste à certaines modifications. Par exemple, changer de distribution statistique des idiosyncrasies ne déplacerait que la valeur $J$ du point critique. Que l'intensité de l'imitation $J$ varie d'une commune à l'autre (i.e. $J\rightarrow J^\aaa$) selon une gaussienne de moyenne $J$, n'altère pas non plus profondément les solutions obtenues avec un $J$ unique. (Un changement majeur peut éventuellement se produire si, à l'instar des verres de spin, $J^\aaa$ prend des valeurs négatives. Mais un tel modèle diffère alors du RFIM basé sur une interaction purement imitative.) De même, des simulations montrent que la fluctuation de l'un des différents paramètres autour de leur précédente valeur ($F\rightarrow F(t)$, ou bien $f^\aaa\rightarrow f^\aaa(t)$, ou bien $J\rightarrow J(t)$) ne modifie pas notablement les conclusions établies à l'annexe~\ref{annexe-refutation-rfim}.

Notons enfin que le champ appliqué, $F$, peut être assimilé à une imitation des agents à l'échelle nationale, i.e. $F\equiv J_0\x\langle m^\aaa\rangle$. Dans le même ordre d'idée, introduire une sorte de réseau $\pab$ des imitations du choix des agents des communes $\beta$ sur ceux des communes $\aaa$, non circonscrit aux plus proches communes (comme dans Eqs.~(\ref{epab-pop}), (\ref{epab})), ne devrait pas non plus modifier substantiellement les solutions précédemment établies si le réseau est construit sur des critères généraux. Nous l'avons vérifié dans un cas particulier et \og réaliste \fg, issu du réseau de type \og gravity model \fg{} observé empiriquement sur les probabilités de connexion et sur la durée des communications téléphoniques entre personnes de différentes communes belges~\cite{lambiotte_geographical, gravity_model}. Dans ce cas, 
\be \label{epab-grav} \pab = \frac{N^\beta}{(r_{\aaa \beta})^d},\ee
avec $d=2$ pour le \og gravity-model \fg. Une distance limite $r_c$, i.e. un cutoff, doit intervenir ici, tel que $\pab=\frac{N^\beta}{(r_c)^d}$ si $r_{\aaa\beta} < r_c$, et $\pab$ donné par Eq.~(\ref{epab-grav}) autrement. Nous avons choisi $r_c=3~km$ qui correspond environ à la distance moyenne entre deux communes.

Bien entendu, un réseau $\pab$ construit de façon \textit{ad hoc} afin de contenir l'ensemble des corrélations $\ti$y compris à longues distances$\ti$ entre différentes communes, ou l'existence \og réelle \fg{} d'un tel réseau $\ti$que nous n'avons pas été en mesure de percevoir$\ti$, pourrait permettre de retrouver les phénomènes à partir du RFIM avec imitation des choix. (Voir~\cite{dorogovtsev_crit_netw} pour une vue d'ensemble sur les modifications engendrées par un réseau sur un phénomène critique.) Mais dans ce cas, la principale information contenue dans le modèle proviendrait de la masse d'informations contenue dans le réseau, et non plus seulement du couple de paramètres $(J,\ell_c)$ du modèle utilisé jusqu'alors, et circonscrit à l'influence du voisinage. Nous nous en tenons là, et préférons pour l'heure que le modèle utilisé pour reproduire les phénomènes contienne le moins d'ingrédients possible. Un modèle dans lequel les corrélations observées à longue distance peuvent résulter des interactions locales à l'échelle microscopique, i.e. à l'échelle des agents. (Nous testerons par la suite la validité d'une influence restreinte au voisinage, et la comparerons avec celle décrite par Eq.~\ref{epab-grav}).) Nous restons par là même fidèle à l'objectif que nous nous sommes fixés~: une interaction simple $\ti$et donc locale$\ti$ peut-elle rendre compte des phénomènes~?  

Élargissons le débat.\\

Nous venons de réfuter le RFIM basé sur l'imitation locale du choix des agents. (Ici, les choix sont binaires : voter en faveur de l'un des deux candidats en présence d'une élection présidentielle, apposer un un oui ou bien un non lors d'un référendum, et enfin participer ou s'abstenir à une élection.) Cette réfutation a été étendue à de légères variantes de ce modèle. Nous venons donc de montrer que, dans le cadre du RFIM, et pour les élections étudiées, les agents n'imitent pas le choix des autres agents de leur voisinage. Autrement dit, en suivant le RFIM et toujours pour les élections analysées, l'imitation d'un agent ne procède pas d'un décompte des deux choix possibles réalisés $\ti$ou ayant l'intention d'être réalisés$\ti$ par les autres agents qu'il rencontre. Tout se passe comme si le nombre d'agents $\ti$\textit{stricto sensu}$\ti$ en faveur ou contre un choix possible n'a pas d'influence majeure\footnote{Le modèle utilisé n'essaie de relater, sous forme mathématique et statistique, que les principaux traits stylisés possibles d'un agent.} sur les choix réalisés. Plus explicitement et avec les précautions d'usage du \og tout se passe comme si \fg{}: qu'il y ait dans l'entourage d'un agent, $50$ et $50\%$ d'agents en faveur des deux choix possibles, ou bien, par exemple, $80$ et $20\%$, n'a pas, en tant que tel, d'importance sur son choix.

Que l'étude des données électorales permette de tirer une telle conclusion constitue en soi un résultat fort. Mais, le caractère inopérant de l'imitation directe du choix des agents ne concerne bien entendu que les élections traitées ici, et surtout, se place dans le cadre interprétatif d'un modèle\footnote{Nous n'exigeons d'un modèle que sa capacité à \og sauver les phénomènes \fg. La question relative à sa véracité ne se pose pas encore à nous, pour l'heur(e).} : le RFIM. Est-il maintenant possible de généraliser cette affirmation à tout autre modèle~? En d'autres termes, existe-t-il un modèle qui puisse rendre compte des phénomènes électoraux ici mesurés, dans lequel l'interaction entre agents ne repose que sur l'imitation locale de leurs choix~? Dit encore autrement, est-il possible qu'un modèle où les spins imitent les autres spins de leur entourage reproduise ces phénomènes électoraux~? (L'interaction locale entre spins seraient alors de type ferromagnétique $\ti$pour restituer l'imitation$\ti$, et les spins ne pourraient prendre que les valeurs $\pm 1$, de manière à transcrire un choix $+1$ ou $-1$.) On peut penser aux \og voter models \fg, aux \og majority rule models \fg, aux modèles de percolation, aux modèles épidémiologiques, etc. (voir notamment~\cite{fortunato_stat_phys, stauffer_sociophys, bettencourt_epidemiological, schweitzer_brownian} et leurs références). Par rapport aux élections étudiées ici, je ne le crois pas. Je ne le crois pas, et pour les mêmes raisons qui nous ont poussé à réfuter le RFIM : à cause de la particularité introduite par le point critique, en contradiction avec la régularité des phénomènes observés, indépendante de la moyenne nationale des résultats (ou des taux de participation). Je me hasarderai donc à croire cette affirmation quelque peu contre-intuitive : effectivement $\ti$i.e. pour n'importe quel modle$\ti$, tout se passe comme si les agents au cours de ces campagnes électorales ne sont pas directement influencés par la proportion des choix, ou des intentions de choix, des autres agents de leur voisinage. Mais par delà le simple fait peu scientifique d'y croire ou de ne pas y croire, la physique sociale gagnerait évidemment à ce qu'elle puisse affirmer une telle assertion.

Ou la réfuter.

\subsection{Imiter autrement}
\label{section-imiter-autrement}

Si l'interaction directe entre agents via leur choix ne peut rendre compte des phénomènes observés, sous quelle autre forme l'influence existe-t-elle~?

Comme nous l'avions annoncé au début de cette section, nous continuerons à adopter le même cadre conceptuel qui avait conduit au RFIM avec imitation des choix. Autrement dit, nous ne repartons pas de zéro dans la modélisation, mais nous voulons modifier l'équation~(\ref{esia-choix}) de manière à la rendre compatible aux données. Modifier, et non pas complexifier en y incluant d'autres ingrédients, puisque nous voulons des modèles simples, voire minimalistes.

Précisons aussi que nous continuons à vouloir répondre à la même question posée plus haut~: l'imitation des agents situés dans le voisinage, permet-elle de retrouver les phénomènes~? Nous venons de voir que l'imitation des choix, ne le permet pas. Voyons maintenant si d'autres formes d'imitation le permettent. 

\subsubsection{Imitation de la conviction}
\label{pt-conviction}
Partons de l'équation~(\ref{esia-choix}) donnant le choix de l'agent $i$ de la commune $\aaa$, écrite sous la forme plus générale~:
\be \label{esia-gal} S_i^\aaa = \operatorname{sign} \big[h_i + f^\aaa + F + J \x \iva \big]~,\ee
où $\iva$ représente l'influence que reçoit un agent d'une commune $\aaa$. (Précédemment, $\iva$ était égal à $\mva$, à cause de l'imitation des choix.) Nous cherchons ici, une façon différente d'exprimer le terme d'influence $\iva$, autrement dit une façon qui ne repose pas sur une imitation du choix des agents.

Dans l'équation ci-dessus, le choix $S_i^\aaa=\pm 1$ de l'agent $i$ dans la commune $\aaa$ résulte d'une réduction de ce que nous appelons sa conviction~:
\be \label{ecia} c_i^\aaa = h_i + f^\aaa + F + J\x \iva ~.\ee
(Si $c_i^\aaa>0$ alors $S_i^\aaa=+1$, ou $-1$ dans le cas contraire.) Notons que ce que nous appelons conviction $\ti$au regard du choix à faire$\ti$ est une grandeur réelle non bornée et à une seule dimension.

Il semble convenable d'imiter non pas les choix $S_j^\beta=\pm 1$ des agents $j$ dans les communes $\beta$, mais leur conviction $c_j^\beta$. Pourquoi~?

Premièrement, à cause de l'élection en tant que telle~: une décision à prendre de façon quasiment synchrone (le même jour pour tous les agents). Les agents ne peuvent alors imiter les choix $\pm 1$ des autres agents, puisqu'ils n'ont pas encore été réalisés, ou ne le connaissent pas. Le caractère secret que le vote peut avoir, renforce la difficulté à imiter les choix lors d'une élection puisqu'ils ne sont pas forcément connus. Tous les électeurs ne disent pas non plus de façon certaine pour qui ils vont voter. Deuxièmement, notamment dans une discussion, une personne ne se réduit pas à un choix $\pm 1$. Il paraît vraisemblable qu'un individu à forte conviction puisse influencer davantage qu'un autre individu peu convaincu, mais faisant tous deux le même choix le jour de l'élection. (Cette vraisemblance n'est pas partagée par les adeptes des modèles à opinion continue et confiance limitée.) Ainsi, l'imitation procède des agents eux-mêmes, i.e. de leur conviction, avant leur réduction en choix $\pm 1$ le jour de l'élection. (Cette réduction de la conviction au choix $\pm 1$ s'apparente à une opération de mesure, à une \og réduction du paquet d'ondes \fg{} se laisseraient à dire les chantres de la mécanique quantique.)\\

Voyons maintenant comment s'écrit le modèle où l'imitation porte sur la conviction des agents, et non plus sur leur choix.

$\iva$ s'obtient de manière identique à ce qui avait été fait pour $\mva$ (cf. Eq.~(\ref{emva-brut})), soit
\be \label{eiva} \iva = \frac{\sum_\beta \sum_{j=1}^{N^\beta} \exp(-\frac{r_{\aaa \beta}}{\ell_c}) \x c_j^\beta}{\sum_\beta \sum_{j=1}^{N^\beta} \exp(-\frac{r_{\aaa \beta}}{\ell_c})}.\ee

Notons $\bar{h}^\aaa$, la conviction moyenne des agents de la commune $\aaa$. Cette grandeur mesurable s'exprime comme
\be \label{eresa-convic} \resa=\pcum[-\bar{h}^\aaa]~,\ee
puisque $\resa = \mathcal{P}[h_i+\bar{h}^\aaa > 0]$. Rappelons que les idiosyncrasies, $h_i$, uniformément réparties sur l'ensemble des communes, ont une moyenne nulle.

Avec $c_i^\aaa=h_i+\bar{h}^\aaa$ mis dans Eq.~(\ref{eiva}), et $c_i^\aaa$ donné par Eq.~(\ref{ecia}), il vient~:
\be \label{econvic} \bar{h}^\aaa = K^\aaa\cdot\bar{h}^{\mathcal{V}_\aaa} + (1-K^\aaa)\frac{f^\aaa + F}{1-J}~,\ee
où
\be \label{ehva-convic} \bar{h}^{\mathcal{V}_\aaa} = \frac{\sum_{\beta\neq\aaa} \pab\cdot\bar{h}^\beta}{\sum_{\beta\neq\aaa} \pab}~,\ee
dans laquelle $\pab$ est défini par l'équation~(\ref{epab-pop}), et
\be \label{eK-convic} K^\aaa = \frac{J}{1+x^\aaa\:(1-J)}\quad \xrightarrow[x^\aaa \ll 1]{}\quad K = J ~,\ee
où $x^\aaa$ traduit le poids relatif des agents de la commune $\aaa$ par rapport à tous les autres, et s'exprime comme
\be \label{xaa} x^\aaa = \frac{p_{\aaa\rightarrow\aaa}}{\sum_{\beta\neq\aaa} \pab}~.\ee

Plus $\ell_c$ augmente, plus $x^\aaa$ diminue et varie peu d'une commune à une autre, en accord avec la plus grande uniformisation imposée par les plus grandes échelles de longueur. Dans la pratique, pour $\ell_c\gtrsim4,5~km$, tous les $K^\aaa$ prennent environ la même valeur $K$. Noter enfin que si $J=0$, alors $K^\aaa=K=0$, et si $J=1$, $K^\aaa=K=1$ et $\frac{1-K^\aaa}{1-J}=1+x^\aaa$. Enfin, le passage de $\bar{h}^\aaa$ au résultat électoral (ou au taux de participation) de la commune utilise l'équation~(\ref{eresa-convic}).

Avant de mieux comprendre ce qu'implique ces équations, voyons une autre façon possible de traduire l'imitation.

\subsubsection{Imitation indirecte via les lieux de résidence}
\label{pt-ressemble-s'assemble}
Dans la partie ci-dessus nous avons écrit le terme d'influence $\iva$ d'une façon différente de celle due à l'imitation des choix. Dans les deux cas, l'imitation des choix ou l'imitation des convictions, $\iva$ peut être considérée comme une influence directe~: celle que reçoit un agent $i$ d'une commune $\aaa$ de par les agents de son voisinage. Nous allons maintenant nous tourner vers une autre forme d'influence, indirecte celle-ci.

Pour cela, nous commençons par simplifier l'équation générale (\ref{esia-gal}) en lui supprimant le terme de l'influence directe des agents du voisinage $\iva$. Le choix d'un agent $i$ qui réside dans la commune $\aaa$ s'écrit alors $S_i^\aaa = \mathcal{P}[h_i + F^\aaa > 0]$, où $F^\aaa=F+f^\aaa$. Ce qui amène tout simplement pour une commune $\aaa$ à 
\be \label{eresa-fa} \resa = \pcum[-F^\aaa] ~.\ee

Essayons maintenant de mieux comprendre le rôle de $F^\aaa=F+f^\aaa$ dans l'équation générale (\ref{esia-gal}). Nous avons vu à la section~\ref{pt-rfim-termes} que la dénommée tendance spécifique, $f^\aaa$, de la commune $\aaa$ pouvait s'interpréter, soit comme un champ appliqué qui se surajoute à un champ global $F$, soit comme un déplacement de la moyenne des idiosyncrasies des habitants de la commune $\aaa$. N'entrons pas dans des querelles de Byzantins pour savoir laquelle de ces deux interprétations convient. Peu importe, rien ne change mathématiquement.

La discussion précédente peut s'étendre de $f^\aaa$ à $F^\aaa=F+f^\aaa$. Ainsi, d'une manière ou d'une autre, $F^\aaa$ traduit mathématiquement la tendance globale des agents de la commune $\aaa$ à se positionner sur le choix demandé. Écrite selon la notation de la section~\ref{pt-conviction}, $F^\aaa=\bar{h}^\aaa$ et s'interprète comme la moyenne des convictions des agents de la commune $\aaa$.  Peu importe ensuite que cette tendance globale, ou moyenne des convictions, puisse être expliquée par l'histoire des gens de la commune, par un état de fait qui résulte de la façon dont la commune est peuplée, par une réponse spécifique des agents de la commune face au choix demandé, par le rôle de certaines personnes, de certains établissements, par intérêt stratégique global, etc. Il est essentiel de remarquer que $F^\aaa$ traduit, certes une tendance globale des agents de la commune (terme flou), mais, surtout, un aspect qui engage l'ensemble des agents de la commune. Une tendance incarnée ou subie (selon les deux interprétations de $f^\aaa$), bref, une tendance partagée par l'ensemble de ses agents. Peu importe ensuite l'origine, ou l'explication, ou encore l'interprétation de $F^\aaa$. Il manifeste à nos yeux l'expression mathématique d'une tendance globale des agents d'une commune. (A ce stade $F^\aaa$ traduirait donc l'hypothétique proverbe~: \og qui s'assemble se ressemble. \fg) 

Il serait des plus surprenants que cette tendance s'arrête aux limites de la commune $\ti$à l'encontre d'un certain nuage radioactif qui n'aurait pas franchi les frontières d'un certain pays, il y a de cela quelques années déjà$\ti$ et que les communes contiguës ne partagent pas, à un certain degré, cette tendance globale. (Un type géographique, rural, urbain, etc., ne se limite pas souvent à une seule commune, semble-t-il.)\\

Traduisons cela en termes mathématiques.

Il paraît vraisemblable que la tendance globale, $F^\aaa$, de la commune $\aaa$ partage, à un bruit près, la même tendance globale que ses voisines, notée $\fva$, ainsi qu'une tendance générale ou nationale, notée $\overline{F}$. Ainsi,
\be k_1\cdot (\fva - F^\aaa) + k_2\cdot(\overline{F}-F^\aaa) = \eta^\aaa ~,\ee
où $k_1$ et $k_2$ sont deux réels positifs (à l'instar des constantes de rappel des ressorts), et $\eta^\aaa$, une variable aléatoire décorrélée $\ti$et toujours pour la même raison, afin que l'écriture minimaliste ci-dessus ne contienne pas d'autre information résiduelle ou cachée.

L'équation ci-dessus s'écrit tout simplement comme
\be \label{efa-gal} F^\aaa = K\cdot\fva + (1-K)\cdot\overline{F} + \eta'^\aaa ~,\ee
où $K=\frac{k_1}{k_1+k_2}$ et $\eta'^\aaa=\frac{-\eta^\aaa}{k_1+k_2}$. Il est important de noter que $\fva$ capture ici les effets dus à la proximité, i.e. les interactions avec le voisinage.\\

Après ces longues tergiversations sur la tendance globale $F^\aaa$ d'une commune, sur son rapport au voisinage, nous voilà avec les mêmes équations que celles obtenues pour l'imitation directe de la conviction (cf. Eqs.~(\ref{eresa-convic}-\ref{econvic}) à comparer avec Eqs.~(\ref{efa-gal}-\ref{eresa-fa})). Tout ça pour rien~?

Il est d'une part rassurant de retrouver les mêmes équations pour les tendances globales des communes que pour celles obtenues à partir de l'imitation de la conviction. La tendance globale $F^\aaa$ peut effectivement provenir d'un partage, d'une imitation, des convictions des agents de la même commune. Mais plus intéressant encore, l'approche par tendance générale permet de comprendre plus naturellement une autre façon d'imiter que celle qui procède de l'imitation directe des convictions.

L'imitation des convictions des agents rend naturellement l'influence d'une commune $\beta$ sur un agent de $\aaa$, proportionnelle à la population de $\beta$. En effet, l'équation~(\ref{ehva-convic}) fait intervenir $\pab$, défini par Eq.~(\ref{epab-pop}), proportionnel à $N^\beta$. Est-il possible d'obtenir un terme d'influence indépendant de la population~?

Oui, savons-nous, si l'agent $i$ de la commune $\aaa$ interagit avec des agents indépendamment de la taille des communes (cf. Eq.(\ref{epab})). Autrement dit, s'il interagit, à distances égales des communes, avec le même nombre d'agents par commune, et ce, quelle que soit leur taille. Qui plus est, avec des agents aux convictions aléatoirement réparties, pour que Eq.~(\ref{ehva-convic}) reste valable. Bien que plausible, cette argumentation paraît assez artificielle. Autrement dit, il paraît plus vraisemblable qu'un modèle reposant sur une imitation de la conviction fasse intervenir la population des communes, plutôt que non~; et donc décrive $\pab$ par Eq.~(\ref{epab-pop}) plutôt que par Eq.~(\ref{epab}).

Les discussions ci-dessus au sujet des tendances globales $F^\aaa$ des communes vont pouvoir maintenant jouer leur rôle, et ce pour mieux appréhender une imitation indépendante du nombre d'agents.\\

Il paraît vraisemblable qu'une tendance globale d'une commune puisse s'étendre à ses voisines indépendamment de sa population. Cette diffusion de la tendance proviendrait alors, davantage de sa localisation, de sa position spatiale, que du nombre de ses habitants. Autrement dit, au regard de cette tendance globale et partagée par l'ensemble des agents, les lieux de résidence importent plus que le nombre d'habitants. Ce genre de tendance globale, incarnée ou subie, par l'ensemble des agents d'une commune s'apparente $\ti$certes, assez lâchement$\ti$ au proverbe bien connu, \og qui se ressemble s'assemble \fg. Il traduit simplement que les habitants d'une zone géographique donnée partagent en commun une même tendance, indépendante du nombre de sa population, et non circonscrite aux limites de la commune. Nous pourrions peut-être y voir des zones rurales, urbaines, résidentielles, d'immigration, etc. De plus, ce partage de tendance globale peut-être considéré non pas comme une influence directe des choix ou des convictions, mais plutôt comme une influence indirecte entre agents $\ti$si le terme d'influence convient encore$\ti$, autrement dit comme un point commun partagé par les agents, sans pour autant qu'ils se convainquent directement. Ce que nous appelons imitation, interaction, influence, se conçoit ici comme l'expression d'une simple similarité $\ti$en rapport aux choix à faire lors des élections$\ti$ des agents, due à la contiguïté des communes dans lesquelles ils résident.

Bref, il ressort des discussions précédentes qu'un comportement commun entre zones voisines, et indépendant du nombre de population, se comprend facilement par l'intermédiaire d'une tendance globale partagée par les habitants qui y résident. Ce comportement commun $\ti$qui donne la primauté aux lieux de résidence sur le nombre de population$\ti$ a l'aspect d'une influence collective entre agents, sans pour autant qu'il y ait une influence directe entre agents de leurs convictions, ou de leurs choix. Cette forme d'influence $\ti$ou d'interaction, ou d'imitation$\ti$ indirecte sera notée, faute de mieux, influence indirecte via les lieux de résidence, et exprime $\pab$ par Eq.~(\ref{epab}), plutôt que par Eq.~(\ref{epab-pop}).

Nous voilà plongés dans une nouvelle querelle de Byzantins, dans de nouvelles exégèses, dans des interprétations sans fin... diriez-vous. Certes. Mais nous voulions distinguer deux types possibles de ce que nous appelions, imitation, influence, interaction. Il est temps maintenant de se résumer et de tester ces genres d'imitation.

\subsection{Modèle à tendances communales statiques}
\label{section-statique}
Un modèle dit à tendances communales statiques, ou plus simplement à tendances statiques, permet d'écrire les deux variantes ci-dessus de ce que nous appelions imitation. Il se rapporte à $F^\aaa$, la tendance globale, ou le champ appliqué global, ou la moyenne des convictions des agents, de la commune $\aaa$ tel que
\be \label{efaa} F^\aaa = K\x\fva + f^\aaa ~,\ee
où le choix binaire de l'agent $i$ de cette commune s'écrit comme $S_i^\aaa=\mathcal{P}[h_i+F^\aaa > 0]$. Le résultat (ou taux de participation) de la commune $\aaa$ est $\resa=\pcum[-F^\aaa]$, comme à l'équation~(\ref{eresa-fa}).

Il est important de noter que les tendances spécifiques aux communes, notées $f^\aaa$, doivent être décorrélées entre elles si nous voulons que les traits saillants de ce modèle reproduisent les données $\ti$i.e. que le modèle ne contienne pas d'information résiduelle qui parasiterait le crédit donné à un éventuel accord du modèle avec les phénomènes observés. Enfin, il paraît vraisemblable que les particularités propres aux communes, $f^\aaa$, puissent être considérées comme stationnaires tout au long du processus qui aboutit à l'élection. En résumé, dans ce modèle, comme dans le précédent avec imitation des choix, les tendances communales sont aux communes ce que les idiosyncrasies sont aux agents~: i.i.d. (et donc non corrélés), et fixes au cours du temps considéré.

Dans ce modèle, $\fva$ contient l'influence $\ti$au sens large$\ti$ du voisinage sur un agent d'une commune $\aaa$, autrement dit $\fva$ exprime l'une des formes d'imitation vue ci-dessus. Il s'écrit comme
\be \label{efva} \fva = \frac{\sum_{\beta\neq\aaa} \pab\x F^\beta}{\sum_{\beta\neq\aaa} \pab}~,\ee
où $\pab$ dénote le poids relatif de l'influence de la commune $\beta$ sur la commune $\aaa$.\\

Remarques~: 
\begin{itemize}
\item nous aurions pu écrire $F^\aaa = K\x\fva + (1-K)\x\overline{F}+f^\aaa$, équation dans laquelle $\langle F^\aaa\rangle = \overline{F}$ et $\langle f^\aaa \rangle = 0$, où $\langle ...\rangle$ signifie la moyenne sur l'ensemble des communes. Mais nous verrons plus bas que modifier la moyenne d'ensemble, $\langle F^\aaa\rangle = \overline{F}$, n'a aucune importance sur les propriétés du système. Par la suite, nous imposerons au système (lors des simulations ou de son analyse théorique) une moyenne nulle de ses tendances spécifiques, soit $\langle f^\aaa\rangle = 0$.
\item nous aurions dû attribuer un coefficient d'interaction, $K^\aaa$, à chaque commune $\aaa$ comme à l'équation~(\ref{econvic}). Mais dès que $\ell_c$ est suffisamment grand ($\ell_c\gtrsim4,5~km$) alors $K^\aaa\simeq K$, à cause du lissage à l'échelle de $\ell_c$ des disparités de la répartition locale de population. Par la suite, $K$ sera assimilé à une valeur uniforme pour toutes les communes $\aaa$, ce qui néglige de la sorte les hétérogénéités locales de la répartition des communes ou des population, à l'échelle de $\ell_c$.   
\end{itemize}
\vspace{0.25cm}

Il faut maintenant distinguer deux variantes du modèle.
\begin{itemize}
\item Dans le cas où l'interaction avec le voisinage provient d'une imitation dite de conviction, $\pab$ prend la forme donnée à l'équation~(\ref{epab-pop}), soit~: $\pab=N^\beta\x\exp(-\frac{r_{\aaa\beta}}{\ell_c})$.
\item Dans le cas où l'influence provient de la dénommée influence indirecte via les lieux de résidence, $\pab$ prend la forme donnée à l'équation~(\ref{epab}), soit~: $\pab=\exp(-\frac{r_{\aaa\beta}}{\ell_c})$.
\end{itemize}
\vspace{0.25cm}

Avant d'analyser l'équation~(\ref{efaa}) correspondant au modèle des tendances communales statiques, il sera intéressant de remarquer le point suivant, qui reliera directement $F^\aaa$ à $\taa$ utilisé tout au long de la première partie de cette thèse. En choisissant de nouveau les idiosyncrasies $h_i$, i.i.d. selon une distribution gaussienne de moyenne nulle et de variance unité, il vient $\resa = \frac{1}{2}\operatorname{erfc}[\frac{-F^\aaa}{\sqrt{2}}]$. Si nous avions utilisé une distribution logistique des idiosyncrasies, de moyenne nulle et de variance unité, nous obtiendrions $\resa = \frac{1}{1+exp(-\pi F^\aaa/\sqrt{3})}$. Ce qui aurait conduit à identifier la grandeur $\taa$ (cf. Eq.~(\ref{etau})) utilisée jusqu'alors à la tendance globale $F^\aaa$ de ce modèle. Plus précisément et avec les conventions adoptées, $\taa$ aurait été égal à $F^\aaa$ à un coefficient $\frac{\pi}{\sqrt{3}}$ multiplicatif près.

Nous maintiendrons néanmoins une distribution gaussienne des idiosyncrasies, en accord avec les considérations de simplicité et de généralité développées précédemment (cf. section~\ref{pt-rfim-ecriture}). Et puisque les distributions gaussienne et logistique ne sont pas trop différentes l'une de l'autre, nous retiendrons que
\be \label{etaa-faa} \taa \simeq \frac{\pi}{\sqrt{3}}\;F^\aaa ~.\ee 

\subsubsection{Analyse et conditions de stabilité du modèle}
\label{pt-statique-analyse}
$K$ exprime l'intensité des effets dus au voisinage. Si $K~=~1$, l'équation précédente se rapproche d'une équation de diffusion en présence d'un champ aléatoire, $f^\aaa$, fixe et localisé dans chaque commune $\aaa$. $K > 0$ signifie une tendance à la concordance entre communes voisines, alors que $K < 0$ impliquerait, au contraire, une tendance à l'opposition.

L'annexe~\ref{annexe-continu-stabilite} étudie les conditions de stabilité de l'équation~(\ref{efaa}) $\ti$en utilisant la transformation de Fourier de cette équation linéaire. Il ressort de cette annexe que les solutions sont toujours stables à temps long~:
\begin{itemize}
\item avec $K<1$;
\item avec $K=0$, mais en ajoutant la condition $\sum_\aaa f^\aaa = 0$.
\end{itemize}
\vspace{0.25cm}

Le modèle des tendances communales statiques, défini par les équations~(\ref{efaa}-\ref{efva}), est un modèle linéaire. Ceci facilitera par la suite son étude analytique. Mais plus intéressant encore, cette linéarité empêche l'existence d'un point critique. Et nous avons déjà vu, lors de la réfutation du modèle RFIM avec imitation des choix, toute l'importance que revêt l'existence ou non d'un point critique. (Ce qui ne signifie pas non plus, bien évidemment, que tout modèle sans point critique est, ici, un bon modèle.)

Le système comporte \textit{a priori} trois paramètres, ajoutées à cela les conditions initiales de $F^\aaa$. Les trois paramètres sont, l'intensité de couplage $K$, la longueur caractéristique $\ell_c$ de prise en compte des distances des communes voisines et l'écart-type $\sigma_f$ des $f^\aaa$. Puisque le système est linéaire, $\sigma_f$ n'a aucun effet sur les corrélations spatiales $C(R)$ engendrées par le système. Les conditions initiales ne jouent aucun rôle $\ti$dans les solutions d'équilibre stable, les seules prises en compte comme nous l'avons déjà discuté dans l'annexe~\ref{annexe-continu-stabilite}$\ti$, hormis un décalage global de la moyenne de l'ensemble des $F^\aaa$. Remarquons au passage que les observations faites sur les résultats (ou les taux de participation) des élections, attestent des caractéristiques indépendantes de la moyenne des $\taa$, qui comme nous l'avons vu, a une très forte accointance avec $F^\aaa$. Enfin, La moyenne des $f^\aaa$ peut être réinjectée, à un coefficient près, dans celle, sans importance, des $F^\aaa$ $\ti$excepté si $K=1$, auquel cas cette moyenne doit être nulle, pour s'assurer de la stabilité du système. Indiquons alors que la moyenne des $f^\aaa$ sera toujours nulle par la suite, dans les simulations que nous ferons ainsi que dans l'étude analytique du modèle.

En résumé, le modèle ci-dessus ne contient que deux paramètres pertinents~: le coefficient de couplage, ou de l'intensité d'interaction avec le voisinage, $K$, et la distance caractéristique de la prise en compte de la proximité, $\ell_c$. Enfin, notons le caractère avantageux d'un modèle à deux paramètres pertinents, puisqu'il facilite grandement son étude et sa comparaison avec les données. 

Essayons maintenant de tester, ne serait-ce qu'imparfaitement, laquelle des deux variantes du modèle convient le mieux aux données. Serait-ce la variante qui provient de l'imitation dite des convictions, ou l'autre qui se réfère à une imitation dite indirecte et via les lieux de résidence~?
Nous profiterons de l'occasion pour tester, ne serait-ce encore qu'imparfaitement, la validité de l'hypothèse utilisée jusqu'alors~: la restriction de l'influence au voisinage géographique.

\subsubsection{Le classement des variantes}
\label{pt-statique-test}
Nous allons tester et classer entre-elles, trois variantes de ce modèle. Chacune d'elles porte sur l'expression de $\pab$ de l'équation~(\ref{efva}) du modèle.

La première variante interprète préférentiellement le modèle en terme d'une imitation dite indirecte et qui passe par les lieux de résidence. Ou peu importe l'interprétation, $\pab$ $\ti$qui représente le poids relatif de l'influence d'une commune $\beta$ sur une autre commune $\aaa$ $\ti$ est donné ici par l'équation~(\ref{epab}), indépendante des populations des communes. En outre, seules les communes $\beta$ proches de la commune $\aaa$ sont prises en compte (cf. la décroissance exponentielle en distance).

Avec la deuxième variante, $\pab$, dont l'expression est donnée par l'équation~(\ref{epab-pop}), est proportionnelle à la population des communes et ne prend encore en compte que le voisinage. Une interprétation en terme d'imitation des convictions des agents du voisinage de $\aaa$ $\ti$et ce, avec une imitation basique, sans aucune particularité$\ti$ semble ici la plus convenable.

Enfin la troisième variante, donnée par l'équation~(\ref{epab-grav}) pour $\pab$, correspond encore à une imitation des convictions $\ti$puisque $\pab$ est proportionnel à la population des communes$\ti$, mais ici, non circonscrite aux agents des communes voisines de $\aaa$. Cette écriture provient d'une extrapolation des observations empiriques~\cite{lambiotte_geographical, gravity_model} des probabilités d'appel, et de la durée des communications téléphoniques, entre personnes de différentes villes belges. $d=2$ a été observé par les auteurs des deux articles ci-dessus. Ici, la valeur de l'exposant $d$ testé peut prendre des valeurs comprises entre $0$ et $10,5$ par pas de $0,5$. $d=0$ signifie que la distance réelle entre deux communes n'a strictement aucune importance, et plus $d$ augmente, plus l'importance des communes voisines de $\aaa$ augmente par rapport aux communes lointaines. Nous testons, par ce biais, la pertinence de l'hypothèse dans laquelle nous nous sommes volontairement cantonnés ici~: restreindre l'imitation, l'influence au voisinage.

L'annexe~\ref{annexe-test} teste et compare ces trois variantes, et ce à partir des résultats (ou des taux de participation) des élections. Le critère à la base du test est l'absence de corrélations des tendances spécifiques $f^\aaa$. Avec un modèle idéal, les corrélations spatiales des $f^\aaa$ devraient être nulles comme nous l'avons écrit plus haut. Dans cette annexe, les $f^\aaa$ sont estimés à partir des données, ce qui permet ensuite de calculer leur corrélation spatiale afin d'évaluer leur degré d'indépendance. Noter enfin que cette méthode convient particulièrement à un modèle linéaire sans point critique.\\

Il ressort principalement de cette analyse les deux points suivants~:
\begin{itemize}
\item Les zones d'imitation paraissent être limitées au voisinage;
\item Le modèle interprété comme imitation indirecte via les lieux d'habitation semble le plus convenable des trois.
\end{itemize}

Bien évidemment, ces conclusions sont d'autant plus catégoriques que le modèle convient aux données. Même s'il semble assez convenable (le rapport des corrélations spatiales des $f^\aaa$ estimés sur celles des résultats réels est de l'ordre de $10^{-2}$), nous verrons par la suite qu'un autre modèle convient mieux aux données. Bref, bien que l'analyse ci-dessus ne soit pas catégorique, elle fournit néanmoins des informations non négligeables quant à la possibilité de restreindre l'influence au voisinage, et au meilleur accord des données avec $\pab$ exprimé par l'équation.~(\ref{epab}).

Un facteur de qualité lié à l'indépendance des $f^\aaa$ estimés, n'est pas la seule source possible de confrontation du modèle avec les données. Il existe une autre voie associée aux corrélations spatiales, et qui se révélera par la suite particulièrement efficace et discriminante.

\subsubsection{Corrélations spatiales, théoriques et numériques}
\label{pt-statique-correl}
Nous commencerons l'étude des corrélations spatiales produites par ce modèle par leur calcul analytique, mais fondé sur une forte simplification. Ensuite, elles seront établies à partir de simulations numériques, mais sans simplification. Nous pourrons alors comparer les formes obtenues des corrélations spatiales avec celles provenant des données, en vue de retenir ou non ce modèle.\\

Les équations~(\ref{efaa}-\ref{efva}) qui caractérisent ce modèle sont linéaires. Elles se prêtent donc aisément à un traitement avec les transformées de Fourier. Le calcul analytique peut aboutir assez facilement à la condition de considérer une répartition uniforme des communes (ou une répartition uniforme des populations dans le cas où Eq.~(\ref{epab-pop}) définit $\pab$).

\begin{figure}[t]
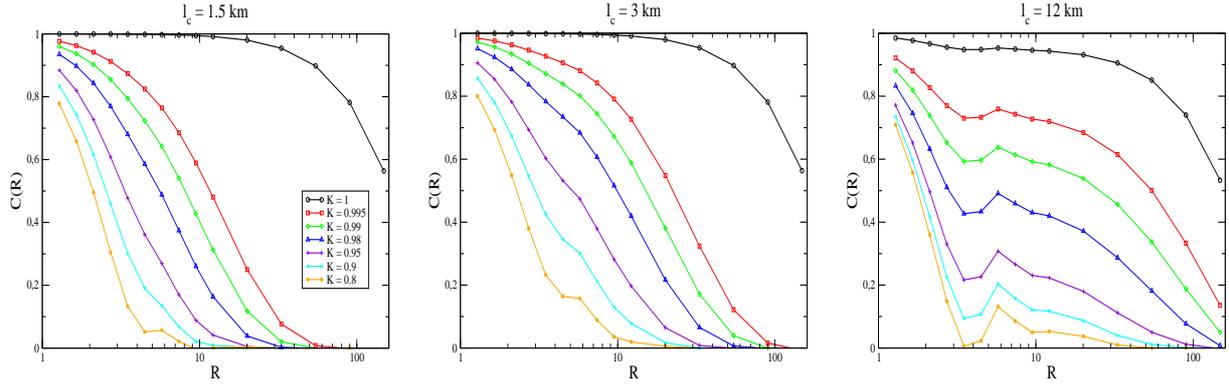

\includegraphics[width=5cm, height=5cm]{cor-statique-th-0.5.eps}\hfill
\includegraphics[width=5cm, height=5cm]{cor-statique-th-1.eps}\hfill
\includegraphics[width=5cm, height=5cm]{cor-statique-th-4.eps}
\caption{\small Corrélations théoriques obtenues avec le modèle des tendances statiques et une répartition uniforme des communes, voire si nécessaire des populations.}
\label{fcor-th-statique}
\end{figure}

L'annexe~\ref{annexe-correl-th} calcule explicitement les corrélations spatiales $C(R)$ du modèle en fonction de ses deux paramètres pertinents, $K$ et $\ell_c$ (cf Eqs~(\ref{ecor-sol}, \ref{eiq-statique})). Les figures~\ref{fcor-th-statique} tracent ces corrélations pour différentes valeurs de $K$ et de $\ell_c$, à partir des équations~(\ref{ecor-sol}, \ref{eiq-statique}). Dans ces figures, la distance caractéristique de la taille du système (i.e. de la France) vaut  $L=500~km$, et la distance caractéristique minimale (i.e. la distance caractéristique de la commune) vaut $a=2~km$. Nous utilisons la même notation que celle de la section~\ref{pt-cor-spatial-res}, à savoir $R=\frac{r}{D}$ (comme aux figures~\ref{fcorrel-res} et \ref{fcorrel-abst}), avec $r$ en $km$ et $D$, la distance moyenne entre deux communes d'un département, qui est constante ici et vaut $2,7~km$.

Bien que les distances qui interviennent dans le tracé des corrélations spatiales théoriques s'appuient sur les distances réelles ($L$, $a$ et $D$) du système étudié, ces figures ne doivent pas être prises en compte rigoureusement. Elles reposent en effet sur une répartition uniforme des communes, voire des populations. Il suffira de se reporter aux figures~\ref{frepartition-taille}, pour se persuader, si besoin est, du caractère non rigoureux de cette hypothèse. (A noter que l'hypothèse de la répartition uniforme de la population est nettement plus fallacieuse que celle de la répartition uniforme des communes $\ti$à cause des lois de puissance qui y interviennent~\cite{repartition_pop}). Néanmoins elles fournissent une bonne indication sur ce que devraient être les corrélations spatiales en l'absence de fortes hétérogénéités dans la répartition des communes, et si nécessaire des populations.

Passons maintenant aux corrélations spatiales provenant des simulations numériques du modèle.\\

\begin{figure}
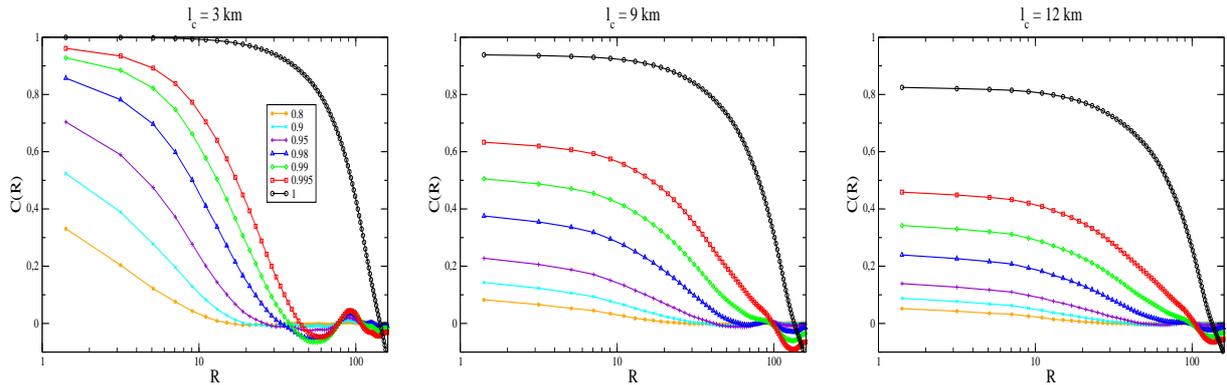

\includegraphics[width=5cm, height=5cm]{cor-statique-1.eps}\hfill
\includegraphics[width=5cm, height=5cm]{cor-statique-3.eps}\hfill
\includegraphics[width=5cm, height=5cm]{cor-statique-4.eps}
\caption{\small Corrélations simulées obtenues avec le modèle des tendances statiques, pour différentes valeurs de $K$ et de $\ell_c$. (Voir Figs.~\ref{fcorrel-res} et \ref{fcorrel-abst} pour les corrélations spatiales des résultats et des taux de participation, ainsi que pour la notation.)}
\label{fcor-statique}
\end{figure}

Les figures~\ref{fcor-statique} tracent les corrélations spatiales obtenues numériquement à partir du modèle. Elles correspondent aux solutions stables du modèle. (Si $K=1$, il faut au préalable contraindre $\sum_\aaa f^\aaa = 0$ pour que les solutions ne divergent pas.) Les tendances très générales dessinées par l'étude théorique se retrouvent ici. A savoir, si $\ell_c$ diminue, alors la portée de corrélations augmente, et la valeur des corrélations avec les proches communes diminue. Et naturellement, si l'intensité d'interaction $K$ augmente, alors augmentent aussi la portée des corrélations ainsi que la valeur des corrélations avec les communes voisines. De plus et en n'étant pas trop exigeant, excepté aux faibles distances, l'allure des courbes simulées s'apparente à celle des courbes théoriques.

Mais aucune de ces courbes $\ti$ou, par extrapolation, de ces genres de courbes$\ti$ ne ressemble suffisamment aux courbes réelles (cf. Figs.~\ref{fcorrel-res} et \ref{fcorrel-abst}), à nos yeux. La valeur des corrélations à courte distance conjuguée à la portée des corrélations pourraient certes convenir, mais leur forme se distingue trop nettement d'une allure quasi-logarithmique observée avec les courbes réelles.

Ce modèle ne peut donc être retenu, puisqu'il produit des corrélation spatiales qui s'écartent trop de celles des données réelles. Faut-il pour autant rejeter en bloc ce modèle~?

Avant de répondre à cette question, voyons ce qu'apporte ce modèle.

\subsubsection{Bilan}
\label{pt-statique-bilan}
Comme nous l'avons vu dans la section consacrée à l'analyse de ce modèle, sa linéarité implique un comportement indépendant de la moyenne globale des $F^\aaa$. Ce qui s'accorde avec les régularités observées et résumées à la section~\ref{pt-sozein-ta-phainomena} $\ti$en se rappelant de la similarité entre $F^\aaa$ et $\taa$. De plus, la distribution des $F^\aaa$ (non montrée ici) est unimodale, sauf quand $K\simeq 1$. De même, la moyenne des $F^\aaa$ en fonction de $F_0^\aaa$ (où $F_0^\aaa$ dénote, à l'instar de $\toa$, la moyenne des $F^\beta$ sur les $n_p=16$ plus proches communes de la commune centrale $\aaa$) se comporte sensiblement comme une droite de pente unité (non représentée ici). Enfin, les corrélations obtenues sont à longue portée. Les valeurs obtenues des corrélations entre proches voisins peuvent convenir avec les résultats réels.

Rappelons que le modèle à tendances communales statiques répond à l'objectif que nous nous étions fixés dans cette partie consacrée à la modélisation~: faire ressortir les phénomènes observés à partir d'une simple imitation, ou influence, due au voisinage.

En résumé, ce modèle a de bonnes raisons d'être retenu, hormis la forme des corrélations spatiales qu'il génère. Ainsi, plutôt que de l'abandonner, nous tenterons de l'amender.

Un dernier point avant de chercher à améliorer le modèle des tendances communales statiques. Les tests et le classement des trois variantes du modèle ne sont pas en toute rigueur exacts, puisque le modèle ne convient pas en toute rigueur. Le test des variantes réalisé ci-dessus nous convie néanmoins à préférer une imitation des agents sous la forme indirecte de ce que nous avions appelé, les lieux de résidence, plutôt que celle émanant d'une imitation directe dite de conviction. En outre, ils accréditent l'hypothèse d'une imitation restreinte au voisinage de la zone d'habitation des agents.

\subsection{Dynamiques tendances}
\label{section-dyn}

Comment modifier le modèle précédent à tendances communales statiques, tout en étant fidèle à nos objectifs $\ti$retrouver les phénomènes à partir d'une forme simple d'imitation, limitée au voisinage~? La modification ne portera pas sur la forme d'imitation puisque l'imitation exprimée par le modèle précédent satisfait déjà aux critères de simplicité. La changer, signifierait la compliquer, la complexifier. (Nous n'entendons pas non plus revenir à une imitation des choix, précédemment invalidée.) Nous ne remettons pas non plus en cause la description $\ti$adoptée jusqu'ici$\ti$ du choix d'un agent $i$ de la commune $\aaa$, en terme d'idiosyncrasie fixe au cours du temps, et tel que $S_i^\aaa=\operatorname{sign}[h_i+F^\aaa]$. En se penchant sur l'équation générale du modèle (Eq.~\ref{efaa}), il ne reste alors plus qu'une possibilité~: modifier les tendances communales statiques, $f^\aaa$.

Comme nous l'avions déjà évoqué, les tendances communales spécifiques doivent être décorrélées, et ce pour que le modèle satisfasse aux objectifs impartis. Dans le cas contraire, les corrélations des $f^\aaa$ masqueraient d'autres informations que celles mises en avant par le modèle, et brouilleraient de surcroît l'action de la seule imitation. En d'autres termes, un modèle avec un ensemble de $f^\aaa$ corrélées et non indépendantes, pourrait certes fournir des solutions en accord avec les données, mais ne pourrait en aucun cas avaliser l'hypothèse d'une imitation responsable des phénomènes $\ti$ou qui permettrait de retrouver les phénomènes. (Précisons que nous ne rejetons pas l'existence des $f^\aaa$ corrélées, ou expliquées par d'autres facteurs quantitatifs, comme par exemple la taille de la population des communes, mais que nous n'en tenons volontairement pas compte~: elle ne permet pas de répondre à la question centrale que nous nous étions posée en début de chapitre.) Comment alors modifier les tendances des communes $f^\aaa$ tout en préservant leur absence de corrélation~?

Il ne reste plus qu'une possibilité~: casser leur aspect statique. Rien n'empêche en effet de considérer les tendances communales comme fluctuantes au cours du temps où s'élabore la prochaine élection. Avec le modèle précédent, elles étaient fixes au cours du temps considéré $\ti$à l'image des idiosyncrasies des agents. Mais il semble plausible qu'elles évoluent, qu'elles changent pendant, ou avant, la campagne électorale. Par exemple, l'effet d'un évènement particulier et local, la réaction à une information pas forcément locale, les changements de population au sein d'une commune, peuvent affecter la dénommée tendance propre de la commune, et la modifier. En revanche, pour que ce modèle réponde à nos objectifs, les variations des tendances communales doivent être également décorrélées les unes des autres. (Si elles étaient parfaitement corrélées, nous retrouverions le modèle précédent avec un champ global $F$ appliqué à l'ensemble des communes, non pas statique mais dynamique.) Nous le choisirons alors comme étant le plus simple possible, i.e. un bruit blanc gaussien, non corrélé.\\

En résumé, le modèle dit à tendances communales dynamiques, ou à tendances dynamiques, écrit au temps $t$ le terme de tendance globale, ou le champ appliqué global, ou encore la moyenne des convictions des agents de la commune $\aaa$, i.e. $F^\aaa$, comme~:
\be \label{efaa-dyn} \dfrac{\mathrm{d}F^\aaa(t)}{\mathrm{d}t} = K\x \fva(t) - F^\aaa(t) + f^\aaa(t)~,\ee
avec
\begin{eqnarray}
\label{efaa-dyn-cor-spa} \langle f^\aaa(t)\x f^\beta(t)\rangle_{_\aaa} & = & \sigma_f^2\;\delta_{\aaa\beta}~,\\
\label{efaa-dyn-cor-t} \langle f^\aaa(t)\x f^\aaa(t')\rangle_{_t} & = & \sigma_f^2\;\delta(t-t')~,
\end{eqnarray}
où $\sigma_f^2$, $\delta_{\aaa\beta}$ et $\delta(t-t')$ dénotent respectivement la variance des tendances dynamiques, le symbole de Kronecker et la fonction de Dirac. $\fva$ garde la même expression de l'équation~(\ref{efaa}), dans laquelle $\pab$ peut provenir de Eq.~(\ref{epab}) ou de Eq.~(\ref{epab-pop}).

La moyenne du membre de gauche de Eq.~(\ref{efaa-dyn-cor-spa}) s'effectue sur l'ensemble des communes, et signifie qu'à un instant donné, les tendances des communes sont non corrélées. Cette équation reste valable pour le modèle des tendances statiques, mais avec des tendances $f^\aaa$ indépendantes du temps. La moyenne du membre de gauche de Eq.~(\ref{efaa-dyn-cor-t}) porte sur l'ensemble des réalisations pour une commune $\aaa$ au cours du temps, et signifie que la tendance d'une commune fluctue sans mémoire, i.e. sans être corrélée à son passé. Pour simplifier, et comme pour le modèle des tendances statiques, nous considérons les moyennes (temporelle ou étendue à l'ensemble des communes à un instant donné) des $f^\aaa(t)$, comme nulles. Notons enfin que ce modèle se classe parmi les processus d'Ornstein-Uhlenbeck.

\subsubsection{Stabilité et paramètres pertinents du modèle}
\label{pt-dyn-analyse}
Le passage du modèle des tendances statiques au modèle à tendances dynamiques, ne modifie que l'expression des tendances communales, qui, de statiques deviennent fluctuantes, $f^\aaa \rightarrow f^\aaa(t)$. L'équation régissant le modèle est donc la même pour les deux $\ti$à l'aspect statique ou dynamique de $f^\aaa$ près$\ti$ comme le montrent les équations~(\ref{efaa-t}, \ref{efaa-dyn}). Il en découle les mêmes conditions de stabilité et le même nombre de paramètres pertinents du modèle.

Ainsi, les solutions du modèle ne divergent pas pour $K<1$ ainsi que pour $K=1$ (mais pour cette dernière, avec la condition supplémentaire d'une moyenne, temporelle ou étendue à l'ensemble des communes à un instant donné, $\langle f^\aaa(t) \rangle = 0$, comme nous l'avions prescrit lors de l'écriture du modèle des tendances communales statiques).

Les paramètres pertinents de ce modèle linéaire sont encore au nombre de deux~: l'intensité de couplage $K$, et la longueur $\ell_c$ caractéristique des distances du voisinage.
 
Voyons maintenant ce qu'une étude théorique $\ti$étayée ensuite par des simulations$\ti$ peut montrer sur les corrélations spatiales.

\subsubsection{Corrélations spatiales théoriques et conséquences}
\label{pt-dyn-correl-th}
Comme précédemment, le calcul analytique qui mène aux corrélations spatiales des solutions du modèle se fonde sur l'hypothèse d'une répartition uniforme des communes, voire de la population si $\pab$ en tient compte. L'annexe~\ref{annexe-correl-th} détaille ces calculs, qui reprennent pour une bonne part ceux du modèle statique. Le caractère fluctuant des tendances communales influe néanmoins sur l'expression des corrélations, comme le montre Eq.~(\ref{eiq-dyn}) comparée à son pendant du modèle statique, Eq.~(\ref{eiq-statique}).

\begin{figure}[t]
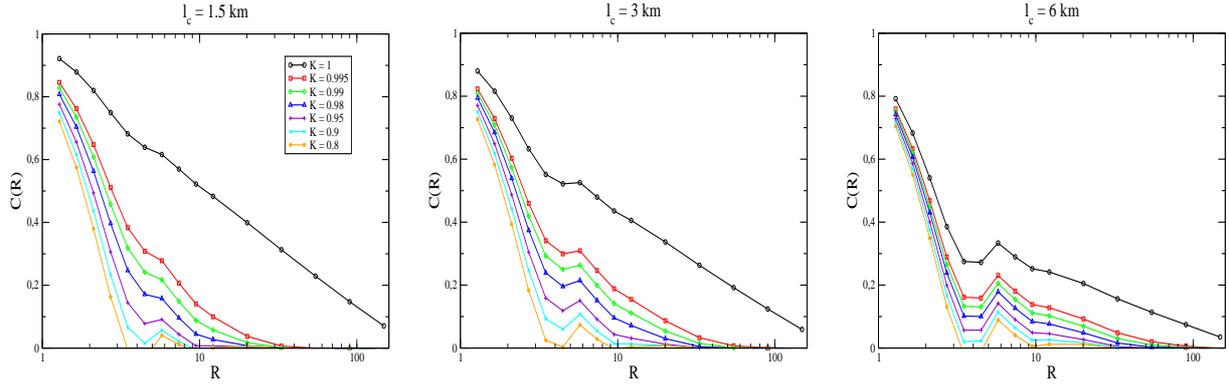

\includegraphics[width=5cm, height=5cm]{cor-dyn-th-0.5.eps}\hfill
\includegraphics[width=5cm, height=5cm]{cor-dyn-th-1.eps}\hfill
\includegraphics[width=5cm, height=5cm]{cor-dyn-th-2.eps}
\caption{\small Moyennes des corrélations théoriques obtenues avec le modèle des tendances dynamiques et une répartition uniforme des communes, voire si nécessaire des populations.}
\label{fcor-th-dyn}
\end{figure}

Les figures~\ref{fcor-th-dyn} montrent, à partir des équations~(\ref{ecor-sol}, \ref{eiq-dyn}), les corrélations spatiales du modèle dans le cadre d'une répartition uniforme des communes ou de la population. Les distances $L$, $a$ et $D$ prises en compte reprennent les valeurs de la section~\ref{pt-statique-correl} qui se voulaient réalistes, i.e. en accord avec les distances réelles du système. Il est important de noter que les corrélations obtenues analytiquement sont des moyennes sur l'ensemble des réalisations. Et de nouveau, seules les solutions obtenues après le régime transitoire sont considérées.

Les corrélations théoriques correspondant à une répartition uniforme des communes $\ti$et des populations si $\pab$ en tient compte$\ti$ ont une décroissance quasi-logarithmique pour la valeur $K=1$. Elles en diffèrent nettement quand $K\neq1$. La forme quasi-logarithmique obtenue avec $K=1$ ne tient plus pour les les plus petites distances, ce qui n'est pas trop gênant puisqu'avec le modèle précédent, l'écart entre corrélations théoriques et numériques résidait essentiellement dans cette zone.

Ce qui amène à se poser les deux questions suivantes~? Premièrement, quel enseignement tirer de la forme appropriée des corrélations spatiales théoriques~? Deuxièmement, quelle signification donner au modèle avec la valeur particulière $K=1$~?\\

Premièrement, l'accord des corrélations spatiales théoriques et réelles, justifie de nouveau la restriction de l'interaction au seul voisinage. L'analyse théorique limite en effet le terme d'influence $\fva$ à la proximité spatiale seule. Rappelons que les tests des trois variantes du modèle des tendances statiques accréditaient également l'hypothèse dans laquelle nous nous étions placés dès le début de cette étude~: l'interaction entre agents se cantonne aux agents du voisinage.

Deuxièmement, avec $K=1$ l'équation~(\ref{efaa-dyn}) du modèle s'apparente à une équation de diffusion en présence d'une source bruitée. Noter que cette équation diffère d'une équation de diffusion au sens propre $\ti$définie avec un Laplacien$\ti$ mais s'en rapproche puisque $(\fva - F^\aaa)$ se comporte comme un Laplacien équivalent. (cf. Eq.~(\ref{ef-fourier}) en annexe, où $(\fva - F^\aaa)$ se comporte comme un Laplacien pour les grandes distances, mais s'en écarte pour les petites distances.) Bref, le modèle des tendances communales dynamiques peut se concevoir comme un modèle de type diffusif en présence d'un bruit blanc. Et de nouveau se pose la question~: quel enseignement en tirer~? Ou plus précisément, quelle interprétation en faire~?

Que le coefficient $K$ soit égal à un, autrement dit que le modèle se ramène à une sorte de diffusion, nous pousse à adopter l'interprétation d'une interaction indirecte via les lieux de résidence, et ce au détriment d'une interaction directe via l'imitation de la conviction. En effet, une diffusion des tendances communales globales $F^\aaa$ se comprend aisément par une interaction indirecte par le biais des lieux de résidence, où seule la distance entre communes intervient. Cette diffusion reflète alors l'homogénéisation locale des résultats (ou des taux de participation) électoraux, par delà le découpage en communes des zones d'habitation relativement homogènes.

Nous pourrions arguer qu'une diffusion de la conviction semble également naturelle. Mais comment justifier à l'échelle microscopique que le coefficient $K$ soit égal à un~? Autrement dit, comment naturellement comprendre que l'intensité de l'imitation de conviction, $J$, soit égal à un, alors qu'il peut prendre n'importe quelle valeur comprise entre $0$ et $1$~? (Rappel~: $J$ se retrouve dans l'équation générale (\ref{esia-gal}), et Eq.~(\ref{eK-convic}) exprime le lien entre $K$ et $J$.)

Précisons que nous n'avons pas la preuve formelle que l'influence procède d'une interaction dite indirecte via les lieux de résidence, mais plutôt un faisceau de convergence. Ce faisceau de convergence provient des deux points suivants. Premièrement, la diffusion des $F^\aaa$ se comprend plus facilement, à l'échelle microscopique, sous la forme d'une homogénéisation des tendances globales des lieux de résidence, plutôt que sous la forme d'une imitation directe des convictions des agents. Deuxièmement, les tests des trois variantes relatives au modèle des tendances statiques penchaient en faveur de l'imitation indirecte par le biais des lieux de résidence $\ti$et renforçaient de plus l'hypothèse d'une influence locale.

Notons enfin que nous acceptons les conclusions de cette analyse théorique, puisque l'hypothèse d'une répartition uniforme des communes ne nous paraît pas complètement farfelue, du mois, à l'échelle locale ou départementale.\\

Pour résumer, l'analyse théorique des corrélations spatiales du modèle des tendances communales dynamiques accrédite les trois points suivants.
\begin{itemize}
\item L'interaction est bornée au voisinage spatial.
\item Le modèle est de type diffusif, i.e. $K=1$. Ainsi, le modèle n'a plus qu'un seul paramètre pertinent à considérer, la longueur caractéristique, $\ell_c$, de prise en compte des distances du voisinage.
\item L'interprétation d'une influence, appelée peut-être à tort, indirecte et au travers des lieux de résidences, est favorisée.
\end{itemize}
Dit en terme mathématique~: l'équation~(\ref{efaa-dyn}) du modèle s'écrit avec $K=1$, et $\pab$ qui intervient dans l'expression de $\fva$ provient de Eq.~(\ref{epab}).

Nous nous placerons par la suite dans ce cadre. Voyons alors ce que nous apprennent les corrélations spatiales numériques, établies non pas à partir de l'hypothèse d'une répartition uniforme des communes, mais à partir de leur position réelle.

\subsubsection{Corrélations spatiales numériques}
\label{pt-dyn-correl-num}
\begin{figure}[t]
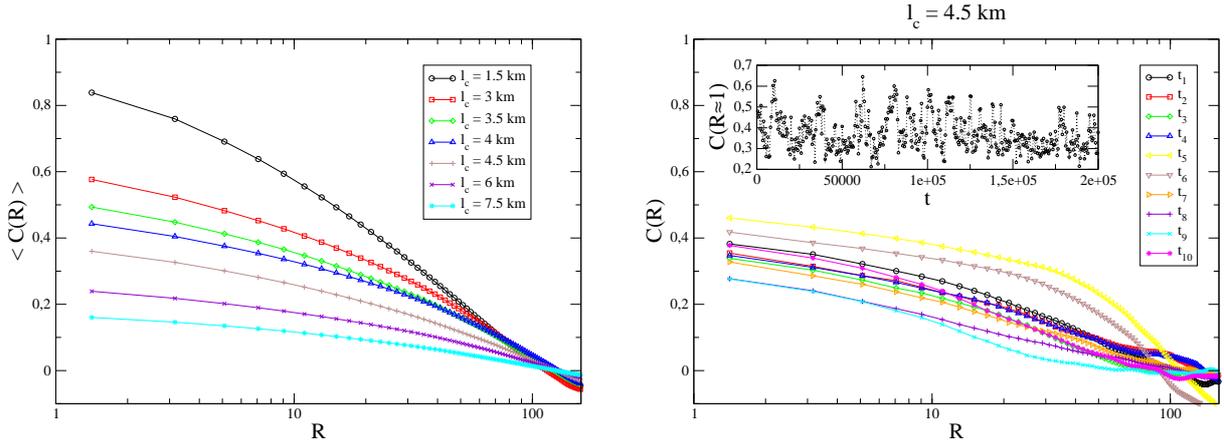

\includegraphics[scale = 0.32]{cor-dyn-multi.eps}\hfill
\includegraphics[scale = 0.32]{cor-dyn-1.5.eps}\hfill
\caption{\small Figure de gauche~: moyennes $\ti$après le régime transitoire$\ti$ des corrélations simulées du modèle des tendances dynamiques, et ce en fonction de l'unique paramètre libre, $\ell_c$. (Ici, le modèle est de type diffusif, i.e. $K=1$.) Figure de droite~: Corrélations spatiales à différents temps ($t_i=3,9\:10^5+500\x i$) après le régime transitoire, avec $\ell_c=4,5~km$. Dans la fenêtre, les valeurs des corrélations des proches voisins, $C(r\leqslant 2\,D)$, à différents temps, et toujours avec $\ell_c=4,5~km$. (Voir Figs.~\ref{fcorrel-res} et \ref{fcorrel-abst} pour les corrélations spatiales des résultats et des taux de participation, ainsi que pour la notation.)}
\label{fcor-dyn}
\end{figure}

La figure~\ref{fcor-dyn} de gauche montre les moyennes, après le régime transitoire, des corrélations spatiales issues du modèle des tendances dynamiques et pour différentes valeurs de $\ell_c$. Les simulations s'effectuent sur un pas de temps élémentaire $\dd t=0,1$, et de façon synchrone $\ti$de manière à respecter le déroulement de l'élection au jour J. Le temps caractéristique du régime transitoire est égal à $\frac{1}{2(\tau_k)^2_{min}}\simeq\frac{1}{3}\left(\frac{L}{\ell_c}\right)^2$ d'après les équations~(\ref{etauk}, \ref{ecor-sol}, \ref{eik-dyn}), où $L=500~km$, caractérise la longueur caractéristique du système, i.e. de la France métropolitaine. Trois fois ce temps (pour que les transitoires comptent moins de $5\%$ de la solution globale) donne $t\simeq 10^5$ dans le cas où $\ell_c$ est le plus petit ($\ell_c=1,5~km$ ici). Par souci d'homogénéité, les moyennes des corrélations spatiales de Fig.~\ref{fcor-dyn} s'obtiennent, pour chaque valeur de $\ell_c$, à des intervalles de temps égaux à $500$ et pour des temps $t\in]10^5\;;2\;10^5]$ (i.e. sur $200$ réalisations différentes pour chaque $\ell_c$ après le régime transitoire).

La figure~\ref{fcor-dyn} de droite montre $10$ réalisations différentes des corrélations spatiales obtenues avec $\ell_c=4,5~km$, et met en évidence les fluctuations $\ti$inhérentes à ce modèle de type diffusif en présence d'une source de bruit blanc$\ti$ des corrélations spatiales. Ces fluctuations sont auto-moyennantes puisqu'il existe une moyenne théorique sur l'ensemble des réalisations. La fenêtre de cette figure le met en évidence qualitativement en représentant les valeurs de la corrélation avec les proches voisins au cours du temps, et toujours pour $\ell_c=4,5~km$. Ces valeurs semblent en effet fluctuer autour d'une valeur moyenne et avec un bruit constant.

Une chose est de tracer numériquement les corrélations spatiales du modèle, autre chose est d'en tirer des conséquences, d'en discuter. La section qui suit s'y attelle.

\subsubsection{Deux ou trois choses que je sais d'elles}
\label{pt-dyn-discussion}
Le premier constat qui sort des figures~\ref{fcor-dyn} est la perte de la forme quasi-logarithmique des corrélations spatiales, par rapport aux corrélations théoriques à grandes distances. Nous attribuons ce point à la non homogénéité de la répartition des $36200$ communes sur le territoire français, comme le montre la figure~\ref{frepartition-taille}.

Deuxième constat~: la forme quasi-logarithmique des corrélations spatiales réelles, se rapproche nettement plus de la forme obtenue avec ce modèle qu'avec le modèle précédent des tendances statiques.

Troisième constat~: la portée des corrélations correspond à peu près à celle des corrélations réelles, et ne dépend pas de la valeur de la corrélation entre proches communes ($C(R\simeq1)$) $\ti$à l'image des corrélations réelles.

Quatrième constat~: les corrélations spatiales fluctuent de façon non négligeable pour différentes réalisations, i.e. pour des temps différents~; rendant par là même difficile une mesure précise du paramètre libre $\ell_c$ à attribuer à chaque élection.

Enfin, cinquième constat~: le modèle fait ressortir naturellement une distance caractéristique du système $\ti$la distance qui sépare deux communes voisines. Autrement dit, pour que les simulations soient comparables aux corrélations réelles, le paramètre pertinent du modèle, $\ell_c$, doit prendre sensiblement la valeur de la distance réelle séparant deux communes voisines (dans les environs de $3~km$).\\

\begin{figure}
 \begin{minipage}[c]{0.5\linewidth}
  \includegraphics[scale=0.32]{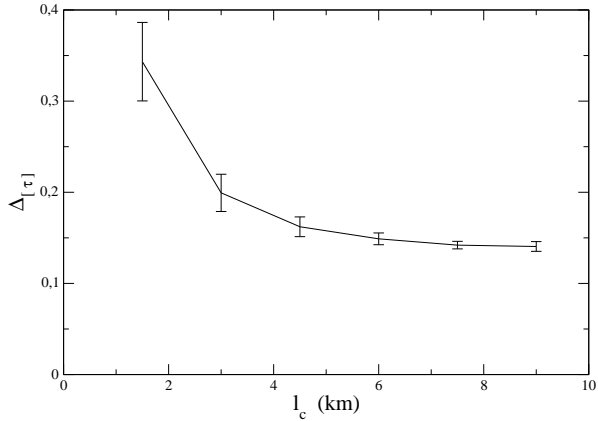}
 \end{minipage} \hfill
 \begin{minipage}[c]{0.45\linewidth}
  \caption{\small Écart-type des $\taa$ sur l'ensemble des communes avec le modèle dynamique en fonction de $\ell_c$, et pour un écart-type des tendances dynamiques $\sigma_f=1$ (cf Eqs.~(\ref{efaa-dyn-cor-spa}, \ref{efaa-dyn-cor-t})). Rappel~: $\taa\simeq\frac{\pi}{\sqrt{3}}\:F^\aaa$ d'après Eq.~(\ref{etaa-faa}). Ces mesures ont été effectuées sur les mêmes réalisations (200 par valeur de $\ell_c$, et après le régime transitoire) de la figure~\ref{fcor-dyn} de gauche.}
 \label{fstat-moy-dyn}
 \end{minipage}
\end{figure}

Essayons de mieux comprendre la longueur caractéristique, $\ell_c$, de prise en compte des distances intercommunales. La figure~\ref{fcor-dyn} de gauche montre que la corrélation entre proches voisins augmente quand $\ell_c$ diminue. La figure~\ref{fstat-moy-dyn} montre quant à elle, que l'écart-type des $F^\aaa$ (ou des $\taa$) augmente quand $\ell_c$ diminue; ce qui est en accord avec l'augmentation des corrélations spatiales entre communes. En comparant avec les corrélations réelles (cf. Figs.~\ref{fcorrel-res}, \ref{fcorrel-abst}), il semble que $\ell_c$ vaille environ $4~km$ pour les résultats électoraux (sauf pour l'élection de 2000-b), et environ $7~km$ pour les taux de participation. (Les valeurs de $\ell_c$, écrites ici à titre indicatif, devraient être vraisemblablement plus faibles si le modèle incorporait des effets dues aux tailles finies $\ti$engendrant de la sorte un bruit additionnel.) Ceci s'accorde aussi avec un écart-type des $\taa$ supérieur pour les résultats électoraux que pour les taux de participation. (Encore faudrait-il plus rigoureusement tenir compte du bruit dû aux tailles finies des communes, point d'autant plus important pour un résultat électoral que la participation est faible, comme lors du référendum de 2000.) Bref, pourquoi cette différence de $\ell_c$~? Ou mieux, comment comprendre qu'à chaque type de donnée électorale corresponde une valeur de $\ell_c$ qui lui est propre~?

Plus $\ell_c$ diminue, plus le poids relatif de la, ou des plus proches communes de la commune centrale $\aaa$ augmente par rapport aux communes plus lointaines, comme l'indique directement l'exponentielle de $\pab$ dans Eq.~(\ref{epab}). Au contraire, plus $\ell_c$ augmente, plus l'influence des plus proches communes se dilue par rapport aux autres communes plus lointaines. Dit encore autrement, quand $\ell_c$ diminue, la finesse de résolution de la prise en compte du voisinage augmente. Ceci nous incite à émettre l'hypothèse suivante~: tout se passe comme si l'importance d'une élection $\ti$au regard des agents$\ti$ pouvait se mesurer à l'aune de la finesse des détails du voisinage pris en compte. Avec l'idée $\ti$potentiellement contestable$\ti$ que l'expression d'un vote semble plus important aux yeux des agents qu'une participation à l'élection, il en résulte que $\ell_c$ est plus petit pour les résultats électoraux que pour les taux de participation. Et dans le rôle de l'exception qui confirme la règle~: le résultat du référendum de 2000. Cette élection, aux taux de participation et de suffrages exprimés les plus faibles des douze étudiées ($\expr/\insc\simeq25\%$ à l'échelle nationale), n'a semble-t-il pas engagé profondément les citoyens français ni fait ressortir de profonds clivages idéologiques. Arrêtons-nous là sur les interprétations d'ordre politique, et contentons-nous après avoir émis cette hypothèse, de constater que $\ell_c$ paraît plus petit pour les résultats électoraux (référendum de 2000 excepté) que pour les taux de participation.\\

Notons enfin que les simulations obtenues avec $\pab$ proportionnel à la population des communes et donné par Eq.~(\ref{epab-pop}) fournissent des corrélations spatiales qui s'éloignent encore plus de la forme quasi-logarithmique. Ce point semble accréditer la discussion précédente qui nous a poussés à préférer l'influence indirecte via les lieux de résidence, au détriment de l'imitation directe de la conviction. Mais qualitativement, ce plus grand écart avec la forme théorique peut aussi se comprendre par une plus grande inhomogénéité de la répartition de la population (y compris sur des zones de distance caractéristique $\ell_c$) comparée à l'hétérogénéité de la répartition des communes.

Arrive maintenant la question fatidique~: ce modèle est-il ou non satisfaisant~?

\subsubsection{Valider ou non le modèle}
\label{pt-dyn-valider}
Avant de savoir si ce modèle peut être validé ou non, essayons de quantifier, ne serait-ce qu'approximativement, ce que nous avions dit jusqu'ici qualitativement et à l'\oe{}il concernant l'allure quasi-logarithmique des corrélations spatiales. Nous utilisons pour cela le coefficient de détermination $R^2$ entre les corrélations spatiales et la forme logarithmique attendue. Autrement dit, nous cherchons comment les corrélations spatiales $C(R)$ peuvent bien s'accorder avec la forme logarithmique voulue. Le coefficient $R^2$ indique en effet la proportion de variance expliquée, i.e. le rapport des variances de l'écart des données à la forme voulue sur les variances des données brutes (cf. Eq.~(\ref{er})). (Plus $R^2$ se rapproche de $1$, meilleur est l'accord entre les données et la forme attendue.) Dans le cas d'un ajustement linéaire, $R$ (du coefficient $R^2$, à distinguer de la distance normalisée et sans dimension $R=r/D$ introduite à la section~\ref{pt-cor-spatial-res} qui intervient ici dans les corrélations spatiales $C(R$)) est simplement égal à la corrélation entre les données et la forme voulue. Ainsi, dans le cas qui nous intéresse, $R$ est égal à la corrélation entre les corrélations spatiales, $C(R)$, et la forme logarithmique $\ln(R)$. (Il correspond à l'ajustement linéaire de $C(R)$ avec $a\:\ln\big(\frac{R}{R_c}\big)$, où $a$ désigne la pente de décroissance, et $R_c$ la longueur ou portée caractéristique des corrélations spatiales.) Notons enfin que les corrélations spatiales $C(R)$ peuvent aussi bien provenir des données réelles que des modèles à tendances communales statiques ou dynamiques.

La table~\ref{tquasilog} évalue les coefficients $R^2$ des corrélations spatiales $C(R)$ avec une forme de décroissance logarithmique. Les corrélations spatiales proviennent des données réelles~: les résultats électoraux et les taux de participation sur l'ensemble des communes. Elles proviennent aussi du modèle à tendances statiques (avec $K=0,995$ et $\ell_c=12~km$), ainsi que du modèle à tendances dynamiques (avec $\ell_c=4~km$). Ces paramètres des modèles reproduisent en effet une portée convenable des corrélations spatiales, conjuguée à une bonne valeur (comparée à celles des résultats électoraux) des corrélations à courte distance, $C(R\simeq 1)$, comme le montrent les figures~\ref{fcor-statique} et \ref{fcor-dyn}. Excepté pour le modèle statique, les valeurs indiquées par la table~\ref{tquasilog} correspondent à des valeurs moyennes (sur $12$ taux d'abstention, $6$ résultats électoraux et $200$ réalisations du modèle dynamique).

La table~\ref{tquasilog} détermine ces coefficients $R^2$ de deux manières différentes. Premièrement, le calcul s'effectue sur $R \leqslant 80$ avec $40$ points~: $R\in]2n\,;2(n+1)]$ où $n=0,1,\ldots,39$. Le choix arbitraire de $80$ correspond environ à la portée des corrélations spatiales des résultats électoraux et des taux de participation. Noter que cette méthode donne plus de poids aux zones des plus grandes valeurs de $R$ (par l'effet de $\ln(R)$, la densité de points pris en compte est supérieure dans les zones à grand $R$, comparée à celle des zones à faible $R$). (Cette première façon de calculer $R^2$ se note \og $1^{\grave{e}re}$ façon \fg{} à la table~\ref{tquasilog}.) La seconde méthode pallie ce problème. Elle considère un espacement constant entre les points $\ln(R)$ considérés. En revanche le calcul s'effectue sur $7$ points à peine, et pour $R \leqslant 66$, tels que $R\in]2\x2^n\,;2\x(2^n+1)]$ où $n=0,1,\dots,6$. (Cette seconde façon de calculer $R^2$ se note, on pouvait s'en douter, \og $2^{nde}$ façon \fg{} à la table~\ref{tquasilog}.)

\begin{table}[h!]
\begin{minipage}[c]{0.6\linewidth}
\begin{tabular}{|c|c|c|}
\cline{2-3}
\multicolumn{1}{c|}{}
 & $1^{\grave{e}re}$ façon & $2^{nde}$ façon \rule[-7pt]{0pt}{22pt}\\ %
\hline
Taux de participation & 0,99 & 0,99\\
Résultats électoraux & 0,99 & 0,98\\
Modèle statique & 0,83 & 0,86\\
Modèle dynamique & 0,95 & 0,93\\
\hline
\end{tabular}
\end{minipage}\hfill
\begin{minipage}[c]{0.4\linewidth}
\caption{\small $R^2$ entre les corrélations spatiales $C(R)$ et une forme logarithmique. Voir texte pour les deux façons de calculer le $R^2$.  Modèle des tendances statiques~: ($K=0,995\,;\ell_c=12~km$). Modèle des tendances dynamiques~: $\ell_c=4~km$.}
\label{tquasilog}
\end{minipage}
\end{table}

La table~\ref{tquasilog} confirme bien tous les jugements qualitatifs précédents. Premièrement les corrélations spatiales des taux de participation et des résultats électoraux ont bien une forme quasi-logarithmique. Deuxièmement, les corrélations spatiales obtenues avec le modèle des tendances dynamiques se rapprochent plus d'une forme quasi-logarithmique, que celles produites par le modèle des tendances statiques. Enfin, en moyenne, la forme des corrélations spatiales issues du modèle des tendances dynamiques diffère, à proprement parler, de la forme des corrélations logarithmique.

Mais la différence entre les corrélations spatiales du modèle des tendances dynamiques, et leur forme attendue (quasi-logarithmique), est-elle suffisante pour invalider le modèle~? Autrement dit, l'écart entre les deux formes $\ti$mesuré de façon un peu plus quantitative à la table~\ref{tquasilog}$\ti$ est-il acceptable ou non~?

Il survient alors l'épineuse et incontournable question du seuil de tolérance. Face à cette embarrassante question, présentons néanmoins notre position.\\

Notre jugement relatif à la validité du modèle des tendances dynamiques est donc mitigé.

Nous ne pouvons rigoureusement pas valider le modèle des tendances dynamiques puisque les corrélations spatiales qu'il engendre diffèrent notablement, et en général, de la forme quasi-logarithmique afférente aux données. Néanmoins, l'écart ne nous semble pas être trop important au point de le rejeter complètement. L'écart à la forme quasi-logarithmique est d'ailleurs moins important qu'avec le modèle des tendances statiques $\ti$ce qui en soi, ne donne aucune justification du modèle dynamique. De plus, le modèle permet de retrouver (non montré ici) tous les autres phénomènes principaux à sauver énumérés à la section~\ref{pt-sozein-ta-phainomena}. Enfin, qu'un modèle à un seul paramètre puisse rendre compte des phénomènes majeurs, à l'exception de l'un d'entre-eux, et encore pas excessivement, nous semble très encourageant.

Précisons aussi que nous n'attendons pas qu'un modèle simpliste, à un seul paramètre pertinent, puisse rendre compte de toutes les particularités relatives à une élection. 

En conclusion, nous ne validons donc pas le modèle des tendances dynamiques, mais le trouvons suffisamment encourageant pour ne pas le rejeter complètement. Plutôt que se focaliser sur la validation, ou non, du modèle, nous trouvons extrêmement encourageant le fait qu'un modèle à un unique paramètre pertinent, et bâti sur des principes microscopiques plausibles, puisse s'approcher à ce point de l'ensemble des phénomènes majeurs à sauver.

Précisons enfin qu'un accord avec l'ensemble des phénomènes majeurs à sauver ne signifierait pas automatiquement la validation du modèle. Encore faudrait-il qu'il puisse rendre compte des phénomènes plus fins $\ti$énumérés à la section~\ref{pt-sozein-ta-phainomena}. La partie suivante illustre une confrontation plus fine du modèle aux données réelles. Elle montre de plus une voie possible d'amélioration du modèle dynamique~: comme pour les solides, liquides et gaz, faire varier le coefficient de diffusion en fonction de l'état du système.

\subsubsection{Tentative d'amélioration et analyse plus fine}
\label{pt-dyn-fine}
Cette partie ne se veut pas rigoureuse puisque le modèle dynamique n'est, à ce stade, pas formellement valide. En revanche, elle indique une voie d'amélioration du modèle dynamique. Ce qui de plus, lui permet de se confronter aux phénomènes plus fins.

Il ne paraît pas absurde de considérer que les tendances des communes fluctuent avec une intensité différente selon les communes. En outre, il paraît judicieux que l'amplitude des fluctuations de la tendance spécifique $f^\aaa(t)$ de la commune $\aaa$, dépende de l'état de ladite commune. Par rapport au modèle dynamique ci-dessus, $f^\aaa(t)\rightarrow f^\aaa(t)=f^\aaa\x \eta^\aaa(t)$, où $f^\aaa$ est fixe au cours du temps et $\eta(t)$ dénote un bruit blanc gaussien $\ti$pris de moyenne nulle et de variance unité$\ti$ qui prend la valeur $\eta^\aaa(t)$ au temps $t$ pour la commune $\aaa$. $\eta$, comme précédemment, n'est ni corrélé en espace (entre différentes communes), ni en temps. Autrement dit, l'équation~(\ref{efaa-dyn}) générale du modèle dynamique reste inchangée, mais les équations~(\ref{efaa-dyn-cor-spa}, \ref{efaa-dyn-cor-t}) deviennent avec les mêmes notations~:
\begin{eqnarray}
\label {efaa-dyn-fin-cor-spa} \langle f^\aaa(t)\x f^\beta(t)\rangle_{_\aaa} & = & (f^\aaa)^2\;\delta_{\aaa\beta}\x (\eta^\aaa(t))^2~,\\
\label {efaa-dyn-fin-cor-t} \langle f^\aaa(t)\x f^\aaa(t')\rangle_{_t} & = & (f^\aaa)^2\;\delta(t-t')~,
\end{eqnarray}
avec $\langle (\eta^\aaa(t))^2 \rangle_{_t} = 1$ et $\langle (f^\aaa)^2 \rangle_{_\aaa} = \sigma_f^2$. (Notons que la moyenne des $f^\aaa$ est encore nulle.)

L'intensité $f^\aaa$ des fluctuations de la tendance propre de la commune $\aaa$ dépend ici de la commune elle-même. Par rapport à l'état de la commune, nous avons choisi $f^\aaa\propto\frac{1}{\sqrt{N^\aaa}}$, certes arbitrairement, mais de manière à refléter aussi le bruit dû aux tailles finies des communes.

L'annexe~\ref{annexe-correl-th} montre que les corrélations spatiales théoriques, selon cette contrainte et sous couvert d'une répartition uniforme des communes, garde la même forme que pour celles du modèle des tendances dynamiques vu précédemment. Les corrélations spatiales issues des simulations ressemblent très fortement $\ti$comme prévu$\ti$ à celles de la figure~\ref{fcor-dyn}.

Illustrons maintenant comment se comporte ce modèle face aux phénomènes plus fins à sauver.\\

Des simulations ont été faites avec $\ell_c=4,5~km$.

Avec $\sigma_f=1$, l'écart-type des $\taa$ vaut $0,16$, à comparer avec des écarts-types autour de $0,49$ et de $0,37$ respectivement pour les résultats électoraux et pour les taux de participation.

\begin{figure}
 \begin{minipage}[c]{0.5\linewidth}
  \includegraphics[scale = 0.32]{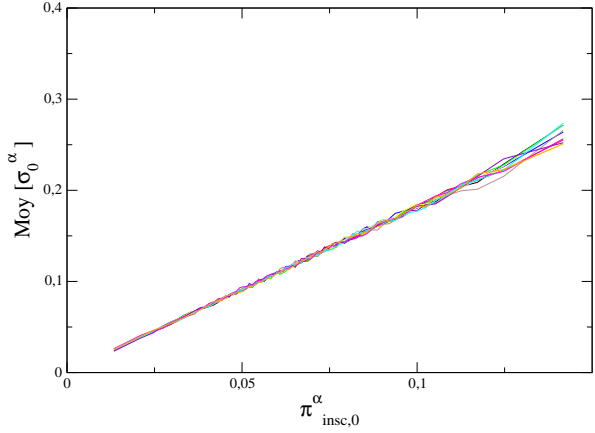}
 \end{minipage} \hfill
 \begin{minipage}[c]{0.45\linewidth}
  \caption{\small Moyenne des $\saa$ en fonction de $\piinsc$, calculé avec $36$ intervalles. Voir Figs.~\ref{fsaa-piinsc-res} et \ref{fsaa-piinsc-abst} pour les résultats et les taux de participation.}
 \label{fsaa-piinsc-dyn-invsqrt}
 \end{minipage}
\end{figure}

La corrélation entre $\saa$ et $\piinsc$ (voir Eq.~(\ref{epiinsc})) vaut $0,85$, à comparer avec celles des résultats et des taux de participation, aux alentours de $0,66$. La figure~\ref{fsaa-piinsc-dyn-invsqrt} trace $\saa$ (évalués ici à partir des grandeurs par commune) en fonction de $\piinsc$. Nous y apercevons une nette différence d'allure par rapport aux courbes réelles des figures~\ref{fsaa-piinsc-res}, et \ref{fsaa-piinsc-abst}, notamment dans les zones à forte population, i.e. à faible $\piinsc$.

Les corrélations temporelles effectuées avec $11$ différentes réalisations donnent $\overline{C}_t(\sigma_0)=0,72$, à comparer avec $0,57$ qui prévaut pour les résultats et les taux de participation des élections. Le \textit{Ratio Significatif Minimum} (voir annexe~\ref{annexe-extraire}) est inférieur à $1\%$, en accord avec une absence d'information purement locale et différente du bruit dû aux tailles finies des communes. Noter que le temps équivalent séparant les $11$ différentes réalisations est suffisamment grand pour qu'il n'y ait plus de mémoire entre les $11$ ensembles des tendances spécifiques $f^\aaa(t)$. Même si ce modèle était valable, il resterait encore à savoir si les ensembles des tendances propres, $f^\aaa(t)$, à deux élections différentes sont complètement indépendants entre eux, ou non.

\begin{figure}[t!]
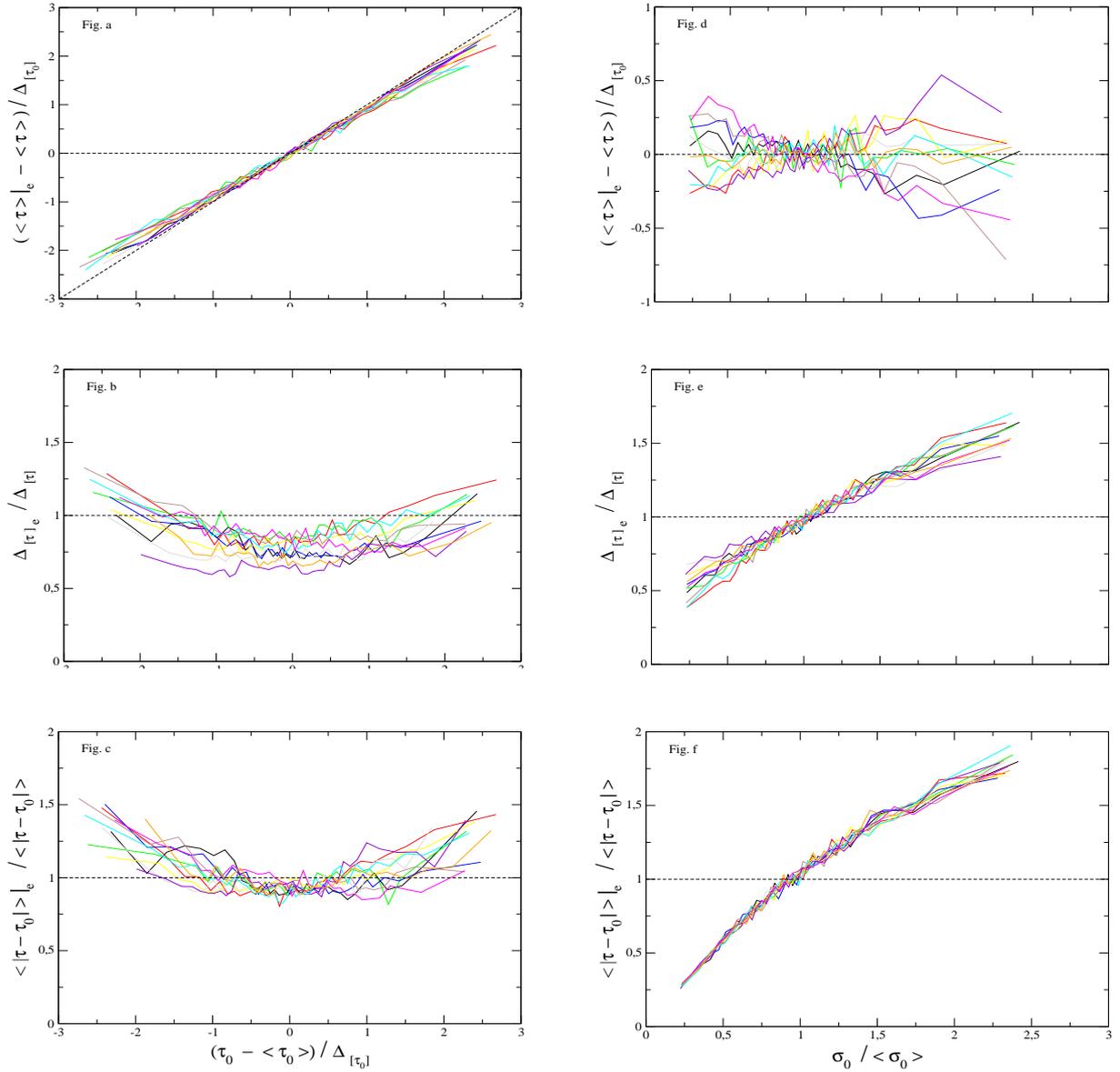

\includegraphics[width=7.5cm, height=5cm]{1-dyn-invsqrt.eps}\hfill
\includegraphics[width=7.5cm, height=5cm]{3-dyn-invsqrt.eps}\vspace{0.25cm}
\includegraphics[width=7.5cm, height=5cm]{2-dyn-invsqrt.eps}\hfill
\includegraphics[width=7.5cm, height=5cm]{4-dyn-invsqrt.eps}\vspace{0.25cm}
\includegraphics[width=7.5cm, height=5cm]{5-dyn-invsqrt.eps}\hfill
\includegraphics[width=7.5cm, height=5cm]{6-dyn-invsqrt.eps}
\caption{\small Effet de l'environnement sur le résultat de la commune centrale. Voir Figs.~\ref{fenvironnement-res} et \ref{fenvironnement-abst} pour une comparaison avec les résultats électoraux et les taux de participation.}
\label{fenvironnement-dyn-invsqrt}
\end{figure}

Enfin, les figures~\ref{fenvironnement-dyn-invsqrt} montrent l'effet de l'environnement des $n_p=16$ communes sur une commune centrale, à l'instar de leurs homologues Figs.~\ref{fenvironnement-res}, et \ref{fenvironnement-abst}. Pour mémoire, ces dernières laissaient apparaître des régularités aux sous-figures, a, e et f.\\


Pour terminer, indiquons deux autres menues modifications, et sans amélioration notable pour chacune d'elles.

Premièrement~: faire dépendre la longueur caractéristique, $\ell_c$, de l'environnement géographique autour de la commune $\aaa$ considérée $\ti$et non pas lui allouer une valeur unique à toutes les communes, comme dans le modèle ci-dessus. Nous avons attribué une longueur caractéristique à chaque département, $\\ell_c\rightarrow \ell_c\x\frac{D}{\langle D \rangle}$, où $D$ représente la distance caractéristique entre communes du département considéré, et $\langle D \rangle$ la distance caractéristique entre communes pour la France métropolitaine entière. Les simulations n'ont montré aucun effet notable.

Deuxièmement, modifier la façon dont décroît la prise en compte du voisinage. Jusqu'ici, cette décroissance était exponentielle (en $\exp(-\,\frac{r}{\ell_c})$). Nous avons pris une autre forme, assez naturelle aussi~: $\exp(-\,\frac{r^2}{2\,(\ell_c)^2})$. Les simulations ont montré un moins bon accord avec les données, i.e. une forme des corrélations spatiales qui s'écarte davantage de la forme quasi-logarithmique que celle obtenue avec le modèle utilisé jusqu'alors

\subsubsection{Bilan}
\label{pt-dyn-bilan}
En résumé, ce modèle à un paramètre pertinent s'accorde relativement bien aux données. Assez correctement, certes, mais pas complètement.

Même si nous ne pouvons le valider, à ce stade, il représente une tentative encourageante.

Demander de plus, à un modèle à un paramètre pertinent, de contenir l'ensemble des caractéristiques et particularités du phénomène électoral, représenterait à nos yeux une gageure démesurée. Une gageure, ou plutôt une erreur d'appréciation de ce que peut apporter un modèle au phénomène étudié.

Même si ce modèle procure incontestablement des encouragements ou des satisfactions, je crois néanmoins que quelque chose $\ti$de profond~?$\ti$ lui échappe encore.\\

Les réponses que nous venons de porter au modèle des tendances communales dynamiques peuvent maintenant s'étendre directement à la question centrale qui animait tous les modèles que nous avons essayés ici. A savoir, une imitation simple et restreinte au voisinage, peut-elle rendre compte des phénomènes~? 

Ainsi, au stade du développement des modèles rencontrés ici $\ti$tous bâtis sur une imitation simple et restreinte au voisinage$\ti$, notre réponse à notre question centrale restera hélas mitigée. (Une réponse claire, quelle qu'elle soit, aurait été à court terme préférable. Préférable à court terme sur le plan scientifique, i.e. avec un point de vue cumulatif.) Bref, nous pouvons retenir les deux points suivants.
\begin{itemize}
\item L'hypothèse d'une imitation simple et restreinte au voisinage géographique s'accorde assez correctement avec les données. Assez correctement mais pas parfaitement.
\item L'imitation doit alors prendre probablement la forme d'une interaction indirecte et via les lieux de résidence.
\end{itemize}
\vspace{0.5cm}

Peut-être n'avons nous pas suffisamment bien développé les modèles basés sur cette question centrale~? A moins qu'il manque encore un autre ingrédient essentiel à prendre en compte~? Mais peut-être nous sommes-nous fourvoyés d'emblée dans une mauvaise question centrale~?  Mais quoi qu'il en soit, les résultats obtenus ici $\ti$et notamment un accord relativement correct entre les données et un modèle à un paramètre pertinent$\ti$ ne peuvent que nous inciter à poursuivre, et à enrichir, la démarche qui a animé tout ce travail.

\clearpage
\section{Conclusion}
\hfill
\begin{minipage}[r]{0.8\linewidth}
\textit{A mon avis pourtant, découvrir une chose parmi celles qui n'ont pas été trouvées, et qui, une fois trouvée, vaille mieux que si elle n'avait pas été découverte, c'est l'ambition et l'\oe{}uvre de l'intelligence, comme aussi accomplir jusqu'à son terme ce qui était à moitié accompli.}\\
Hippocrate, \textit{Art}~\cite{hippocrate}
\end{minipage}
\vspace{0.75cm}

La conclusion de ce travail tiendra en trois points. Nous récapitulerons, dans un premier temps, les principaux résultats dégagés de cette étude. Le fait de se confronter à des domaines et à des données, actuellement assez exotiques à la physique, ne peut nous laisser indifférent et nous incite à nous questionner. Pour cela, le deuxième point de la conclusion explicitera la démarche que nous avons volontairement suivie tout au long de ce travail. Enfin, nous ne pouvons nous dérober à de nécessaires remises en cause, et autres critiques que cette étude suscite.

\subsection{Bilan}
\label{pt-conclusion-bilan}
Nous avons étudié séparément les résultats électoraux de $6$ élections aux choix binaires, et $12$ taux de participation, tous deux pour des élections françaises entre $1992$ et $2007$, de même choix (ou assimilable) à l'échelle nationale. Les données électorales ont été analysées à l'échelle de la commune, voire du bureau de vote (avec environ $36000$ communes et $62000$ bureaux de vote pour la France métropolitaine). Nous avons de plus tenu compte de leur localisation spatiale, ou plus précisément de la position en coordonnées XY de la mairie de chaque commune. Ce sont les seules données non électorales dont nous avons tenu compte, ici.\\

L'étude des données a dégagé les points essentiels suivants~:
\begin{itemize}
\item la distribution des $\tau$ par commune (ou par bureau de vote) des taux de participation semble avoir un caractère permanent d'une élection à l'autre~;
\item l'hétérogénéité locale, $\sigma_0$, des résultats électoraux ou des taux de participation se révèle importante $\ti$au regard des régularités temporelles et spatiales, ou en terme d'information dite positive~; 
\item une commune centrale vote (ou s'abstient) en moyenne comme les communes de son environnement ($\tau=\tau_0$), et l'écart à son entourage, $|\tau-\tau_0|$, augmente en moyenne régulièrement avec la dispersion, $\sigma_0$, des communes de son environnement~;
\item les corrélations spatiales des résultats électoraux (ou des taux d'abstention) sont relativement fortes à faible distance (voire au sein des bureaux de vote d'une même ville), à longue portée ($\simeq 200~km$) et de décroissance spatiale quasi-logarithmique.
\end{itemize}

Les résultats électoraux montrent, généralement, de plus fortes corrélations intercommunales que les taux d'abstention. En revanche, les taux d'abstention des communes témoignent d'une plus grande permanence des valeurs (à un décalage de la moyenne nationale près) entre différentes élections.

Des régularités quantitatives $\ti$plus clairement prononcées pour les taux de participation$\ti$ sont ainsi apparues. Et, il est important de le signaler, des régularités indépendantes de la nature des élections, ainsi que du résultat (ou du taux de participation) moyen de l'élection.

Nous avons donc fait $3$ prédictions quantitatives concernant les taux de participation des européennes de juin $2009$. Chacune d'elles a été vérifiée à moins d'un \textit{sigma} (à moins d'un écart-type autour de la moyenne des mesures liées aux élections précédentes).\\

De plus, l'étude des régularités a le grand mérite de rendre significatives, les quelques irrégularités qui s'en détachent. Or, nous n'avons vu aucune irrégularité surgir $\ti$d'un contexte de régularités$\ti$, ni dans des zones à faible ou moyenne population, ni pour les taux de participation. Autrement dit, le surgissement d'irrégularité $\ti$au sein des régularités ambiantes$\ti$ traduit l'existence d'un autre, ou d'un plus intense, phénomène, appelé ici, faute de mieux, le \og fait politique \fg. Et ledit \og fait politique \fg, à l'aune des élections étudiées, ne peut alors se manifester que dans des zones à fortes population et pour l'expression des votes de certaines élections (et particulièrement ici, le référendum de $2005$).\\

Fort de cette analyse de données, nous avons voulu ensuite retrouver les principaux phénomènes à partir des modèles simples et reposant sur une dynamique plausible à l'échelle microscopique (ici, un comportement plausible des agents). De plus, partant des seules données électorales et de leur localisation spatiale, les modèles que nous utilisons se fondent sur une interaction entre agents telle que~:
\begin{itemize}
\item l'interaction n'est qu'une imitation très simple, et ne concerne indistinctement que les agents voisins~;
\item l'interaction entre agents n'est fonction que de la distance qui sépare leur lieu de résidence, et décroît exponentiellement en fonction de cette distance.
\end{itemize}
Autrement dit nous posons aux données la même question centrale~: une imitation simple (et à déterminer la forme) entre agents géographiquement voisins permet-elle de s'accorder aux phénomènes~? Enfin, concernant la formalisation du choix des agents, nous nous sommes servis du cadre conceptuel proposé par le modèle RFIM qui, par ailleurs, a pu relativement bien s'accorder à d'autres phénomènes sociaux.\\

Le premier point fort qui ressort de la confrontation d'un modèle aux données... est une réfutation. Nous avons montré l'impossibilité d'utiliser un modèle RFIM basé sur l'imitation des choix entre agents voisins pour reproduire les phénomènes électoraux. Tout se passe comme si, avec ce modèle, la proportion des agents du voisinage ayant fait un choix par rapport à l'autre (voter en faveur de l'un des deux candidats de la présidentielle, d'un oui ou d'un non lors d'un référendum, participer ou s'abstenir à une élection) n'avait aucun effet notable sur le choix à réaliser par l'agent. Nous avons ensuite laissé en question ouverte l'impossibilité pour tout modèle basé sur une simple imitation du choix des agents du voisinage (à l'instar d'une interaction ferromagnétique des spins), à pouvoir s'accorder avec ces données électorales.\\

Parmi les modèles correspondant à notre questionnement central, celui qui s'accorde le mieux aux données $\ti$mais pas encore complètement$\ti$ est un modèle de type diffusif, et de surcroît à un seul paramètre pertinent (la longueur caractéristique d'atténuation de la prise en compte du voisinage). Ce modèle possède trois caractéristiques. Premièrement, tout se passe comme si la tendance globale des communes, ou la conviction moyenne des agents (fortement apparente à $\tau$ d'une commune), diffuse. (Dans le modèle à seuil utilisé, le choix binaire pris par l'agent résulte de la position par rapport à un seuil de sa tendance, ou de sa conviction $\ti$une grandeur à une dimension réelle et non bornée.) Deuxièmement, les spécificités propres aux communes (l'équivalent des sources internes de chaleur dans un modèle de diffusion thermique) fluctuent sans corrélation et sans mémoire. Enfin, troisièmement, l'intensité de l'interaction entre deux communes voisines ne dépend pas de leur nombre d'habitants.  Elle ne dépend ici que de la distance qui les sépare. Ceci nous permet de privilégier l'interprétation d'une interaction indirecte des agents (via les tendances des lieux où ils résident), au détriment d'une interaction directe entre agents (via leur conviction).

\subsection{Démarche adoptée}
\label{pt-conclusion-demarche}
\og Ce que la physique peut apporter aux sciences sociales, c'est sa démarche expérimen\-tale \fg{}\cite{roehner_oral} déclare Bertrand Roehner $\ti$qui le met en application, notamment dans~\cite{roehner_econophysics, roehner_interaction}. (A ce sujet, les deux thèses~\cite{gallo_thesis, salganik_thesis_final}, ne manquent pas non plus d'intérêt.)

J'aurai tendance à adhérer à cette position, bien qu'à mon grand dam je ne connaisse pas encore suffisamment les sciences sociales. Je ferai donc dévier la discussion des sciences sociales aux phénomènes sociaux.

Cette thèse n'est pas à proprement parler une thèse expérimentale. En revanche, elle se contraint d'adopter la rigueur de la méthodologie scientifique, et celle de la physique en particulier. Cette démarche semble d'autant plus impérative que les données manipulées et les thèmes abordés s'écartent des objets et concepts usuellement traités en physique. Revenons donc succinctement sur la démarche ici suivie.\\

Les grandeurs électorales, spatialement localisées, ont ainsi été considérées et manipulées à l'instar de toutes autres grandeurs physiques mesurables. Peu importe donc qu'elles proviennent de l'activité humaine, aussi diverse et variée soit-elle. Plus important encore, les données électorales se sont départies de toute une quantité de \og détails \fg{} et d'interprétations. Des \og détails \fg{} liés au contexte social, politique et historique dans lequel se produit une élection, ainsi qu'aux candidats, partis politiques, etc., engagés dans l'élection, et, enfin, au comportement d'un même électeur à différentes élections. Nous pensions, par ce biais, atteindre plus facilement des régularités et des caractéristiques du phénomène électoral. En un mot, nous nous sommes efforcés à la simplicité.

Nous reconnaissons en revanche l'aspect limité et restreint de cette étude. La prise en compte des données non électorales (de type social, économique, politique, historique, etc.) pourraient, et devraient, enrichir ce travail. Et plus important encore, nous reconnaissons le risque qu'occasionnerait la prise en compte des données non électorales~: la réfutation de certains pans de cette étude. (Cf. le \og modèle zéro \fg{} infirmé par la prise en compte de la localisation spatiale des données électorales.) Mais de telles infirmations constitueraient-elles véritablement un problème, scientifiquement parlant~?\\

Sur un plan théorique, les modèles utilisés ici devaient naturellement s'accorder aux données, i.e. aux principaux traits dégagés de l'étude empirique des données. Les modèles devaient aussi représenter une dynamique plausible à l'échelle microscopique (ici, les agents). Ils devaient aussi être simples. Volontairement simples, de manière à ce que leurs traits saillants ressortent facilement. Leur confrontation aux données peut alors se comprendre comme une question posée aux données. A savoir, ce que disent les données sur les points saillants du modèle.

Précisons néanmoins que l'accord d'un modèle aux données n'exclut pas nécessairement d'autres approches possibles. De plus, une recherche empirique des régularités ne s'établit pas \textit{ex nihilo}. Elle sous-tend nécessairement, et en filigrane, des idées $\ti$même vagues$\ti$ sous-jacentes. Ici, nous avons été guidé par la recherche d'une relation entre une commune centrale et son environnement, et, de façon plus large, par la volonté de comprendre ce que traduit, et ce qu'induit, l'environnement d'une commune.\\

On peut être surpris de constater ici à quel point la volonté de simplicité prédomine tout au long de ce travail, tant sur son volet empirique que théorique. La réponse $\ti$ pour une fois$\ti$ est peut-être claire. C'est l'un des moyens, coutumier à la science, d'acquérir une emprise sur les phénomènes étudiés.

\subsection{Critique}
\label{pt-conclusion-critique}
La science part, usuellement, des phénomènes les plus simples pour aller aux phénomènes plus compliqués, ou plus complexes. Or, ici, on pourrait être surpris d'avoir mis sur un plan d'égalité les phénomènes liés aux résultats électoraux et ceux associés aux taux de participation. Ces derniers témoignent en effet, sans conteste, d'une plus grande régularité. On pourrait donc s'étonner de n'avoir pas commencé l'étude des données électorales par les taux de participation, et de n'avoir pas élaboré un modèle qui se préoccupe des seules données relatives à l'abstention.

La critique est valable. Je n'ai rien à y redire, si ce n'est quelques précisions supplémen\-taires.

Le grand mérite de l'étude des résultats électoraux est d'avoir, d'emblée, discerné l'importance de $\sigma_0$ (l'hétérogénéité des résultats électoraux, ou des taux d'abstention, dans l'environnement autour d'une commune centrale). En commençant l'étude des données par l'abstention, l'importance de $\sigma_0$ ne se serait pas détachée aussi nettement, et serait apparue au sein d'autres régularités, notamment celles liées à la variation du taux de participation en fonction de la population de la commune.

Quant aux modèles, j'ai cru $\ti$et continue à le croire$\ti$ que l'élément clé est la similarité, d'une élection à l'autre, de la distribution du taux de participation $\tau$ sur l'ensemble des communes. De par son importance même, cet élément devrait constituer le point de départ de toute modélisation des phénomènes électoraux (en France). La distribution des taux de participation $\tau$, et plus particulièrement sa dissymétrie (skewness), provient très probablement de la décroissance de la participation à une élection quand le nombre d'habitants de la commune augmente. J'ai essayé de partir de ce point de départ (un taux d'abstention qui n'est fonction, à un bruit près, que de la population de la commune, et notamment une décroissance en fonction de $1\big/\sqrt{\insc}$, cf. aussi les formes des corrélations spatiales de $1/\sqrt{\insc}$ et de $\pio$ comparées à celles de $\tau$ et de $\tau_0$), y ai passé du temps, et n'ai pas réussi.\\

La science, habituellement, se construit en se basant sur les connaissances précédentes $\ti$et ce, de manière cumulative, ou au contraire, parfois, sous forme de rupture. Bref, la physique n'a pas pour vocation de réinventer l'eau chaude, ni de retrouver des bribes d'un autre savoir déjà bien constitué. Pourtant, ici, on pourrait constater qu'il n'a été fait mention des sciences sociales qu'avec une très grande parcimonie. On devrait donc trouver risible, pour le moins, que cette étude réalisée sur les données électorales se soit construite presque indépendamment des sciences sociales.

La critique est de nouveau valable. Elle appelle néanmoins de nouvelles remarques, en plus de l'avertissement initial mis en introduction.

Je ne veux pas rentrer dans le débat concernant l'inutilité, à court terme, de la recherche scientifique $\ti$en me référant notamment, à la fulgurante progression des \og sciences \fg\, grecques comparées aux \og sciences \fg{} millénaires égyptiennes et mésopotamiennes~\cite{pichot}, ou bien à l'importance historique de la chute des corps dans l'élaboration de la science moderne. Bref, je ne parle pas ici de l'intérêt pratique de ce travail, mais du manque apparent d'intérêt à retrouver par la physique ce que les sciences sociales auraient préalablement établi. On pourrait alors considérer ce travail comme un jeu stérile, une pure perte de temps. Peut-être. Mais ce serait méconnaître l'enrichissement que peut procurer la diversité des angles de vues. (Mais vouloir multiplier sans arrêt les approches, ne présente guère d'intérêt non plus.) Il serait maintenant enrichissant de se confronter aux sciences sociales. Des points soulevés dans ce travail, lesquels ont été déjà trouvés, lesquels éventuellement entrent en contradiction, lesquels sont suraccentués, sous-accentués, etc.~? Bref, par la comparaison, par le décalage, produits par des voies d'investigations différentes, ce travail devrait probablement donner du relief à l'objet étudié ici (les données électorales). Mais cette comparaison, ce décalage, nécessitent préalablement des études menées quasiment indépendamment l'une de l'autre.

D'autre part, je me laisserais porter à croire qu'une problématique exclusivement physique de cette étude présente un avantage supplémentaire. (Il existe bien évidemment, bien d'autres questionnements possibles, et purement physiques.) Ce travail s'est efforcé de rendre explicite la problématique, le parti pris, dans lesquel il s'est engagé. Or, plutôt que de s'embarquer directement avec ses propres outils sur la problématique d'une autre discipline, un véritable travail pluridisciplinaire gagne, me semble-t-il, à s'appuyer sur différentes problématiques clairement identifiées. (Mais vouloir multiplier indéfiniment les problématiques, ne présente guère d'intérêt non plus.) La rencontre de plusieurs problématiques, indépendantes entre elles, sur le même objet d'étude pourrait $\ti$du moins, espérons-le$\ti$ s'avérer fécond. Il serait maintenant enrichissant de mieux connaître les problématiques et les outils issus des sciences sociales en vue de poursuivre ce travail. Puisse donc cette étude de physique faciliter et participer à une construction plus large à venir.\\

La science, aux dires de certains, serait une insatisfaction permanente et structurée. Les phénomènes sociaux ont été étudiés ici, avec une démarche entièrement issue de la physique. On pourrait se gausser à constater que cette étude, au thème exotique, n'ait suscité aucun questionnement $\ti$hormis l'affirmation de se positionner dans une démarche purement physique$\ti$ ni aucune remise en cause.

La critique est encore valable, et nécessite quelques explications supplémentaires.

Premièrement, je ne veux pas rentrer dans un débat philosophique sur le politique qui, par exemple, le définit comme \og un espace entre les hommes \fg, et le conçoit comme le \og miracle de la liberté \fg{} reposant sur l'agir de l'homme et sa capacité à créer un \og nouveau commencement \fg $\ti$incalculable donc~\cite{arendt}. Le politique n'a pas été étudié ici~: seules l'ont été les données électorales. Quant à savoir si les phénomènes humains sont, par nature, complètement séparés des phénomènes naturels $\ti$à l'instar des mondes sublunaire et supralunaire de la physique aristotélicienne$\ti$, je m'en tiendrai à une position naïve. Étudions et on verra. Et l'étude ne pourra qu'enrichir chacun des deux domaines si elle parvient $\ti$ce serait une sorte de Graal$\ti$ à montrer qu'ils restent distincts (de par l'affirmation de leur distinction), ou à faire naître un monde plus large, englobant les deux domaines anciennement distincts.

Enfin, j'inclinerais à croire que l'étude des phénomènes sociaux représente, pour la physique, une chance à saisir. Non seulement pour lui permettre d'affirmer sa démarche expérimentale $\ti$bien attestée par ailleurs$\ti$, ou pour appliquer des outils très $\ti$voire trop~?$\ti$ efficaces, comme ceux de la physique statistique. (Des outils trop efficaces dans le sens où, grisé par la réussite, ils n'incitent pas au premier abord à une réflexion critique.) Mais surtout, pour lui permettre de remettre en cause quelques-uns de ses concepts actuels. Je pense notamment au principe de séparabilité, i.e. la possibilité de séparer, de distinguer, une grandeur de son environnement, ou de permettre à une grandeur, à un objet, une existence isolée.

Mais avant de penser à un potentiel élargissement, encore fallait-il connaître ce qu'une étude classique pouvait dire sur le domaine. Ainsi s'achève cette thèse.

\clearpage
\renewcommand{\thesection}{A}
\renewcommand{\theequation}{A-\arabic{equation}}
\setcounter{equation}{0}  
\renewcommand{\thefigure}{A-\arabic{figure}}
\setcounter{figure}{0}
\renewcommand{\thetable}{A-\arabic{table}}
\setcounter{table}{0}
\section{Figures et tables supplémentaires}
\label{annexe-fig}
\vspace{0.25cm}

\begin{figure}[h]
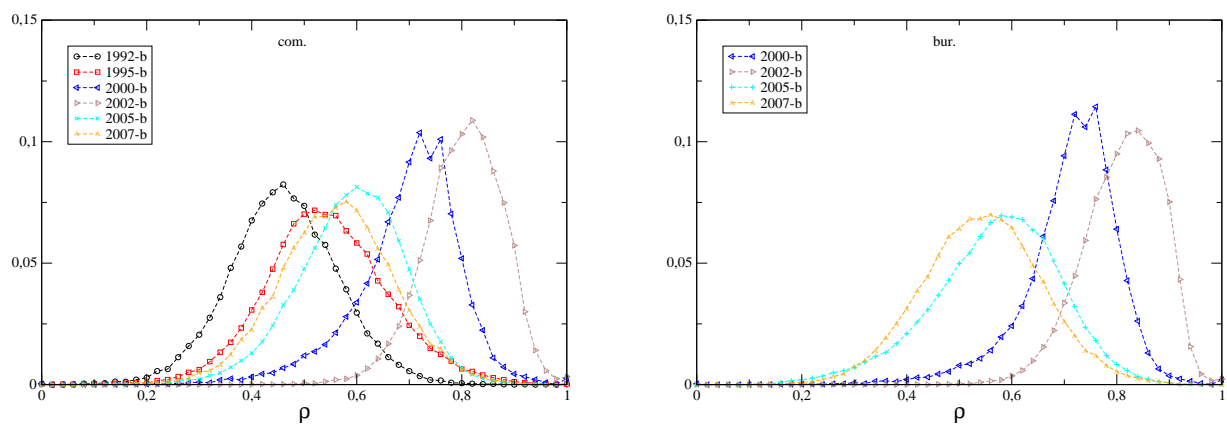

\includegraphics[scale = 0.32]{histo-dens-res.eps}\hfill
\includegraphics[scale = 0.32]{histo-dens-bvot-res.eps}
\caption{\small Histogrammes comparables à la Fig.~\ref{fhisto-log-res} des résultats $\res = \frac{\gagn}{\expr}$.}
\label{fhisto-dens-res}
\end{figure}

\begin{table}[h]
\begin{tabular}{|c||c|c|c|c|c||c|c|c|c|c|}
\hline
Année & \textbf{com.} & moy & éc-typ & skew & kurt & \textbf{bur.} & moy & éc-typ & skew & kurt\\
\hline
1992 & 36186 & 0,461 & 0,104 & 0,024 & 0,524 & & & & & \\ 
1995 & 36197 & 0,543 & 0,118 & 0,093 & 0,478 & & & & & \\ 
2000 & 36202 & 0,697 & 0,102 & -0,999 & 2,80 & 62253 & 0,713 & 0,091 & -1,20 & 3,99\\
2002 & 36215 & 0,803 & 0,075 & -0,476 & 0,728 & 62352 & 0,811 & 0,074 & -0,577 & 0,526\\
2005 & 36220 & 0,589 & 0,102 & -0,315 & 0,447 & 62775 & 0,569 & 0,121 & -0,410 & 0,279\\
2007 & 36219 & 0,560 & 0,111 & -0,091 & 0,455 & 63516 & 0,543 & 0,115 & -0,028 & 0,200\\
\hline
\end{tabular}
\caption{\small Statistiques de Tab.~\ref{tstat-log-res} pour les résultats $\res = \frac{\gagn}{\expr}$.}
\vspace*{0.5cm}
\label{tstat-dens-res}
\end{table}

\begin{figure}[!h]
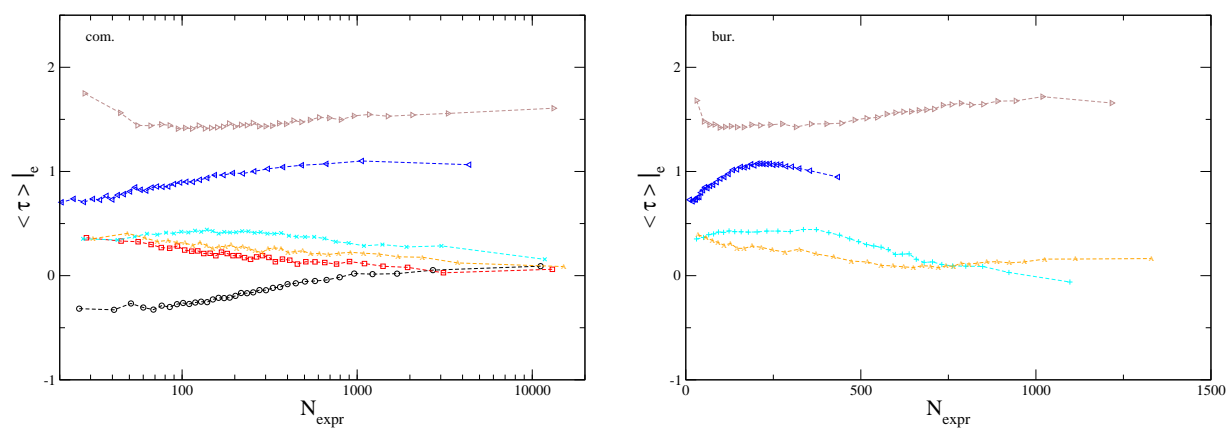

\includegraphics[scale=0.32]{moy-taille-log-res.eps}\hfill
\includegraphics[scale=0.32]{moy-taille-log-bvot-res.eps}
\caption{\small Moyenne des résultats $\tau$ calculés à l'intérieur de $36$ intervalles sur $\expr$ ; pour les communes à gauche et pour les bureaux de vote à droite.}
\vspace{0.5cm}
\label{fmoy-taille-res}
\end{figure}

\begin{figure}
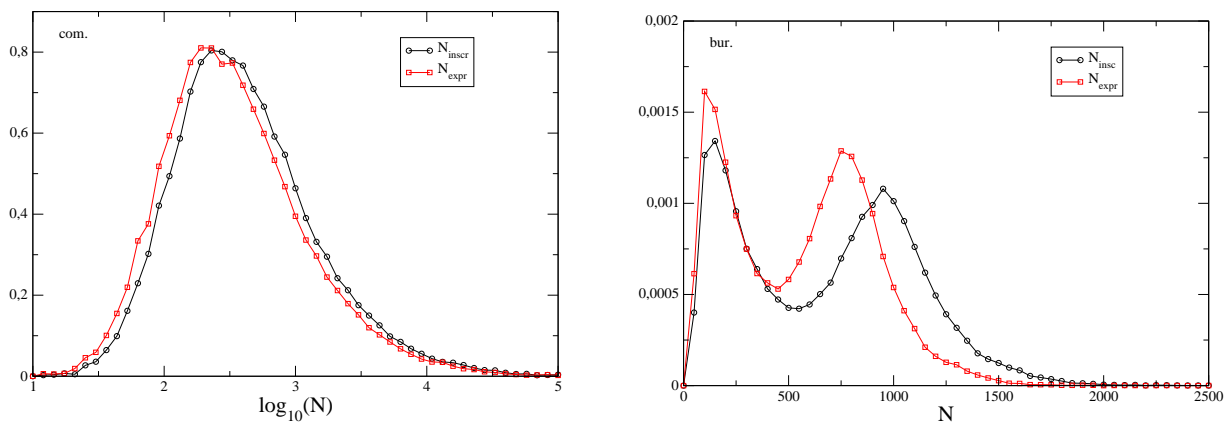

\includegraphics[scale = 0.32]{histo-insc-expr.eps}\hfill
\includegraphics[scale = 0.32]{histo-insc-expr-bvot.eps}
\caption{\small Histogrammes pour l'élection 2007-b des nombres d'électeurs inscrits $\insc$ et de suffrages exprimés $\expr$, par commune à gauche et par bureau de vote à droite. }
\vspace{0.5cm}
\label{fhisto-insc-expr}
\end{figure}

\begin{figure}
\includegraphics[scale = 0.32]{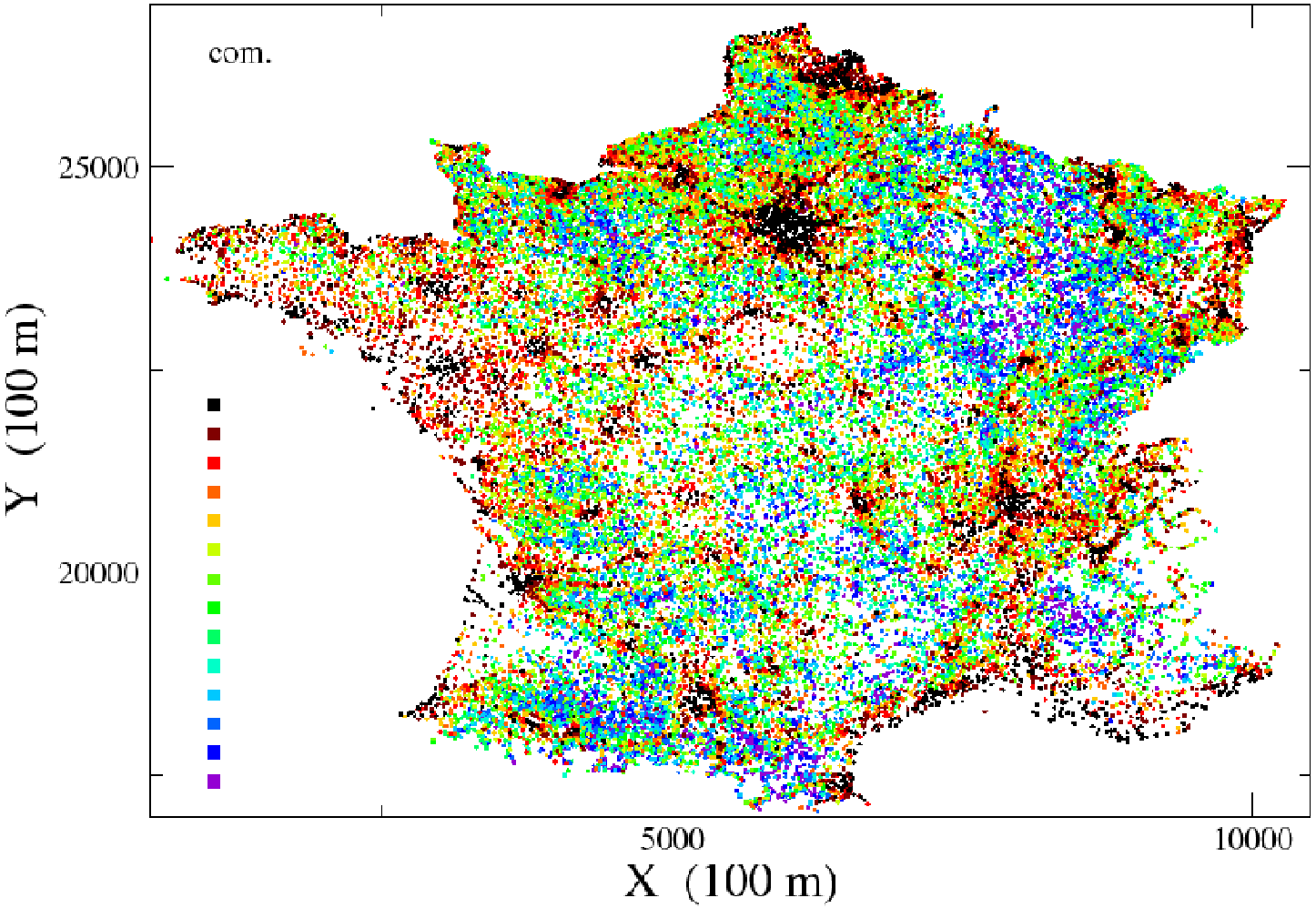}\hfill
\includegraphics[scale = 0.32]{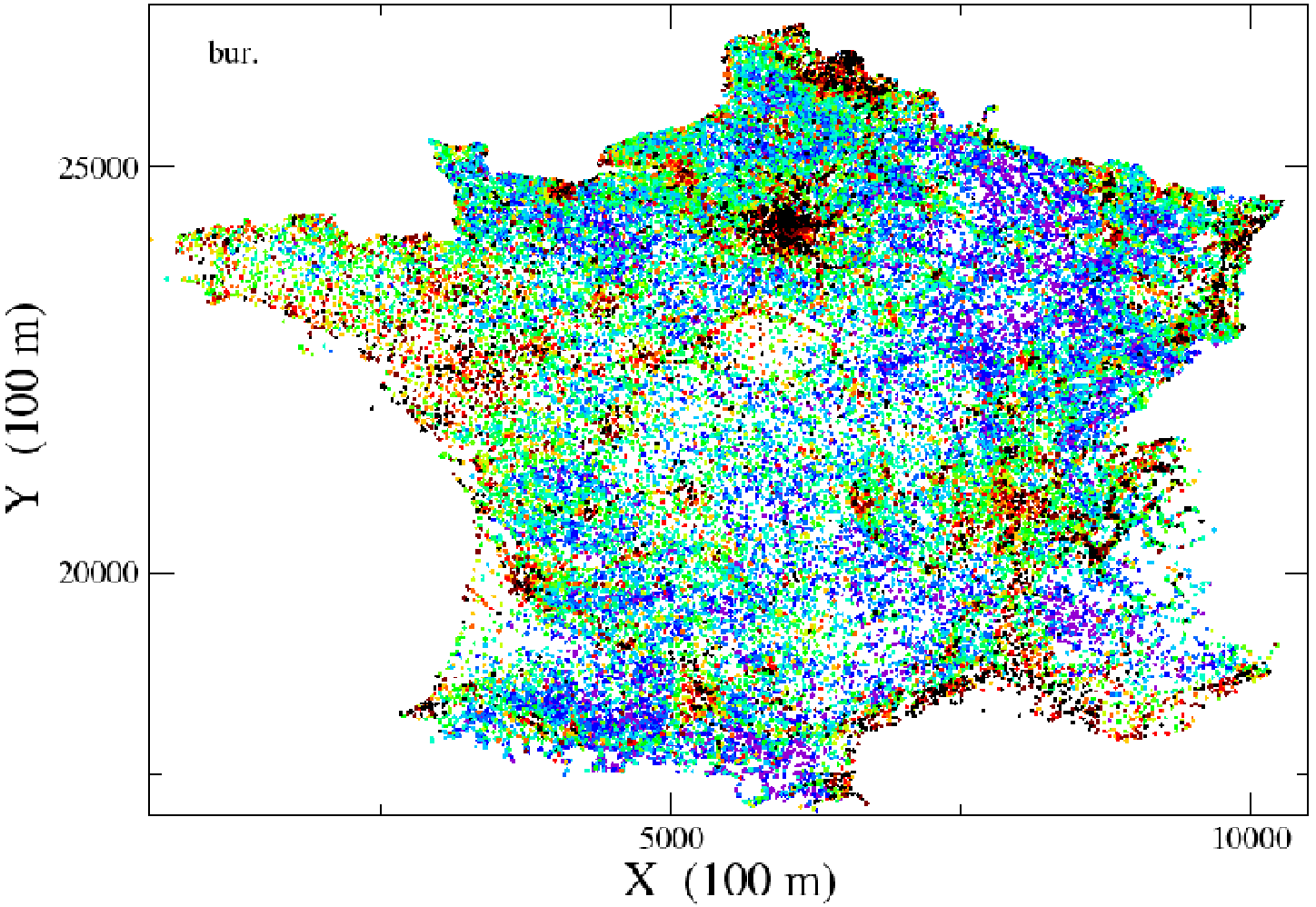}
\caption{\small Coordonnées XY des communes selon le nombre d'inscrits à l'élection 2007-b, des communes (à gauche), ou des bureaux de vote (à droite). Les nombres d'inscrits s'échelonnent sur $14$ intervalles, partant par ordre croissant du violet jusqu'au noir\protect\footnotemark.}
\vspace{0.5cm}
\label{frepartition-taille}
\end{figure}
\footnotetext{À l'origine, le choix des couleurs devait beaucoup à l'ardeur de Hugues Borghesi et de Léna Borghesi, de par leurs intenses réflexions, leurs débordements d'imagination, leurs jugements critiques, leurs réflexions amusées, leurs engouements passionnés, etc., et je les en remercie.}

\begin{figure}
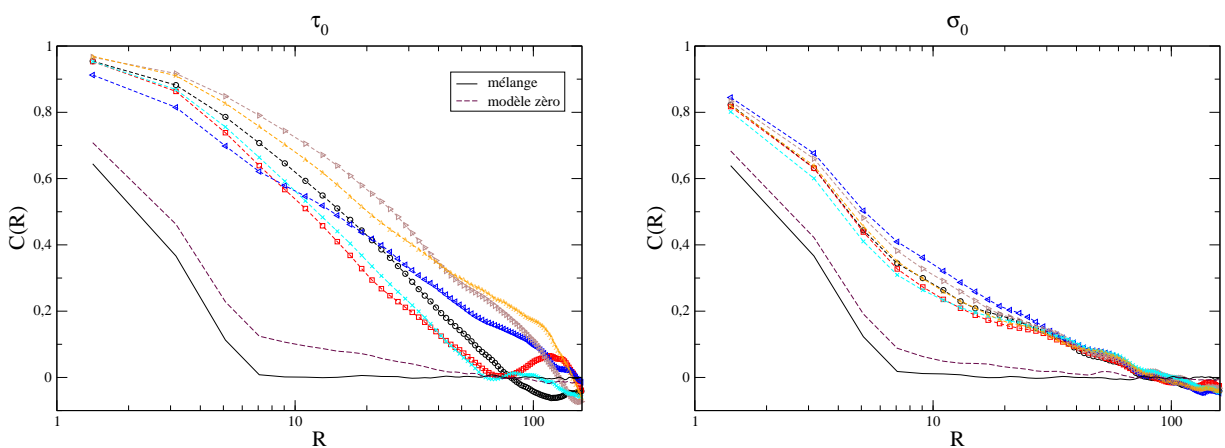

\includegraphics[scale=0.32]{correl-tppv-res.eps}\hfill
\includegraphics[scale=0.32]{correl-saa-res.eps}
\caption{Corrélations spatiales des $\toa$ (à gauche) et des $\saa$ (à droite) des résultats électoraux par commune. Le mélange aléatoire des résultats par commune et le \og modèle zéro \fg{} s'appliquent aux données de l'élection 2007-b.}
\vspace{0.5cm}
\label{fcorrel-toa-saa-res}
\end{figure}

\begin{figure}
\includegraphics[scale = 0.32]{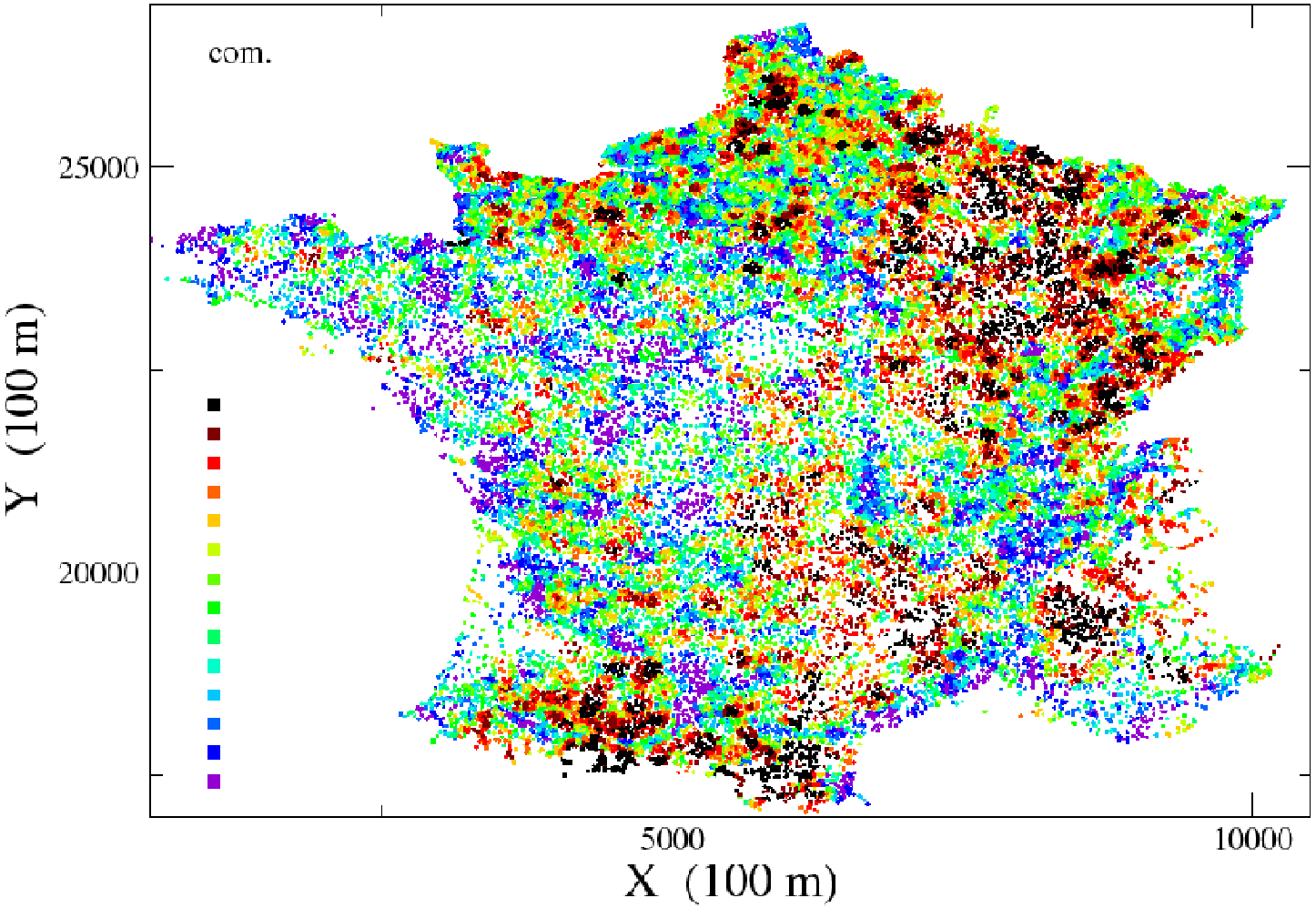}\hfill
\includegraphics[scale = 0.32]{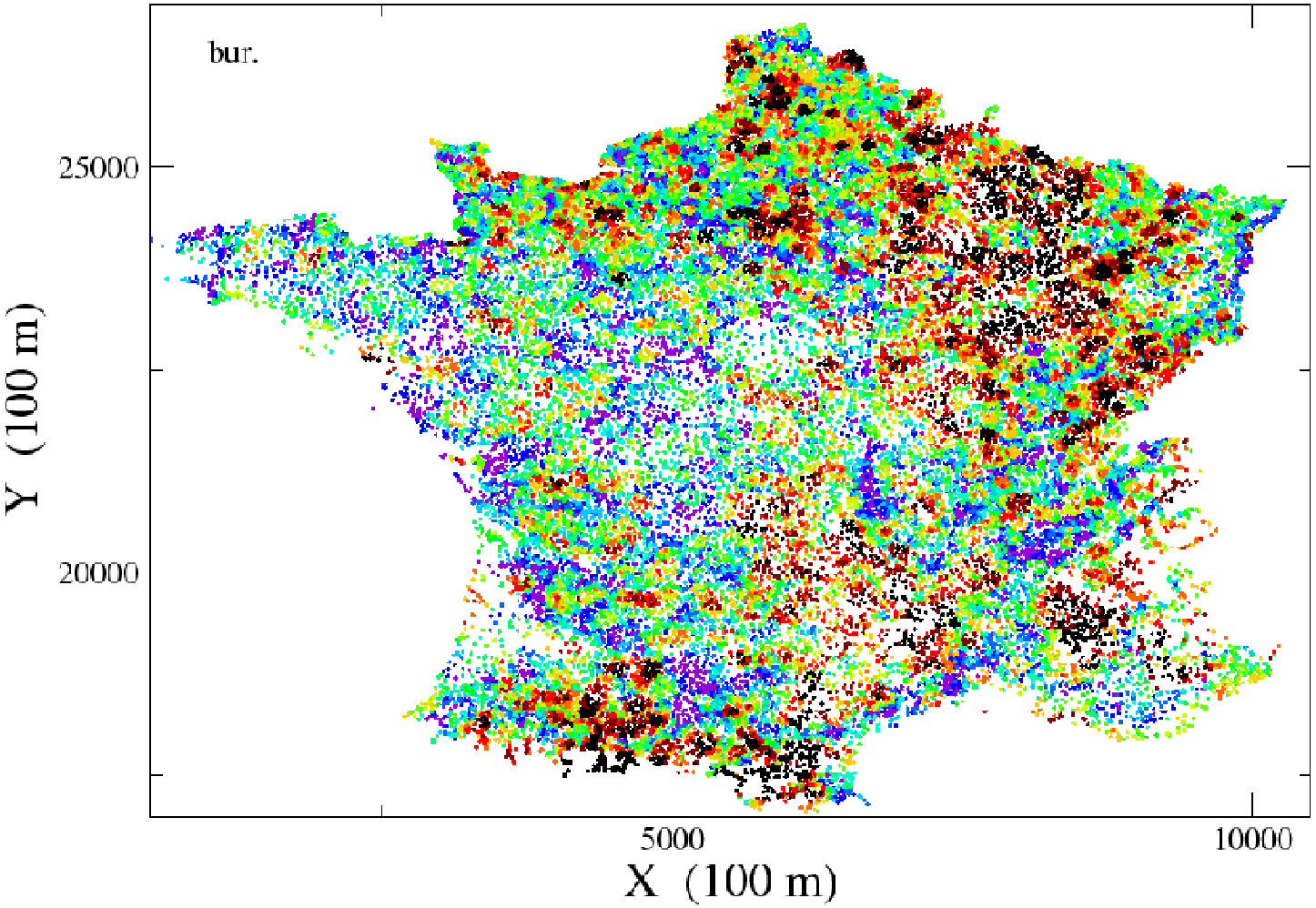}
\caption{\small Coordonnées XY des communes centrales sur lesquelles s'évaluent $\saa$ des résultats de l'élection 2007-b, en fonction des grandeurs par commune (à gauche), ou par bureau de vote (à droite). Les valeurs de $\saa$ se répartissent sur $14$ intervalles, partant du violet pour les plus faibles, jusqu'au noir pour les plus grandes.}
\vspace{0.5cm}
\label{frepartition-saa-res}
\end{figure}

\begin{figure}
\includegraphics[width=5cm, height=5cm]{histo-tppv-res.eps}\hfill
\includegraphics[width=5cm, height=5cm]{histo-sppv-res.eps}\hfill
\includegraphics[width=5cm, height=5cm]{histo-delta-res.eps}
\caption{\small Histogrammes de $\toa$, de $\saa$ et de $(\taa - \toa)$}
\vspace{0.5cm}
\label{fhisto-environnement-res}
\end{figure}

\begin{figure}
\includegraphics[scale = 0.32]{s0eqt0-res.eps}\hfill
\includegraphics[scale = 0.32]{t0eqs0-res.eps}
\caption{\small Eventuel lien entre les deux grandeurs caractérisant l'environnement : $\saa = f(\toa)$ et $\toa = f(\saa)$.}
\vspace{0.5cm}
\label{fs0t0-res}
\end{figure}

\begin{figure}
 \begin{minipage}[c]{0.55\linewidth}
  \includegraphics[scale=0.32]{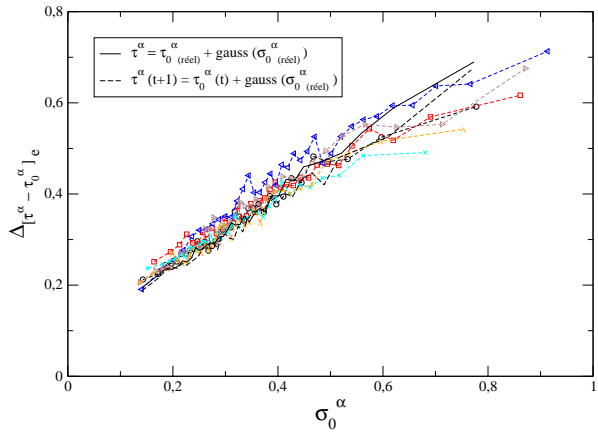}
 \end{minipage} \hfill
 \begin{minipage}[c]{0.45\linewidth}
  \caption{\small Ecart-type de $(\taa - \toa)$ en fonction de $\saa$, calculé avec $36$ intervalles. Les simulations (voir texte), en trait continu et en tirets, s'établissent à partir des valeurs réelles de l'élection 2007-b. gauss(X) signifie une distribution gaussienne de moyenne nulle et d'écart-type X.}
 \label{fdelta-saa-res}
 \end{minipage}
\end{figure}

\begin{table}
\begin{minipage}[c]{0.6\linewidth}
\begin{tabular}{|c|c|c|c|c|c|}
\hline
\backslashbox{$t_i$}{$t_j$} & 1995 & 2000 & 2002 & 2005 & 2007\\
\hline
1992 & -0,189 & 0,347 & 0,318 & -0,353 & -0,180\\
1995 & & -0,290 & -0,037 & -0,516 & 0,734\\
2000 & & & 0,253 & -0,028 & -0,329\\
2002 & & & & -0,308 & -0,339\\
2005 & & & & & -0,417\\
\hline
\end{tabular}
\end{minipage}\hfill
\begin{minipage}[c]{0.36\linewidth}
\caption{\small Corrélation temporelle $C_{t_i,t_j}(\tau)$ des $\taa$ sur chaque couple d'élections $(t_i,t_j)$. Pour les valeurs absolues : moyenne $\overline{C}_t(\tau) = 0,309$ ; écart-type $\Delta_{[C_t(\tau)]} = 0,175$ ; $\frac{\Delta_{[C_t(\tau)]}}{\overline{C}_t(\tau)} = 0,564$.}
\label{ttempo-res}
\end{minipage}
\end{table}

\begin{table}
\begin{minipage}[c]{0.6\linewidth}
\begin{tabular}{|c|c|c|c|c|c|}
\hline
\backslashbox{$t_i$}{$t_j$} & 1995 & 2000 & 2002 & 2005 & 2007\\
\hline
1992 & -0,075 & 0,574 & 0,367 & -0,535 & -0,109\\
1995 & & -0,421 & -0,069 & -0,644 & 0,795\\
2000 & & & 0,487 & -0,084 & -0,525\\
2002 & & & & -0,311 & -0,533\\
2005 & & & & & -0,458\\
\hline
\end{tabular}
\end{minipage}\hfill
\begin{minipage}[c]{0.36\linewidth}
\caption{\small Corrélation temporelle $C_{t_i,t_j}(\tau_0)$ des $\toa$ sur chaque couple d'élections $(t_i,t_j)$. Pour le valeurs absolues : moyenne $\overline{C}_t(\tau_0) = 0,399$ ; écart-type $\Delta_{[C_t(\tau_0)]} = 0,226$ ; $\frac{\Delta_{[C_t(\tau_0)]}}{\overline{C}_t(\tau_0)} = 0,567$.}
\label{ttempo-toa-res}
\end{minipage}
\end{table}

\begin{figure}
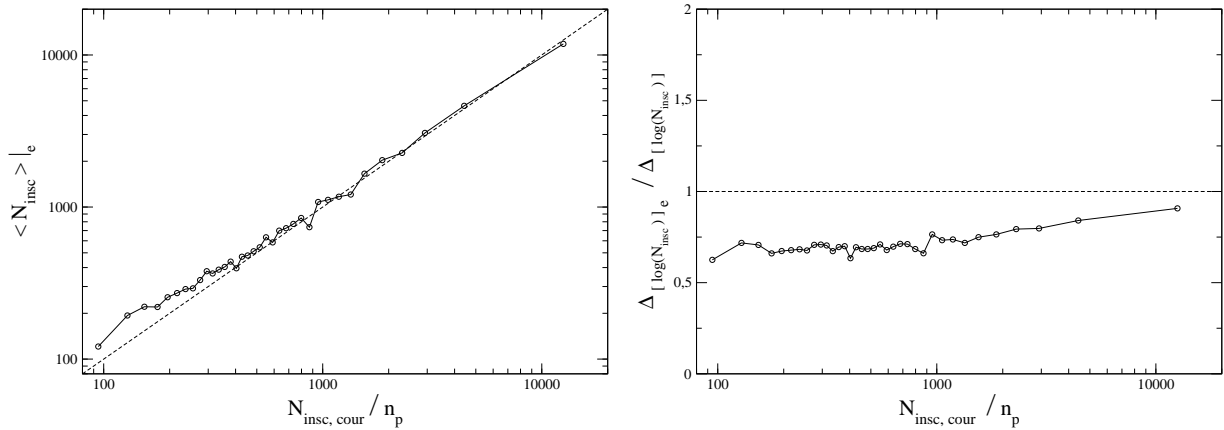

\includegraphics[scale = 0.32]{moy-inscrit=inscrit_0.eps}\hfill
\includegraphics[scale = 0.32]{sig-inscrit=inscrit_0.eps}
\caption{\small Le nombre d'inscrits $\insc^\aaa$ par commune centrale $\aaa$, en fonction de l'ensemble des inscrits, $\inscoa$, dans les $n_p$ communes de son environnement (cf. Eq.~(\ref{einscoa})). Le calcul des écarts-types relatifs, qui indique l'intensité de la polarisation, s'effectue à partir de $\log(\insc)$ afin d'éluder les problèmes liés aux lois de puissance.}
\label{fenvironnement-inscrit}
\end{figure}

\begin{figure}
 \begin{minipage}[c]{0.55\linewidth}
  \includegraphics[scale=0.32]{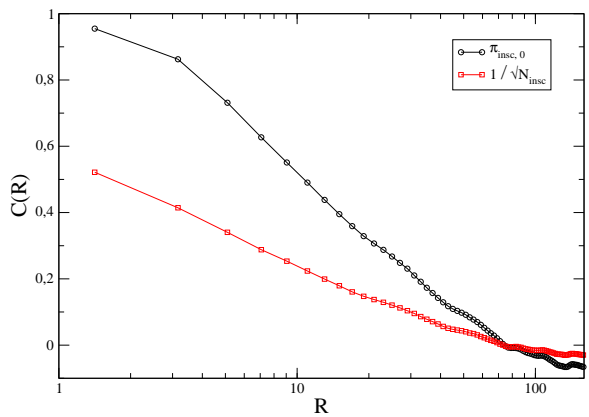}
 \end{minipage} \hfill
 \begin{minipage}[c]{0.45\linewidth}
  \caption{\small Corrélations spatiales de $\pio$ et de $\frac{1}{\sqrt{\insc}}$ correspondant à l'élection 2007-b. Noter la ressemblance $\ti$peut-être fortuite ~?$\ti$ des corrélations spatiales de $\frac{1}{\sqrt{\insc}}$ avec celles des résultats électoraux $\tau$ (excepté ceux de l'élection de $2000$) et, à un bruit supplémentaire près, celles des taux de participation (cf. Figs.~\ref{fcorrel-res} et \ref{fcorrel-abst}). Noter aussi la similarité des corrélations spatiales de $\pio$ avec celles des $\toa$ des résultats électoraux, et plus particulièrement encore avec celles des $\toa$ des taux de participation (cf. Figs.~\ref{fcorrel-toa-saa-res} et \ref{fcorrel-toa-saa-abst}).}
 \label{fcorrel-piinsc}
 \end{minipage}
\end{figure}

\clearpage
\begin{figure}[h]
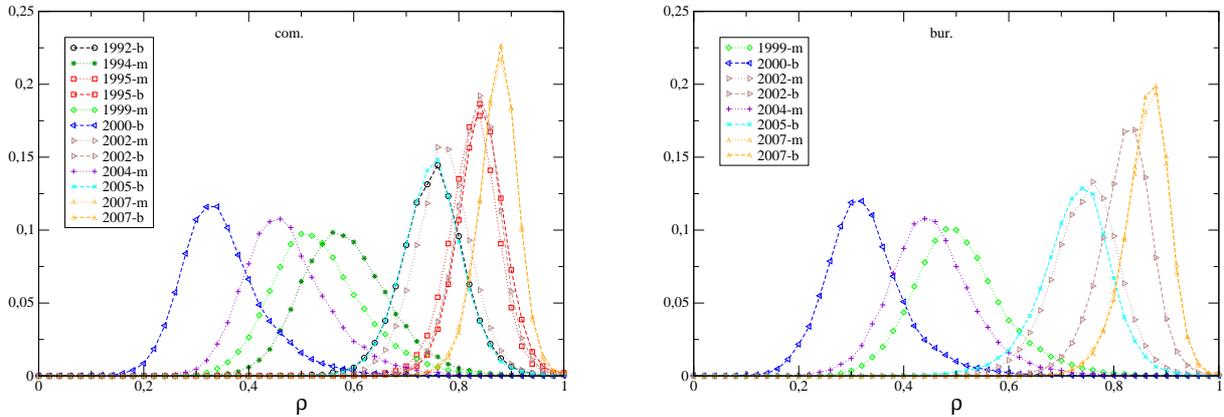

\includegraphics[scale = 0.32]{histo-dens-abst.eps}\hfill
\includegraphics[scale = 0.32]{histo-dens-bvot-abst.eps}
\caption{\small Histogrammes comparables à la Fig.~\ref{fhisto-log-abst} des taux de participation $\res = \frac{\vot}{\insc}$.}
\vspace*{0.25cm}
\label{fhisto-dens-abst}
\end{figure}

\begin{table}
\begin{tabular}{|c||c|c|c|c|c||c|c|c|c|c|}
\hline
Élection & \textbf{com.} & moy & éc-typ & skew & kurt & \textbf{bur.} & moy & éc-typ & skew & kurt\\
\hline
1992-b & 36186 & 0,750 & 0,061 & -0,148 & 1,20 & & & & &\\
1994-m & 36196 & 0,585 & 0,087 & 0,254 & 1,57 & & & & &\\
1995-m & 36195 & 0,827 & 0,050 & -0,841 & 6,60 & & & & &\\
1995-b & 36197 & 0,842 & 0,047 & -0,229 & 1,70 & & & & &\\
1999-m & 36209 & 0,534 & 0,090 & 0,565 & 1,07 & 62005 & 0,507 & 0,090 & 0,558 & 1,27\\
2000-b & 36212 & 0,352 & 0,083 & 1,082 & 3,34 & 62267 & 0,330 & 0,081 & 1,06 & 3,66\\
2002-m & 36217 & 0,770 & 0,056 & -0,255 & 2,64 & 62356 & 0,745 & 0,065 & -0,359 & 1,63\\
2002-b & 36217 & 0,837 & 0,045 & -0,296 & 1,28 & 62358 & 0,820 & 0,052 & -0,628 & 2,67\\
2004-m & 36221 & 0,476 & 0,086 & 0,870 & 1,88 & 62775 & 0,455 & 0,085 & 0,669 & 1,94\\
2005-b & 36223 & 0,750 & 0,059 & 0,004 & 0,93 & 62778 & 0,727 & 0,069 & -0,452 & 1,34\\
2007-m & 36219 & 0,873 & 0,040 & -0,768 & 7,51 & 63514 & 0,861 & 0,046 & -0,844 & 3,79\\
2007-b & 36219 & 0,874 & 0,039 & -0,397 & 1,39 & 63516 & 0,862 & 0,044 & -0,633 & 1,70\\
\hline
\end{tabular}
\caption{\small Statistiques de Tab.~\ref{tstat-log-abst} pour les taux de participation $\res = \frac{\vot}{\insc}$.}
\vspace*{0.25cm}
\label{tstat-dens-abst}
\end{table}

\begin{figure}[!h]
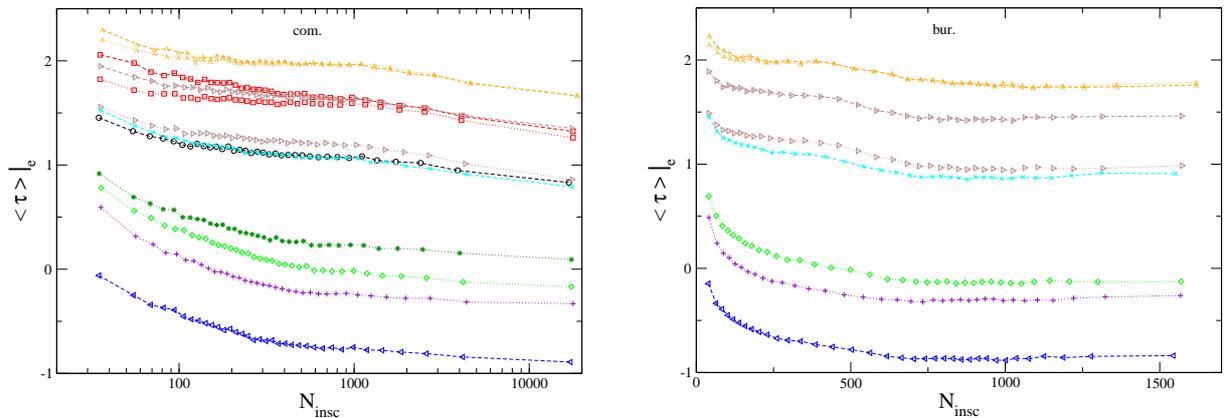

\includegraphics[scale=0.32]{moy-taille-log-abst.eps}\hfill
\includegraphics[scale=0.32]{moy-taille-log-bvot-abst.eps}
\caption{\small Moyenne des taux de participation $\tau$ calculés à l'intérieur de $36$ intervalles sur $\insc$ ; pour les communes à gauche et pour les bureaux de vote à droite.}
\vspace*{0.5cm}
\label{fmoy-taille-abst}
\end{figure}

\begin{figure}
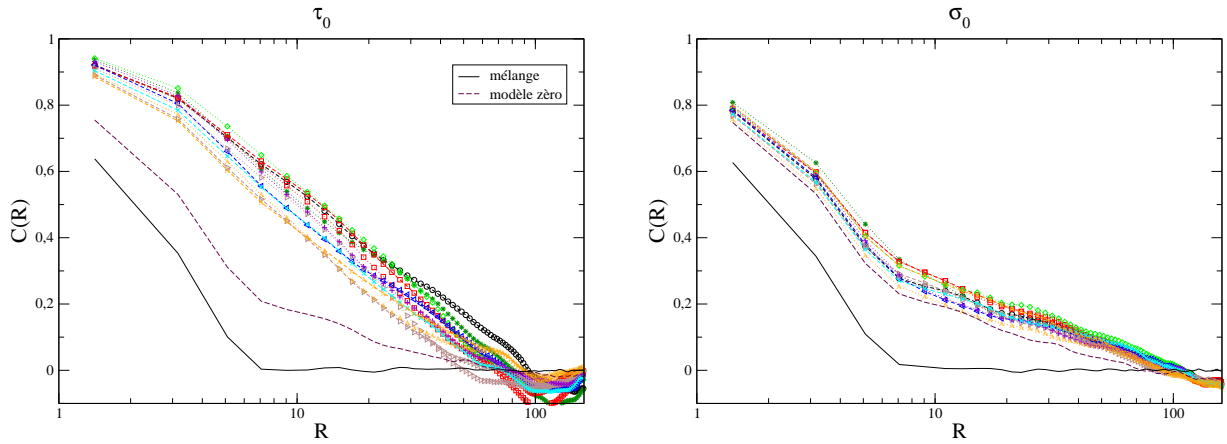

\includegraphics[scale=0.32]{correl-tppv-abst.eps}\hfill
\includegraphics[scale=0.32]{correl-saa-abst.eps}
\caption{Corrélations spatiales des $\toa$ (à gauche) et des $\saa$ (à droite) des taux de participation par commune. Le mélange aléatoire des résultats par commune et le \og modèle zéro \fg{} s'appliquent aux données de l'élection 2007-b.}
\vspace{0.5cm}
\label{fcorrel-toa-saa-abst}
\end{figure}

\begin{figure}
\includegraphics[scale = 0.32]{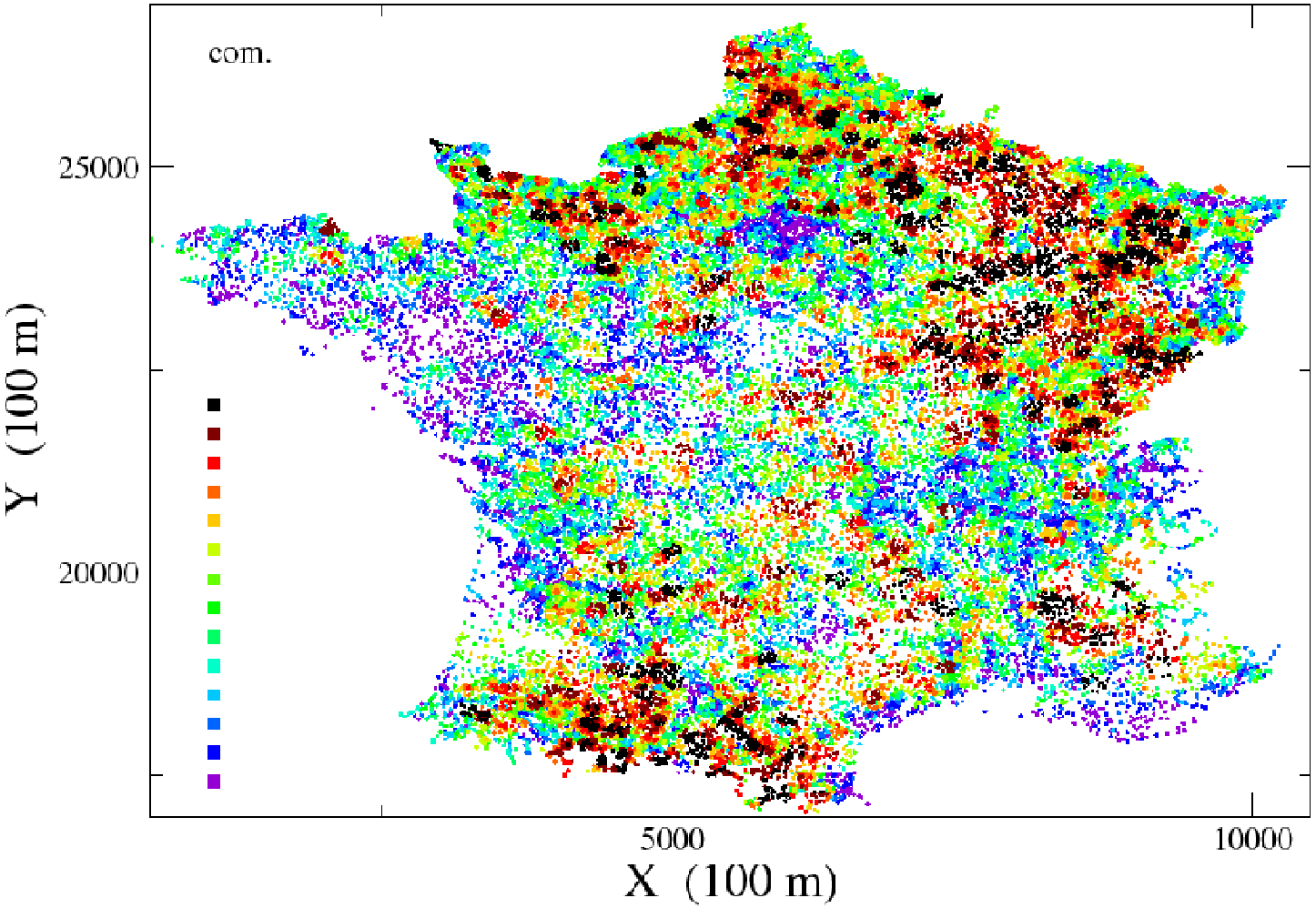}\hfill
\includegraphics[scale = 0.32]{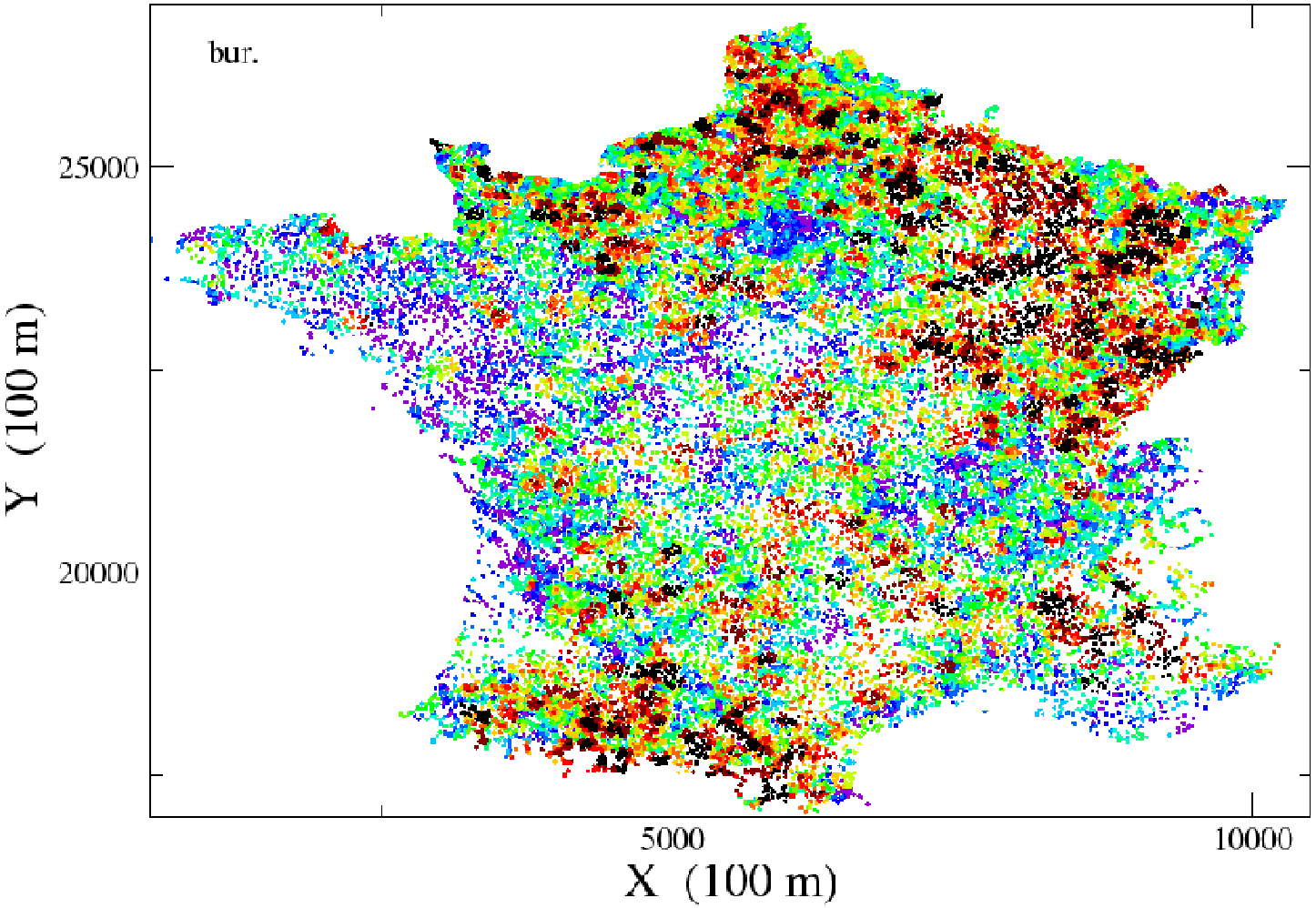}
\caption{\small Coordonnées XY des communes centrales sur lesquelles s'évaluent $\saa$ des taux de participation de l'élection 2007-b, en fonction des grandeurs par commune (à gauche), ou par bureau de vote (à droite). Les valeurs de $\saa$ se distribuent sur $14$ intervalles, partant du violet pour les plus faibles, jusqu'au noir pour les plus grandes.}
\vspace{0.5cm}
\label{frepartition-saa-abst}
\end{figure}

\begin{figure}
\includegraphics[width=5cm, height=5cm]{histo-tppv-abst.eps}\hfill
\includegraphics[width=5cm, height=5cm]{histo-sppv-abst.eps}\hfill
\includegraphics[width=5cm, height=5cm]{histo-delta-abst.eps}
\caption{\small Histogrammes de $\toa$, de $\saa$ et de $(\taa - \toa)$ des taux de participation.}
\vspace{0.5cm}
\label{fhisto-environnement-abst}
\end{figure}

\begin{figure}
\includegraphics[scale = 0.32]{s0eqt0-abst.eps}\hfill
\includegraphics[scale = 0.32]{t0eqs0-abst.eps}
\caption{\small Eventuel lien entre les deux grandeurs caractérisant l'environnement : $\saa = f(\toa)$ et $\toa = f(\saa)$.}
\vspace{0.5cm}
\label{fs0t0-abst}
\end{figure}

\begin{table}
\small
\begin{tabular}{|c|c|c|c|c|c|c|c|c|c|c|c|}
\hline
\backslashbox{$t_i$}{$t_j$} & 94-m & 95-m & 95-b & 99-m & 00-b & 02-m & 02-b & 04-m & 05-b & 07-m & 07-b\\
\hline
1992-b & 0,601 & 0,585 & 0,570 & 0,513 & 0,453 & 0,458 & 0,442 & 0,405 & 0,457 & 0,363 & 0,369\\ 
1994-m & & 0,549 & 0,578 & 0,663 & 0,576 & 0,487 & 0,439 & 0,562 & 0,488 & 0,345 & 0,373\\ 
1995-m & & & 0,740 & 0,471 & 0,395 & 0,523 & 0,488 & 0,347 & 0,442 & 0,408 & 0,410\\ 
1995-b & & & & 0,549 & 0,453 & 0,505 & 0,500 & 0,447 & 0,462 & 0,398 & 0,441\\ 
1999-m & & & & & 0,687 & 0,562 & 0,506 & 0,700 & 0,574 & 0,400 & 0,444\\ 
2000-b & & & & & & 0,485 & 0,451 & 0,628 & 0,496 & 0,351 & 0,387\\ 
2002-m & & & & & & & 0,736 & 0,525 & 0,595 & 0,510 & 0,527\\ 
2002-b & & & & & & & & 0,479 & 0,575 & 0,531 & 0,549\\ 
2004-m & & & & & & & & & 0,640 & 0,434 & 0,477\\ 
2005-b & & & & & & & & & & 0,595 & 0,607\\ 
2007-m & & & & & & & & & & & 0,764\\ 
\hline
\end{tabular}
\normalsize
\caption{\small Taux de participation : corrélation temporelle $C_{t_i,t_j}(\tau)$ des $\taa$ sur chaque couple d'élections $(t_i,t_j)$. Moyenne $\overline{C}_t(\tau) = 0,507$ ; écart-type $\Delta_{[C_t(\tau)]} = 0,099$ ; $\frac{\Delta_{[C_t(\tau)]}}{\overline{C}_t(\tau)} = 0,200$.}
\label{ttempo-abst}
\end{table}

\begin{table}
\small
\begin{tabular}{|c|c|c|c|c|c|c|c|c|c|c|c|}
\hline
\backslashbox{$t_i$}{$t_j$} & 94-m & 95-m & 95-b & 99-m & 00-b & 02-m & 02-b & 04-m & 05-b & 07-m & 07-b\\
\hline
1992-b & 0,733 & 0,748 & 0,705 & 0,654 & 0,656 & 0,649 & 0,631 & 0,473 & 0,679 & 0,486 & 0,526\\
1994-m & & 0,636 & 0,724 & 0,805 & 0,745 & 0,639 & 0,590 & 0,693 & 0,697 & 0,431 & 0,529\\
1995-m & & & 0,806 & 0,545 & 0,522 & 0,773 & 0,712 & 0,380 & 0,618 & 0,560 & 0,585\\
1995-b & & & & 0,747 & 0,680 & 0,734 & 0,750 & 0,651 & 0,715 & 0,572 & 0,681\\
1999-m & & & & & 0,828 & 0,673 & 0,645 & 0,854 & 0,772 & 0,485 & 0,620\\
2000-b & & & & & & 0,601 & 0,594 & 0,770 & 0,699 & 0,436 & 0,548\\
2002-m & & & & & & & 0,826 & 0,593 & 0,762 & 0,604 & 0,706\\
2002-b & & & & & & & & 0,598 & 0,767 & 0,755 & 0,806\\
2004-m & & & & & & & & & 0,759 & 0,486 & 0,630\\
2005-b & & & & & & & & & & 0,696 & 0,784\\
2007-m & & & & & & & & & & & 0,857\\
\hline
\end{tabular}
\normalsize
\caption{\small Taux de participation : corrélation temporelle $C_{t_it_j}(\tau_0)$ des $\toa$ sur chaque couple d'élections $(t_i,t_j)$. Moyenne $\overline{C}_t(\tau_0) = 0,661$ ; écart-type $\Delta_{[C_t(\tau_0)]} = 0,110$ ; $\frac{\Delta_{[C_t(\tau_0)]}}{\overline{C}_t(\tau_0)} = 0,167$.}
\label{ttempo-toa-abst}
\end{table}


\clearpage
\renewcommand{\thesection}{B}
\renewcommand{\theequation}{B-\arabic{equation}}
\setcounter{equation}{0}  
\renewcommand{\thefigure}{B-\arabic{figure}}
\setcounter{figure}{0}
\renewcommand{\thetable}{B-\arabic{table}}
\setcounter{table}{0}
\section{Taux de croissance de population et hétérogénéité locale $\mathbf{\sigma_0}$}
\label{annexe-croiss-pop}

Cette annexe esquisse une façon de mesurer la présence éventuelle d'une grandeur, d'une information, dans $\sigma_0$, lorsque cette grandeur s'évalue sur la même zone géographique que $\sigma_0$. Ici, $\sigma_0$ peut aussi bien se calculer à partir des données par commune que par bureau de vote, et concerne soit des résultats d'élections, soit des taux de participation.

Cette annexe se consacre à la connexion entr'aperçue précédemment (voir section~\ref{pt-sarko} et Tab.~\ref{tsarko}) entre les différents $\sigma_0$ de l'élection de 2007-b et le taux d'augmentation de population entre 1992 et 2007. (Remarque au passage~: n'ayant aucune certitude concernant le sens de la causalité entre $\sigma_0$ et le taux d'augmentation de population, ou de leur éventuelle rétroaction, nous préférons employer les termes de connexion, ou de présence, plutôt que celui d'influence. Ces deux premiers termes se mesurent d'ailleurs correctement par des coefficients de corrélation, sans oublier l'adage bien connu \og corrélation ne fait pas causalité \fg.)\\

Précisons les grandeurs utilisées pour décrire le taux d'augmentation de population dans la zone géographique, $\vaa$, d'évaluation de $\sigma_0$. De nouveau, la population se réduit et se mesure, ici, par le nombre de personnes inscrites sur les listes électorales des communes ou des bureaux de vote. Les données électorales que nous possédons s'échelonnent entre 1992 et 2007, ce qui incite à définir le taux d'augmentation de population sur le plus large intervalle possible, soit ici entre 1992 et 2007. N'ayant pas la connaissance des données par bureau de vote de l'année 1992, nous ne pouvons accéder qu'au taux d'augmentation de population par commune. Notons comme à la section~\ref{pt-sarko},
\be p^\aaa=\ln\big(\frac{\insc^\aaa(2007)}{\insc^\aaa(1992)}\big)~,\ee
le taux d'augmentation de population de la commune $\aaa$. Sur l'environnement $\vaa$ autour de la commune centrale $\aaa$, le taux d'augmentation de population peut alors se déterminer comme
\be \label{epoa} \poa = \frac{1}{n_p}\sum_{\beta\in\vaa} p^\beta~,\ee
ou comme
\be \label{epcour} \pcour = \ln\Big(\frac{\sum_{\beta\in\vaa} \insc^\beta(2007)}{\sum_{\beta\in\vaa} \insc^\beta(1992)}\Big)~,\ee
où $\beta$ désigne l'une des $n_p=16$ plus proches communes de $\aaa$.

La grandeur $\poa$, non pondérée par la taille de la population, offre plus de compatibilité avec $\saa$ (grandeur non pondérée avec la population), que $\pcour$ avec $\saa$. Par contre, la grandeur $\pcour$ s'accorde aussi bien à un traitement des données par commune que par bureau de vote, tandis que $\poa$ se restreint à celui des communes. En résumé, $\poa$ et $\pcour$ définissent ici le taux d'augmentation de population dans la zone géographique de $\saa$ ; avec une prédilection pour $\poa$ s'il s'agit de travailler avec $\saa$ calculé à l'aide des résultats électoraux (ou des taux de participation) par commune, et $\pcour$, faute de mieux, si $\saa$ se détermine avec des données par bureau de vote.\\

Chercher la présence de $\poa$ (ou de $\pcour$) dans $\saa$ par une simple corrélation $\ti$comme nous l'avions fait entre $p^\aaa$ et $\saa$ à la section~\ref{pt-sarko} et notamment à la table~\ref{tsarko}$\ti$ ne suffit pas. En effet, la taille de la population se connecte à la fois avec $\saa$ et avec le taux d'augmentation de la population. Une corrélation entre $\saa$ et $\poa$ (ou $\pcour$) ne signifie pas alors nécessairement que le taux d'augmentation de population se relie directement à $\sigma_0$. Par exemple, la corrélation sur l'ensemble des communes de $X^\aaa=\saa$ des résultats par commune de 2007-b et de $Y^\aaa=\piinsc$ (cf. Eq.~(\ref{epiinsc})) vaut $C_{XY}\simeq0,66$ (voir Tab.~\ref{tsaa-piinsc-res})~; et pour $X^\aaa=\poa$ et $Y^\aaa=\piinsc$, $C_{XY}\simeq-0,51$ (i.e. en moyenne, plus la population augmente, plus le taux de croissance de population, augmente aussi). La corrélation $C_{XY}\simeq-0,42$ entre $X^\aaa=\saa$ et $Y^\aaa=\poa$ ne permet pas alors d'affirmer une connexion directe entre ces deux dernières grandeurs, du fait de leur connexion commune avec $\piinsc$.

Nous devons donc avoir recours à une régression multiple, entre d'un côté $\saa$, et d'autre côté $\piinsc$ et $\poa$ (ou $\pcour$). De plus, pour des raisons de simplicité et de cohérence par rapport aux corrélations calculées dans le corps du texte, nous utiliserons une régression linéaire\footnote{L'idée d'une régression linéaire multiple entre $\saa$, une grandeur liée à la population et une autre au taux de croissance de population revient entièrement à Jean-Philippe Bouchaud.}. Ainsi, après avoir centré et réduit $\saa$, $\piinsc$ et $\poa$ (ou $\pcour$), nous pouvons déterminer par régression les paramètres estimer, $C_\pi$ et $C_p$, tels que :
\be \label{eregr-multicor} \tilde{\sigma}_0^\aaa = C_\pi.\tilde{\pi}_{insc,0}^\aaa + C_p.\tilde{p}_0^\aaa + \epsilon^\aaa~.\ee
($\tilde{X}$ désigne une variable $X$ de moyenne nulle et de variance unité. $\tilde{p}_{cour}^\aaa$ pourra remplacer $\tilde{p}_0^\aaa$ dans l'équation ci-dessus.) Le paramètre recherché, $C_p$, mesure la relation linéaire entre $\saa$ et le taux d'augmentation de population, $\poa$ (ou $\pcour$), non entachée de relation indirecte à la taille de la population $\ti$caractérisée ici par $\piinsc$. De même, $C_\pi$ exprime une relation linéaire entre $\piinsc$ et $\saa$, dégagée d'une relation indirecte avec le taux d'augmentation de population$\ti$mesuré par $\poa$ ou $\pcour$.

La qualité de la régression se retrouve dans le coefficient de corrélation multiple, $R$, écrit ici comme 
\be \label{er} R=\sqrt{1-\frac{\sum_{\aaa=1}^n (\epsilon^\aaa)^2}{n}}~,\ee 
où $n$ désigne le nombre de communes centrales $\aaa$ prises en compte. Plus $R$ se rapproche de $1$, plus la pertinence de la régression linéaire, entre d'une part $\saa$ et d'autre part $\piinsc$ et $\poa$ (ou $\pcour$), s'accroît.\\

La table~\ref{tregr-multicor} fournit les paramètres estimés, $C_\pi$ et $C_p$ ainsi que le coefficient de corrélation multiple $R$ pour les $\saa$ de l'élection 2007-b. $\saa$ et $\piinsc$ s'appliquent de façon concomitante sur les mêmes types de données~: celles issues des bureaux de vote ou bien celles issues des communes. Lorsque $\saa$ dérive des résultats électoraux (ou des taux de participation) des bureaux de vote, $\piinsc$ provient à son tour du nombre d'inscrits de ces bureaux de vote. Par contre, si $\saa$ découle des résultats électoraux (ou des taux de participation) par communes, $\piinsc$ se calcule aussi à partir du nombre d'inscrits par commune. Rappelons que $\piinsc$ se réfère ici aux inscrits de l'élection 2007-b. Remarque~: le taux d'augmentation de population exprimé par $\pcour$ s'applique à la fois, avec $\saa$ calculé via les données des bureaux de vote, et via celles des communes, afin de mieux comparer la régression linéaire appliquée à ces deux échelles de mesures différentes.

Par la suite, quand $\saa$ s'évaluera à l'aide des données issues des communes, le taux d'accroissement de population sur la zone géographique correspondante s'estimera uniquement par $\poa$. Par contre, si $\saa$ s'évalue à l'aide des données issues des bureaux de vote, $\pcour$ mesurera le taux d'augmentation de population correspondant.

\begin{table}
\begin{tabular}{|c c|c|c|c|}
\cline{3-5}
\multicolumn{2}{c|}{} & $C_\pi$ & $C_p$ & $R$\\
\hline
Résultats : & $\saa$ par com. et $\poa$ & 0,608 & -0,109 & 0,67\\
& $\saa$ par com. et $\pcour$ & 0,621 & -0,099 & 0,67\\
& $\saa$ par bur. et $\pcour$ & 0,554 & -0,110 & 0,61\\
\hline
Taux de participation : & $\saa$ par com. et $\poa$ & 0,707 & +0,085 & 0,67\\
& $\saa$ par com. et $\pcour$ & 0,681 & +0,042 & 0,66\\
& $\saa$ par bur. et $\pcour$ & 0,644 & +0,036 & 0,63\\
\hline
\end{tabular}
\caption{\small Paramètres $C_\pi$, $C_p$, et le coefficient $R$ pour différents $\saa$ de l'élection 2007-b. Pour mémoire, la corrélation $C_{\sigma\pi}$ entre $\saa$ et $\piinsc$ (voir Tabs.~\ref{tsaa-piinsc-res} et \ref{tsaa-piinsc-abst}) vaut, pour les résultats : $C_{\sigma\pi}\simeq 0,66$ à l'échelle des communes, $C_{\sigma\pi}\simeq 0,60$ à l'échelle des bureaux de vote ; et pour les taux de participation : $C_{\sigma\pi}\simeq 0,66$ à l'échelle des communes, $C_{\sigma\pi}\simeq 0,63$ à l'échelle des bureaux de vote. Par comparaison, pour un tirage binomial de probabilité $\overline{\rho}$ par commune, ou par bureau de vote, avec les données des résultats de l'élection 2007-b : $\tilde{\sigma}_0^\aaa\simeq0,85\tilde{\pi}_{insc,0}^\aaa$ (i.e. $R=C_{\sigma\pi}\simeq0,85$).}
\label{tregr-multicor}
\end{table}

Interprétons succinctement la table~\ref{tregr-multicor}, en s'aidant des corrélations $C_{\sigma\pi}$ entre $\saa$ et $\piinsc$ auxquelles elle se rapporte. Premièrement, le taux de croissance de population, sans sa connexion à la taille de la population, ou plus précisément ici à $\piinsc$, contribue faiblement à expliquer $\saa$. En effet, $C_\pi$ dépasse de beaucoup $C_p$ en valeur absolue. De plus, le coefficient de corrélation multiple $R$ n'augmente que très faiblement la corrélation simple $C_{\sigma\pi}$ entre $\saa$ et $\piinsc$. Ceci confirme la prédominance de $\piinsc$, comparée à $\poa$ (ou à $\pcour$), dans $\saa$. Deuxièmement, l'échelle de mesure des bureaux de vote pour les $\saa$ et les $\piinsc$ ne modifie que légèrement l'analyse effectuée à l'échelle des communes. Enfin, la connexion entre le taux de croissance de la population et $\saa$, diffère selon que $\saa$ se réfère aux résultats ou aux taux de participation. Cette connexion, mesurée ici par $C_p$, prend une valeur petite et positive pour les taux de participation, et signifie par là même une légère augmentation de la dispersion des taux de participation à l'intérieur d'une zone à forte augmentation de population. Par contre, pour les résultats électoraux, $C_p$ a une valeur négative et en valeur absolue supérieure à celle des taux de participation. Ce qui s'interprète selon les deux points suivants~: d'une part, la connexion entre $\saa$ et taux d'augmentation de population se marque davantage pour les résultats électoraux que pour les taux de participation, et, d'autre part, en moyenne plus une zone géographique connaît une forte augmentation de population, plus l'hétérogénéité des résultats à l'intérieur de cette zone diminue. Voilà ce qui constitue à nos yeux l'enseignement majeur de l'analyse faite ci-dessus recherchant la présence du taux d'augmentation de la population dans les divers $\saa$ de l'élection 2007-b.\\

Mais nous pouvons encore enrichir cette étude. Tout d'abord, à titre anecdotique, la table~\ref{tregr-multicor-region} fournit les paramètres $C_\pi$, $C_p$ et $R$, non plus sur toutes les communes centrales $\aaa$ de la France métropolitaine comme à la table~\ref{tregr-multicor}, mais pour chacune des $21$ régions métropolitaines. En termes de résultats électoraux, la région Île-de-France dépareille des autres régions. Ceci nous pousse tout naturellement à savoir comment les coefficients $C_\pi$, $C_p$ et $R$, varient en fonction du nombre d'inscrit $\inscoa$ (cf Eq.~(\ref{einscoa})) à l'intérieur de $\vaa$, et ce indépendamment de la localisation de la commune centrale $\aaa$.

\begin{figure}[t]
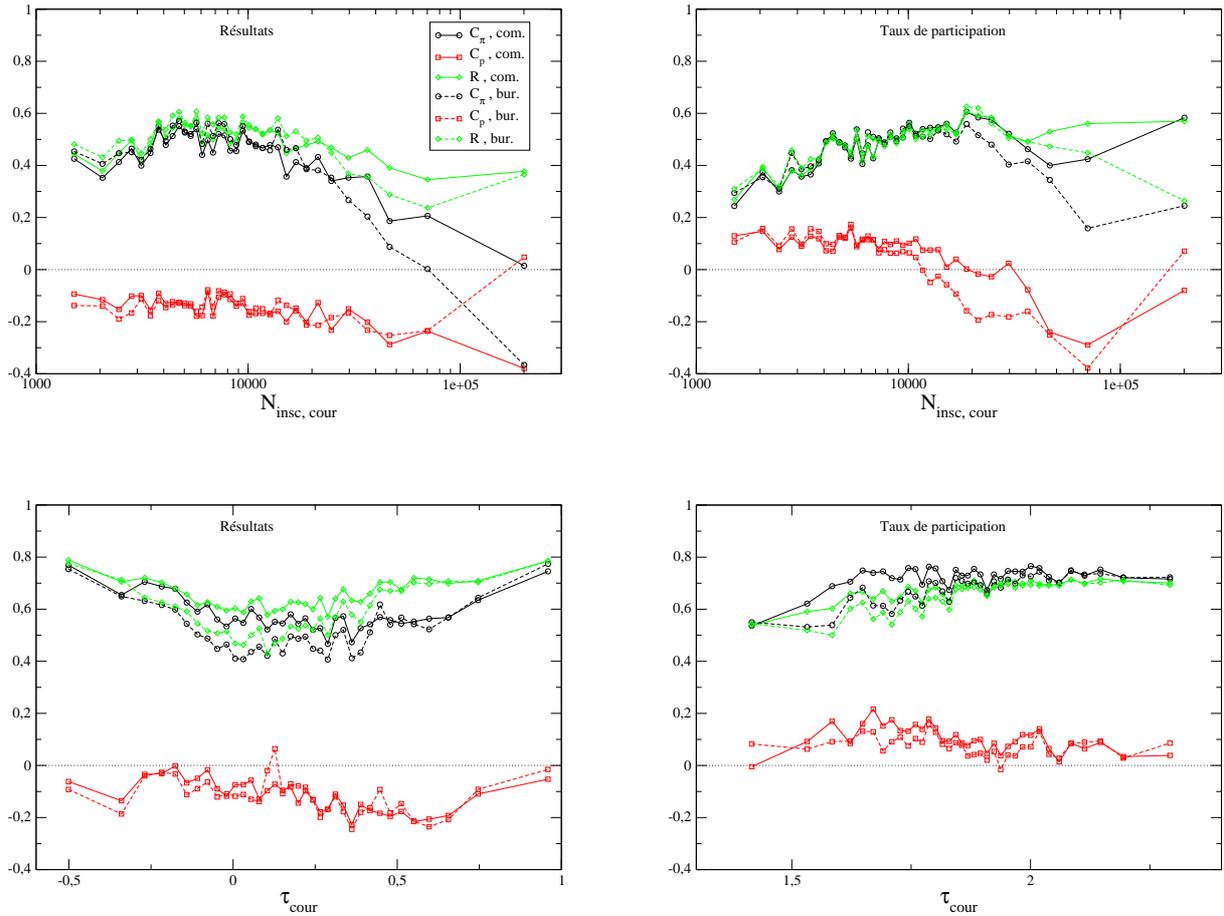

\includegraphics[scale = 0.32]{n0insc-multicor-res.eps}\hfill
\includegraphics[scale = 0.32]{n0insc-multicor-abst.eps}\vspace{1cm}
\includegraphics[scale = 0.32]{tcour-multicor-res.eps}\hfill
\includegraphics[scale = 0.32]{tcour-multicor-abst.eps}
\caption{\small Paramètres $C_\pi$, $C_p$ et le coefficient $R$ pour les résultats (à gauche) et pour les taux de participation (à droite) de l'élection 2007-b, calculés à l'intérieur de $36$ échantillons en $\inscoa$ (en haut) et en $\tau_{cour}$ (en bas). $\saa$ calculé à l'aide des données par commune (com.) utilise $\poa$, et à l'aide des données par bureau de vote (bur.), $\pcour$.}
\label{fmulticor}
\end{figure}

Et puisque nous nous posons ce genre de question, voyons aussi comment ces trois coefficients varient en fonction du résultat électoral (ou du taux de participation) global à l'intérieur de $\vaa$, et de nouveau pour l'ensemble des communes métropolitaines. Le résultat global, $\tau_{cour}^\aaa$, à l'intérieur de $\vaa$, se détermine à l'instar de $\taa$ (voir Eq.~\ref{etau}) comme
\be \label{etaucour} \tau_{cour}^\aaa = \ln\Big(\frac{\sum_{\beta\in\vaa} \gagn^\beta}{\sum_{\beta\in\vaa} (\expr^\beta - \gagn^\beta}\Big)~.\ee
Le nombre de suffrages exprimés $\expr$ et de suffrages $\gagn$ en faveur du choix gagnant à l'échelle nationale concernent évidemment ceux de l'élection 2007-b. Pour évaluer le taux de participation global $\tau_{cour}^\aaa$ sur cette zone, il suffit de remplacer $\expr$ par le nombre de personnes inscrites $\insc$, et $\gagn$ par le nombre de participants $\vot$. Notons que $\tau_{cour}^\aaa$, comme $\pcour$, s'applique indifféremment à une analyse à l'échelle des bureaux de vote ou à l'échelle des communes. Les figures~\ref{fmulticor} tracent $C_\pi$, $C_p$ et $R$ sur $36$ échantillons en $\inscoa$ ou en $\tau_{cour}$.

Terminons cette analyse en commentant succinctement les figures~\ref{fmulticor}. La première déduction tirée de ces figures se rapporte à la variation des paramètres $C_p$, $C_\pi$ et $R$, essentiellement sur les zones à forte population $\inscoa$. La plus grande modification concerne non pas les taux de participation mais les résultats électoraux. Interpréter les valeurs de ces trois coefficients, pour les résultats électoraux, et pour l'échantillon de la plus forte population, se révèle d'autant plus périlleux que $C_\pi$ et $C_p$ diffèrent selon l'échelle de mesure, i.e. selon que $\saa$ se calcule à partir des résultats électoraux des communes ou des bureaux de vote. Notons toutefois que $\saa$ des résultats électoraux croît quand $\inscoa$ augmente pour les zones à forte population (cf. Fig.~\ref{fsaa-taille-res}). Le second résultat $\ti$fort, celui-ci$\ti$ qui découle des figures~\ref{fmulticor}, concerne l'indépendance des paramètres $C_\pi$, $C_p$ et $R$ en fonction du résultat électoral (ou du taux de participation) global sur la zone d'évaluation de $\saa$. Autrement dit le fait qu'une zone vote plus ou moins pour un candidat (ou participe plus ou moins à l'élection) n'a aucun lien, en moyenne, sur la présence du taux de population et de $\piinsc$ dans $\saa$.\\

Nous avons voulu présenter dans cette annexe des méthodes pour extirper de l'information de $\saa$. En s'attaquant à d'autres données, autres que celle dont nous disposions $\ti$du taux d'augmentation de population, qui fournit, admettons-le, relativement peu d'information$\ti$ peut-être obtiendrait-on plus de renseignements~?

\begin{table}
\begin{tabular}{|c|c c c c|c c c c|}
\hline
\backslashbox{Régions}{Résultats}& com. & $C_\pi$ & $C_p$ & $R$ & bur. & $C_\pi$ & $C_p$ & $R$\\
\hline
Alsace& &0,42&-0,25&0,53& &0,16&-0,33&0,38\\
Aquitaine& &0,63&-0,00&0,64& &0,58&+0,04&0,55\\
Auvergne& &0,53&-0,22&0,70& &0,63&-0,13&0,70\\
Bourgogne& &0,56&-0,19&0,68& &0,49&-0,23&0,62\\
Bretagne& &0,26&-0,22&0,43& &0,08&-0,16&0,21\\
Centre& &0,57&+0,01&0,56& &0,33&-0,11&0,38\\
Champagne-Ardenne& &0,55&-0,05&0,58& &0,64&+0,07&0,61\\
Franche-Comté& &0,62&-0,12&0,69& &0,56&-0,17&0,64\\
Île-de-France& &-0,09&-0,34&0,37& &-0,25&-0,36&0,47\\
Languedoc-Roussillon& &0,66&-0,19&0,81& &0,63&-0,21&0,78\\
Limousin& &0,72&+0,07&0,68& &0,64&+0,05&0,61\\
Lorraine& &0,61&-0,11&0,67& &0,60&-0,18&0,67\\
Midi-Pyrénées& &0,65&-0,14&0,73& &0,66&-0,16&0,75\\
Nord-Pas-de-Calais& &0,44&-0,14&0,51& &0,26&-0,21&0,37\\
Basse-Normandie& &0,49&-0,09&0,54& &0,34&-0,16&0,42\\
Haute-Normandie& &0,27&-0,24&0,41& &0,12&-0,32&0,35\\
Pays de la Loire& &0,42&-0,08&0,47& &0,21&-0,15&0,32\\
Picardie& &0,39&-0,34&0,59& &0,30&-0,34&0,50\\
Poitou-Charentes& &0,45&-0,28&0,69& &0,36&-0,25&0,56\\
Provence-Alpes-Côte d'azur& &0,71&+0,01&0,70& &0,56&-0,15&0,64\\
Rhône-Alpes& &0,57&-0,21&0,72& &0,51&-0,20&0,63\\
\hline
\hline
\backslashbox{Régions}{Taux de participation}& com. & $C_\pi$ & $C_p$ & $R$ & bur. & $C_\pi$ & $C_p$ & $R$\\
\hline
Alsace& &0,37&+0,26&0,41& &0,23&-0,01&0,23\\
Aquitaine& &0,45&-0,08&0,51& &0,39&-0,17&0,51\\
Auvergne& &0,58&+0,01&0,57& &0,48&+0,03&0,46\\
Bourgogne& &0,59&+0,03&0,57& &0,51&-0,02&0,52\\
Bretagne& &0,39&+0,06&0,36& &-0,05&-0,13&0,12\\
Centre& &0,63&+0,15&0,58& &0,44&-0,04&0,45\\
Champagne-Ardenne& &0,65&+0,30&0,49& &0,51&+0,14&0,46\\
Franche-Comté& &0,59&+0,04&0,57& &0,55&-0,04&0,57\\
Île-de-France& &0,65&+0,00&0,65& &0,53&+0,01&0,53\\
Languedoc-Roussillon& &0,83&+0,08&0,77& &0,73&+0,04&0,71\\
Limousin& &0,62&+0,21&0,52& &0,48&+0,22&0,38\\
Lorraine& &0,67&+0,03&0,66& &0,63&-0,02&0,64\\
Midi-Pyrénées& &0,75&+0,14&0,69& &0,68&+0,07&0,65\\
Nord-Pas-de-Calais& &0,53&-0,06&0,56& &0,47&-0,15&0,53\\
Basse-Normandie& &0,66&+0,22&0,58& &0,48&+0,03&0,47\\
Haute-Normandie& &0,36&-0,10&0,40& &0,33&-0,10&0,35\\
Pays de la Loire& &0,76&+0,14&0,68& &0,50&-0,06&0,54\\
Picardie& &0,52&-0,03&0,53& &0,49&-0,04&0,50\\
Poitou-Charentes& &0,45&-0,08&0,52& &0,26&-0,17&0,40\\
Provence-Alpes-Côte d'azur& &0,59&+0,08&0,54& &0,66&+0,10&0,63\\
Rhône-Alpes& &0,73&+0,02&0,72& &0,63&-0,06&0,66\\
\hline
\end{tabular}
\caption{\small Paramètres $C_\pi$, $C_p$ et le coefficient $R$ par région, pour les résultats (en haut) et pour les taux de participation (en bas) de l'élection 2007-b ; avec $\saa$ évalué à partir des communes (com.) et $\poa$, ou bien à partir des bureaux de vote (bur.) et $\pcour$.}
\label{tregr-multicor-region}
\end{table}

\clearpage
\renewcommand{\thesection}{C}
\renewcommand{\theequation}{C-\arabic{equation}}
\setcounter{equation}{0}  
\renewcommand{\thefigure}{C-\arabic{figure}}
\setcounter{figure}{0}
\renewcommand{\thetable}{C-\arabic{table}}
\setcounter{table}{0}
\renewcommand{\thesubsection}{C. \arabic{subsection}}
\setcounter{subsection}{0}
\section{L'hétérogénéité locale $\mathbf{\sigma_0}$ et l'information locale dite positive}
\label{annexe-extraire}

Cette annexe s'attache à déceler la trace présente dans $\saa$, d'une information locale, permanente et non entachée de bruit dû aux tailles finies.

Dans un premier temps, nous essaierons de mesurer $\ti$y compris \textit{a minima}$\ti$ dans quelle proportion cette partie pérenne intervient dans la permanence globale de $\saa$ à diverses élections. Ensuite, nous appréhenderons cette partie pérenne par l'intermédiaire des corrélations inter-communales, que nous tenterons indirectement de mesurer. Nous localiserons ensuite les zones à plus ou moins forte partie pérenne. Puis, nous discuterons des méthodes et des résultats obtenus.

Tous les calculs et discussions s'effectueront à partir des résultats électoraux. Nous appliquerons ensuite, directement, les mêmes méthodes aux taux d'abstention. La richesse apportée par la comparaison, ou par la multiplication des données, nous permettra de mieux comprendre les différentes notions développées pour les résultats électoraux.

\subsection*{Avec les résultats électoraux}
\label{annexe-extraire-res}

\subparagraph{$\bullet$ Extraction par mélange\\}
\label{pt-extraire-mel}
Nous essayons ici de ne pas tenir compte du bruit statistique dû aux tailles finies dans la permanence temporelle globale des $\saa$, et ce de façon purement empirique. Nous voulons par ce biais détecter, et mesurer, l'information locale dite positive que contient $\saa$ (cf. section~\ref{pt-separer-res}).

Supposons que l'information locale dite positive se trouve présente dans $\saa$ de manière additive. Autrement dit, supposons que nous puissions écrire pour une élection à la date $t$~:
\be \label{esaa-a+b} \saa(t) = a^\aaa(t) + b^\aaa(t)~,\ee 
où $a^\aaa$ contient l'information locale dite positive ou pérenne de $\saa$, et $b^\aaa$ renferme tout le reste, y compris le bruit statistique dû aux tailles finies.

Noter au passage que les figures~\ref{fsaa-piinsc-res} ne tracent que la moyenne de $\saa$ en fonction de $\piinsc$, mais ne fournissent aucune indication quant à la position, pour une commune $\aaa$ donnée, de $\saa$ par rapport à cette moyenne. Dit autrement ces figures n'indiquent pas si $\saa$ d'une commune $\aaa$ donnée se trouve, par exemple, constamment au-dessus de la moyenne, ou fluctue par rapport à celle-ci d'une élection à l'autre. 

Il reste maintenant à savoir comment filtrer $a^\aaa$ de $\saa$~?

Il convient d'utiliser un filtre qui se base sur $\piinsc$ afin d'éliminer assez correctement le bruit de type binomial dû aux tailles finies des communes (ou des bureaux de vote). $\piinsc$ intervient en effet directement dans $\saa$ d'une binomiale pure et aussi dans $\saa$ des résultats réels (cf. section~\ref{pt-bruit-res}, Fig.~\ref{fsaa-piinsc-res} et Tab.~\ref{tsaa-piinsc-res}). Nous utilisons alors $n_e$ échantillons classés en fonction de $\piinsc$, dans lesquels se répartissent les $\saa$ avec un nombre sensiblement égal de $\saa$ par échantillon. Nous considérons donc qu'à l'intérieur de chaque échantillon, l'intensité du bruit dû aux tailles finies présent dans $\saa$ prend la même valeur $\ti$aux fluctuations près.

La technique d'extraction que nous utiliserons se base sur un échantillonnage en $\piinsc$, voire un mélange aléatoire à l'intérieur de chaque échantillon, et repose donc sur les deux hypothèses suivantes~:
\begin{itemize}
\item à l'intérieur d'un échantillon, l'amplitude du bruit dû aux tailles finies des communes (ou des bureaux de vote) prend sensiblement la même valeur pour tous les $\saa$~;
\item $\saa(t)$ peut s'écrire comme $\saa = a(t) + b(t)$, où $a(t)$ exprime la partie permanente de $\saa$ $\ti$à laquelle se rattache l'information locale dite à caractère positif$\ti$, et $b(t)$ dénote tout le reste, dont la partie provenant du bruit statistique des tailles finies.
\end{itemize}
Nous discuterons ultérieurement de ces deux hypothèses.\\

Notons 
\be \saa=a^\aaa_e(t)+b^\aaa_e(t)~,\ee
l'expression de $\saa$ de la commune $\aaa$ appartenant à l'échantillon $e$ lors de l'élection à la date $t$. Écrivons ensuite \begin{eqnarray}
                                                  a^\aaa_e(t)=\overline{a}_e(t)+\epsilon^\aaa_{a,\:e}(t)~, \\
                                                  b^\aaa_e(t)=\overline{b}_e(t)+\epsilon^\aaa_{b,\:e}(t)~,                      \end{eqnarray}
où $\overline{a}_e(t)$ et $\overline{b}_e(t)$ désignent respectivement les moyennes des $a^\aaa_e(t)$ et des $b^\aaa_e(t)$ à l'intérieur de l'échantillon $e$. Bien évidemment, selon cette écriture et pour une commune centrale $\aaa$ donnée, les $\epsilon^\aaa_{a,\:e}$ n'ont pas de corrélation avec les $\epsilon^\aaa_{b,\:e}$, de même les $\epsilon^\aaa_{b,\:e}(t_i)$ et $\epsilon^\aaa_{b,\:e}(t_j)$ à deux dates différentes $t_i$ et $t_j$. Les seules corrélations existantes, pour $\aaa$ fixé, concernent la partie permanente $\epsilon^\aaa_{a,\:e}(t_i)$ et $\epsilon^\aaa_{a,\:e}(t_j)$. Ainsi, 
\be \label{emoye} \langle\saa(t_i).\saa(t_j)\rangle\big|_e = \langle\epsilon^\aaa_{a,\:e}(t_i).\epsilon^\aaa_{a,\:e}(t_j)\rangle\big|_e +\big(\overline{a}_e(t_i)+\overline{b}_e(t_i)\big).\big(\overline{a}_e(t_j)+\overline{b}_e(t_j)\big)~,\ee
où $\langle...\rangle\big|_e$ dénote la moyenne à l'intérieur de l'échantillon $e$.

L'équation précédente se simplifie en faisant intervenir la covariance (donnée par l'expression $cov(X,Y)=\langle X.Y\rangle -\langle X\rangle.\langle Y\rangle$). En reprenant la notation des corrélations temporelles (cf. section~\ref{pt-tempo-res}), notons $cov_{t_i,t_j}(\sigma_0)$, la covariance sur l'ensemble des communes centrales $\aaa$, des $\saa$ des résultats d'élections aux dates $t_i$ et $t_j$. Restreinte aux communes centrales $\aaa$ d'un échantillon $e$, la covariance s'écrira comme $cov_{t_i,t_j}(\sigma_0)\big|_e$. L'équation~(\ref{emoye}) s'écrit alors plus simplement comme~:
\be \label{ecove} cov_{t_i,t_j}(\sigma_0)\big|_e = \langle\epsilon^\aaa_{a,\:e}(t_i).\epsilon^\aaa_{a,\:e}(t_j)\rangle\big|_e =cov_{t_i,t_j}(a)\big|_e.\ee

Ainsi, la covariance de la partie permanente $a^\aaa$ contenue dans $\saa$ découle directement d'un simple calcul de covariance à l'intérieur d'un échantillon $\ti$dans lequel l'amplitude du bruit dû aux tailles finies prend, par hypothèse, sensiblement la même valeur pour tous les $\saa$. En d'autres termes, nous avons réalisé une sorte de filtre qui extrait le bruit dû aux tailles finies dans la covariance de $\saa$.

La covariance sur l'ensemble des communes de $a^\aaa$, qui, rappelons-le, contient la partie pérenne de $\saa$, s'écrit alors comme~:
\be \label{ecov} cov_{t_i,t_j}(a)=\sum_e cov_{t_i,t_j}(\sigma_0)\big|_e~.\ee
\vspace{0.25cm}

Un autre variante, qui nous semble plus familière que la précédente, permet d'obtenir facilement le terme $cov_{t_i,t_j}(a)$.

Mélangeons aléatoirement les valeurs des $\saa$ à l'intérieur d'un échantillon $e$. Autrement dit, $\sigma^\aaa_{0,\:mel}$ prend la valeur $\sigma^\beta_0$ de la commune centrale $\beta$, où les communes $\aaa$ et $\beta$ appartiennent obligatoirement au même échantillon $e$. En brisant de la sorte la partie permanente $a^\aaa$ de $\saa$ $\ti$ou plus précisément la partie $\epsilon^\aaa_{a,\:e}$, et ce sans atteindre à la moyenne $\overline{a}_e$ $\ti$, il vient~:
\be \label{ecove-mel} cov_{t_i,t_j}(\sigma_{0,\:mel})\big|_e = 0~.\ee
Ce qui donne un résultat simple pour le calcul de la covariance étendu à l'ensemble des communes $\aaa$ :
\be \label{ecov-mel} cov_{t_i,t_j}(\sigma_{0,\:mel})= cov_{t_i,t_j}(\sigma_0) - cov_{t_i,t_j}(a)~.\ee
Puisque le mélange à l'intérieur d'un échantillon $e$ n'affecte en rien la variance des $\saa$, l'équation précédente permet d'utiliser comme de coutume les corrélations $C_{t_i,t_j}$ des $\sigma_0$ et des $\sigma_{0,\:mel}$ pour aboutir au résultat suivant :
\be \label{eratio} \frac{cov_{t_i,t_j}(a)}{cov_{t_i,t_j}(\sigma_0)} = 1-\frac{C_{t_i,t_j}(\sigma_{0,\:mel})}{C_{t_i,t_j}(\sigma_0)}~.\ee

Cette opération se répète ensuite à tous les couples d'élections afin d'obtenir les moyennes des covariances et des corrélations, notées respectivement $\overline{cov}_t$ et $\overline{C}_t$. Les valeurs de $\piinsc$ variant peu d'une élection à l'autre, le même échantillonnage en $\piinsc$, tiré des inscrits de l'élection $2007$, sert à tous les couples d'élections. Nous pouvons alors définir le \textit{ratio significatif minimal}, (ou $R.S.M.$) comme :
\be \label{ersm-def} R.S.M. \;\stackrel{def}{=} \;\frac{\overline{cov}_t(a)}{\overline{cov}_t(\sigma_0)}~.\ee
Le \textit{ratio significatif minimal} fournit la proportion minimale occasionnée par la composante spécifique, permanente de $\saa$ dans la covariance temporelle des $\saa$. Autrement dit, le \textit{R.S.M.} procure par valeur minimale, la contribution due à la partie permanente, $a^\aaa$,  de $\saa$ dans la covariance temporelle globale de $\saa$. Le terme \og significatif \fg{} doit se comprendre au sens d'une information pérenne, attachée à la zone géographique sur laquelle se calcule $\saa$ $\ti$comme nous l'avions discuté à la section~\ref{pt-separer-res}. \og Minimal \fg, puisque la démarche précédente ne permet pas d'extraire la totalité de l'information pérenne, exprimée par $a^\aaa$. En effet, pour un couple d'élections $t_i$, $t_j$, la technique utilisée  ne permet d'accéder qu'à $cov_{t_i,t_j}(a)=\langle\epsilon^\aaa_{a}(t_i).\epsilon^\aaa_{a}(t_j)\rangle$, sans pour autant avoir accès aux moyennes de $a^\aaa$ (comme par exemple $\overline{a}_e(t_i)$, etc.).

D'après l'équation~(\ref{eratio}), il vient :
\be \label{ersm} R.S.M. \simeq 1-\frac{\overline{C}_t(\sigma_{0,\:mel})}{\overline{C}_t(\sigma_0)}~.\ee
Pour écrire l'égalité des deux termes de l'équation ci-dessus, il faudrait que les écarts-types des $\saa(t)$ prennent une valeur constante pour toutes les élections. Mais comme elle varie peu d'une élection à l'autre, l'écart entre les valeurs des membres de droite des équations~(\ref{ersm-def}, \ref{ersm}) reste faible~: inférieur à $1\%$ pour toutes les mesures qui suivront. Autrement dit, ici, nous appliquerons indifféremment l'une des deux équations donnant le \textit{R.S.M}.

Éprouvons maintenant la bonne qualité du filtre, qui, dans la négative, rendrait caduque cette méthode. Pour cela, nous usons comme précédemment des $\saa$ calculés à partir des $\taa$ simulés. La simulation consiste en un tirage binomiale de moyenne $\overline{\rho}$ sur $\expr^\aaa$ évènements. (La valeur $\overline{\rho}$ (voir Tab.~\ref{tstat-dens-res}) et l'ensemble des $\expr^\aaa$ par commune, ou par bureau de vote, proviennent encore des valeurs réelles obtenues pour l'élection concernée.) Avec un nombre d'échantillon $n_e = 72$ en $\pio$, le \textit{ratio significatif minimal}, idéalement nul avec un filtrage parfait, devient pour les $\saa$ d'origine binomiale $\ti$par commune ou par bureau de vote$\ti$ inférieure à $1\%$. Ce qui nous satisfait. Nous avons opté pour $n_e=72$ (ce qui correspond à environ $500$ communes centrales $\aaa$ par échantillon) plutôt que pour $n_e=36$ employé jusqu'alors, afin d'améliorer le procédé utilisé et obtenir d'un $R.S.M.$ théoriquement nul, une valeur inférieure à la barre symbolique de $1\%$. A titre indicatif, avec un mauvais filtrage, constitué par exemple avec $n_e=72$ échantillons en $\inscour$ (cf. Eq.~(\ref{einscoa})), le \textit{ratio significatif minimal} des $\saa$ d'origine binomiale par commune vaut $17\%$.  Nous avons répété l'opération avec des $\taa$ simulés selon le \og modèle zéro \fg{} et avons obtenu les mêmes conclusions, à savoir qu'avec $n_e$ échantillons en $\piinsc$, le \textit{R.S.M} prend une valeur inférieur à $1\%$ $\ti$tandis qu'avec un mauvais filtrage, avec $\inscour$ par exemple, le \textit{R.S.M} avoisinerait les $80\%$. Notons néanmoins que les $\taa$ qui ont servi à vérifier la qualité du filtre proviennent tous deux d'évènements indépendants (i.e. des votes indépendants) au niveau de la commune elle-même, voire entre les $n_p$ communes de l'environnement pour le \og modèle zéro \fg.\\

La table~\ref{trsm-res} fournit les différentes valeurs conduisant au \textit{ratio significatif minimal} dans les $\saa$ des résultats électoraux, calculés par commune ou par bureau de vote. Une partie pérenne de $\saa$, associée à une information locale dite positive, contribue donc à hauteur minimale de $15\%$ dans la permanence temporelle globale des $\saa$.

\begin{table}[h]
\begin{minipage}[c]{0.2\linewidth}
\begin{tabular}{|c|c|}
\cline{2-2}
\multicolumn{1}{c|}{}
 & \textit{R.S.M.} \\ 
\hline
com. & 0,15\\
bur. & 0,15\\
\hline
\end{tabular}
\end{minipage}\hfill
\begin{minipage}[c]{0.75\linewidth}
\caption{\small \textit{Ratio significatif minimal} (voir texte) obtenu pour les $\saa$ calculés à partir des résultats électoraux par commune ou bien par bureau de vote.}
\label{trsm-res}
\end{minipage}
\end{table}

La table~\ref{trsm-res} calcule les corrélations sur chaque couple d'élections différentes de $(\sigma_0 - \overline{\sigma}_{0,\:e})$, où $\overline{\sigma}_{0,\:e}$ dénote la moyenne des $\saa$ à l'intérieur de l'échantillon $e$ en $\pio$. Comme $cov_{t_i,t_j}(\sigma_0 - \overline{\sigma}_{0,\:e})=cov_{t_i,t_j}(a)$ d'après l'équation~(\ref{ecov}), cette corrélation concerne uniquement $\epsilon^\aaa_{a,\:e}+\epsilon^\aaa_{b,\:e}$ et fait mieux ressortir les corrélations dues à l'information locale, à caractère dit positif, de $\saa$ (ou plus précisément à $\epsilon^\aaa_{a,\:e}$). Pour mieux comprendre, cette corrélation a une valeur nulle, ou inférieure à $1\%$, avec des $\saa$ provenant des tirages binomiaux ou du \og modèle zéro \fg. Bref, la table~\ref{trsm-res} montre plus spécifiquement les corrélations temporelles d'une information locale contenue dans $\saa$. Comparée à la table~\ref{ttempo-saa-res} qui relate les corrélations temporelles globales de $\saa$, $C_{t_i,t_j}(\sigma_0 - \overline{\sigma}_{0,\:e})$ prend des valeurs inférieures à $C_{t_i,t_j}(\sigma_0)$ et surtout fluctue davantage d'une élection à l'autre. Les corrélations $C_{t_i,t_j}(\sigma_0 - \overline{\sigma}_{0,\:e})$ deviennent par la même plus significatives que les $C_{t_i,t_j}(\sigma_0)$ $\ti$puisqu'elles possèdent une moindre part de bruit dû aux tailles finies.

\begin{table}
\begin{minipage}[c]{0.5\linewidth}
\begin{tabular}{|c|c|c|c|c|c|}
\hline
\backslashbox{$t_i$}{$t_j$} & 1995 & 2000 & 2002 & 2005 & 2007\\
\hline
1992 & 0,19 & 0,11 & 0,16 & 0,12 & 0,14\\
1995 &  & 0,11 &  0,13 & 0,32 & 0,54\\
2000 &  &  & 0,12 & 0,05 & 0,09\\
2002 &  &  &  & 0,14 & 0,12\\
2005 &  &  &  &  & 0,30\\
\hline
\end{tabular}
\end{minipage}\hfill
\begin{minipage}[c]{0.45\linewidth}
\caption{\small Corrélation temporelle $C_{t_i,t_j}(\sigma_0 - \overline{\sigma}_{0,\:e})$ (voir texte) sur chaque couple d'élections $(t_i,t_j)$, et pour les résultats des élections. A comparer aux corrélations $C_{t_i,t_j}(\sigma_0)$ à la table~\ref{ttempo-saa-res}, dans lesquelles contribue le bruit dû aux tailles finies.}
\label{ttempo-separer-res}
\end{minipage}
\end{table}

\begin{figure}[t]
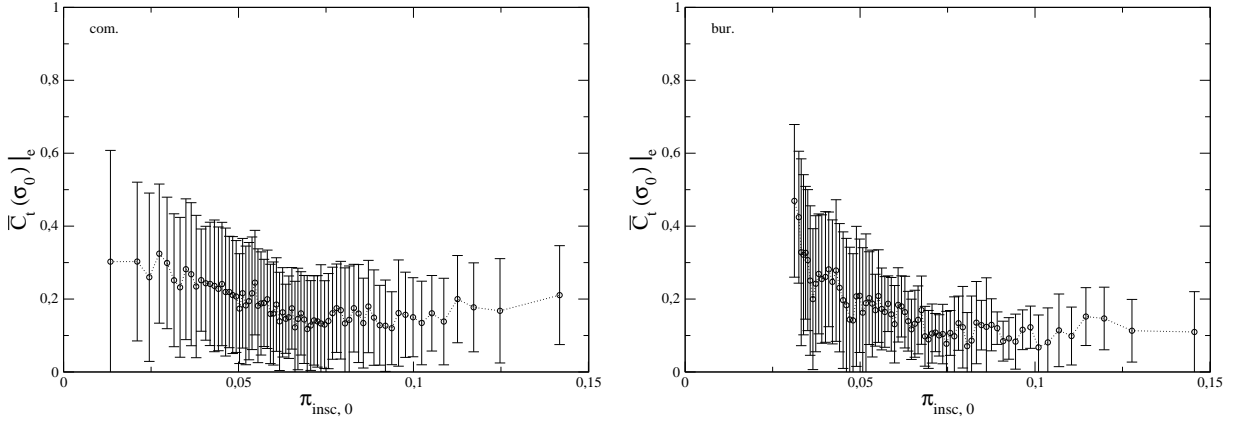

\includegraphics[scale = 0.32]{tempo-piinsc-saa-res.eps}\hfill
\includegraphics[scale = 0.32]{tempo-piinsc-saa-res-bvot.eps}
\caption{\small A l'intérieur des $72$ intervalles en $\pio$ de $2007$~: moyenne $\overline{C}_t(\sigma_0)\big|_\pi$, avec la dispersion standard en barre d'erreur, des corrélations temporelles des $\saa$ des résultats électoraux sur l'ensemble des couples d'élections. A gauche, les $\saa$ proviennent des résultats par commune, et à droite, par bureau de vote. Pour mémoire, sur l'ensemble des communes (voir Tab.~\ref{ttempo-saa-res}), $C_t(\sigma_0) = 0,579 \pm 0,060$, et sur l'ensemble des bureaux de vote, $C_t(\sigma_0) = 0,519 \pm 0,085$.}
\label{ftempo-piinsc-res}
\end{figure}

Les figures~\ref{ftempo-piinsc-res} tracent quant à elles les valeurs moyennes des corrélations $\overline{C}_t$ à l'intérieur des $n_e = 72$ échantillons en $\pio$. Elles montrent que l'information pérenne contenue dans $\saa$ se trouve davantage présente dans les zones à forte population (i.e. à faible $\piinsc$), et se décèle mieux encore à partir des résultats par bureaux de vote. Remarquons néanmoins que les résultats électoraux connus à l'échelle des bureaux de vote concernent uniquement quatre des six élections traitées.

Notons que cette méthode de filtrage n'affecte pas véritablement la grandeur $\toa$, peu sensible en effet au bruit dû aux tailles finies de par sa définition même. En remplaçant $\saa$ par $\toa$ dans les équations ci-dessus, il vient (cf. Eqs.~(\ref{ecov}, \ref{ersm-def})) pour l'équivalent du \textit{R.S.M.} des $\toa$~: $\frac{\sum_e \overline{cov}_t(\tau_0)\big|_e}{\overline{cov}_t(\tau_0)}=95\%$. (La moyenne sur tous les couples d'élections s'effectue ici avec les valeurs absolues $|cov_{t_i,t_j}(\tau_0)|$ de manière à prendre en compte toutes les élections, et ce indépendamment de leur choix gagnant à l'échelle nationale). Un \textit{R.S.M.} équivalent pour les $\toa$  proche de $1$ signifie, conformément à ce dont nous nous attendions, que le bruit dû aux tailles finies ne contribue quasiment pas à la permanence temporelle des $\toa$.\\

Nous venons de montrer qu'un minimum de $15\%$ environ de la corrélation temporelle des $\saa$ des résultats électoraux par commune (ou par bureau de vote), provient d'une information attachée à la zone géographique de $\saa$, et dénuée de bruit dû au tailles finies des communes (ou des bureaux de vote). En outre, cette information locale, dite à caractère positif, intervient davantage dans les zones à forte population que dans les autres zones géographiques moins peuplées, et semble plus apparente avec des résultats électoraux mesurés à l'échelle du bureau de vote.

Nous avons suivi deux hypothèses pour établir les points dégagés ci-dessus. La première considère la partie pérenne $a^\aaa$, présente dans $\saa$ de manière additive~; et la seconde pose que le bruit dû aux tailles finies dépend de $\piinsc$ uniquement. (Nous discuterons ultérieurement de ces deux hypothèses.)

Nous n'avons pas essayé de comprendre sous quelle forme apparaît l'information locale à caractère positif, présente dans $\saa$. Nous avons considéré ici $\saa$ à l'image d'une boîte noire dans laquelle intervenaient notamment un mélange et un échantillonnage. Voyons maintenant une façon plus physique de concevoir l'information à caractère positif associée à $\saa$.

\subparagraph{$\bullet$ $\mathbf{\saa}$ et les corrélations intercommunales\\}
\label{pt-extraire-intercom}
L'information à caractère positif, d'ordre sociologique ou politique, contenue dans $\saa$ peut aussi se comprendre $\ti$parmi d'autres manières$\ti$ comme l'état interne des corrélations des communes (ou des bureaux de vote) de la zone géographique, $\vaa$, sur laquelle s'évalue $\saa$. Nous avions abordé ce point à la section~\ref{pt-separer-res}, et abandonné cet angle d'attaque à cause de l'absence de modèle de vote à l'échelle communale. Nous tentons ici de mesurer l'état des corrélations intercommunales des communes (ou des bureaux de vote) du voisinage $\vaa$ de la commune centrale $\aaa$, en contournant le problème crucial d'absence de modèle. Rappelons aussi que nous nous soucions non seulement de l'état des corrélations des communes (ou des bureaux de vote) à l'intérieur de $\vaa$ à une élection donnée, mais aussi de la permanence de cet état de corrélation au cours de chacune des élections. Plus clairement, un même état de corrélation des communes (ou des bureaux de vote) de $\vaa$ à chaque élection témoigne d'une information à caractère positif sur la zone géographique considérée~; une information que nous nous efforçons de déceler.

L'idée de base du procédé utilisé s'appuie sur une comparaison entre $\saa$ réel et une autre de valeur, notée $\saam$, que pourrait prendre la déviation standard des $n_p$ communes autour de la commune centrale $\aaa$ en l'absence de corrélation interne. La connaissance de ces deux valeurs nous permettra d'évaluer le taux de corrélation des $\tau$ des $n_p$ communes, $r^\aaa$, par le biais de la connexion de ces $n_p$ résultats $\tau$ à une même grandeur locale. Il suffira ensuite de calculer les corrélations temporelles des $r^\aaa$ sur l'ensemble des communes centrales $\aaa$, pour évaluer l'information à caractère dit positif contenue dans $\saa$. Mais rappelons que l'intérêt majeur de cette partie réside dans une compréhension plus physique de l'information locale à caractère positif, appréhendée ici comme l'état des corrélations intercommunales.\\

Supposons que les résultats $\tau$ des $n_p$ plus proches communes de la commune centrale $\aaa$ dépendent d'une grandeur locale, $l^\aaa$, assignée à la zone géographique autour de la commune $\aaa$. Notons $r^\aaa$, le taux de corrélation, ou le rapport de corrélation, qui exprime le degré de corrélation entre un résultat $\tau$ et la grandeur locale $l^\aaa$. (Si le coefficient de corrélation entre deux variables aléatoires $Y$ et $X$, vaut $C$, alors $Y = \pm\sqrt{r}\:X + \sqrt{1-r}\:\eta$, où $\eta$ dénote un bruit sans corrélation, et $r$ le taux de corrélation tel que $C^2=r$. Dans le cas d'une corrélation positive (i.e. $C>0$), l'équation précédente prend le signe $+$, et inversement dans le cas d'une anti-corrélation. Dans l'écriture précédente, les variables aléatoires $X$, $Y$ et $\eta$ ont une moyenne nulle et une variance unité.) Par extrapolation et avec une corrélation positive, la commune $\beta$, parmi les $n_p$ communes en question, a un résultat $\tau^\beta$ qui peut alors s'écrire comme~:
\be \label{etaub} \tau^\beta=(\sqrt{r^\aaa}\:l^\aaa + \sqrt{1-r^\aaa}\:\eta^\beta)\sigma_\tau ~,\ee
où $l$ et $\eta$ dénotent respectivement l'information locale et un bruit additionnel, considérés tous deux comme de moyenne nulle et de variance unité à l'intérieur d'un ensemble donné. (Par la suite, les communes centrales $\aaa$ à l'intérieur d'un échantillon $e$ en $\pio$ constitueront l'ensemble des valeurs considérées.) Pour simplifier, les $\tau$ ont aussi une moyenne nulle sur cet ensemble, et $\sigma_\tau$ exprime l'écart-type des $\tau$ en l'absence de corrélation.

L'écart-type de ces $n_p$ communes qui en résulte devient alors
\be \label{esaa-cor} \saa = \sqrt{1-r^\aaa}\:\sigma_\tau ~.\ee
Le cas des $n_p$ communes complètement corrélées $\ti$ou anticorrélées$\ti$ (i.e. $r^\aaa=1$) à la variable locale donnerait évidemment $\saa=0$, alors que le cas des communes indépendantes ($r^\aaa=0$) fournirait $\saa=\sigma_\tau$.

L'équation~(\ref{esaa-cor}) permet donc d'évaluer le taux de corrélation intercommunale $r^\aaa$. Mais pour cela, il faut nécessairement connaître à la fois, $\saa$ et un intermédiaire calculatoire, $\sigma_\tau$~: l'écart-type qu'auraient les résultats de ces $n_p$ communes en l'absence d'inter-corrélation à une variable locale. $\saa$ se mesure directement à partir des données (cf. Eq.~(\ref{esaa})). Il reste donc à déterminer $\sigma_\tau$, ce que nous allons tenter de faire.\\

Comme nous venons de le voir, sans connexion des résultats à une variable locale (i.e. $r^\aaa=0$), $\saa=\sigma_\tau$. De plus, $\sigma_\tau$ doit préserver le bruit dû aux tailles finies des $n_p$ communes. Une manière de  déterminer $\sigma_\tau$ consiste alors à évaluer l'écart-type des résultats électoraux $\tau$ des $n_p$ communes, non pas à partir des communes de la zone $\vaa$ où se calcule $\saa$ et qui ont une connexion à la même grandeur locale, mais à partir d'autres communes ayant le même bruit dû aux tailles finies et connectées avec des grandeurs locales différentes. Dit autrement, nous déterminons un nouvel écart-type, noté $\saam$, obtenu à partir des résultats électoraux de $n_p$ communes, de même population que chacune des $n_p$ plus proches communes de $\aaa$, mais situées à des endroits différents $\ti$ce qui leur interdit de partager la même connexion à la même variable locale. L'opération consiste alors à mélanger aléatoirement les résultats des communes en fonction de leur taille de population, $\expr$, et de calculer ensuite un nouvel écart-type $\saam$. Ainsi $\saam$ possède, d'une part, l'avantage d'avoir le même bruit dû aux tailles finies que $\saa$, et, d'autre part, une absence d'inter-corrélation des résultats électoraux des communes prises en compte $\ti$puisque les $n_p$ communes prises en compte ne partagent pas leur connexion à la même grandeur locale. En revanche, cette méthode assigne arbitrairement un comportement commun à toutes les communes de la France entière ayant la même population. En d'autres termes, $\sigma_\tau$ doit posséder la même valeur, aux fluctuations près, pour toutes les communes centrales ayant la même valeur de $\pio$. Nous reconnaissons volontiers cet arbitraire, mais sans modèle de vote fin, à l'échelle microscopique (de l'individu), voire à l'échelle intra ou intercommunale, nous n'avons pas trouvé d'autre façon de casser les corrélations intercommunales $\ti$exprimées ici par le biais d'une connexion à une variable locale$\ti$ tout en préservant le bruit dû aux tailles finies inclus dans $\saa$.

Plus rigoureusement, en notant $\langle ... \rangle\big|_e$ la moyenne sur l'ensemble des communes centrales $\aaa$ d'un même échantillon $e$, ayant la même valeur de $\pio$, et en supposant naturellement l'absence de corrélation entre $r^\aaa$, $l^\aaa$ et $\eta^\beta$ dans l'équation~(\ref{etaub}), il vient d'après ce qui précède~:
\be \label{esaam} (\saam)^2\simeq \big[\langle r \rangle\big|_e \; \langle l^2 \rangle\big|_e + (1-\langle r \rangle\big|_e)\; \langle \eta^2 \rangle\big|_e\;\big].(\sigma_\tau)^2~=~(\sigma_\tau)^2 ~.\ee
\vspace{0.25cm}

Concrètement, nous avons pour chaque élection mélangé aléatoirement les résultats des communes (ou bien des bureaux de vote) à l'intérieur de chacun des $72$ intervalles en $\expr$. Ceci a permis de calculer $\saam$ pour chaque commune centrale $\aaa$. Pour amoindrir l'effet des fluctuations des $\saam$, qui présentent la même intensité due aux tailles finies pour les mêmes valeurs de $\pio$, nous calculons leur moyenne à l'intérieur de $72$ intervalles de $\pio$. Des équations~(\ref{esaa-cor}) et (\ref{esaam}) découle alors le taux de corrélation recherché $r^\aaa$, qui s'écrit comme~:
\be \label{eraa} r^\aaa = 1-\frac{(\saa)^2}{\langle(\sigma'_0)^2\rangle\big|_\pi}~,\ee
où $\langle(\sigma'_0)^2\rangle\big|_\pi$ signifie la moyenne des $(\sigma'_0)^2$ de l'intervalle en $\pio$ dans lequel se situe $\piinsc$. 

Une fois obtenu le taux de corrélation $r^\aaa$ pour chaque commune centrale $\aaa$, et pour chaque élection, il ne reste plus qu'à calculer comme précédemment les corrélations temporelles $C_{t_,t_j}(r)$ des $r^\aaa$. L'information que nous appelons locale et à caractère positif se manifeste, selon les discussions précédentes (voir section~\ref{pt-separer-res}) par la permanence temporelle des $r^\aaa$ $\ti$à un éventuel décalage global près, dû à la moyenne des $r^\aaa$ de l'élection considérée$\ti$, et se mesure par $\overline{C}_t(r)$. Précisons pour lever toute ambiguïté, que l'information locale dite positive ne réside pas dans la valeur de $l^\aaa$, potentiellement variable d'une élection à l'autre, mais dans l'état des corrélations intercommunales $r^\aaa$. Et plus encore, dans leur permanence temporelle.\\

Nous nous attendons qu'en l'absence de corrélation intercommunale $r^\aaa$, i.e. en l'absence de cette fameuse information locale à caractère positif dont nous cherchons tant la trace, $\overline{C}_t(r)$ prenne une valeur nulle. Nous le vérifions pour les... $r'^\aaa$ dérivés des $\saam$ $\ti$qui ne possèdent évidemment pas, par construction même, d'information locale à caractère dit positif. ($\overline{C}_t(r')\simeq 0,6\%$, inférieur donc au $1\%$ symbolique.) Notons toutefois que, si la moyenne $\langle(\sigma'_0)^2\rangle$ de l'équation~(\ref{eraa}) se déterminait à l'intérieur des échantillons en $N_{insc,\:cour}$ au lieu des échantillons en $\pio$, et avec encore $\saam$ au numérateur de la même équation, on obtiendrait $\overline{C}_t(r')\simeq 13\%$. Ce qui confirme de nouveau la pertinence des échantillons en $\pio$ pour éliminer le bruit dû aux tailles finies, du moins, dans le cas des évènements (ou des résultats électoraux par commune ou par bureau de vote) indépendants. Notons enfin que cette vérification présente une faille~: les $r'^\aaa$, par leur simple fluctuation, prennent ici des valeurs tant positives que négatives, alors qu'ils devraient rigoureusement prendre uniquement des valeurs positives.

\begin{figure}[t]
\includegraphics[scale = 0.32]{caa-res.eps}\hfill
\includegraphics[scale = 0.32]{caa-res-bvot.eps}
\caption{\small Taux de corrélation intercommunale $r^\aaa$ calculés avec l'équation~(\ref{eraa}), en fonction de $\piinsc$~; évalués à partir des $\saa$ des résultats, par commune à gauche, et par bureau de vote à droite. Les $72$ intervalles en $\pio$ proviennent de l'élection 2007-b.}
\label{fraa-res}
\end{figure}

Bref, les figures~\ref{fraa-abst} tracent la moyenne des $r^\aaa$ en fonction de $\pio$, en utilisant l'équation~(\ref{eraa}), et pour les $\saa$ déterminés à partir des résultats par commune ou bien par bureau de vote. (La démarche précédente établie à partir des résultats par commune, s'étend sans aucune difficulté aux bureaux vote, avec comme seule différence, un nombre de bureaux de vote non uniforme pour l'ensemble des $\vaa$.) Excepté éventuellement pour les zones à plus forte population (i.e. pour celles à plus faible $\pio$), $r^\aaa$ diminue quand la population augmente (i.e. quand $\pio$ augmente). Noter les plus faibles valeurs des $r^\aaa$ de l'élection 2000-b par rapport aux cinq autres élections ; ce qui s'accorde avec les plus faibles corrélations spatiales entre proches communes ($C(R\simeq 1$) de cette élection en comparaison aux cinq autres élections (voir Fig~\ref{fcorrel-res}).

La table~\ref{traa-res} fournit $\overline{C}_t(r)$ à partir des $\saa$ calculés à partir des communes, ou bien à partir des bureaux de vote. Chacune de ces valeurs dépasse les $20\%$, et dépasse le \textit{ratio significatif minimal} calculé précédemment à la table~\ref{trsm-res}. Nous discuterons plus loin de cette différence. 

\begin{table}[h]
\begin{minipage}[c]{0.2\linewidth}
\begin{tabular}{|c|c|}
\cline{2-2}
\multicolumn{1}{c|}{}
 & $\overline{C}_t(r)$ \rule[-7pt]{0pt}{22pt}\\
\hline
com. & 0,24\\
bur. & 0,22\\
\hline
\end{tabular}
\end{minipage}\hfill
\begin{minipage}[c]{0.75\linewidth}
\caption{\small Moyenne $\overline{C}_t(r)$ des corrélations temporelles sur l'ensemble des couples d'élections des taux de corrélation intercommunale $r^\aaa$, issues des résultats par commune ou par bureau de vote. $r^\aaa$ se détermine à partir des $\saa$ et en utilisant l'équation~(\ref{eraa}).}
\label{traa-res}
\end{minipage}
\end{table}

En résumé, nous venons de déterminer le taux de corrélation des communes à une grandeur locale, appelé également état de corrélation intercommunale, $r^\aaa$. Nous l'avons effectué $\ti$en l'absence de modèle de vote à l'échelle infra ou intercommunale$\ti$ de manière indirecte, sous l'hypothèse d'un bruit potentiel, $\sigma_\tau$, similaire pour toutes les zones ayant le même $\pio$. Il apparaît sous cette condition que l'état des corrélations, $r^\aaa$, diminue quand les populations diminuent, excepté dans les zones aux très fortes populations, et prend des valeurs sensiblement similaires à l'exception de l'élection de 2000. L'information locale à caractère positif se comprend physiquement ici par l'intermédiaire de l'état des corrélations intercommunales. Et plus particulièrement par leur permanence temporelle, $\overline{C}_t(r)$, ici de l'ordre de $20\%$.

Ces résultats concernent l'ensemble des communes (ou des bureaux de vote) de la France entière. Essayons, certes de manière anecdotique, de cartographier les zones à plus ou moins forte information locale pérenne.

\subparagraph{$\bullet$ Une visualisation de la persistance temporelle de $\mathbf{\saa}$\\}
\label{pt-extraire-visualisation}
Essayons, à titre indicatif, de visualiser les zones à plus ou moins forte persistance temporelle des $\saa$. 

Jusqu'ici nous nous avons déterminé des corrélations (des $\saa$ ou des $r^\aaa$) sur toutes les communes pour des couples d'élections différentes. Afin de connaître les communes centrales $\aaa$ pour lesquelles $\saa$ varie plus ou moins fortement, nous devons procéder autrement. Nous devons engager une seule commune à la fois, et au travers de plusieurs élections.

Notons $\Delta_{[\saa]_t}$ l'écart-type que prend $\saa$ sur toutes les élections traitées. Ne s'occuper que de $\Delta_{[\saa]_t}$ pour déterminer si $\saa$ fluctue plus ou moins fortement d'une élection à l'autre n'a pas véritablement de sens. Il faut aussi tenir compte de la moyenne $\langle\saa\rangle_t$ des $\saa$ d'une commune $\aaa$ sur toutes les élections. Autrement, par exemple les zones à fort bruit dû aux tailles finies, et donc à fort $\saa$, auraient d'emblée plus facilement une forte variation de leurs $\saa$. Il convient donc d'utiliser le coefficient de variation, $\cva$, des valeurs que prend $\saa$ d'une commune $\aaa$ donnée, à chacune des élections~:
\be \label{ecva} \cva=\frac{\Delta_{[\saa]_t}}{\langle\saa\rangle_t}~.\ee

\begin{figure}
\begin{minipage}[c]{0.55\linewidth}
\includegraphics[scale = 0.32]{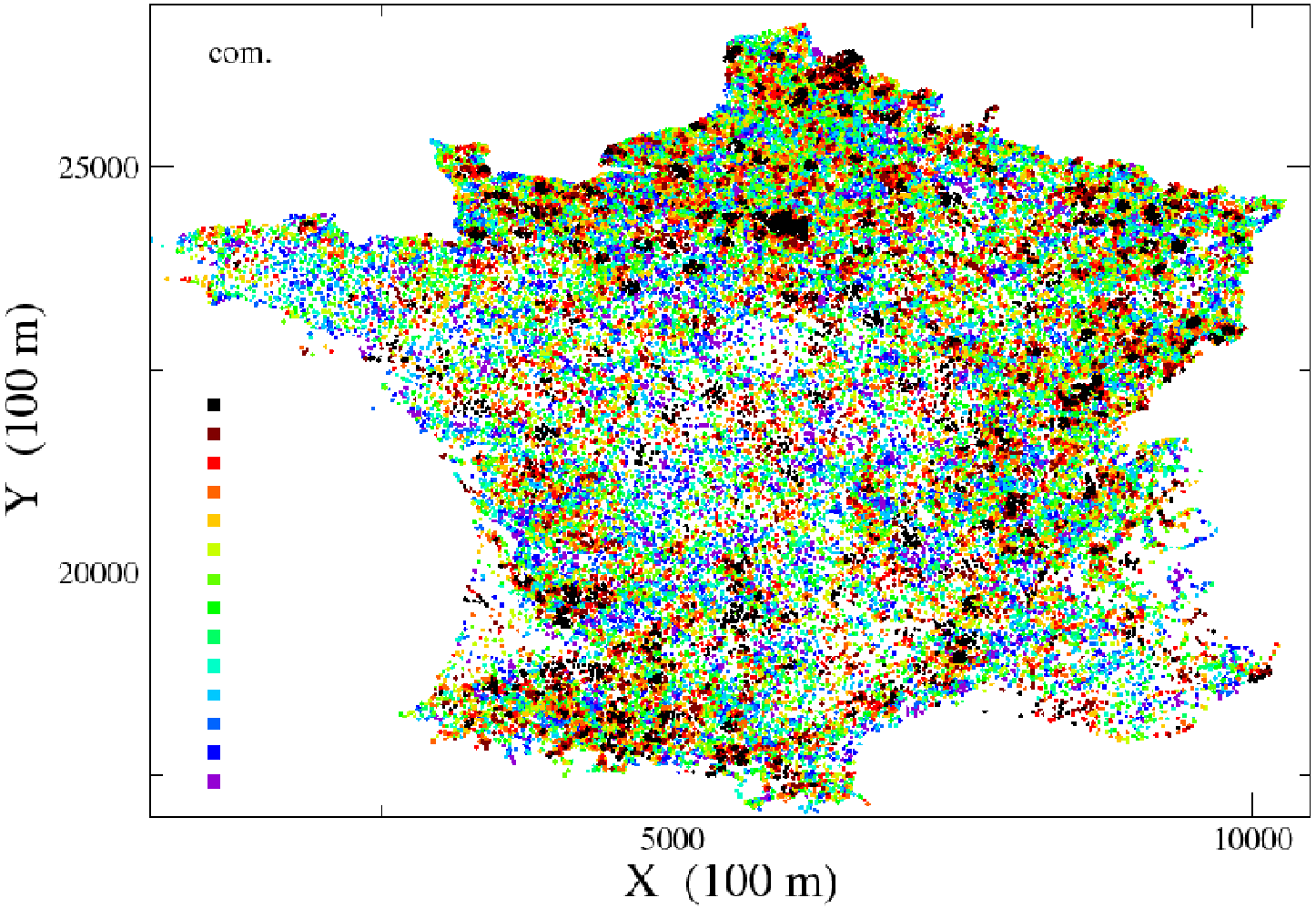}
\end{minipage}\hfill
\begin{minipage}[c]{0.45\linewidth}
\caption{\small Position XY des communes centrales des coefficients de variation de $\saa$. Les $\saa$ proviennent des résultats des communes. Les $\cva$ (cf. Eq.~(\ref{ecva})) se classent en $14$ intervalles, partant du violet pour les plus faibles valeurs, au noir pour les plus fortes. Les zones à faible $\cva$ indiquent une forte information locale positive pérenne. Noter les différences avec la carte de la répartition des $\saa$ à la figure~\ref{frepartition-saa-res}.}
\label{fpermanence-res}
\end{minipage}
\end{figure}

La figure~\ref{fpermanence-res} représente la position des communes centrales $\aaa$, selon la répartition en $14$ niveaux des $\cva$. Les $\saa$ ne proviennent sur cette figure que des résultats par commune. Les résultats par bureaux de vote n'auraient guère de sens avec à peine $4$ élections connues à cette échelle.

$6$ valeurs pour calculer un coefficient de variation n'a rien de vraiment glorieux non plus, diriez-vous. Voilà pourquoi nous n'avons pas abordé le problème de la permanence des $\saa$ par ce biais. Mais puisque nous aimons à croiser différents angles de vue d'un même phénomène, ne gâchons surtout pas notre plaisir en indiquant ce qu'aurait apporté cette méthode.

Comme précédemment, nous pouvons comparer les écarts-types $\Delta_{[\saa]_t}$ à ceux issus d'un mélange à l'intérieur d'un même échantillon en $\pio$, et ce pour casser la permanence temporelle (à la moyenne d'ensemble près) attachée  $\saa$. (L'argumentation reprend les mêmes termes, notations et remarques précédents.) En passant ensuite à l'ensemble des communes $\aaa$~: $\frac{\langle \Delta_{[\sigma^\aaa_{0,\:mel}]_t}\rangle}{\langle \Delta_{[\saa]_t}\rangle}=1,09$, où $\langle ...\rangle$ dénote la moyenne sur l'ensemble des communes centrales $\aaa$. Ainsi, la grandeur qui s'apparente au \textit{ratio significatif minimal} $\ti$dans le sens où il se préoccupe aussi de permanence temporelle des $\saa$\--- vaut par cette façon environ $9\%$, au lieu de $15\%$ comme précédemment. Nous accordons moins de crédit à la façon entreprise dans cette partie, à cause du plus fort bruit qu'elle implique à la base (i.e. dans le calcul de $\Delta_{[\saa]_t}$). Constatons néanmoins la concordance approximative des deux calculs.

\subparagraph{$\bullet$ Résumé\\}
\label{pt-extraire-resume}
En résumé, nous avons vu que $\saa$ d'une même commune centrale $\aaa$ contient une partie pérenne d'une élection à l'autre, et indépendante du bruit dû aux tailles finies des communes (ou des bureaux de vote). Cette partie permanente exprime très probablement une information spécifiquement locale à teneur dite positive, i.e. toute particularité locale, à l'exception de ce qu'implique la taille de la population par le biais du bruit statistique des tailles finies. Cette partie pérenne de $\saa$ contribue à hauteur minimale de $15\%$ dans la permanence temporelle globale des $\saa$ $\ti$mesurée par $\overline{C}_t(\sigma_0)$.

L'une des manières de comprendre cette information locale à caractère positif réside dans l'état des corrélations des résultats électoraux entre communes voisines (ou entre les bureaux de vote voisins). Nous avons essayé de mesurer indirectement ces corrélations intercommunales en les comparant à ce qu'aurait fourni une absence d'interconnexion.

L'état des corrélations intercommunales et la partie pérenne de $\saa$ se réfèrent toutes deux à la présence d'une information locale à caractère positif. Néanmoins ces deux notions ne s'identifient pas, à proprement parler, l'une à l'autre. Ceci explique pourquoi les mesures des tables~\ref{trsm-res} et \ref{traa-res} se différencient à ce point. D'autant plus que l'une des mesures, le \textit{R.S.M.} concerne une mesure \textit{a minima}. Par contre, ces deux notions indiquent une présence de l'information locale et positive globalement plus forte dans les zones à forte population.

Enfin la carte~\ref{fpermanence-res} représente les zones à plus ou moins fort coefficient de variation des $\saa$ d'une même commune centrale et sur l'ensemble des élections analysées. Cette carte fait apparaître $\ti$heureusement$\ti$ des différences avec la carte des plus ou moins grandes valeurs de $\saa$ lors de l'élection de 2007-b (cf. Fig.~\ref{frepartition-saa-res}).

Toutes les conclusions ci-dessus chancelleraient si les hypothèses sur lesquelles elles reposent ne tenaient pas. Discutons-les donc.

\subparagraph{$\bullet$ Discussion des hypothèses\\}
\label{pt-extraire-discussion}
Selon la première méthode utilisée, la recherche de la partie pérenne $a^\aaa$ présente dans $\saa$ à diverses élections se fonde sur les deux hypothèses suivantes~:
\begin{itemize}
\item l'amplitude du bruit statistique dû aux tailles finies des communes (ou des bureaux de vote) ne dépend que de $\piinsc$~;
\item $\saa= a^\aaa + b^\aaa$, i.e. $\saa$ se décompose entre, d'une part $a^\aaa$ qui contient la partie permanente recherchée de $\saa$, et d'autre part tout le reste, $b^\aaa$, comprenant notamment le bruit statistique des tailles finies.
\end{itemize}

Quant à la recherche de l'information locale positive au travers de l'état des corrélations intercommunales, celle-ci se base sur l'hypothèse d'un bruit statistique global (comprenant notamment le bruit dû aux tailles finies) identique à l'intérieur d'un même échantillon en $\piinsc$. Selon cette seconde méthode d'investigation, l'état des corrélations, $r^\aaa$, intercommunales se détermine ensuite par l'équation~(\ref{eraa}).\\

Dans les deux cas, la grandeur recherchée ($a^\aaa$ ou $r^\aaa$) se réfère à un bruit statistique à l'intérieur d'un échantillon (celui de $b^\aaa$, ou de $\langle(\sigma'_0)^2\rangle\big|_\pi$). La grandeur recherchée issue de $\saa$, se rapporte au bruit de l'échantillon~: dans le premier cas, de façon additive, et dans le second cas, de façon multiplicative.

Précisons de nouveau que les deux méthodes précédemment développées ne se prétendent pas rigoureuses. Nous les avons néanmoins utilisées afin de pallier l'absence de modèle de vote. Cette différence d'approche, de méthode, dans la façon d'utiliser le bruit statistique de l'échantillon ne constitue nullement un handicap. Bien au contraire, ces deux méthodes différentes permettent toutes deux de déceler et de mesurer $\ti$y compris \textit{a minima}$\ti$ la présence d'une information locale positive dans $\saa$. Bien que non rigoureuses, elles attestent donc, par une sorte de faisceau de convergence, l'existence d'une information locale dite positive dans $\saa$.

Quelques remarques complémentaires néanmoins. En utilisant la première méthode, non pas avec $\saa$ mais avec $(\saa)^2$ (i.e. $(\saa)^2=a'^\aaa + b'^\aaa$, où $a'^\aaa$ renferme la partie pérenne de $(\saa)^2)$), nous obtenons les mêmes \textit{R.S.M.} que ceux de $\saa$, à $1\%$ près. Notons que l'égalité $\ti$prévisible$\ti$ des \textit{R.S.M.} de $(\saa)^2$ et de $\saa$ ne dément pas la méthode utilisée $\ti$sans pour autant la confirmer pleinement. Enfin, en l'absence de modèle de vote, nous n'avons pu établir comment le bruit binomial dû aux tailles finies intervient dans $\saa$. (Nous n'avons pu, ni le retrancher, ni le diviser, etc. de $\saa$.) Nous avons simplement considéré son intensité comme constante à l'intérieur d'un échantillon en $\piinsc$.\\

Passons maintenant à l'épreuve la seconde hypothèse, relative dans les deux méthodes au bruit statistique des $\saa$ à l'intérieur des échantillons.

Avec la première méthode, les $\saa$ d'un échantillon $e$ doivent avoir le même niveau de bruit dû aux tailles finies $\ti$aux fluctuations près, évidemment. Notons que l'échantillon peut aussi contenir une information locale pérenne, moyenne à l'ensemble des $\saa$ (i.e. $\overline{a}_e$), mais non prise en compte dans la mesure minimale du \textit{R.S.M.}.

Selon la seconde méthode, non seulement l'échantillon $e$ doit avoir pour les $\saa$ le même niveau de bruit statistique dû aux tailles finies, mais aussi le même niveau \og d'agitation \fg{} des résultats $\tau$ ($\sigma_\tau=\sqrt{\langle(\sigma'_0)^2\rangle\big|_\pi}$~) lors d'une hypothétique absence de corrélations intercommunales.

Rappelons une fois de plus qu'en l'absence de modèle de vote, nous n'avons pas trouvé d'autre moyen pour mesurer, dans $\saa$ purgé de son bruit statistique dû aux tailles finies, la présence d'une information locale dite positive.

Nous avons précédemment vu, pour les deux méthodes, l'importance de la grandeur servant à l'échantillonnage (cf. un échantillonnage basé sur $\inscoa$, plutôt que sur $\piinsc$ pour le tirage binomial, ou pour le \og modèle zéro \fg). Il importe donc dans les deux méthodes que la grandeur, sur laquelle se bâtit l'échantillonnage, exprime le bruit statistique dû aux tailles finies.

Dans le cas d'un vote indépendant, ce bruit statistique procède de $\piinsc$ (ou une grandeur assimilée à $\piinsc$ si nous tenions compte des suffrages exprimés plutôt que du nombre d'inscrits sur les listes électorales) comme nous l'avions vu à la section~\ref{pt-bruit-res}. Or, les résultats réels attestent incontestablement l'existence d'un vote non indépendant, i.e. corrélé, et à l'intérieur des communes (cf. Annexe~\ref{annexe-communautes}), et entres différentes communes voisines (cf. Annexe~\ref{annexe-communautes} ou Figs.~\ref{fcorrel-res} et \ref{fcorrel-abst}). Et en l'absence de modèle de vote, nous ne savons pas quelle grandeur exprime le bruit statistique dû aux tailles finies. Nous avions simplement constaté la similarité d'allure des $\saa$ réels, et ceux issus d'un tirage binomial de probabilité uniforme (cf. sections~\ref{pt-comprendre-res} et \ref{pt-bruit-res}). Ce qui nous a conduit à considérer que $\piinsc$ exprime également le bruit statistique dû aux tailles finies dans les $\saa$ des résultats électoraux réels.

Or, l'absence de modèle de vote implique que nous ne pouvons pas tester rigoureusement l'hypothèse que nous avons utilisée, à savoir un bruit statistique dû aux tailles finies, exprimé par $\piinsc$. Nous n'allons pas nous décourager pour autant, et encore moins abandonner la discussion à ce stade. D'autant plus que cette annexe, rappelons-le, ne peut avoir de caractère rigoureux. Bref, nous n'allons pas tester rigoureusement l'hypothèse ci-dessus, mais sa robustesse. Nous voulons voir si une modification de la grandeur ($\piinsc$ selon l'hypothèse retenue), qui exprime le bruit statistique  dû aux tailles finies, modifie notablement les précédentes mesures.

L'expression de $\piinsc$ provient d'un bruit dû aux tailles finies par commune proportionnel à $\frac{1}{\sqrt{N}}$ (cf. section~\ref{pt-bruit-res}), où $N$ dénote la taille de la population. Nous allons donc considérer la variante de ce bruit, pour les résultats réels, proportionnel à $\frac{1}{N^a}$, où l'exposant $a$, précédemment égal à $0,5$, s'étend de $0,1$ à $1$. Avec cet hypothétique bruit statistique dû aux tailles finies dans chaque résultat $\taa$, et en suivant la même démarche qu'à la section~\ref{pt-bruit-res}, le bruit dû aux tailles finies s'exprimerait dans $\saa$ par~:
\be \label{epiinsc-a} \pi^\aaa(a) = \sqrt{\langle\frac{1}{(\insc^\beta)^{2\,a}}\rangle_{_{\beta\in\vaa}}} = \sqrt{\frac{1}{n_p} \sum_{\beta\in\vaa} \frac{1}{(\insc^\beta)^{2\,a}}}~.\ee

Nous obtenons de la sorte, une nouvelle grandeur, $\pi^\aaa(a)$, qui englobe la grandeur $\piinsc$ retenue jusqu'alors. Nous testons ensuite la robustesse de l'hypothèse d'un bruit dû aux tailles finies et exprimé par $\piinsc$ (i.e. par $a=0,5$ dans l'équation ci-dessus) en faisant varier l'exposant $a$ autour de $a=0,5$.

\begin{figure}
\begin{minipage}[c]{0.55\linewidth}
\includegraphics[scale = 0.32]{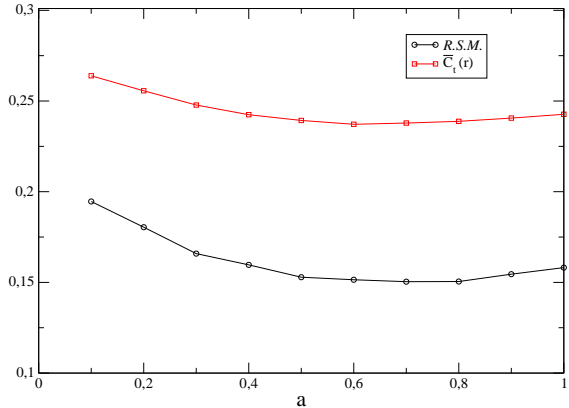}
\end{minipage}\hfill
\begin{minipage}[c]{0.45\linewidth}
\caption{\small \textit{R.S.M.} et $\overline{C}_t(r)$ calculés comme précédemment, mais avec des échantillons en $\pi^\aaa(a)$ (cf. Eq.~(\ref{epiinsc-a})). Remarque~: $\piinsc$ s'obtient avec $a=0,5$.}
\label{frobustesse}
\end{minipage}
\end{figure}

La figure~\ref{frobustesse} trace le \textit{R.S.M.} et $\overline{C}_t(r)$ de la même façon que précédemment, mais en fonction des échantillons en $\pi^\aaa(a)$, au lieu de $\piinsc$. Il ressort que le coefficient $a=0,6$ conviendrait peut-être mieux que $a=0,5$ à la base de $\piinsc$. Mais plus important encore, nous voyons que les résultats ne se modifient guère par modification de l'exposant $a$ autour de la valeur que nous avions précédemment utilisée, à savoir $a=0,5$.

En conclusion, même si nous n'avons pu tester rigoureusement l'hypothèse d'un bruit statistique dû aux tailles finies exprimé par $\piinsc$ dans $\saa$, nous avons cependant vérifié la robustesse de cette hypothèse en modifiant l'expression de $\piinsc$. (Il existe probablement d'autres façons de modifier $\piinsc$, mais nous nous contenterons de celle-ci. Elle nous paraît en effet suffisamment générale, en l'absence de modèle de vote, pour s'en satisfaire.)

\subsection*{Avec les taux de participation}
\label{annexe-extraire-abst}

Recherchons maintenant la présence d'une information locale dite positive dans les $\saa$ des taux de participation. Pour cela nous appliquerons, directement et sans les discuter, les précédentes méthodes établies pour les $\saa$ des résultats électoraux. Nous noterons, quand l'occasion s'en fera ressentir, les différences avec les mesures précédentes. Et une fois de plus, la comparaison entre données différentes (ici, les résultats électoraux et les taux de participation) permettra, espérons-le, d'enrichir la compréhension des notions développées.

Précisons au préalable que les deux informations locales à caractère positif, présentes dans les $\saa$ soit issus des résultats électoraux soit issus des taux de participation, ne s'identifient pas obligatoirement l'une à l'autre. En effet, les choix que sollicitent l'expression des votes ne s'identifient pas forcément aux choix impliqués par une participation ou non à une élection.\\

La table~\ref{trsm-abst} fournit les \textit{R.S.M.} des $\saa$ calculés à partir des taux de participation par commune ou par bureau de vote. Que les \textit{R.S.M.} des taux de participation prennent des valeurs plus grandes que celles correspondant aux résultats électoraux n'a rien de surprenant. Les choix relatifs à la participation aux différentes élections semblent en effet plus homogènes entre eux, que ceux relatifs à l'expression du vote en faveur de l'un des deux candidats des différentes présidentielles, ou lors d'un des trois référendums analysés. L'aspect pérenne de $\saa$ se marque alors naturellement davantage pour les taux de participation, comparés aux résultats électoraux.

\begin{table}[h]
\begin{minipage}[c]{0.2\linewidth}
\begin{tabular}{|c|c|}
\cline{2-2}
\multicolumn{1}{c|}{}
 & \textit{R.S.M.} \\ 
\hline
com. & 0,27\\
bur. & 0,36\\
\hline
\end{tabular}
\end{minipage}\hfill
\begin{minipage}[c]{0.75\linewidth}
\caption{\small \textit{Ratio significatif minimal} (voir texte) obtenu pour les $\saa$ calculés à partir des taux de participation par commune ou bien par bureau de vote.}
\vspace{1.5cm}
\label{trsm-abst}
\end{minipage}
\end{table}

\begin{figure}
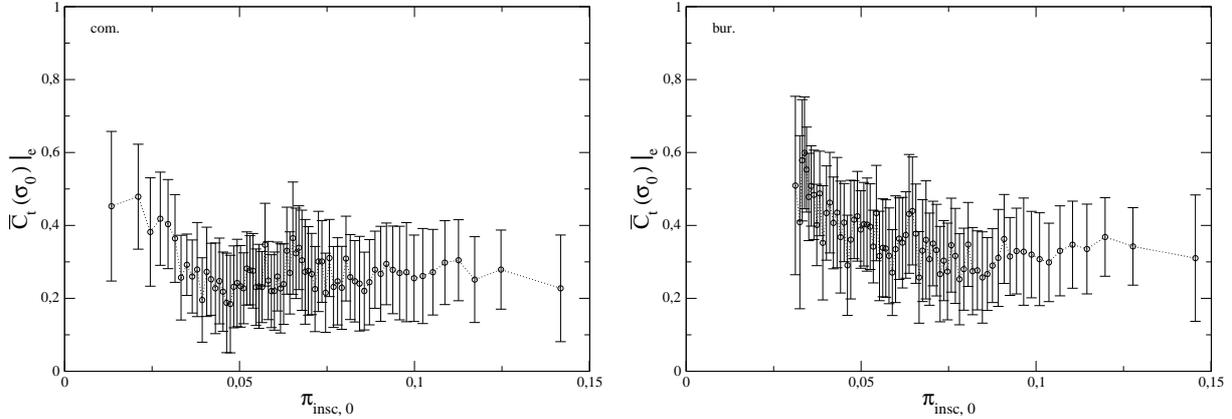

\includegraphics[scale = 0.32]{tempo-piinsc-saa-abst.eps}\hfill
\includegraphics[scale = 0.32]{tempo-piinsc-saa-abst-bvot.eps}
\caption{\small A l'intérieur des $72$ intervalles en $\pio$ de $2007$~: moyenne $\overline{C}_t(\sigma_0)\big|_\pi$, avec la dispersion standard en barre d'erreur, des corrélations temporelles des $\saa$ des taux de participation sur l'ensemble des couples d'élections. A gauche, les $\saa$ proviennent des résultats par commune, et à droite, par bureau de vote. Pour mémoire, sur l'ensemble des communes (voir Tab.~\ref{ttempo-saa-abst}), $C_t(\sigma_0) = 0,567 \pm 0,058$, et sur l'ensemble des bureaux de vote, $C_t(\sigma_0) = 0,586 \pm 0,067$.}
\label{ftempo-piinsc-abst}
\end{figure}

La table~\ref{ttempo-separer-abst} fournit pour tous les couples d'élections, les corrélations $C_{t_i,t_j}(\sigma_0-\overline{\sigma}_{0,\:e})$ des taux de participation. Le bruit statistique dû aux tailles finies se marque moins que pour les corrélations $C_{t_i,t_j}(\sigma_0)$ des $\saa$ bruts, donnés à la table~\ref{ttempo-saa-abst}. Nous pouvons noter que certains couples d'élections (entre le premier et le second tour de la même présidentielle) se détachent nettement plus des autres couples, dans la table~\ref{ttempo-separer-abst} comparée à la table~\ref{ttempo-saa-abst}. Ceci s'explique par une contribution moins prononcée du bruit statistique dû aux tailles finies dans la table~\ref{ttempo-separer-abst}.

\begin{table}[h]
\begin{tabular}{|c|c|c|c|c|c|c|c|c|c|c|c|}
\hline
\backslashbox{$t_i$}{$t_j$} & 94-m & 95-m & 95-b & 99-m & 00-b & 02-m & 02-b & 04-m & 05-b & 07-m & 07-b\\
\hline
92-b & 0,40 & 0,32 & 0,33 & 0,32 & 0,24 & 0,18 & 0,20 & 0,20 & 0,20 & 0,16 & 0,17\\
94-m & & 0,34 & 0,32 & 0,47 & 0,32 & 0,22 & 0,19 & 0,31 & 0,24 & 0,13 & 0,15\\
95-m & &  & 0,53 & 0,31 & 0,20 & 0,25 & 0,22 & 0,19 & 0,21 & 0,17 & 0,17\\
95-b & &  &  &  0,31 & 0,19 & 0,22 & 0,23 & 0,20 & 0,20 & 0,16 & 0,17\\
99-m & &  &  &  & 0,46 & 0,31 & 0,27 & 0,46 & 0,32 & 0,18 & 0,20\\
00-b & &  &  &  &  & 0,24 & 0,21 & 0,37 & 0,22 & 0,12 & 0,15\\
02-m & &  &  &  &  &  & 0,50 & 0,28 & 0,28 & 0,19 & 0,21\\
02-b & &  &  &  &  &  &  & 0,23 & 0,29 & 0,23 & 0,26\\
04-m & &  &  &  &  &  &  &  & 0,39 & 0,20 & 0,22\\
05-b & &  &  &  &  &  &  &  &  & 0,30 & 0,32\\
07-m & &  &  &  &  &  &  &  &  &  & 0,50\\
\hline
\end{tabular}
\caption{Corrélation temporelle $C_{t_i,t_j}(\sigma_0 - \overline{\sigma}_{0,\:e})$ (voir texte) sur chaque couple d'élections $(t_i,t_j)$, et pour les taux d'abstention. A comparer aux corrélations $C_{t_i,t_j}(\sigma_0)$ à la table~\ref{ttempo-saa-abst}, dans lesquelles contribue le bruit dû aux tailles finies.}
\label{ttempo-separer-abst}
\end{table}

Les figures~\ref{ftempo-piinsc-abst} montrent les corrélations temporelles des $\saa$ des taux de participation, à l'intérieur de chaque intervalle en $\pio$. Conformément à leurs homologues concernant les résultats électoraux (cf. Fig.~\ref{ftempo-piinsc-res}), l'information locale, dite positive ou pérenne, se situe davantage dans les zones à forte population, et se manifeste aussi davantage avec les données recueillies à l'échelle des bureaux de vote.

\begin{figure}
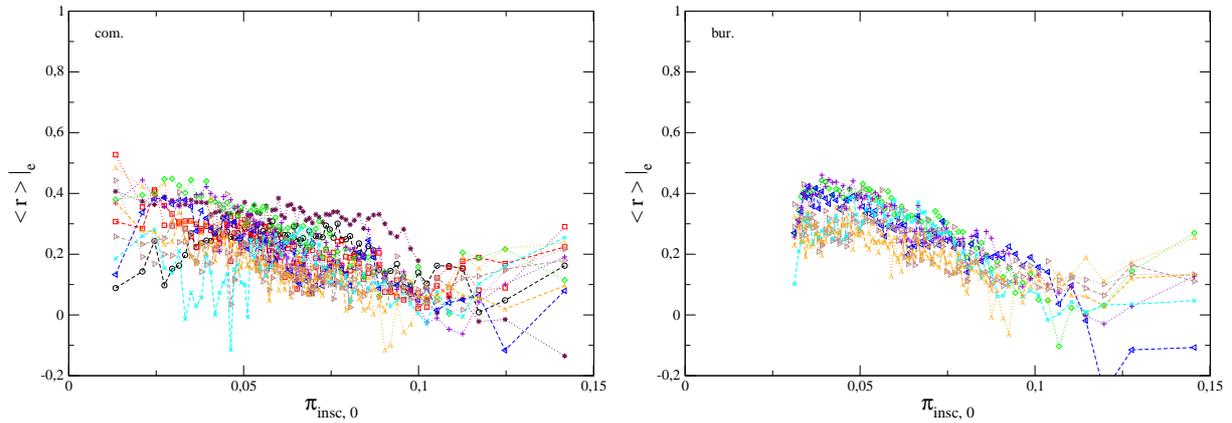

\includegraphics[scale = 0.32]{caa-abst.eps}\hfill
\includegraphics[scale = 0.32]{caa-abst-bvot.eps}
\caption{\small Taux de corrélation intercommunale $r^\aaa$ calculés avec l'équation~(\ref{eraa}), en fonction de $\pio$~; évalués à partir des $\saa$ des taux de participation, par commune à gauche, et par bureau de vote à droite. Les $72$ intervalles en $\pio$ proviennent de l'élection 2007-b.}
\label{fraa-abst}
\end{figure}

Les figures~\ref{fraa-abst} représentent les taux de corrélation intercommunale, $r^\aaa$, en fonction de $\piinsc$, pour chacune des élections considérée pour son taux de participation. Les valeurs des $r^\aaa$ prennent, comme pour les résultats électoraux, de plus grandes valeurs dans les zones à forte population que dans les zones faiblement peuplées. En revanche, nous remarquons de plus faibles valeurs des $r^\aaa$, comparées à leurs homologues des résultats électoraux (cf. Fig.~\ref{fraa-res}). Ce dernier point s'accorde avec de plus grandes corrélations entre communes voisines des $\tau$ des résultats électoraux, comparées à celles des taux de participation (voir Figs.~\ref{fcorrel-res} et \ref{fcorrel-abst}).

La table~\ref{traa-abst} fournit les corrélations temporelles $\overline{C}_t(r)$ des $r^\aaa$, issus des $\saa$ des taux de participation à l'échelle des communes ou des bureaux de vote. Bien que les $r^\aaa$ des taux de participation prennent des valeurs plus faibles que ceux des résultats électoraux, ils présentent une corrélation temporelle sensiblement égale à ceux des résultats électoraux. Ceci provient probablement d'une plus grande homogénéité des choix relatifs aux taux d'abstention, que ceux relatifs aux résultats électoraux $\ti$comme nous l'avions déjà évoqué plus haut, notamment à la table~\ref{trsm-abst}.

\begin{table}[h]
\begin{minipage}[c]{0.2\linewidth}
\begin{tabular}{|c|c|}
\cline{2-2}
\multicolumn{1}{c|}{}
 & $\overline{C}_t(r) \rule[-7pt]{0pt}{22pt}$\\
\hline
com. & 0,19\\
bur. & 0,26\\
\hline
\end{tabular}
\end{minipage}\hfill
\begin{minipage}[c]{0.75\linewidth}
\caption{\small Moyenne $\overline{C}_t(r)$ des corrélations temporelles sur l'ensemble des couples d'élections des corrélations intercommunales $c^\aaa$, issues des taux de participation par commune ou par bureau de vote. $r^\aaa$ se détermine à partir des $\saa$ et en utilisant l'équation~(\ref{eraa}).}
\label{traa-abst}
\end{minipage}
\end{table}

\begin{figure}
\begin{minipage}[c]{0.55\linewidth}
\includegraphics[scale = 0.32]{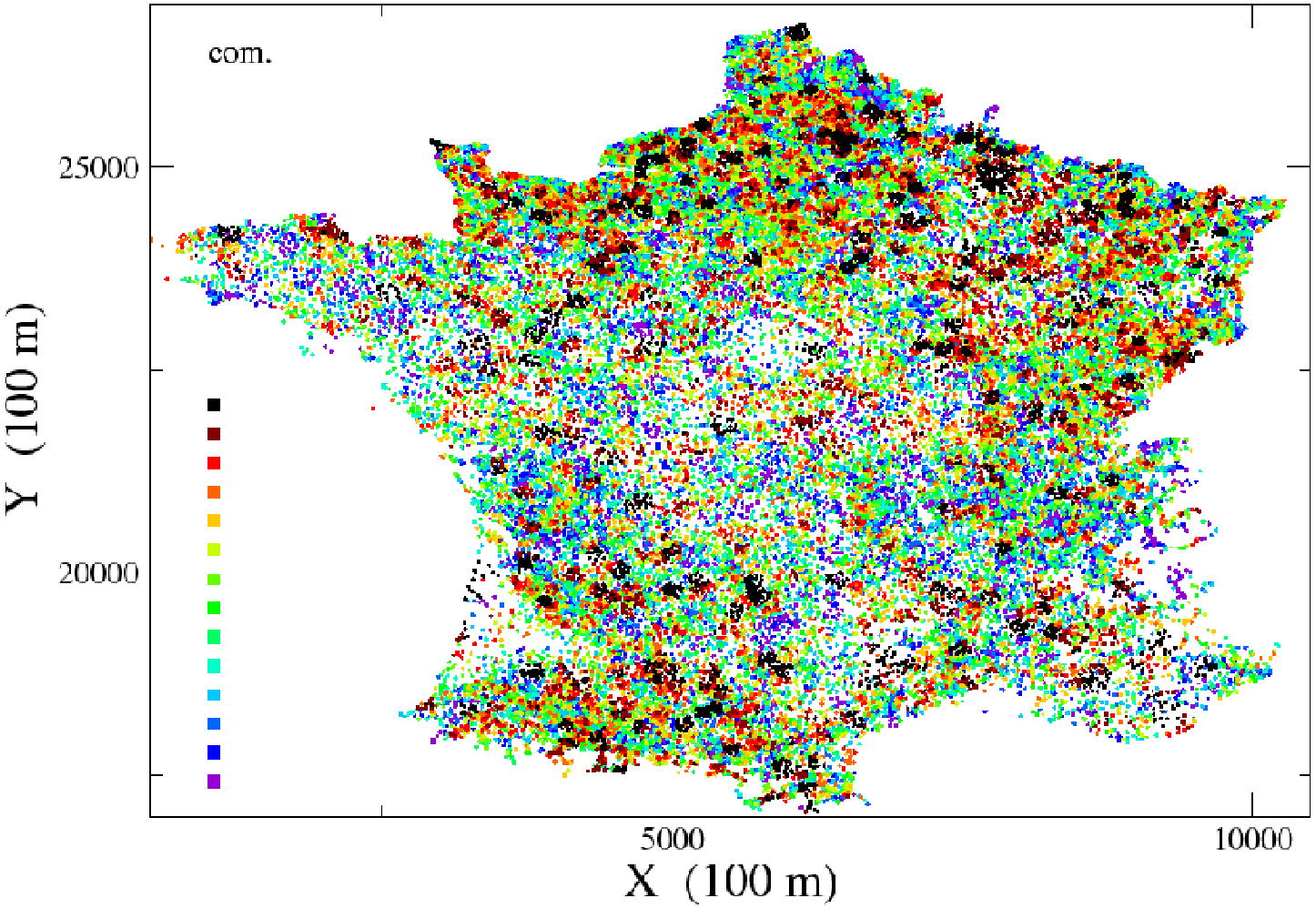}
\end{minipage}\hfill
\begin{minipage}[c]{0.45\linewidth}
\caption{\small Position XY des communes centrales des coefficients de variation de $\saa$. Les $\saa$ proviennent des taux de participation des communes. Les $\cva$ (cf. Eq.~(\ref{ecva})) se classent en $14$ intervalles, partant du violet pour les plus faibles valeurs, au noir pour les plus fortes. Les zones à faible $\cva$ indiquent une forte information locale positive pérenne. Noter les différences avec la carte de la répartition des $\saa$ à la figure~\ref{frepartition-saa-abst}.}
\label{fpermanence-abst}
\end{minipage}
\end{figure}

Enfin, la figure~\ref{fpermanence-abst} indique, avec un classement selon leurs valeurs, la position géographique des coefficients de variation, $\cva$ (cf Eq.~(\ref{ecva})), des $\saa$ issus des taux de participation par commune. Les faibles valeurs de $\cva$ indiquent un voisinage $\vaa$ d'une commune centrale $\aaa$, pour lequel les $\saa$ varient relativement peu d'une élection à l'autre~; et témoignent de la sorte d'une information locale pérenne relativement forte. Les taux de participation par commune concernent $12$ élections, à la différence des résultats par commune qui n'en considèrent qu'un nombre plus restreint d'élections --$6$. Les coefficients de variation d'une commune se noient donc un peu moins dans le bruit statistique dû aux petits échantillons, que ceux des résultats électoraux. De plus, avec les taux de participation, $\frac{\langle \Delta_{[\sigma^\aaa_{0,\:mel}]_t}\rangle}{\langle \Delta_{[\saa]_t}\rangle}=1+20\%$. Ce qui s'accorde mieux avec le \textit{R.S.M.} issu des taux de participation par commune.

\vspace{0.5cm}
En conclusion, ces petits calculs et artifices permettent d'établir que $\saa$ ne mesure pas uniquement un bruit statistique dû aux tailles finies des communes (ou des bureaux de vote), mais aussi une information stable, inhérente à sa zone d'évaluation, vraisemblablement à caractère politique, sociologique, historique, comportementale, etc. $\ti$dite tout simplement, positive. Nous n'avons pu $\ti$en l'absence de modèle de vote$\ti$ mesurer rigoureusement son importance, mais avons pu néanmoins, au travers de plusieurs méthodes différentes, déceler sa présence et mesurer approximativement son importance.

\clearpage
\renewcommand{\thesection}{D}
\renewcommand{\theequation}{D-\arabic{equation}}
\setcounter{equation}{0}  
\renewcommand{\thefigure}{D-\arabic{figure}}
\setcounter{figure}{0}
\renewcommand{\thetable}{D-\arabic{table}}
\setcounter{table}{0}
\section{De l'indépendance des bureaux de vote des zones urbaines}
\label{annexe-communautes}
\vspace{0.5cm}

\begin{figure}[h!]
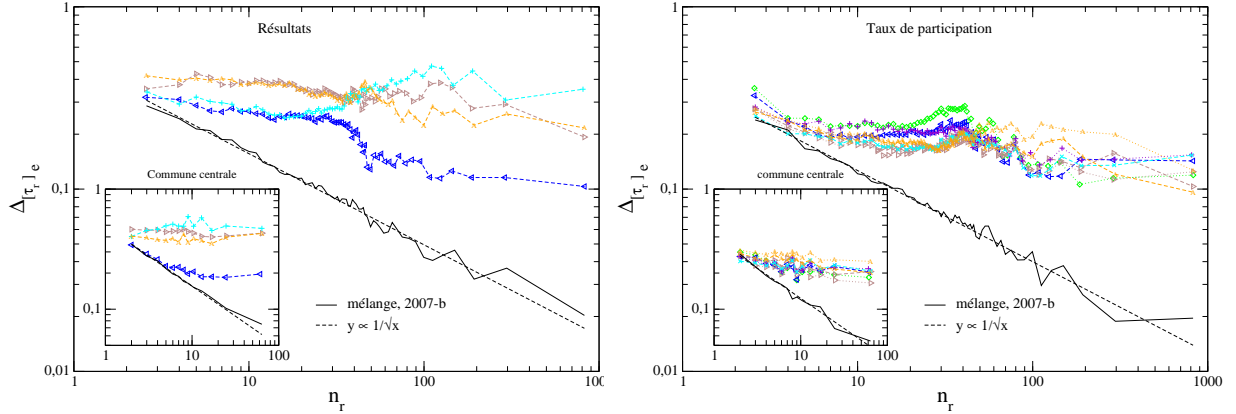

\includegraphics[scale = 0.32]{sig-t0-bvot-plein-dist-nb-res.eps}\hfill
\includegraphics[scale = 0.32]{sig-t0-bvot-plein-dist-nb-abst.eps}
\caption{\small Écarts-types des $\tau_r^\aaa$ par bureaux de vote, $\Delta_{[\tau_r]}\big|_{n_r}$ (notés $\Delta_{[\tau_r]_{\,e}}$ sur ces figures), en fonction du nombre $\nra$ de bureaux de vote des communes situées à l'intérieur d'un cercle de rayon $r_d~=~8,0~km$ centré sur une commune centrale. A gauche, les $\tau_r^\aaa$ proviennent des résultats électoraux et, à droite, des taux de participation. Les fenêtres se restreignent aux bureaux de vote de la seule commune centrale. En ligne continue, la courbe issue du mélange par bureau de vote des valeurs de l'élection de 2007-b ; et en tirets, la droite $y \propto 1/\sqrt{x}$ de l'équation~(\ref{eindep}) adéquate aux processus indépendants.}
\label{fcommunautes}
\end{figure}

Cette annexe s'inspire du bel article de Rozenfeld et al.~\cite{stanley_pop_growth}, qui cherche notamment à savoir comment les taux des croissances des cellules constituant les villes se relient entre eux, et infirme dans leur cas la loi de Gibrat~\cite{gibrat}. Quant à nous, nous cherchons à savoir si, en terme de données électorales, les zones urbaines peuvent se concevoir comme des cellules indépendantes ou non. La cellule élémentaire s'identifie alors à l'espace géographique que circonscrit le bureau de vote considéré. Cette étude analyse plus finement les corrélations spatiales entre bureaux de vote à courte distance (voir figs.~\ref{fcorrel-res} et \ref{fcorrel-abst}), voire même aux bureaux à l'intérieur d'une seule commune, et convient particulièrement de par la méthode utilisée aux zones fortement peuplées.\\

Pour des raisons de clarté, les calculs s'effectuent ici sur des disques de rayon $r_d~=~8,0~km$ autour d'une commune centrale $\aaa$, plutôt que sur des zones $\vaa$ constituées jusqu'alors de ses $n_p = 16$ plus proches communes. (Par souci de cohérence avec les sections précédentes, nous convenons de la distance $r_d~=~8,0~km$, puisque cette distance donne en moyenne $16,6$ communes autres que la commune centrale à l'intérieur de ces disques.) De plus, par souci de clarté, ici les résultats électoraux (ou les taux de participation) de la commune centrale $\aaa$ se rajoutent à ceux de ses plus proches communes.

Chaque zone considérée autour d'une commune $\aaa$ contient $\nra$ bureaux de vote~: la somme des bureaux de vote de la commune centrale et des bureaux de vote des communes à une distance $r \leqslant r_d$ de la commune centrale. Rappelons au passage que les données dont nous disposons n'indiquent pas les coordonnées spatiales des bureaux de vote, mais uniquement la position de la mairie de chaque commune. La surface réelle d'une zone définie comme il précède, qui s'identifie à la superposition des surfaces des $\nra$ bureaux de vote, diffère donc en toute rigueur d'un disque de rayon $r_d$. Notons enfin que les communes à population relativement faible ne contiennent qu'un seul bureau de vote, ce qui représente environ $83\%$ des cas.

Revenons à notre problème. La moyenne des résultats (ou des taux de participation) des bureaux vote à l'intérieur de la zone considérée de commune centrale $\aaa$ s'écrit alors comme
\be \label{etra} \tra = \frac{1}{\nra} \sum_{\beta=1}^{\nra} \tbb ~,\ee
où $\beta$ désigne l'un des $\nra$ bureaux de vote, et $\tbb$ son résultat électoral (ou son taux de participation).

Calculons l'écart-type des $\tra$ ayant sensiblement le même nombre $\nra$ de bureaux de vote. Autrement dit, à l'intérieur d'un échantillon des $\tau_r^\aaa$ ayant environ les mêmes $\nra$ déterminons l'écart-type des $\tau_r^\aaa$, noté $\Delta_{[\tau_r]}\big|_{n_r}$. En écrivant $\tbb = \toa + \eta^\beta$, il vient 
\be \label{edelta2} \big(\Delta_{[\tau_r]}\big|_{n_r}\big)^2 = \langle (\frac{1}{\nra})^2 \sum_{\beta=1}^{\nra} \eta^\beta . \sum_{\gamma=1}^{\nra} \eta^\gamma \rangle \Big|_{n_r} ~.\ee

Parmi les différentes possibilités, traitons au préalable les deux cas extrêmes~: le cas des bureaux de votes indépendants et le cas des bureaux de vote complètement corrélés.\\

$\bullet$ \textbf{Bureaux de vote indépendants}\\
L'indépendance des bureaux de vote peut s'écrire simplement comme $\langle \eta^\beta . \eta^\gamma\rangle = \sigma^2_\eta\: \delta_{\beta\gamma}$, où $\delta_{\beta\gamma} = 1$ si $\beta=\gamma$ et $0$ autrement, et $\langle ... \rangle$ représente la moyenne sur l'ensemble des $\tra$. L'équation~(\ref{edelta2}) devient alors, 
\be \label{eindep} \Delta_{[\tau_r]}\big|_{n_r} \propto \frac{1}{\sqrt{n_r}}\ .\ee
\vspace{0.25cm}

$\bullet$ \textbf{Bureaux de vote complètement corrélés}\\
Cette condition s'écrit ici comme $\langle \eta^\beta . \eta^\gamma\rangle = \sigma^2_\eta$ pour tous $\beta$ et $\gamma$. Il vient directement de l'équation~(\ref{edelta2}) : 
\be \label{egibrat} \Delta_{[\tau_r]}\big|_{n_r} = cste~.\ee
En terme d'accroissement de population, ce cas correspondrait à la loi de Gibrat, où la distance séparant deux cellules élémentaires n'a aucune importance.\\

Préoccupons-nous désormais d'un cas intermédiaire qui peut rendre compte des corrélations à longue portée (voir figs.~\ref{fcorrel-res} et \ref{fcorrel-abst}).

$\bullet$ \textbf{Décroissance de la corrélation des cellules élémentaires en loi de puissance}\\
Ce cas s'écrit alors avec un exposant $d$ de la loi de puissance, comme
\be \label{ecorpuiss} \langle \eta^\beta . \eta^\gamma\rangle \sim \frac{\sigma^2_\eta}{(r_{\beta\gamma})^d} ~\ee
où $r_{\beta\gamma}$ représente la distance qui sépare les bureaux de vote $\beta$ et $\gamma$. (Il faudrait aussi tenir compte d'une longueur de coupure pour éviter des divergences pour les petites distances.) Pour les zones urbaines à forte population, i.e. ayant un grand nombre $\nra$ de bureaux de vote, nous pouvons utiliser le passage au continu de l'équation~(\ref{edelta2}), qui devient alors pour $\delta < 2$, en négligeant les effets de bord et en considérant la zone de surface $S_d$ comme un disque de rayon $r_d$~:
\be \big(\Delta_{[\tau_r]}\big|_{n_r}\big)^2 \sim \frac{1}{n_r}\int_{S_d} \frac{\sigma_\eta^2\: \mu\: \dd r \: \dd \theta}{r^d} \sim \frac{1}{n_r}\frac{\sigma^2_\eta\: \mu\: (r_d)^{2-d}}{2-d} ~,\ee
où $\mu = \frac{n_r}{S_d} \propto \frac{n_r}{(r_d)^2}$ désigne la densité moyenne de bureau de vote par unité de surface. Il vient alors que~:
\be \label{edeltapuiss} \Delta_{[\tau_r]}\big|_{n_r} \sim \frac{1}{(n_r)^{d/4}} ~.\ee
Notons que les deux cas extrêmes précédents se retrouvent ici : le cas des bureaux de vote complètement corrélés, avec $d = 0$, et le cas des bureaux de vote indépendants, avec $d = 2$ $\ti$i.e. la dimension d'espace du problème considéré, ici une surface de dimension spatiale 2. Avec un exposant $d > 2$, les corrélations deviennent non relevantes et se confondent alors avec le cas indépendant.\\

Les figures~\ref{fcommunautes} tracent $\Delta_{[\tau_r]}\big|_{n_r}$ en fonction de $n_r$ pour les résultats électoraux (et pour les taux de participation), avec un minimum d'environ $180$ $\tra$ par échantillon. En outre, leur fenêtre visualise $\Delta_{[\tau_r]}\big|_{n_r}$ en fonction de $n_r$ pour les bureaux de vote de la commune centrale uniquement. Les $6000$ communes environ concernées pour le tracé des fenêtres $\ti$les communes qui possèdent plus d'un bureau de vote$\ti$ se répartissent alors dans des échantillons de taille d'environ $120$ au minimum. Afin de mieux percevoir le cas des bureaux de vote indépendants, nous ajoutons une courbe issue du mélange des résultats électoraux (ou des taux de participation) des bureaux de vote de l'élection de 2007-b. Qu'en déduire donc~?\\

Les taux de participation des bureaux de vote, de la seule commune centrale ou des communes à l'intérieur d'un disque de rayon $r_d~=8,0~km$, montrent une forte corrélation $\ti$bien que différente du cas complètement corrélé. Un peu plus quantitativement, pour les bureaux de vote de la seule commune centrale où l'équation~(\ref{edeltapuiss}) s'accorde relativement bien aux données, l'exposant de la décroissance des corrélations des bureaux de vote (voir Eq.~(\ref{ecorpuiss})) vaut en moyenne $d \simeq 0,5$ sur l'ensemble des $8$ élections. (Le cas complètement corrélé aurait donné $d = 0$, et le cas complètement indépendant, $d = 2$.)

Quant aux résultats électoraux des bureaux de vote à l'intérieur d'une seule commune centrale, ils vérifient relativement bien le cas complètement corrélé, excepté pour l'élection de $2000$. Les résultats électoraux des bureaux de vote à l'intérieur d'un disque de rayon $r_d~=8,0~km$ s'éloignent sensiblement du cas complètement corrélé pour se rapprocher légèrement du cas indépendant, excepté pour l'élection de $2005$. Cette dernière élection présente, de nouveau, un aspect nettement non trivial et en dehors des schèmes usuels.

\clearpage
\renewcommand{\thesection}{E}
\renewcommand{\theequation}{E-\arabic{equation}}
\setcounter{equation}{0}  
\renewcommand{\thefigure}{E-\arabic{figure}}
\setcounter{figure}{0}
\renewcommand{\thetable}{E-\arabic{table}}
\setcounter{table}{0}
\section{Réfutation du RFIM avec imitation du choix des agents}
\label{annexe-refutation-rfim}

\begin{figure}[h!]
\includegraphics[width=5cm, height=5cm]{pente.eps}\hfill
\includegraphics[width=5cm, height=5cm]{th1.eps}\hfill
\includegraphics[width=5cm, height=5cm]{th2.eps}
\caption{\small $m^\aaa=f(\mva)$ d'après Eq.~(\ref{erfim-choix-gauss}) avec un champ uniforme $F^\aaa=F$. A gauche, $J$ en fonction de $F$ donnant un équilibre stable en $\meq=0,6$, ainsi que la pente $\dfrac{d m^\aaa}{d \mva}=f'(\meq)$ en ce point d'équilibre. Le couple charnière avec lequel $\meq=0,6$ passe d'un équilibre simple à un équilibre multiple est $(F\simeq0,025; J\simeq1,36)$. La pente vaut $1$ pour le couple $(F\simeq -0,23; J\simeq 1,79)$. Figures du milieu et de droite :  en pointillé, les droites $y=x$, $x=0$ et $y=0$ ; les étoiles indiquent une position d'équilibre instable, alors que les ronds symbolisent une position d'équilibre stable. Dans la figure de droite avec $F=0,01$, la position d'équilibre stable $\meq=0,6$ est atteinte pour $J\simeq 1,39$.}
\label{fth-j}
\end{figure}

Nous allons voir ici que les données permettent de rejeter le modèle RFIM basé sur l'imitation du choix des agents et exprimé par les équations~(\ref{erfim-choix}) ou (\ref{erfim-choix-gauss}).
Nous l'établissons en nous servant d'une valeur particulière, $<\resa>\simeq0,8$ soit $<m^\aaa>\simeq0,6$, obtenue pour les résultats de l'élection 2002-b, ou pour des taux de participation de plusieurs élections (voir Tabs.~\ref{tstat-dens-res} et \ref{tstat-dens-abst}). (Rappelons que $m=2\res-1$ ou voir Eqs.~(\ref{eres},~\ref{emaa}).) Les simulations seront faites en utilisant $\pab$ donné par Eq.~(\ref{epab}), plutôt que par Eq.~(\ref{epab-pop}), afin de se rapprocher davantage de l'analyse théorique~; autrement dit, pour que la disparité de la répartition de la population ne modifie pas trop ce que prévoit une étude théorique. Les conclusions générales établies avec $\pab$ donné par Eq.~(\ref{epab}) seront ensuite vérifiées avec $\pab$ exprimé par Eq.~(\ref{epab-pop}).\\

Considérons pour commencer un champ uniforme, $F^\aaa = F$ sur l'ensemble des communes $\aaa$. Les équations~(\ref{erfim-choix}) ou (\ref{erfim-choix-gauss}) fournissent alors une valeur uniforme $m = m^\aaa$ sur l'ensemble des communes $\aaa$. Cette valeur, notée $\meq$, indiquera par la suite qu'elle se réfère à un champ uniforme où $f^\aaa=0$. La figure~\ref{fth-j}-a trace à partir de l'équation~(\ref{erfim-choix-gauss}) les couples $(F; J)$ (et avec $J\geqslant 0$) qui permettent l'existence d'un équilibre en $m^\aaa=\mva=\meq=0,6$, ainsi que la pente de $m^\aaa=f(\mva)$ en ce point d'équilibre. Cette pente vaut $f'(\mva)=\dfrac{d m^\aaa}{d \mva}=2J.p[-F-J.\mva]$, où $p$ exprime la densité de probabilité des idiosyncrasies $h_i$ $\ti$ici, une gaussienne centrée réduite. Cette pente est inférieure à un en valeur absolue pour un équilibre stable. (Rappelons qu'au point critique, en $F=0$, et $m^\aaa=\mva=0$, cette pente vaut un, ce qui implique avec la distribution choisie des $h_i$, un $J$ critique égale à $J_c=\sqrt{\pi/2}\simeq 1,25$.)\\

$\bullet$ \textbf{Équilibre unique $\mathbf{\meq=0,6}$, avec $\mathbf{J\geqslant 0}$}\\
Plaçons-nous dans le cas où les paramètres $(F; J)$ (voir Fig.~\ref{fth-j}-a) assurent l'existence d'un unique équilibre $\meq=0,6$, et notons $f'(\meq)$ la valeur de la pente au point d'équilibre. Que se passe-t-il alors en présence des fluctuations $f^\aaa$~? Mieux, comment savoir si ce modèle avec ces paramètres s'accorde avec les données réelles des résultats électoraux et des taux de participation~?

Pour répondre à cette question, nous utiliserons la valeur de la pente au point d'équilibre $f'(m^*)$. En effet, elle permet de connaître qualitativement si les corrélations sont à longue ou courte portée : plus la pente s'approche de un, plus les corrélations spatiales doivent avoir une longue portée, comme au point critique. D'autre part, les pentes des courbes empiriques donnant les valeurs moyennes de $\taa$ en fonction de $\toa$ (voir Figs.~\ref{fenvironnement-res}-a et ~\ref{fenvironnement-abst}-a) devraient être approximativement égales à la valeur de $f'(\meq)$.

Pour le démontrer faisons un développement limité au premier ordre de $\mva$ autour de $\meq$. Avec $\mva = \meq + \delta\mva$ et Eq.~(\ref{erfim-choix}), il vient $m^\aaa=\meq+\delta m^\aaa\simeq2J.p[-F-J\meq-f^\aaa].\delta\mva$. En prenant la moyenne des $m^\aaa$ sur un échantillon ayant le même $\mva$ et en considérant des fluctuations $f^\aaa$ pas trop grandes (i.e. $\sigma_f$ assez petit), il en découle
\be <\delta m^\aaa>\big|_{\mva} \simeq f'(\meq).\delta\mva~.\ee
Or les pentes empiriques valent approximativement un, alors qu'avec un équilibre unique en $<m^\aaa>\simeq\meq=0,6$, $f'(\meq)\lesssim0,76$ (voir Fig.~\ref{fth-j}-a). (Le passage des grandeurs $m=2\res-1$ aux grandeurs $\tau=\ln(\frac{m+1}{1-m})$ n'affecte pas la valeur de la pente.) Des simulations numériques, non montrées ici, confirment nos affirmations et mettent en évidence des corrélations des $m^\aaa$ faibles et à courte portée.

Ainsi, dans le cas de la présence d'un équilibre unique en $\meq=0,6$, le modèle du RFIM avec imitation du choix des agents ne permet pas de retrouver les mesures empiriques. Étudions alors le cas où existent des équilibres multiples, dont l'un proche ou égal à $<m^\aaa> = 0,6$.\\

$\bullet$ \textbf{Équilibre double avec une pente $\mathbf{f'(\meq=0,6)=1}$}\\
Pour $F\simeq-0,23$, $J\simeq1,79$ et en l'absence de fluctuation $f^\aaa$, au point d'équilibre $\meq=0,6$ la pente de $m^\aaa=f(\mva)$ prend la valeur convenable et précédemment recherchée de $f'(\meq)\simeq1$ (voir Fig.~\ref{fth-j}-a). Mais ce point d'équilibre est clairement instable (voir figure~\ref{fth-j}-b), à moins que toutes les communes vérifient initialement $m^\aaa(t=0)>0,6$, condition initiale, semble-t-il, peu réaliste. L'équilibre stable se situe quant à lui à une valeur négative proche de $-1$ (cf. Fig.~\ref{fth-j}-b). Même si les conditions initiales drastiques étaient respectées, pour oblitérer la présence de l'autre équilibre à $m$ négatif, la présence des $f^\aaa\neq 0$ favoriserait la manifestation de l'autre équilibre à $m<0$, et déstabiliserait de la sorte l'équilibre $\meq=0,6$. (Il en irait quasiment de même avec des paramètres proches de ceux-ci.) Bref, la faible stabilité de $\meq=0,6$ avec ce jeu de paramètres ne permet pas à ce modèle de sauver les phénomènes.\\

$\bullet$ \textbf{Équilibre double, avec $\mathbf{F \geqslant 0}$}\\
Terminons alors par le cas où il existe un équilibre double, dont un, proche de $\meq=0,6$, et ce avec $F>0$ de manière à favoriser naturellement cet équilibre stable plutôt que l'autre à $\meq<0$. La plage de $F$ qui réalise ces conditions s'étend de $F=0$ à $F\simeq0,025$ en l'absence de $f^\aaa$ (voir Fig.~\ref{fth-j}-a). Dans un premier temps, analysons en détail ce qui se produit en $F=0,01$. Les conclusions dégagées seront dans un second temps confirmées avec d'autres valeurs de $F$ encadrant $F=0,01$. Notons qu'avec $F=0,01$, l'équilibre stable $\meq=0,6$ s'obtient pour $J\simeq1,39$. Enfin, la figure~\ref{fth-j}-c trace $m^\aaa=f(\mva)$ avec un champ uniforme $F^\aaa=0,01$ pour différents $J$.

Le cas de l'existence d'un équilibre unique ayant été traité plus haut, la question qui se pose ici est de savoir comment la présence du second équilibre stable influe sur les corrélations et sur la distribution des $m^\aaa$. Pour répondre à cette question, nous faisons des simulations de l'équation~(\ref{erfim-choix-gauss}) avec $F=0,01$, $J=1,4$, mais avec différentes conditions initiales et écarts-types de $f^\aaa$. En effet, plus l'écart-type $\sigma_f$ des $f^\aaa$ augmente, plus le second équilibre stable $\meq<0$ se manifeste. Pour le comprendre qualitativement, d'après l'équation~(\ref{erfim-choix-gauss}), il suffit de remplacer $\mva$ sur l'axe des abscisses de la figure~\ref{fth-j}-c par $(\mva+f^\aaa)$. Quant aux conditions initiales des $m^\aaa$, elles peuvent favoriser un équilibre $\meq$ au détriment de l'autre. Nous choisissons ici deux types de conditions initiales afin de mieux mettre en lumière le rôle de $\sigma_f$ : la première où les $m^\aaa(t=0)$ sont tous nuls, et la seconde où les $m^\aaa(t=0)$ résultent d'un tirage aléatoire i.i.d. de moyenne nulle et d'écart-type $\sigma_{m(t=0)}$ donné, égal à $0,4$ en l'occurrence. (Rappelons que la moyenne des $f^\aaa$ est nulle.) Le second type de conditions initiales rend évidemment plus manifeste la présence de l'autre équilibre $\meq<0$. Les figures~\ref{fj1,4} reportent les corrélations spatiales $C(R)$ des $m^\aaa$ ainsi que les histogrammes en grandeur $\res=\frac{m+1}{2}$. Ces histogrammes en grandeur $\rho$ peuvent directement se comparer avec ceux des résultats électoraux (voir l'élection 2002-b par commune de Fig.~\ref{fhisto-dens-res}) ou des taux de participation (voir Fig.~\ref{fhisto-dens-abst}). 

\begin{figure}
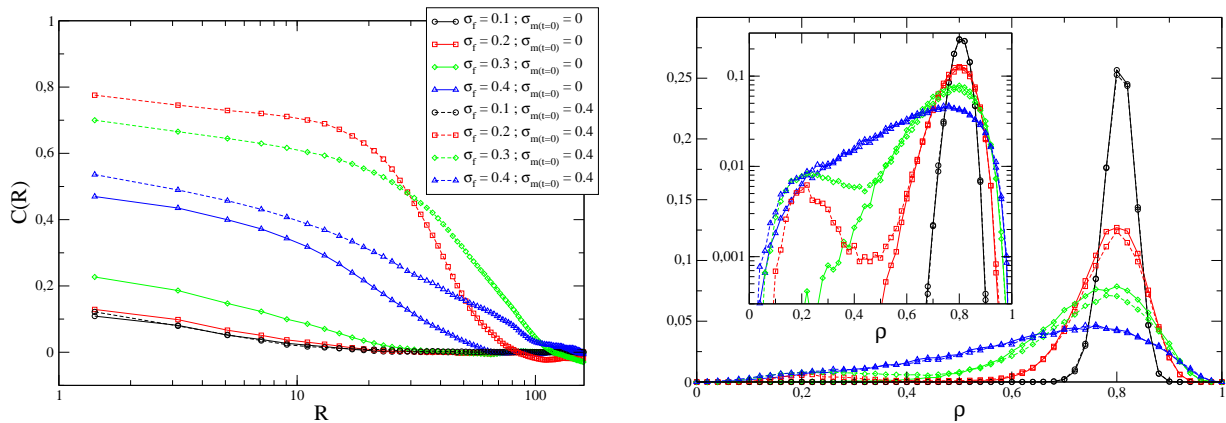

\includegraphics[scale = 0.32]{cor-1.4-multi.eps}\hfill
\includegraphics[scale = 0.32]{histo-1.4-multi.eps}
\caption{\small Simulations avec $F=0,01$, $J=1,4$ dans Eq.~(\ref{erfim-choix-gauss}), avec différents écarts-types $\sigma_f$ des $f^\aaa$ et différentes conditions initiales (soit toutes nulles, soit de moyenne nulle et d'écart-type $\sigma_{m(t=0)}=0,4$). A gauche, les corrélations spatiales $C(R)$ et à droite les histogrammes en grandeur $\rho$ des $m^\aaa$ simulés.}
\label{fj1,4}
\end{figure}

Que nous apprennent principalement les simulations représentées à la figure~\ref{fj1,4}~? Premièrement, quand s'exerce la présence d'un seul équilibre $\meq$ (cf. les histogrammes de Fig.~\ref{fj1,4} avec $\sigma_f=0,1$ ou $\sigma_f=0,2$, avec pour ce dernier les conditions initiales $m^\aaa(t=0)=0$), les corrélations spatiales $C(R)$ sont peu élevées et surtout de courte portée $\ti$ce qui confirme nos dires précédents relatifs au cas d'existence d'un équilibre unique $\meq$. Deuxièmement, dès que la présence du second équilibre $\meq<0$ se manifeste (cf. les histogrammes), les corrélations spatiales deviennent non seulement importantes entre proches voisins (cf. $C(R\simeq1)$) mais aussi à longue portée.

Comment comprendre ce dernier point~? Quand les deux équilibres $\meq$ se manifestent, cela signifie qu'il existe des domaines dans lesquels tous les $m^\aaa$ prennent des valeurs autour de l'équilibre $\meq>0$, et d'autres domaines dans lesquels tous les $m^\aaa$ prennent des valeur autour du second équilibre $\meq<0$. Ceci explique la forte valeur des corrélations entre proches voisins quand les domaines sont bien distincts.

Nous pouvons retenir de cette analyse que le RFIM avec imitation de choix exprimé par Eq.~(\ref{erfim-choix}) peut engendrer des corrélations à longue portée et relativement fortes entre proches voisins pour une moyenne des $\resa$ bien éloignée de $\rho=0,5$ (comme par exemple $<\resa>\simeq0,7$ pour $\sigma_f=0,3$ et $\sigma_{m(t=0)}=0,4$), mais en faisant apparaître de façon concomitante une distribution bimodale ou très élargie.\\

\begin{figure}
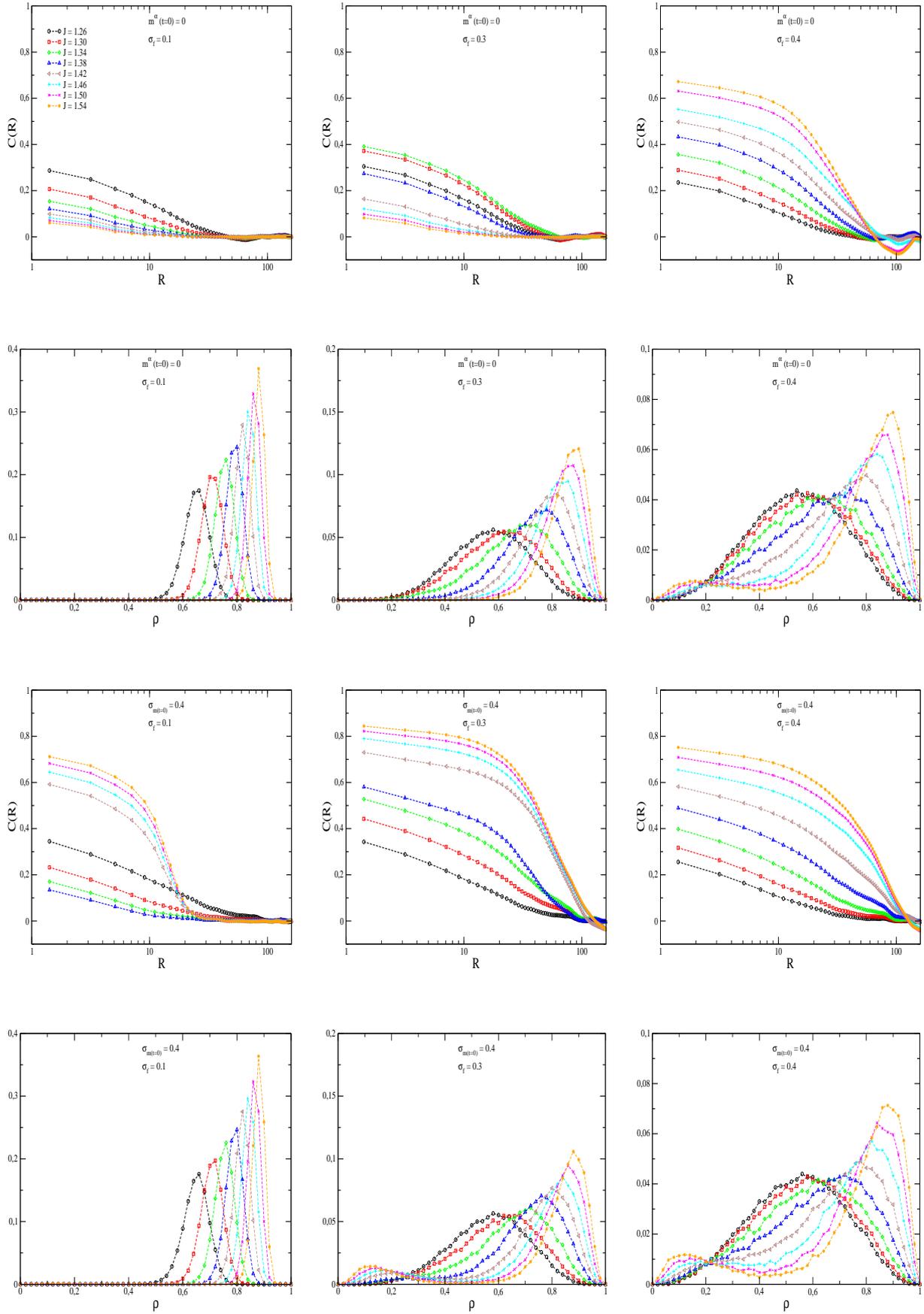

\includegraphics[width=5cm, height=5cm]{cor-J-0.01-0.1-ci0.eps}\hfill
\includegraphics[width=5cm, height=5cm]{cor-J-0.01-0.3-ci0.eps}\hfill
\includegraphics[width=5cm, height=5cm]{cor-J-0.01-0.4-ci0.eps}\vspace{1cm}
\includegraphics[width=5cm, height=5cm]{histo-J-0.01-0.1-ci0.eps}\hfill
\includegraphics[width=5cm, height=5cm]{histo-J-0.01-0.3-ci0.eps}\hfill
\includegraphics[width=5cm, height=5cm]{histo-J-0.01-0.4-ci0.eps}\vspace{1cm}
\includegraphics[width=5cm, height=5cm]{cor-J-0.01-0.1-ci0.4.eps}\hfill
\includegraphics[width=5cm, height=5cm]{cor-J-0.01-0.3-ci0.4.eps}\hfill
\includegraphics[width=5cm, height=5cm]{cor-J-0.01-0.4-ci0.4.eps}\vspace{1cm}
\includegraphics[width=5cm, height=5cm]{histo-J-0.01-0.1-ci0.4.eps}\hfill
\includegraphics[width=5cm, height=5cm]{histo-J-0.01-0.3-ci0.4.eps}\hfill
\includegraphics[width=5cm, height=5cm]{histo-J-0.01-0.4-ci0.4.eps}
\caption{\small Corrélations spatiales et histogrammes tirés des simulations avec $F=0,01$ dans Eq.~(\ref{erfim-choix-gauss}), avec $J$ allant de $1,26$ à $1,56$, $\sigma_f$ égal à $0,1$, $0,2$ et $0,4$, et les conditions initiales des $m^\aaa$, soit tous nuls, soit de moyenne nulle et d'écart-type $\sigma_{m(t=0)}=0,4$.}
\label{fmulti-j}
\end{figure}

Confirmons le point fort dégagé par l'analyse précédente $\ti$l'apparition conjointe d'une distribution bimodale ou très élargie et des corrélations à longue portée$\ti$ par d'autres simulations, encore à $F=0,01$ mais avec différentes valeurs de $J$. En l'absence de $f^\aaa$, les courbes théoriques $m^\aaa=f(\mva)$ se voient à la figure~\ref{fth-j}-c. Les corrélations spatiales et les distributions obtenues par simulations numériques sont représentées à la figure~\ref{fmulti-j} pour plusieurs valeurs de $J$ allant de $1,26$ à $1,54$, pour $\sigma_f$ égal à $0,1$, $0,3$ et $0,4$, et les deux types de conditions initiales utilisées plus haut. Notons qu'à $F=0,01$, les équilibres multiples de $m^\aaa=f(\mva)$ en l'absence de $f^\aaa$ apparaissent quand $J\gtrsim1,31$, et que les histogrammes obtenus pour $\sigma_f=0,1$ avec comme conditions initiales $\sigma_{m(t=0)}=0,4$ en coordonnées logarithmiques sur l'axe des ordonnées témoignent de la présence du second équilibre quand $J\gtrsim1,42$. (Les histogrammes des résultats ou des taux de participation des figures~\ref{fhisto-dens-res} et \ref{fhisto-dens-abst} ne laissent toujours pas percevoir de second équilibre en coordonnées logarithmiques sur l'axe des ordonnées.)

Sans exploiter en détail les courbes de Fig.~\ref{fmulti-j}, elles corroborent l'assertion précédente : les corrélations à longue portée apparaissent pour une moyenne $<\resa>$ bien éloignée de $\rho=0,5$ avec la manifestation du second équilibre, élargissant par là même la distribution des $\resa$, la rendant clairement bimodale, ou bien très élargie. Remarquons néanmoins qu'avec une moyenne d'ensemble proche de $\resa=0,5$, les corrélations à longue portée apparaissent sans nécessairement faire intervenir de deuxième équilibre, ou d'élargissement de la distribution des $\resa$. Il suffit en effet de se placer près du point critique ($J\simeq1,25$, $F\simeq0$ donnant $<\resa>\simeq0,5$). Mais toute cette étude se polarise sur les cas où les moyennes d'ensembles $<\resa>$ ne sont pas proches de $\rho=0,5$, et valent aux alentours de $\rho=0,8$.\\

Terminons cette partie en considérant comme nous l'avions dit plus haut, d'autres valeurs de $F$ dans la plage de $0$ à $0,025$. Plus $F$ se rapproche de $F=0$, plus la présence du second équilibre se manifeste facilement, et plus la distribution des $\resa$ devient facilement large ou bimodale $\ti$avec son corollaire de corrélations plus facilement à longue portée et à plus grande valeur entre proches voisins. Ici aussi il devient impossible de générer $<\resa>\simeq0,8$ avec des corrélations à longue portée sans distribution des $\resa$ bimodale ou large. Des simulations effectuées pour $F=0,0001$ (non montrées ici) le confirment. Par contre, quand $F$ augmente, la portée des corrélations tend à s'estomper par rapport à celles obtenues pour $F=0,01$, ce qui s'explique aisément par un plus grand éloignement du point critique. Idem, des simulations réalisées pour $F=0,02$, non montrées ici, l'attestent.\\

$\bullet$ \textbf{Conclusion}\\
Le modèle RFIM avec imitation du choix des agents écrit à l'équation~(\ref{erfim-choix-gauss}) contient quatre paramètres libres : l'intensité de l'imitation $J$, le champ global appliqué $F$, l'intensité des fluctuations des tendances propres des communes $\sigma_f$ et la longueur caractéristique $\ell_c$ de la portée des imitations. En plus de ces quatre paramètres, les conditions initiales peuvent influer sur les solutions fournies par le modèle. Nous avons fait agir dans cette annexe chacun de ces éléments excepté $\ell_c$. Mais son influence est minime, comme le confirment d'autres simulations. Nous pouvons arrêter ici les tentatives pour obtenir des simulations en accord avec les résultats (ou les taux de participation) des élections de moyenne nationale $<\resa>\simeq 0,8$.

Les simulations faites avec $\pab$ donné par Eq.~(\ref{epab-pop}) confirment les conclusions ci-dessus. Par contre, elles font apparaître plus facilement $\ti$à des valeurs de $J$ plus faibles que les précédentes et à champ $F$ égal$\ti$ des histogrammes très élargis ou bimodaux.

Nous pouvons donc conclure l'impossibilité au modèle RFIM avec imitation du choix des agents de générer simultanément : des corrélations à longue portée des $\resa$, une moyenne d'ensemble proche $\rho=0,8$ et une distribution des $\resa$ non bimodale (ni très large). L'explication provient de l'éloignement de la moyenne finale $<\resa>\simeq0,8$ du point critique du modèle en $\rho=0,5$.

Ce modèle ne peut sauver les phénomènes, il doit donc être rejeté.

\clearpage
\renewcommand{\thesection}{F}
\renewcommand{\theequation}{F-\arabic{equation}}
\setcounter{equation}{0}  
\renewcommand{\thefigure}{F-\arabic{figure}}
\setcounter{figure}{0}
\renewcommand{\thetable}{F-\arabic{table}}
\setcounter{table}{0}
\renewcommand{\thesubsection}{F-\arabic{subsection}}
\setcounter{subsection}{0}
\section{Passage au continu et conditions de stabilité du modèle à tendances communales statiques}
\label{annexe-continu-stabilite}

L'équation~(\ref{efaa}) résulte de l'état d'équilibre de l'équation linéaire
\be \label{efaa-t} \dfrac{\mathrm{d}F^\aaa(t)}{\mathrm{d}t}=K\x\fva(t) - F^\aaa(t) + f^\aaa~.\ee

Notons que l'étude du modèle correspond aux états d'équilibre du modèle $\ti$comme nous l'avions fait pour le précédent modèle RFIM avec imitation des choix$\ti$ sur lesquels une analyse générale peut être établie. Tenir compte des solutions transitoires, avec en prime les conditions initiales qui les sous-tendent, alors qu'un équilibre existe, compliquerait énormément l'étude et la perdrait dans les méandres de l'ensemble des cas particuliers.

L'état d'équilibre du système correspond à l'ensemble de l'équation~(\ref{efaa-t}) appliquée à toutes les communes centrales $\aaa$ ; ce qui pourrait donner lieu à une écriture matricielle. Mais l'ensemble de ces équations se prête à une écriture relativement simple en utilisant la grandeur spatiale continue $\rr$. Il est important de remarquer que l'espace, le territoire français sur lequel vivent les habitants répartis en diverses communes, est considéré comme une surface plane de dimension deux, et localisé comme précédemment par les coordonnées XY (longitude et latitude). Nonobstant l'hétérogénéité de la répartition de l'ensemble des communes, le passage au continu se simplifie encore. En adoptant alors une répartition homogène des communes si $\pab$ provient de Eq.~(\ref{epab}) (avec une répartition également homogène de la population si $\pab$ l'équation~(\ref{epab-pop}) définit $\pab$), et en utilisant l'équation~(\ref{efva}) pour $\fva$, l'équation~(\ref{efaa-t}) s'écrit alors comme
\be \label{ef-continu} \dfrac{\partial F(\rr,t)}{\partial t}=K\frac{\int \mathrm{d}\rr'\, e^{-\frac{|\rr-\rr'|}{\ell_c}}\,F(\rr',t)}{\int \mathrm{d}\rr'\, e^{-\frac{|\rr-\rr'|}{\ell_c}}} - F(\rr,t) + f(\rr)~.\ee

L'expression linéaire ci-dessus fait intervenir un produit de convolution. Il convient donc d'utiliser les transformées de Fourier pour la traiter plus facilement. Question notation, $\hat{g}(\kk,t)$ désigne de façon générale la transformée de Fourier de $g(\rr,t)$ à deux dimensions, avec la convention suivante~:
\begin{eqnarray}
g(\rr,t) = \frac{1}{2\pi} \int \mathrm{d}\kk\, e^{i\kk.\rr}\,\hat{g}(\kk,t)~,\\ 
\hat{g}(\kk,t) = \frac{1}{2\pi} \int \mathrm{d}\rr\, e^{-i\kk.\rr}\,g(\rr,t)~.
\end{eqnarray}
L'équation~(\ref{ef-continu}) s'écrit avec les transformées de Fourier, comme
\be \label{ef-fourier} \dfrac{\partial \hat{F}(\kk,t)}{\partial t} = \big[\frac{K}{(1+\qq^2)^{3/2}}-1\big]\,\hat{F}(\kk,t)+\hat{f}(\kk)~,\ee
où $\qq=\kk\:\ell_c$ est une grandeur sans dimension.

L'équation précédente se comprend plus facilement par le traitement des cas limites. En notant $A=\big[\frac{K}{(1+\qq^2)^{3/2}}-1\big]$, il vient $A\underset{q\ll 1}{\longrightarrow}-\frac{3}{2}\qq^2\,K-(1-K)$ et $A\underset{q\gg 1}{\longrightarrow}-1$. Nous retrouvons donc pour les grandes distances à l'échelle de $\ell_c$, i.e. pour les petits vecteurs d'onde ($k\ll\frac{1}{\ell_c}$), le Laplacien de $F(\rr,t)$. (La transformée de Fourier du Laplacien à deux dimensions donne $-\kk^2$.) Ce résultat était prévisible, au coefficient $\ti$inopérant ici$\ti$ $\frac{3}{2}\ell_c^2$ près. En effet, le membre de droite de l'équation~(\ref{efaa-t}) s'écrit comme $K(\fva-F^\aaa)-(1-K)F^\aaa + f^\aaa$. Or, $\fva$ réalise une sorte de moyenne des $F^\beta$ au voisinage de la commune centrale $\aaa$, ce qui permet de concevoir à son tour $(\fva-F^\aaa)$ comme une sorte de Laplacien. Quant aux petites distances à l'échelle de la distance caractéristique $\ell_c$, i.e. pour les grands vecteurs d'onde ($k\gg\frac{1}{\ell_c}$), le voisinage de $F(\rr,t)$ n'intervient pas. Ce qui impliquerait à lui seul et en remontant à l'équation discrète (\ref{efaa-t}), que $F^\aaa$ soit égal à $f^\aaa$.

\subparagraph{$\bullet$ Conditions de stabilité\\}
\label{pt-stabilite}
Revenons à la stabilité du modèle, qui ne dépend pas de la façon dont les communes (ou la population) se répartissent sur le territoire. Les solutions de l'équation~(\ref{ef-fourier}) convergent à temps long quand 
\be \label{estable} \frac{K}{(1+\qq^2)^{3/2}}-1 < 0~.\ee
\vspace{0.25cm}

\begin{wrapfigure}[13]{r}{0.52 \textwidth}
  \centering
  \includegraphics[scale=0.32]{stabilite-laplacien.eps}
\end{wrapfigure}

Ainsi, les solutions sont toujours stables pour $K<1$. L'instabilité ne peut se produire que pour $K \geqslant 1$, et pour les modes adimensionnés $q\leqslant q^* = \sqrt{K^{2/3}\,-1}$. Le premier mode instable apparaît à $K = 1$ avec $q=0$, i.e. pour une solution $F(\rr,t)$ spatialement uniforme. Enfin, pour $K>1$, le mode uniforme $q=0$ reste d'ailleurs le mode le plus instable (celui pour lequel $\big[\frac{K}{(1+\qq^2)^{3/2}}-1\big]$ est positif et de plus grande valeur) comme le montre la figure ci-contre. Remarque : L'assimilation de ($\fva-F^\aaa$) à un Laplacien effectif, partage les mêmes critères de stabilité ci-dessus, à la valeur de $k^*$ près.

Considérons maintenant le cas limite $K=1$ avec $k=0$. Puisque $(\fva-F^\aaa)$ peut être assimilé à une sorte de Laplacien (cf. discussion ci-dessus), l'équation~(\ref{efaa}) se comporte comme une sorte de diffusion en présence de sources fixes : l'ensemble des $f^\aaa$. La solution à temps long n'est alors stable que si
\be \label{estable-diff} \sum_\aaa f^\aaa = 0.\ee
Cette condition de stabilité peut se concevoir par analogie de la diffusion thermique en présence de sources internes de chaleur, l'ensemble des $f^\aaa$ en l'occurrence. Ceci se retrouve mathématiquement. Avec $K=1$ et $k=0$, l'équation~(\ref{ef-fourier}) devient $\dfrac{\partial\hat{F}(0,t)}{\partial t}=\hat{f}(0)$. Or $\hat{f}(\kk) = \frac{1}{2\pi}\sum_\aaa f^\aaa\x e^{-i\kk.\rr^\aaa}$ puisque $f(\rr) = \sum_\aaa f^\aaa \x\delta(\rr - \rr^\aaa)$, où $\delta$ dénote la fonction de Dirac et $\rr^\aaa$ la position de la commune $\aaa$. $\hat{F}(0,t)$ ne diverge pas seulement si $\hat{f}(0)=\sum_\aaa f^\aaa$ est nul.

En résumé. L'équation~(\ref{efaa}) du modèle discuté, admet des solutions stables pour $K<1$. Les solutions deviennent instables pour $K>1$, et avec le mode spatial uniforme (i.e. $\kk=\mathbf{0}$) comme mode le plus instable. Pour $K=1$, l'équation devient de type diffusif, et le système n'est stable que si $\sum_\aaa f^\aaa = 0$.


\clearpage
\renewcommand{\thesection}{G}
\renewcommand{\theequation}{G-\arabic{equation}}
\setcounter{equation}{0}  
\renewcommand{\thefigure}{G-\arabic{figure}}
\setcounter{figure}{0}
\renewcommand{\thetable}{G-\arabic{table}}
\setcounter{table}{0}
\renewcommand{\thesubsection}{G-\arabic{subsection}}
\setcounter{subsection}{0}
\section{L'interaction entre communes testée avec le modèle des tendances communales statiques} 
\label{annexe-test}

Cette annexe se propose de comparer trois types d'interaction entre agents de différentes communes, et ce sur la base du modèle des tendances communales statiques.

Nous exposons dans un premier temps une méthode utilisée pour tester le modèle à tendances statiques, puis nous l'illustrons avec l'une des trois formes de connexion entre communes, avant de comparer entre elles les trois formes d'interaction entre communes. Enfin, nous terminerons cette annexe par une brève discussion.

\subparagraph*{$\bullet$ Méthode suivie\\}
\label{pt-annexe-test-methode}
Le modèle statique donné par l'équation~(\ref{efaa}) attribue la valeur $F^\aaa$ d'une commune en fonction des valeurs d'autres communes (regroupées dans $\fva$), d'un coefficient $K$ et d'une tendance spécifique fixe dans le temps, $f^\aaa$, inhérente à la commune $\aaa$. $F^\aaa$ est une grandeur mesurable puisqu'elle est directement liée au résultat électoral (ou au taux de participation) de la commune $\aaa$ comme le montre Eq.~(\ref{eresa-fa}), d'où la relation~:
\be \label{efrho} F^\aaa = -\sqrt{2}\: \operatorname{erfc}^{(-1)}(2\resa)~,\ee
avec une distribution gaussienne des idiosyncrasies individuelles $h_i$. (Comme nous l'avons déjà évoqué, avec une distribution des $h_i$ de nature logistique, $F^\aaa\propto \taa$.) $\fva$ dépend de la façon dont les agents de la commune $\aaa$ sont connectés $\ti$en termes d'influence sur le choix du vote ou de la participation, à l'élection considérée$\ti$ par les agents d'autres communes $\beta$. Dit autrement, la manière dont une commune $\beta$ influence les agents d'une commune $\aaa$ est transcrite par le coefficient $\pab$, et se retrouve dans $\fva$. Le modèle est satisfaisant s'il se suffit à lui-même, autrement dit s'il parvient à retrouver les phénomènes observés avec un minimum d'ingrédients. Dans ce cas, le modèle est satisfaisant si les tendances propres des communes sont décorrélées entre elles. Le cas contraire extrême, où chaque $f^\aaa$ dériverait directement du seul résultat électoral (ou du taux de participation) $\resa$, n'aurait aucune valeur explicative puisque toute l'information serait contenue dans le champ des $f^\aaa$ $\ti$i.e. du champ des $\resa$ en l'occurrence. En d'autres termes, un modèle convenable parvient à retrouver les phénomènes réels avec un minimum d'information, soit ici avec l'ensemble des $f^\aaa$ le plus simple possible, i.e. le moins corrélé possible.

Voyons maintenant comment mettre à profit les commentaires précédents, de manière à comparer différentes expressions de $\pab$. La pertinence du modèle réside ici dans la corrélation spatiale des $f^\aaa$~: nulle pour un modèle parfait, égale à la corrélation des $\resa$ avec un modèle qui ne sert à rien. Tout ce qui suit consistera donc à estimer la tendance spécifique $\ffa$ par commune $\aaa$ à partir de l'ensemble des résultats électoraux (ou des taux de participation), puis de calculer les corrélations spatiales des $\ffa$.\\

Considérons une élection donnée. Le résultat électoral (ou le taux de participation) réel $\resa$ détermine $F^\aaa$ de chaque commune en utilisant l'équation~(\ref{efrho}). De l'ensemble des $F^\beta$ des communes $\beta\neq\aaa$ et de l'expression de $\pab$, il se déduit $\fva$ d'après l'équation~(\ref{efva}). Noter que $\pab$ dépend d'un paramètre, $\ell_c$, s'il ne dépend que du voisinage (cf. Eqs.~(\ref{epab}, \ref{epab-pop}), ou $d$ avec la connexion de type gravité étendue (cf. Eq.~(\ref{epab-grav})). Ainsi, $\fva=\fva(\ell_c)$ (ou $\fva=\fva(d)$). Ensuite, $\ffa$ s'évalue d'après l'équation~(\ref{efaa}) en fonction de $F^\aaa$, de $\fva(\ell_c)$ (ou de $\fva(d)$) et de $K$. Autrement dit, les données réelles des élections permettent d'évaluer la tendance propre $\ffa(K,\ell_c)$ (ou $\ffa(K,d)$) associée à chaque commune $\aaa$, en fonction des deux paramètres libres du modèle adopté, $K$ et $\ell_c$ (ou $K$ et $d$). Ceci permet ensuite de déterminer, parmi tous les couples testés, le couple de paramètres qui minimise la corrélation spatiale des $\ffa$. Ce qui détermine le couple de paramètres du modèle avec la forme de $\pab$ voulue, qui convient le mieux à une élection donnée.

Notons que cette méthode, qui consiste dans un premier temps à calculer les tendances spécifiques estimées en fonction des paramètres du modèle et des données réelles,  puis, dans un second temps, à mesurer le degré d'indépendance des tendances spécifiques estimées, convient à ce modèle. En effet ce modèle linéaire ne comporte ni de solutions doubles, ni de point critique. Ainsi, $\ffa$ s'obtient de façon univoque à partir des paramètres du modèle et des données réelles. L'inexistence de point critique évite d'introduire un biais possible dans la valeur des corrélations.

Élaborons maintenant une sorte de facteur de qualité $Q$, qui nous permettra de comparer les trois formes de connexion $\pab$ testées.

\subparagraph*{$\bullet$ Facteur de qualité\\}
\label{pt-annexe-test-Q}
Prenons l'exemple du cas où $\pab$ provient de l'équation~(\ref{epab}). Appelons $\tilde{C}_{(K,\ell_c)}(r)$ la corrélation spatiale de $\ffa(K,\ell_c)$ à la distance $r$ réelle (cf. Eq.~(\ref{ecorrel-spatiale}) et section~\ref{pt-cor-spatial-res}). La corrélation s'effectue à l'intérieur des couronnes $n$, telles que $r\in]2\,n\,D;2\,(n+1)\,D]$, et se note $\tilde{C}_{(K,\ell_c)}(n)$. (La distance $D=3~km$, fixe ici, correspond environ à la distance séparant deux plus proches communes. A noter que dans Eq.~(\ref{ecorrel-spatiale}), $D$ dépend du département où se situe la commune $\aaa$ et vaut en moyenne environ $2,7~km$.)

La corrélation spatiale des tendances spécifiques évaluées, $\ffa(K,\ell_c)$ doit être comparée à la corrélation des données réelles $F^\aaa$, et ce afin de mesurer la qualité apportée par le modèle sur l'élection considérée avec le couple de paramètres libres $(K,\ell_c)$. Notons $C(n)$ la corrélation spatiale des $F^\aaa$ à l'intérieur de la couronne $n$. Définissons le facteur de qualité $Q_{(K,\ell_c)}(r_{max})$, la somme sur $n_{max}$ couronnes (où $r_{max}=2\,n_{max}\,D$) des valeurs absolues du rapport entre les corrélations estimées $\tilde{C}_{(K,\ell_c)}(n)$ et les corrélations des données brutes $C(n)$, soit~:
\be \label{eQ} Q_{(K,\ell_c)}(r_{max}) = \sum_{n=1}^{n_{max}}\big|\frac{\tilde{C}_{(K,\ell_c)}(n)}{C(n)}\big|~.\ee
$Q_{(K,\ell_c)}(r_{max})=1$, signifie que le modèle n'apporte rien. Plus $Q_{(K,\ell_c)}(r_{max})$ est petit, meilleure est la qualité du modèle pour l'élection considérée, et avec le couple de paramètres $(K,\ell_c)$ pris en compte.

\begin{figure}[t]
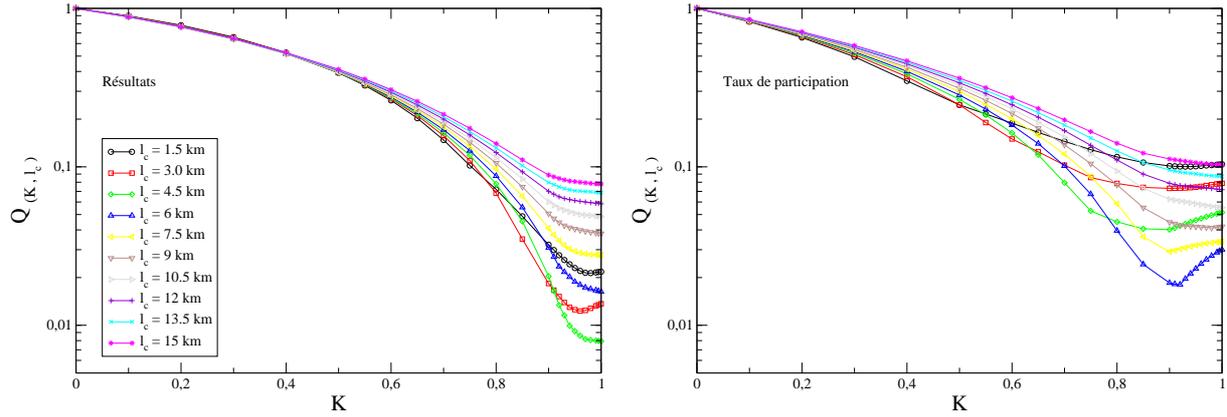

\includegraphics[scale = 0.32]{qa1-res-2007.eps}\hfill
\includegraphics[scale = 0.32]{qa1-abst-2007.eps}
\caption{\small Facteur de qualité $Q_{(K,\ell_c)}$ obtenu pour $r_{max}=96~km$ et pour l'élection 2007-b, avec le modèle à tendances statiques, et avec $\pab$ défini par l'équation~(\ref{epab}). Ici, $\ell_c$ prend des valeurs jusqu'à $15~km$, par multiples entiers de $1,5~km$ quand $\ell_c > 3~km$. Le couple qui convient au mieux à l'élection donnée, sous couvert du modèle considéré, correspond à celui qui minimise $Q_{(K,\ell_c)}$. A gauche, l'élection 2007-b considérée pour ses résultats par commune, et à droite, pour ses taux de participation.}
\label{fQ2007}
\end{figure}

La figure~\ref{fQ2007} montre $Q_{(K,\ell_c)}(r_{max})$ pour les résultats et les taux d'abstention de l'élection 2007-b, avec $n_{max}=16$ (i.e. jusqu'à une distance $r_{max}=96~km$), et avec le type d'influence d'une commune $\beta$ sur un agent de la commune $\aaa$ défini par Eq.~(\ref{epab}). Le couple $(K,\ell_c)$ qui minimise $Q_{(K,\ell_c)}(r_{max})$ est celui qui explique au mieux l'élection considérée, en adoptant le modèle statique et la forme de $\pab$ voulue.

Nous avons testé la fiabilité de la méthode, en simulant les $F^\aaa$ à partir du modèle statique, et avec des couples $(K,\ell_c)$ connus. Les $F^\aaa$ simulés ont été ensuite bruités, autrement dit un bruit additif $\eta^\aaa$, i.i.d., s'est rajouté aux $F^\aaa$ simulés. Nous avons cherché ensuite le couple qui minimise le facteur de qualité. Ce couple correspond à celui attendu (celui qui a servi à la simulation) dès lors que $r_{max}\gtrsim 30~km$.

\begin{figure}[t]
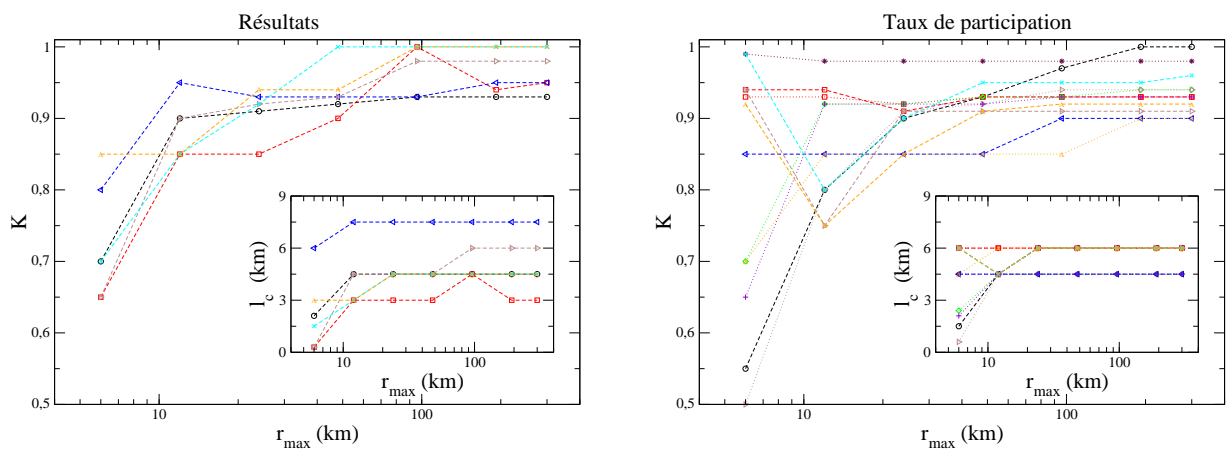

\includegraphics[scale = 0.32]{K-lc-res.eps}\hfill
\includegraphics[scale = 0.32]{K-lc-abst.eps}
\caption{\small Couple $(K,\ell_c)$ du modèle statique, avec $\pab$ défini par Eq.~(\ref{epab}), le plus approprié pour chaque élection, et à une distance maximale $r_{max}$ de prise en compte des corrélations des tendances spécifiques estimées, $\ffa(K,\ell_c)$.  A gauche, les couples obtenus pour les résultats des élections, et à droite, pour les taux de participation.}
\label{ftest-K-lc}
\end{figure}

La figure~\ref{ftest-K-lc} montre le couple $(K,\ell_c)$ du modèle statique (avec $\pab$ défini par Eq.~(\ref{epab})) qui s'accorde au mieux à chaque élection, i.e. qui minimise $Q_{(K,\ell_c)}$ à une distance maximale $r_{max}$ donnée. Les couples $(K,\ell_c)$ restent à peu près les mêmes pour une élection donnée, si la distance maximale pour laquelle les corrélations des $\ffa(K,\ell_c)$ pris en compte est telle que $r_{max}\gtrsim 50~km$.

Comparons maintenant trois manières possibles d'exprimer l'influence d'une commune $\beta$ sur un agent d'une commune $\aaa$, et toujours sous le couvert du modèle des tendances communales statiques.

\subparagraph*{$\bullet$ Comparaison des types d'interaction entre communes\\}
\label{pt-annexe-test-comparaison}
Plutôt que de chercher le couple particulier $(K,\ell_c)$ qui correspond au mieux à une élection donnée selon le modèle utilisé, nous cherchons ici à comparer entre eux différents modèles. Ou mieux, nous testons trois variantes, qui portent sur le type d'interaction entre communes, au sein du modèle dit à tendances communales statiques. En d'autres termes, $\pab$, qui exprime la manière dont les agents d'une commune $\aaa$ interagissent avec les agents de la commune $\beta$, pourra prendre trois formes différentes à l'intérieur de $\fva$.

Peu nous importera donc de déterminer, selon le type d'interaction adopté, le couple particulier $(K,\ell_c)$ qui minimise $Q_{(K,\ell_c)}(r_{max})$ à $r_{max}$ donné. Notons alors $Q(r_{max})$, la valeur minimale de $Q_{(K,\ell_c)}(r_{max})$ obtenue pour le type d'interaction $\pab$ considéré et à $r_{max}$ donné. Autrement dit, $Q(r_{max})$, à $r_{max}$ donné, procure le meilleur facteur de qualité que la variante du modèle prise en compte puisse fournir à l'élection traitée. Plus petit sera $Q(r_{max})$, meilleure sera la variante du modèle, i.e. meilleure sera l'indépendance des $\ffa$ $\ti$estimés selon la variante du modèle$\ti$ pour l'élection traitée. La comparaison entre ces différentes variantes devient dès lors possible.

Les figures~\ref{fQ} fournissent $Q(r_{max})$ pour trois types d'interaction $\pab$, obtenus avec les résultats et les taux de participation des élections étudiées. Les interaction testées, exprimées par la forme de $\pab$ dans Eq.~(\ref{efaa}), proviennent, soit de Eq.~(\ref{epab}), soit de Eq.~(\ref{epab-pop}), soit enfin de Eq.~(\ref{epab-grav}).

Il ressort de ces figures que le mode d'interaction entre deux communes défini par l'équation~(\ref{epab}) procure le meilleur résultat au modèle des tendances communales statiques. Et ce meilleur accord concerne aussi bien les résultats électoraux, que les taux de participation. (Seules les valeurs de $Q$ obtenues pour $r_{max}\gtrsim50~km$ sont significatives, comme il en a été déjà question ci-dessus, ainsi qu'aux figures~\ref{fQ}.)\\

Pour vérifier la robustesse du test effectué, nous avons aussi calculé les corrélations des $\ffa_s=\operatorname{sign}[\ffa - \langle \tilde{f} \rangle]$, qui prennent la valeur $\pm 1$ selon que la tendance spécifique estimée $\ffa$ d'une commune $\aaa$ est au-dessus ou au-dessous de la moyenne sur toutes les communes des $\ffa$. Ces nouveaux tests $\ti$avec la corrélation spatiale des $\ffa_s$ plutôt que celle des $\ffa$ comme précédemment$\ti$ confirment les conclusions précédentes.

Nous avons aussi fait un autre test en tenant compte des coefficients $K^\aaa$ non lissés, tels qu'ils apparaissent dans les équations~(\ref{econvic} et \ref{eK-convic}). Ces tests confirment de nouveau les conclusions précédentes.

Faisons enfin le bilan de ces tests.

\begin{figure}[t]
\includegraphics[width=5cm, height=5cm]{q-res.eps}\hfill
\includegraphics[width=5cm, height=5cm]{q-pop-res.eps}\hfill
\includegraphics[width=5cm, height=5cm]{q-grav-res.eps}\vspace{1cm}
\includegraphics[width=5cm, height=5cm]{q-abst.eps}\hfill
\includegraphics[width=5cm, height=5cm]{q-pop-abst.eps}\hfill
\includegraphics[width=5cm, height=5cm]{q-grav-abst.eps}
\caption{\small Facteur de qualité $Q(r_{max})$ obtenu à partir de trois formes d'interaction, $\pab$, entre  communes, et toujours selon le modèle à tendances statiques. $\pab$ exprimé par Eq.~(\ref{epab}) procure les figures de gauche. Les figures du milieu proviennent de $\pab$ écrit par Eq.~(\ref{epab-pop}). $\pab$ prend la forme de  Eq.~(\ref{epab-grav}) dans les figures de droite. Les figures du haut découlent des résultats des élections, tandis que celles du bas, des taux de participation.}
\label{fQ}
\end{figure}

\subparagraph*{$\bullet$ Résumé et discussion\\}
\label{pt-annexe-test-discussion}
Fait notable, les couples de paramètres $(K,d)$ qui conviennent le mieux aux données (aux résultats électoraux et aux taux de participation) selon la variante afférente à l'équation~(\ref{epab-grav}) pour $\pab$, sont $(K=1\,;\,d=10,5)$. Le fait que $d$ soit égal à $10,5$ (le plus grand exposant testé) est particulièrement remarquable puisque nous testions indirectement l'hypothèse d'une interaction restreinte au voisinage géographique. $d=10,5$ est l'exposant testé qui accorde le plus de poids, ou d'importance, aux communes proches de celle où réside l'agent considéré. Ainsi, l'hypothèse d'une interaction locale n'est pas contredite à ce stade.

Nous avons aussi comparé les facteurs de qualité $Q$ obtenus à partir d'une distribution logistique, et ceux obtenus à partir d'une distribution gaussienne $\ti$utilisée jusqu'alors$\ti$ et avec $\pab$ provenant de Eq.~(\ref{epab}), mais nous n'y avons rien vu de notable.

En résumé, les tests effectués sur le modèle des tendances communales statiques ont mis en évidence les deux points suivants. Premièrement, des trois variantes testées, celle qui convient le mieux correspond à l'équation~(\ref{epab}) pour $\pab$, et est appelée influence indirecte via les lieux de résidence. Deuxièmement, l'hypothèse d'une interaction locale ne semble pas être contredite.\\

Néanmoins, ces conclusions sont sujettes à caution puisqu'elles ne sont rigoureusement valables que dans le cadre du modèle des tendances statiques. Bien que le facteur de qualité soit très petit $\ti$de l'ordre de $10^{-2}$, ce qui nous semble encourageant pour un modèle simpliste$\ti$, nous ne pouvons tenir compte rigoureusement de ces conclusions que si le modèle convient. Quoi qu'il en soit, ces tests fournissent, \textit{a minima}, des indications fortes dont nous tiendrons compte par la suite.

\clearpage
\renewcommand{\thesection}{H}
\renewcommand{\theequation}{H-\arabic{equation}}
\setcounter{equation}{0}  
\renewcommand{\thefigure}{H-\arabic{figure}}
\setcounter{figure}{0}
\renewcommand{\thetable}{H-\arabic{table}}
\setcounter{table}{0}
\renewcommand{\thesubsection}{H-\arabic{subsection}}
\setcounter{subsection}{0}
\section{Corrélations spatiales théoriques}
\label{annexe-correl-th}

Cette annexe se donne pour tâche de tracer les corrélations spatiales des modèles à tendances communales statiques ou dynamiques, et ce analytiquement. Elle se fonde sur l'approximation d'une répartition homogène des communes, voire si nécessaire (dans le cas où $\pab$ est défini par l'équation~(\ref{epab-pop})), d'une répartition homogène de la population. Les corrélations proviennent du passage au continu des équations~(\ref{efaa}-\ref{efva}) déjà traité à l'annexe~\ref{annexe-continu-stabilite}.

Cette annexe reprend les mêmes notations et conventions concernant la transformation de Fourier que celles de l'annexe~\ref{annexe-continu-stabilite}.

\subparagraph*{$\bullet$ Modèle statique\\}
\label{pt-correl-th-statique}

L'équation~(\ref{ef-fourier}) admet pour solution~:
\be \label{ef-sol-statique} \hat{F}(\kk,t)=\int_0^t \mathrm{d}t'\x e^{-(t-t')/\tau_k}\x\hat{f}(\kk)\;+\;C.I.~,\ee
où \textit{C.I.} désigne les conditions initiales, et $\tau_k$ le temps caractéristique du mode $\kk$, défini par~:
\be \label{etauk} \frac{1}{\tau_k} = 1-\frac{K}{(1+(k\,\ell_c)^2)^{3/2}}~,\ee
qui contient les deux paramètres libres du système~: $K$ et $\ell_c$. Remarquons au passage que les équations ci-dessus sont invariantes par rotation, ce qui est en accord avec l'isotropie du modèle. ($\fva$, de par sa définition, ne fait apparaître aucune direction privilégiée, et les communes sont uniformément réparties.) Nous nous plaçons dans la situation où les conditions initiales ne jouent plus aucun rôle afin d'étudier le modèle dans sa généralité $\ti$comme nous l'avions déjà discuté à l'annexe~\ref{annexe-continu-stabilite}. 

La corrélation spatiale $C(\rr,t)$ au temps $t$ s'écrit comme~:
\be \label{ecor-th-def} C(\rr,t)=\int \mathrm{d}\rr_1\langle F(\rr_1,t)\,F(\rr_2=\rr_1-\rr,t)\rangle~,\ee
où $\langle ...\rangle$ désigne la moyenne sur toutes les réalisations.

En utilisant les transformées de Fourier, 
\be \label{efr1-fr2} \langle F(\rr_1,t)\,F(\rr_2,t)\rangle = \frac{1}{(2\pi)^2}\int \mathrm{d}\kk_1,\mathrm{d}\kk_2\x e^{i(\kk_1,\rr_1\,+\,\kk_2,\rr_2)}\x \langle\hat{f}(\kk_1)\,\hat{f}(\kk_2)\rangle\x I(k_1,k_2,t)~,\ee
avec
\be \label{eit-statique} I(k_1,k_2,t)=\int_0^t \mathrm{d}t_1 \int_0^t \mathrm{d}t_2\x  e^{-(t-t_1)/\tau_{k_1}}\x e^{-(t-t_2)/\tau_{k_2}}~,\ee
d'après Eq.~(\ref{ef-sol-statique}).

Les tendances spécifiques des communes s'écrivent comme $f(\rr)=\sum_\aaa f^\aaa\x \delta(\rr-\rr^\aaa)$, où $\delta$ désigne la fonction de Dirac et $\rr^\aaa$ la position de la commune $\aaa$. La décorrélation des tendances propres des communes implique que $\langle f^\aaa\,f^\beta\rangle=(f^\aaa)^2\,\delta_{\aaa\beta}$, où $\delta_{\aaa\beta}$ dénote le symbole de Kronecker. Ainsi,
\be \label{ek1-efk2} \langle\hat{f}(\kk_1)\,\hat{f}(\kk_2)\rangle = \frac{1}{(2\pi)^2}\sum_\aaa (f^\aaa)^2\x e^{-i(\kk_1+\kk_2).\rr^\aaa}~,\ee
ce qui permet d'obtenir $C(\rr,t)$, après simplification, comme~:
\be \label{ecor-simplifie} C(\rr,t) \propto \sum_\aaa (f^\aaa)^2\int \mathrm{d}\kk\x e^{-i\kk.\rr}\x I(k,k,t)~,\ee
où, d'après Eq.~(\ref{eit-statique}), 
\be \label{eik-statique} I(k,k,t)= (\tau_k)^2\x (1-e^{-t/\tau_k})^2 \ \underset{t\gg\tau_k}{\longrightarrow}\ I_k=(\tau_k)^2~.\ee
Notons une fois de plus l'isotropie de la corrélation spatiale, évidemment en accord avec l'isotropie du modèle.

De nouveau, nous nous nous plaçons après le régime transitoire. Ensuite, il ne reste plus qu'à normaliser la corrélation spatiale, soit plus précisément à diviser $C(\rr)$ ci-dessus par $C(\rr=\mathbf{0})$. Il faut enfin tenir compte de la physique du problème, i.e. des dimensions réelles du système étudié : les communes de la France. La distance maximale qui intervient est de l'ordre de $L=500~km$. La distance minimale sur laquelle porte une tendance spécifique uniforme, est de l'ordre de grandeur de la dimension de la commune, soit $a=2~km$. Il en résulte, avec un changement de variable faisant intervenir le mode adimensionné $\qq=\kk\,\ell_c$, que~:
\be \label{ecor-sol} C(r)=\frac{\int \mathrm{d}\theta \int_{\ell_c/L}^{\ell_c/a} \mathrm{d}q\x q\x e^{-iq\frac{r}{\ell_c}\cos\theta}\x I_q}{\int \mathrm{d}\theta \int_{\ell_c/L}^{\ell_c/a} \mathrm{d}q\x q\x I_q}~,\ee
où 
\be \label{eiq-statique} I_q=(\tau_q)^2 = \big[1-\frac{K}{(1+q^2)^{3/2}}\big]^{-2} ~.\ee
Dans l'équation~(\ref{ecor-sol}) ci-dessus, la distance $r$, possède la même unité que les distances $\ell_c$, $L$ et $a$.

En résumé, dans l'approximation d'une répartition uniforme des communes, voire d'une répartition uniforme de la population si nécessaire, la corrélation spatiale $C(r)$ s'obtient analytiquement par Eq.~(\ref{ecor-sol}), où $I_q$ provient avec ce modèle de l'équation~(\ref{eiq-statique}). (Les figures~\ref{fcor-th-statique} tracent ces corrélations pour différentes valeurs de $K$ et de $\ell_c$.)

\subparagraph*{$\bullet$ Modèle dynamique\\}
\label{pt-correl-th-dynamique}

Cette partie suivra la même démarche que celle qui traitait du modèle statique.

La différence par rapport au modèle précédent concerne les tendances propres des communes~: précédemment fixes et décorrélées entre elles, ici elles deviennent non seulement décorrélées entre elles, mais aussi fluctuantes au cours du temps (cf Eqs.~(\ref{efaa-dyn-cor-spa}-\ref{efaa-dyn-cor-t})).
Autrement dit, les tendances spécifiques deviennent du modèle statique au modèle dynamique, $f^\aaa\rightarrow f^\aaa(t)$, et en transformée de Fourier, $\hat{f}(\kk)\rightarrow \hat{f}(\kk,t)$.

Ainsi, $\hat{F}(\kk,t)$ devient en modèle dynamique, et en négligeant les conditions initiales~:
\be \label{ef-sol-dyn} \hat{F}(\kk,t)=\int_0^t \mathrm{d}t'\x e^{-(t-t')/\tau_k}\x\hat{f}(\kk,t')~,\ee
où $\tau_k$ garde la même forme que celle donnée à l'équation~(\ref{etauk}). Ceci implique que l'équation~(\ref{eit-statique}) devienne avec le modèle dynamique~:
\be \label{eit-dyn} I(k_1,k_2,t)=\int_0^t \mathrm{d}t_1 \int_0^t \mathrm{d}t_2\x  e^{-(t-t_1)/\tau_{k_1}}\x e^{-(t-t_2)/\tau_{k_2}}\x \delta(t_1-t_2)~,\ee
d'où,
\be \label{eik-dyn} I(k,k,t)= \frac{1}{2} \tau_k \x (1-e^{-2t/\tau_k}) \ \underset{t\gg\tau_k}{\longrightarrow}\ I_k\propto\tau_k ~.\ee

Ainsi, la corrélation $C(r)$ s'obtient de nouveau par l'équation~(\ref{ecor-sol}), mais avec ici~:
\be \label{eiq-dyn} I_q=\tau_q = \big[1-\frac{K}{(1+q^2)^{3/2}}\big]^{-1} ~.\ee

En résume, avec l'approximation d'une répartition uniforme des communes, ou d'une répartition uniforme de la population si nécessaire, la corrélation spatiale $C(r)$ s'obtient analytiquement de nouveau par Eq.~(\ref{ecor-sol}), mais avec $I_q$ qui provient de Eq.~(\ref{eiq-dyn}) en modèle dynamique. (Les figures~\ref{fcor-th-dyn} tracent ces corrélations pour différentes valeurs de $K$ et de $\ell_c$.)\\

Notons enfin que si l'amplitude moyenne des fluctuations des tendances spécifiques d'une commune, dépend de la commune considérée (voir Eqs.~(\ref{efaa-dyn-fin-cor-spa}-\ref{efaa-dyn-fin-cor-t}) pour leur forme mathématique), l'équation~(\ref{eit-dyn}) est inchangée. Ainsi, la corrélation spatiale $C(r)$ de ce cas, garde la même forme que celle du modèle dynamique traité ci-dessus.

\clearpage
\addcontentsline{toc}{section}{Bibliographie}   

\clearpage


\addcontentsline{toc}{section}{Postface}
\section*{Postface}
\hfill
\begin{minipage}[r]{0.8\linewidth}
\textit{Ne vous faites pas de souci pour le ciel et la Terre, ne craignez pas leur subversion, pas plus que celle de la philosophie [...]. La philosophie elle-même ne peut que tirer avantage de nos disputes~: si nos pensers sont vrais, nous y aurons gagné quelque chose ; s'ils sont faux, les doctrines antérieures n'en seront que mieux confirmées par leur rejet. Ayez plutôt souci de certains philosophes et cherchez à les aider et à les soutenir~: la science, elle, ne peut qu'avancer.}\\
Galilée, \textit{Dialogue sur les deux grands systèmes du monde}~\cite{galilee_1}
\end{minipage}
\vspace{0.75cm}

Revenir une année après la rédaction de la thèse offre l'opportunité de compléter et de développer quelques points. (Il eût été périlleux de modifier le corps du texte sans rompre son unité, son cœur, et donc, certes selon des yeux tout subjectifs, l'une de ses forces~; d'autant plus que les modifications, inscrites dans le flux dynamique de la recherche, n'en seraient que provisoires.) Le premier point concerne les développements apportés par l'article~\footnote{C. Borghesi and J.-P. Bouchaud, \textit{Spatial correlations in vote statistics: a diffusive field model for decision-making}, Eur. Phys. J. B. {\bf 75}, 395-404 (2010)} écrit avec Jean-Philippe Bouchaud $\ti$et dont les principales extensions lui sont dues. Le deuxième permet de situer cette étude de physique par rapport aux sciences sociales et, plus particulièrement, par rapport à la géographie électorale. Enfin, le troisième et dernier point se penche encore et encore $\ti$mais cette fois-ci brièvement$\ti$ sur la méthode d'investigation que nous avons volontairement utilisée tout au long de ce travail.\\ 

{\it Spatial correlations in vote statistics...} s'appuie sur une partie du travail de cette thèse et l'enrichit, et ce avec une vision épistémologiquement plus réaliste ou moins sceptique. (Comme quoi le scepticisme seul ne suffit pas $\ti$si besoin était encore d'en parler$\ti$, mais encore resterait-il à savoir par où le faire commencer et, surtout, par où le faire terminer.) Bref, voyons quels sont les principaux développements amenés par cet article.
\begin{enumerate}
\item L'hypothèse que les $\taa$ de la participation électorale sur l'ensemble des communes (et centrée sur la moyenne nationale) constituent la même distribution, permanente d'une élection à l'autre, ne peut être rejetée selon le test de Kolmogorov-Smirnov. (À noter que cette hypothèse nous avait permis d'émettre la première prédiction, cf. sections~\ref{pt-predictions} et \ref{pt-addenda}.)
\item Une recherche des \og clones \fg, i.e. des agents qui votent ou participent à l'identique, est effectuée au sein des communes.
\item L'idée qui paraît conceptuellement la plus novatrice ou la plus féconde, par rapport à la thèse, est de concevoir l'ensemble des tendances dynamiques communales, $F^\aaa$, comme un unique champ qui évolue au cours du temps, et ce indépendamment de la présence ou non d'une élection. Dit autrement, chacune des élections ne réaliserait en quelque sorte qu'une photo instantanée de ce même champ, mais à des dates différentes. Ce champ se retrouve alors promu à une existence concrète et porte le nom $\ti$certes quelque peu fourre-tout$\ti$ de \og champ culturel \fg. (À noter que les tendances dynamiques, $F^\aaa$, qui s'appliquaient indépendamment d'une élection à l'autre, étaient perçues dans cette thèse comme un objet utile, avec comme seul objectif celui de \og sauver les phénomènes\fg\cite{duhem}.)
\item L'équation qui gouverne l'évolution temporelle du \og champ culturel \fg{} correspond à une équation de diffusion avec sources fluctuantes (en anglais, random diffusion equation, comme ici l'équation~(\ref{efaa-dyn}) du modèle des tendances dynamiques), et, nouveauté par rapport à la thèse, elle appartient à une classe d'universalité $\ti$ce qui la rend robuste à de mineures modifications du modèle. Ceci permettrait d'accorder le primat à l'équation elle-même, plutôt qu'au modèle microscopique permettant d'y parvenir (à l'instar, peut-être~? des équations de Maxwell bâties sur des rouages d'éther et qui, par la suite, ont permis de bannir l'éther de l'électromagnétisme classique).
\item Des arguments, liés aux distances réelles du système, tendent à valider la diffusion d'un champ aléatoire (le \og champ culturel \fg) dans un espace à deux dimensions (cf. Eq.~(\ref{efaa-dyn})). La forme des corrélations spatiales produites par ce genre d'équation est effectivement logarithmique et, de plus, leurs portées est bien de l'ordre de la taille du système (soit ici de l'ordre de $500~km$ pour la France). La longueur de corrélation, $\ell_c$, du bruit ou des sources fluctuantes est, de plus, de l'ordre de la distance caractéristique inter-communale (soit pour la France de l'ordre de $3~km$). En outre, ce type d'équation s'accorde avec une distribution stable (pour différents temps) des $F^\aaa(t)$ sur l'ensemble des communes (en accord avec la permanence temporelle de la distribution des $\taa$ de la participation électorale).
\item Il faut, pour décrire de façon plus complète les phénomènes mesurés, introduire dans l'expression de la conviction d'un agent un autre terme, un champ idiosyncratique propre à la commune où l'agent réside. Ce champ serait de faible longueur caractéristique et de grand temps caractéristique. À noter que le poids relatif du \og champ culturel \fg{} dans la prise de décision est plus grand lors du vote (en faveur de l'un des deux choix proposés par l'élection) comparé à celui qui prévaut lors de la participation ou non à l'élection (sauf encore pour le résultat électoral du référendum de 2000).
\item Enfin, cerise sur le gâteau, le modèle produit une propriété émergente et à laquelle nous ne nous attendions pas~: une persistance temporelle de longue durée (de l'ordre du siècle avec des paramètres en ordre de grandeur réalistes) pour les phénomènes électoraux. Ceci concorde avec la permanence des paysages électoraux, bien connue et de longue date en géographie électorale (cf. {\it Le granite vote \`a droite, le calcaire vote \`a gauche} attribué à Siegfried~\cite{siegfried}). Notons pour terminer que la corrélation temporelle des taux de participation sur l'ensemble des communes, $C_{t_i,t_j}(\tau)$ (cf. section~\ref{pt-tempo-res} et Tab~\ref{ttempo-abst}), exprimée en fonction du temps $|t_i - t_j|$, s'accorde correctement avec ce que prévoit le modèle.\vspace{0.5cm}
\end{enumerate}

Il est amusant et instructif de constater après coup que ce travail aboutit $\ti$sur le plan de la modélisation$\ti$ à retrouver, certes avec une certaine nuance, ce qui était déjà bien connu en sciences sociales et plus particulièrement en géographie électorale. (La partie qui suit se restreint à une légère discussion sur le modèle diffusif obtenu ici, et faisant écho à la diffusion d'idées $\ti$ou contagion spatiale d'opinions$\ti$ que l'on trouve en géographie électorale, notamment dans le beau livre de Michel Bussi~\footnote   {Michel Bussi, \textit{\'El\'ements de g\'eographie \'electorale \`a travers l'exemple de la France de l'Ouest}, Ed. P. U. de Rouen (1998)}.) De plus $\ti$et sans s'étendre sur ce point$\ti$, il est rassurant pour la pratique scientifique que, sans aucune connaissance préalable en géographie électorale, les seules données nous aient amenées (forcées~?) à établir des résultats en accord avec cette discipline. (Peut-être aussi, est-ce partiellement dû au type de données que nous utilisons : uniquement des données électorales spatialement localisées, i.e. géoréférencées.)

Mais ne nous berçons pas pour autant dans un angélisme béat du genre \og tout se vaut \fg. Il persiste des différences essentielles, et enrichissantes, entre les deux approches (celle de la géographie électorale et la nôtre) concernant les faits électoraux tels qu'ils sont observés, mesurés, décrits et modélisés. Et bien évidemment $\ti$le contraire eût été surprenant$\ti$ ici encore se retrouve confirmé l'adage à la Poincaré selon lequel l'échelle d'observation crée le phénomène~\footnote{\og La façon dont ces cellules sont agencées et d’où résulte l’unité de l’individu, n’est-elle pas aussi une réalité, beaucoup plus intéressante que celle des éléments isolés, et un naturaliste, qui n’aurait jamais étudié l’éléphant qu’au microscope, croirait-il connaître suffisamment cet animal ?\fg{} Henri Poincaré, {\it La valeur de la science}}.

La géographie électorale analyse finement, et avec moult détails, les faits électoraux géoréférencés, leurs pose de nombreuses questions avec de belles mises en perspectives $\ti$comme par exemple l'effet notabilitaire, l'effet dû à la longévité d'un élu, la recherche de centres rayonnants, la cohésion spatiale selon les partis politiques, le comportement électoral urbain par rapport au rural, etc. Rappelons pour mémoire que nous avions, dès le départ de ce travail, pris le pari de nous abstraire de ce que nous appelions $\ti$peut-être à tort$\ti$ détail (cf. l'Introduction de ce manuscrit). Il ressort par comparaison que notre étude est bien pauvre en quantité de phénomènes observés, d'histoires racontées et de questions posées, mais, en revanche $\ti$et peut-être grâce à cette pauvreté~?$\ti$, elle accède à une description et à une compréhension quantitative. 

Le modèle que nous obtenons pour s'accorder aux phénomènes électoraux (tels que nous les mesurons) n'est, au final, qu'un simple modèle de diffusion avec sources fluctuantes~; ce qui rappelle la notion de diffusion d'opinion ou de contagion spatiale à l'œuvre en géographie électorale. Et il n'est pas rien de pouvoir déterminer l'ordre de grandeur de la longueur de corrélation des sources fluctuantes. Et, plus important encore, il n'est pas rien non plus de pouvoir affirmer le genre de diffusion en action. Par exemple, nous savons que tout se passe comme s'il ne pouvait exister de diffusion de la seule décision binaire, $\pm 1$ (par rapport au vote d'une élection à deux choix ou par rapport au choix de participer ou non à l'élection, cf. section~\ref{pt-refutation-rfim}), ni de diffusion directe de la conviction (liée au nombre d'individus dans les communes voisines, cf. section~\ref{pt-dyn-discussion}), ni d'une diffusion avec sources statiques (cf. section~\ref{section-statique}). (Peut-être que la diffusion, comme auparavant pour la notion de température et de chaleur avant leurs mesures et leurs définitions précises, est un concept trop vague qui peut tout et rien dire en même temps, bref qui manque de précision dès lors qu'on n'y prend garde.) Et, encore plus important, n'oublions pas que le modèle (ou l'équation de diffusion) obtenu n'est qu'un modèle (ou une équation) à l'interprétation délicate et, surtout, un modèle (ou une équation) perfectible et falsifiable.\\

Enfin, il peut paraître pour le moins surprenant que notre façon de poser le problème électoral n'ait nullement tenu compte d'un aspect citoyen~\footnote{Je tiens à remercier Aurélien Raccah pour avoir soulevé ces questions et surtout pour les avoir pertinemment développées ; ce dont bénéficie la discussion qui suit.}, voire d'un aspect proprement humain $\ti$à supposer qu'il soit, sur le plan conceptuel ou opératoire, facilement saisissable. Par exemple, nous n'avons pris en considération ni la question soumise au vote, ni les candidats, ni même la nature de l'élection. Au contraire, nous avons tenu compte de si peu d'informations qu'une élection pourrait être traitée à l'instar d'un vulgaire champ de température dont nous connaissons la valeur pour chaque commune. Nous aimerions succinctement revenir sur ce parti pris qui, $\ti$comme on commence peut-être à trop en parler$\ti$ est complètement délibéré de notre part. (À noter que nous ne rentrons pas dans le débat de savoir ce qu'apporte, ou retire, ce travail à la vision théorique de la démocratie actuelle.)

Parce que ce parti pris laissait présager une relative simplicité du phénomène étudié, on pouvait s'attendre à extirper assez facilement des régularités empiriques (voire des traits universels avec un peu plus de chance) sur lesquelles des modèles théoriques pourraient ensuite se fonder. Mais nous avions déjà discuté de cela, en rappelant d'ailleurs que la pratique de la réduction d'informations est usuelle en physique $\ti$comme l'illustre par exemple le rôle de la chute des corps dans l'établissement de la mécanique. (Notons en passant que la participation électorale, phénomène {\it a priori} plus simple que l'expression du vote, a effectivement fourni davantage de régularités que ce dernier.)

Ajoutons à cela $\ti$et c'est peut-être moins glorieux$\ti$ une certaine facilité à se satisfaire de si peu de données, à ne pas s'échiner à obtenir, à dépouiller, à quantifier, des données et des notions relatives aux domaines socio-économiques, politiques, historiques, etc. Et d'ailleurs, quel aspect socio-politique choisir~? pourquoi un aspect plutôt qu'un autre~? surtout si l'on est novice dans l'étude des phénomènes sociaux et que, scientifiquement aussi, \og Ce dont on ne peut parler, il faut le taire.\fg{} Pour ne pas commettre d'erreur, pour ne pas se laisser embarquer dans un problème d'emblée trop compliqué (car à trop grand nombre de variables), le plus simple n'eût-il pas été de ne privilégier aucun aspect socio-politique~?

Ajoutons encore à cela $\ti$et c'est peut-être plus amusant, plus jouissif$\ti$ l'aspect ludique et stimulant d'un défi à relever. Puisque \og le bon sens est la chose du monde la mieux partagée \fg, des discussions au comptoir d'un café jusqu'aux réflexions plus ou moins alambiquées des milieux autorisés, et que bon nombre de facteurs socio-politiques paraissent, évidemment, jouer un rôle de la plus haute importance~: peut-on en prendre le contre-pied~? Soit plus naïvement, pourquoi la physique n'aurait-elle pas son mot à dire sur la question~? D'où la gageure~: que peut dire la physique, seule (et donc, pour le moment, sans données socio-politiques), sur les élections~? 

Venons-en maintenant à ce qui nous semble le plus important $\ti$par rapport au parti pris dans lequel nous nous sommes engagés. Et là encore je dirai que le plus important n'est pas tant de savoir si la problématique est pertinente ou non, mais de savoir ce qu'en disent les données. L'essentiel étant qu'il y ait une réponse nette $\ti$et qu'importe que la réponse soit positive ou négative. Ainsi, notre démarche, intransigeante sur la méthode, sans aucune complaisance $\ti$i.e. en ne faisant jamais appel à d'autres explications que celles provenant des seules données électorales spatialement localisées$\ti$, a l'avantage $\ti$du moins nous l'espérons$\ti$ de violenter suffisamment les données afin d'en obtenir des réponses claires. En effet, si de nettes régularités empiriques (correctement calibrées) surgissent, alors elles pourraient être utiles à d'autres connaissances théoriques et empiriques à venir. À l'inverse, si les irrégularités restent tenaces ou bien si quelques irrégularités se détachent d'un halo de régularités (voir section~\ref{pt-discussions}) alors, soit nous n'avons pas bien fait notre travail $\ti$du moins nous ne l'espérons pas$\ti$, soit nous pouvons affirmer qu'il manque un ingrédient essentiel à la problématique. Et là encore, il ressortirait forcément de cette lacune un aspect profitable, et exploitable. Encore resterait-il par la suite à trouver ce critère absent, cet ingrédient manquant, et ce, afin de poursuivre le processus cumulatif de la science~; car la science, elle, ne devrait qu'avancer.

\end{document}